\documentclass[openany,titlepage,12pt,letterpaper]{book}

\newcommand{\nocontentsline}[3]{}
\newcommand{\tocless}[2]{\bgroup\let\addcontentsline=\nocontentsline#1{#2}\egroup}

\usepackage{amsthm}
\usepackage{enumerate}
\usepackage{float}
\usepackage{color}
\usepackage{amsmath}
\usepackage{amssymb}
\usepackage{algorithm}
\usepackage{algorithmic}
\usepackage{subfig}
\usepackage{verbatim,tikz}
\usetikzlibrary{arrows}
\usepackage{color}
\usepackage{enumitem}
\usepackage{makeidx}
\usepackage[font=small]{idxlayout}
\usepackage{latexsym,rotating,pstricks,times,framed,xparse}

\makeatletter
\renewcommand\mainmatter{%
    \clearpage
  \@mainmattertrue
  \pagenumbering{arabic}}
\makeatother

\title{\textbf{Interactive Sensing and Decision Making in Social Networks}}

\author{
\vspace{1cm}\\
Vikram Krishnamurthy\\
Omid Namvar Gharehshiran \\
Maziyar Hamdi\vspace{2cm}\\
University of British Columbia\\
Vancouver, Canada
\date{}\vspace{4cm}\\
{\normalsize This monograph appears in {\em Foundation and Trends in Signal Processing},}\\
{\normalsize DOI: 10.1561/2000000048, Now Publsihers.}\\
{\normalsize \copyright 2014 Vikram Krishnamurthy, Omid Namvar Gharehshiran and Maziyar Hamdi}
}
\makeindex

\begin{document}

\setboolean{@twoside}{false}
\frontmatter
\thispagestyle{empty}
\maketitle

\setboolean{@twoside}{true}
\tableofcontents

\mainmatter

\chapter*{Preface}
\addcontentsline{toc}{chapter}{Preface}
The proliferation of social media such as real time microblogging and  online reputation systems  facilitate real time sensing of social patterns and behavior. In the last decade, sensing and decision making in social networks have witnessed significant  progress in  the electrical engineering, computer science, economics, finance, and sociology research communities. Research in this area involves the interaction of dynamic random graphs, socio-economic analysis, and statistical inference algorithms. This monograph provides a survey, tutorial development, and discussion  of four highly stylized examples: social learning for interactive sensing; tracking the degree distribution of social networks; sensing and information diffusion; and coordination of decision making via game-theoretic learning. Each of the four examples   is motivated by practical examples, and comprises of a literature survey together with careful problem formulation and mathematical analysis. Despite being highly stylized, these examples  provide a rich variety of models, algorithms and analysis tools that are readily accessible to a signal processing, control/systems theory, and applied mathematics audience.
\csname @openrightfalse\endcsname
\clearpage

\chapter*{Acknowledgements}
Some of the work in this monograph is based on papers we have co-authored with Professor George Yin, Department of Mathematics, Wayne State University. We are grateful for his expertise in weak convergence analysis of stochastic approximation algorithms. Chapter~2 is based on a recent paper with Professor Vincent Poor, Department of Electrical Engineering, Princeton University. We are also grateful to Professor Christophe Chamley, Department of Economics, Boston University, for discussions that sparked our interest in social learning.

The research in this monograph was supported in part by the  the Canada Research Chairs Program, Natural Sciences \& Engineering Research Council of Canada, and Social Sciences \& Humanities Research Council of Canada.

\csname @openrigh\endcsname
\newcommand{\degreeseq}{\mbox{${\bf d}$}} 
\newtheorem{theorem}{Theorem}[section]
\newcommand{\trace}{\mathop{\mathrm{trace}}}
\newcommand{\bdd}{\hspace*{-0.08in}{\bf.}\hspace*{0.05in}}
\newcommand{\bq}[1]{\begin{equation} \label{#1}}
\newcommand{\eq}{\end{equation}}
\newcommand{\bed}{\begin{displaymath}}
\newcommand{\eed}{\end{displaymath}}
\newcommand{\bea}{\bed\begin{array}{rl}}
\newcommand{\eea}{\end{array}\eed}
\newcommand{\ad}{&\!\!\!\disp}
\newcommand{\aad}{&\disp}
\newcommand{\bE}{{\mathbf{E}}}
\newcommand{\tr}{{\hbox{tr}}}
\def\disp{\displaystyle}
\newcommand{\e}{\varepsilon}
\newcommand{\wdt}{\widetilde}
\newcommand{\barray}{\begin{array}{ll}}
\newcommand{\earray}{\end{array}}
\newcommand{\M}{{\cal M}}
\newcommand{\initialprob}{\pi_0}
\newcommand{\sizemc}{M}
\newcommand{\sizegraph}{N}
\newcommand{\cd}{(\cdot)}
\newcommand{\rr}{{\mathbb R}}
\renewcommand{\hat}{\widehat}
\renewcommand{\tilde}{\widetilde}
\renewcommand{\bar}{\overline}
\newcommand{\graph}{G}      
\newcommand{\pdupstep}{r}       
\newcommand{\pdel}{q}       
\newcommand{\pdup}{p}          
\newcommand{\E}{\mathbf{E}}
\newcommand{\esa}{\varepsilon} 
\newcommand{\etrue}{\mu}    
\newcommand{\emc}{\rho}   
\newcommand{\bg}{\bar{g}}   
\newcommand{\obs}{y}        
\newcommand{\noise}{e}      
\newcommand{\g}{g}          
\newcommand{\hg}{\hat{\g}}  
\newcommand{\degree}{f}     
\newcommand{\transition}{L} 
\newcommand{\A}{A}          
\newcommand{\tim}{n}        
\newcommand{\ct}{t}         
\newcommand{\baf}{\bar{f}}  
\newcommand{\mc}{\theta}    
\newcommand{\ttrue}{B_{\sizegraph_0}}      
\newcommand{\diag}{\mathop{\mathrm{diag}}}
\newcommand{\tg}{\tilde{\g}}
\newcommand{\Et}{\E_\tim}   
\newcommand{\ser}{\nu}      
\newcommand{\cov}{\Sigma}   
\newcommand{\s}{{N_0}}          
\newcommand{\ind}{\chi}       
\newcommand{\delay}{\lambda}
\newcommand{\de}{\bar{d}_1}
\newcommand{\dd}{\bar{d}_2}
\newcommand{\noisee}{\omega}
\newcommand{\PLn}{n}
\newcommand{\parentnode}{u}
\newcommand{\newnode}{v}
\newcommand{\delnode}{w}
\newcommand{\identity}{I}
\newcommand{\generatormatrix}{Q}
\newcommand{\generatorelement}{q}
\newcommand{\transitionelem}{l}
\newcommand{\utilityfunc}{U}
\newcommand{\ben}{b}
\newcommand{\costmc}{v}
\newcommand{\dynamics}{\mathcal{D}}
\newcommand{\argmaxx}{\operatornamewithlimits{argmax}}
\newcommand{\sizedynamics}{\sizemc}
\newcommand{\costleave}{c}
\newcommand{\salgebH}{\mathcal{H}}
\newcommand{\indicatorit}{I}

\newtheorem{corollary}{Corollary}[chapter]
\newtheorem{lemma}{Lemma}[chapter]
\newtheorem{Assumption}{Assumption}[chapter]
\newtheorem{definition}{Definition}[chapter]
\newtheorem{example}{Example}[chapter]
\newtheorem{remark}{Remark}[chapter]

%

\newcommand{\argmin}{\operatornamewithlimits{argmin}}
\newcommand{\argmax}{\operatornamewithlimits{argmax}}

\newcommand{\degreeGC}{d}
\newcommand{\GCnode}{n}
\newcommand{\GCnodeinitial}{i}
\newcommand{\GCnodefinal}{k}
\newcommand{\setneighbors}{\mathcal{A}}
\newcommand{\GCneighbor}{N}
\newcommand{\rightmost}{A_R}


\renewcommand{\top}{\prime}
\newcommand{\game}{\mathcal{G}}
\newcommand{\plyrset}{\mathcal{N}}
\newcommand{\actset}{\mathcal{A}}
\newcommand{\plyrind}{n}
\newcommand{\fplyr}{N}
\newcommand{\utilityk}{U^n}
\newcommand{\utilitykn}{U^n_k}
\newcommand{\utilitykL}{U^n_L}
\newcommand{\utilitykG}{U^n_G}
\newcommand{\utilitykGn}{U^n_{G,k}}
\newcommand{\utilitykGt}{U^n_{G,t}}
\newcommand{\utilitykGtau}{U^n_{G,\tau}}
\newcommand{\fact}{A^n}
\newcommand{\act}{\textmd{a}}
\newcommand{\actk}{\textmd{a}^n}
\newcommand{\actprof}{\mathbf{a}}
\newcommand{\mixedstrat}{\mathbf{p}}
\newcommand{\mixedstratind}{p}
\newcommand{\CE}{\mathcal{C}}
\newcommand{\CEe}{\mathcal{C}_\epsilon}
\newcommand{\CEdistance}{\epsilon}
\newcommand{\regmatplyr}{R^n}
\newcommand{\regmatplyrinterpol}{R^{n,\stepsize}\ctimeet}
\newcommand{\regmatplyrinterpolcdot}{R^{n,\stepsize}(\cdot)}
\newcommand{\regmatplyrl}{R^l}
\newcommand{\regmatplyrdiff}{\bar{R}^n}
\newcommand{\regmatplyrdiffl}{\bar{R}^l}
\newcommand{\regmatplyrL}{R^{L,n}}
\newcommand{\regmatplyrG}{R^{G,n}}
\newcommand{\regmatplyrm}{R^{m,n}}
\newcommand{\regmatplyrLinterpolcdot}{R^{L,n,\stepsize}(\cdot)}
\newcommand{\regmatplyrGinterpolcdot}{R^{G,n,\stepsize}(\cdot)}
\newcommand{\regplyrji}{r^n_{ji}}
\newcommand{\regplyrij}{r^n_{ij}}
\newcommand{\regplyrijinterpol}{r^{n,\stepsize}_{ij}}
\newcommand{\regplyrLji}{r^{L,n}_{ji}}
\newcommand{\regplyrLij}{r^{L,n}_{ij}}
\newcommand{\regplyrLijinterpol}{r^{L,n,\stepsize}(i,j)}
\newcommand{\regplyrGijinterpol}{r^{G,n,\stepsize}(i,j)}
\newcommand{\regplyrGji}{r^{G,n}_{ji}}
\newcommand{\regplyrGij}{r^{G,n}_{ij}}
\newcommand{\dtimee}{k}
\newcommand{\dtimenext}{k+1}
\newcommand{\ctimeet}{(t)}
\newcommand{\inertia}{\mu}
\newcommand{\stepsize}{\varepsilon}
\newcommand{\indicatori}{I\left\lbrace \act^\plyrind_\dtimee = i\right\rbrace}
\newcommand{\indicatoritau}{I\left\lbrace \act^\plyrind_\tau = i\right\rbrace}
\newcommand{\indicatorj}{I\left\lbrace \act^\plyrind_\dtimee = j\right\rbrace}
\newcommand{\indicatorjtau}{I\left\lbrace \act^\plyrind_\tau = j\right\rbrace}
\newcommand{\neighborkc}{\mathcal{N}^n_c}
\newcommand{\neighborko}{\mathcal{N}^n_o}
\newcommand{\weightkl}{w_{nl}^s}
\newcommand{\weightmat}{W^s}
\newcommand{\Cmat}{Q^s}
\newcommand{\Cmatkron}{\mathbf{Q}^s}
\newcommand{\Cmatkl}{q_{nl}^s}
\newcommand{\diffstep}{\mu}
\newcommand{\identityK}{I_{N}}
\newcommand{\gamegraph}{G}
\newcommand{\gamevertex}{V}
\newcommand{\gameedge}{E}
\newcommand{\neighborhoodk}{N^n}
\newcommand{\cneighborhoodk}{N^n_c}
\newcommand{\nonneighbork}{S^n}
\newcommand{\oneK}{\mathbf{1}_N}
\newcommand{\oneM}{\mathbf{1}_{\fstate}}
\newcommand{\zeroK}{\mathbf{0}_N}
\newcommand{\zeroM}{\mathbf{0}_{\fstate}}
\newcommand{\explor}{\delta}
\newcommand{\communitys}{C^s}
\newcommand{\fcom}{S}
\newcommand{\comind}{s}
\newcommand{\comset}{\mathcal{S}}
\newcommand{\globbehav}{\boldsymbol{z}}
\newcommand{\unitvec}{\boldsymbol{e}}
\newcommand{\lyap}{V}
\newcommand{\statdistk}{\boldsymbol{\psi}^n}
\newcommand{\statdistind}{\psi}
\newcommand{\statdistindi}{\psi^n_i}
\newcommand{\statdistindj}{\psi^n_j}
\newcommand{\globinterpol}{\boldsymbol{z}^\stepsize}
\newcommand{\globinterpolind}{z^{n,\stepsize}\big(i,\actprof^{-\plyrind}\big)(t)}
\newcommand{\globinterpolt}{\boldsymbol{z}^\stepsize\ctimeet}
\newcommand{\diffinclsocial}{H^n\big(\regmatplyr\big)}
\newcommand{\diffinclglobal}{\mathbf{H}}
\newcommand{\diffinclsocialL}{H^{L,n}\big(\regmatplyrL,\regmatplyrG\big)}
\newcommand{\diffinclsocialG}{H^{G,n}\big(\regmatplyrL,\regmatplyrG\big)}
\newcommand{\diffinclsocialind}{h^n_{ij}\big(\regmatplyr\big)}
\newcommand{\diffinclsocialLind}{h^{n,L}_{ij}\big(\regmatplyrL,\regmatplyrG\big)}
\newcommand{\diffinclsocialGind}{h^{n,G}_{ij}\big(\regmatplyrL,\regmatplyrG\big)}
\newcommand{\simplexminus}{\Delta\actset^{-\plyrind}}
\newcommand{\neibmixedstrat}{\boldsymbol{\nu}^{-\plyrind}}
\newcommand{\simplexneighbor}{\Delta\actset^{N^n}}
\newcommand{\statdistindgroup}{\boldsymbol{\sigma}^\plyrind\big(\regmatplyr\big)}
\newcommand{\statdistgroupnor}{\boldsymbol{\sigma}^\plyrind}
\newcommand{\statdistindgroupind}{\sigma_i^\plyrind\big(\regmatplyr\big)}
\newcommand{\statdistindgroupnor}{\sigma_i^\plyrind}
\newcommand{\globalregret}{\mathbf{R}^s}
\newcommand{\globalregretinterpol}{\mathbf{R}^{s,\stepsize}}
\newcommand{\markovgame}{\theta}
\newcommand{\statespacegame}{\mathcal{M}}
\newcommand{\fstate}{M}
\newcommand{\markovspeed}{\rho}
\newcommand{\contmarkovmat}{Q}
\newcommand{\contmarkovmatind}{q_{ij}}
\def\lb{\left[}
\def\rb{\right]}
\def\lbr{\left\lbrace}
\def\rbr{\right\rbrace}

\newcommand{\btheta}{\mbox{\boldmath $\theta$}}
\newcommand{\Canalyte}{X}
\newcommand{\Y}{\mathbb{Y}} \newcommand{\X}{\mathbb{X}}
\newcommand{\sss}{\Bbb S}
\def\H{{\mathbf H}}
\newcommand{\wdh}{\widehat}

\newcommand{\tp}{P}

\newcommand{\belief}{{\pi}}
\newcommand{\tbelief}{\pi^0}

\newcommand{\trust}{\mathbf{L}}

\newcommand{\eSL}{\varepsilon}
\newcommand{\priv}{\eta}
\newcommand{\salpha}{\mbox{\boldmath $\alpha$}}
\newcommand{\ASL}{\mathcal{A}}
\newcommand{\conc}{\mathcal{M}}
\renewcommand{\c}{X}
\newcommand{\beq}{\begin{equation}}
\newcommand{\eeq}{\end{equation}}
\newcommand{\ones}{\mathbf{1}}

\newcommand{\history}{\mathcal{H}}
\newcommand{\full}{\mathcal{F}}

\newcommand{\bA}{\mathbf{A}}
\newcommand{\Bs}{R^\pi}
\newcommand{\ta}{\tilde{a}}
\newcommand{\sigs}{\sigma}
  \def\1{{\mathbf 1}}
\newcommand{\reals} {\Bbb{R}}
\newcommand{\nn}{\nonumber}
\newcommand{\Mu}{\boldsymbol{\mu}}
\newcommand{\pizero}{\pi_0}
\newcommand{\R}{\mathcal{R}}
\newcommand{\gr}{\geq_r}
\newcommand{\lr}{\leq_r}
\newcommand{\gs}{\geq_s}
\newcommand{\ls}{\leq_s}
\newcommand{\Ep}{\E^\mu_{\pi_0}}

\newcommand{\Pp}{\mathbf{P}^\mu_{\pi_0}}

\newcommand{\lbelief}{l}
\newcommand{\Epzero}{\E^\mu_{\pi_0}}
\newcommand{\Tp}{T}
\newcommand{\sigp}{\sigma}
\newcommand{\discount}{\rho}
\newcommand{\Cb}{\bar{C}}
\newcommand{\lbar}{\overline}
    \def\cd{(\cdot)}
    \newcommand{\Ts}{T}
    \newcommand{\ca}{c_a}
\newcommand{\p}{\prime}
    \def\I{{\Pi}} 
  \def\Ind{{\mathcal{I}}} 


\newcommand{\stepsizeDiff}{\epsilon}
\newcommand{\func}{F}

\renewcommand{\P}                 {\Bbb{P}}

\newcommand{\minc}{w}

\newcommand{\failprob}{p_F}

\NewDocumentCommand{\weight}{gg}
  {\IfNoValueTF{#2}
     {\IfNoValueTF{#1}
        {W}
        {W{(#1)}}%
     }
     {W{(#1, #2)}}%
  }

\NewDocumentCommand{\oprob}{gg}
  {\IfNoValueTF{#2}
     {\IfNoValueTF{#1}
        {B}
        {B_{#1}}%
     }
     {B_{#1 #2}}%
  }

\newcommand{\sigmaf}{\mathcal{F}}

\newcommand{\aindex}{a}

\newcommand{\ole}{\stackrel{\text{defn}}{=}}

\NewDocumentCommand{\tpdiff}{gggg}
     {\IfNoValueTF{#4}
     {\IfNoValueTF{#3}
         {\IfNoValueTF{#2}
        {{p}_{#1} }
        {{p}_{#1}{(#2)} } }%
     {{p}_{#1}(#2,#3) }}%
     {p_{#1}(#2,#3,#4)}
  }

\NewDocumentCommand{\atp}{gggg}
     {\IfNoValueTF{#4}
     {\IfNoValueTF{#3}
         {\IfNoValueTF{#2}
          {\IfNoValueTF{#1}
          {\bar{p}}
        {{\bar{p}}_{#1} } }
        {{\bar{p}}_{#1}{(#2)} } }%
     {{\bar{p}}_{#1}(#2,#3) } }%
     {{\bar{p}_{#1}(#2,#3,#4) }}
  }

\NewDocumentCommand{\sentiment}{gggg}
     {\IfNoValueTF{#4}
     {\IfNoValueTF{#3}
         {\IfNoValueTF{#2}
        {{z}_{#1} }
        {{z}_{#1}{(#2)} } }%
     {{z}_{#1}(#2,#3) }}%
     {z_{#1}(#2,#3,#4)}
  }

\NewDocumentCommand{\onoise}{gggg}
     {\IfNoValueTF{#4}
     {\IfNoValueTF{#3}
         {\IfNoValueTF{#2}
        {{v}_{#1} }
        {{v}_{#1}{(#2)} } }%
     {{v}_{#1}(#2,#3) }}%
     {v_{#1}(#2,#3,#4)}
  }

\NewDocumentCommand{\var}{gggg}
     {\IfNoValueTF{#4}
     {\IfNoValueTF{#3}
         {\IfNoValueTF{#2}
        {\sigma^2{(#1)} }
        {\sigma^2{(#1, #2)}  } }%
     {\sigma^2_{#1}(#2,#3) }}%
     {\sigma^2_{#1}(#2,#3,#4)}
  }

\NewDocumentCommand{\steady}{g}
  {
     {\IfNoValueTF{#1}
        {\pi}
        {\pi{(#1)}}%
     }
     }

\NewDocumentCommand{\cost}{g}
  {
     {\IfNoValueTF{#1}
        {c}
        {c^{(#1)}}%
     }
     }

\newcommand{\reward}{r}

\newcommand{\costpdf}{P_{\cost,\target{\dtime}}}

\NewDocumentCommand{\target}{g}
  {
     {\IfNoValueTF{#1}
        {s}
        {s_{#1}}%
     }
     }

\NewDocumentCommand{\targetobs}{g}
  {
     {\IfNoValueTF{#1}
        {o}
        {o_{#1}}%
     }
     }

\NewDocumentCommand{\targetoprob}{gg}
  {\IfNoValueTF{#2}
     {\IfNoValueTF{#1}
        {\bar{B}}
        {\bar{B}_{#1}}%
     }
     {\bar{B}_{#1 #2}}%
  }

\newcommand{\targetobsdim}{O}

\NewDocumentCommand{\tptarget}{gg}
  {\IfNoValueTF{#2}
     {\IfNoValueTF{#1}
        {A}
        {A_{#1}}%
     }
     {A_{#1 #2}}%
  }

\newcommand{\tstate}{s}
\newcommand{\tstatep}{s'}



\newcommand{\indicator}{I}

\newcommand{\obsdiff}{y}

\newcommand{\network}{G}
\newcommand{\Vertexset}{V}
\newcommand{\vertexset}{\{1,2,\ldots,\vertexnum\}}
\newcommand{\edgeset}{E}

\NewDocumentCommand{\infectdist}{gg}
  {\IfNoValueTF{#2}
     {\IfNoValueTF{#1}
        {\rho}
        {\rho_{#1}}%
     }
     {\rho_{#1}(#2)}%
  }

\NewDocumentCommand{\minfectdist}{gg}
  {\IfNoValueTF{#2}
     {\IfNoValueTF{#1}
        {\bar{\rho}}
        {\bar{\rho}_{#1}}%
     }
     {\bar{\rho}_{#1}(#2)}%
  }

\NewDocumentCommand{\tinfectdist}{gg}
  {\IfNoValueTF{#2}
     {\IfNoValueTF{#1}
        {\tilde{\rho}^\vertexnum}
        {\tilde{\rho}^{\vertexnum}_{#1}}%
     }
     {\tilde{\rho}^{\vertexnum}_{#1}(#2)}%
  }

\NewDocumentCommand{\pa}{gg}
  {\IfNoValueTF{#2}
     {\IfNoValueTF{#1}
        {\theta}
        {\theta_{#1}}%
     }
     {\theta_{#1}^{#2}}%
  }

\NewDocumentCommand{\nodeobs}{gg}
  {\IfNoValueTF{#2}
     {\IfNoValueTF{#1}
        {y}
        {y_{#1}}%
     }
     {y_{#1}^{(#2)}}%
  }

\newcommand{\nodeobsdim}{Y}

\NewDocumentCommand{\sample}{g}
  {
     {\IfNoValueTF{#1}
        {\alpha}
        {\alpha{(#1)}}%
     }
     }

\NewDocumentCommand{\obsm}{g}
  {
     {\IfNoValueTF{#1}
        {H}
        {H{(#1)}}%
     }
     }

\NewDocumentCommand{\feedforward}{g}
  {
     {\IfNoValueTF{#1}
        {D}
        {D{(#1)}}%
     }
     }

\newcommand{\degreediff}[1]{D^{(#1)}}
\newcommand{\nbhood}[1]{\mathcal{N}^{(#1)}}

\newcommand{\dtime}{k}
\newcommand{\ctime}{t}
\newcommand{\stime}{\tau}
\newcommand{\sdtime}{k'}
\newcommand{\finaltime}{T}

\newcommand{\seq}{l}

\NewDocumentCommand{\nodem}{g}
  {
     {\IfNoValueTF{#1}
        {m}
        {m_{#1}}%
     }
     }

\newcommand{\noden}{n}
\renewcommand{\deg}{d}
\newcommand{\degmax}{\bar{D}}

\NewDocumentCommand{\state}{gg}
  {\IfNoValueTF{#2}
     {\IfNoValueTF{#1}
        {x}
        {x_{#1}}%
     }
     {x_{#1}^{(#2)}}%
  }

\NewDocumentCommand{\vertexnum}{g}
  {
     {\IfNoValueTF{#1}
        {N}
        {N{(#1)}}%
     }
     }

\newcommand{\normal}{\mathbf{N}}

\NewDocumentCommand{\degdist}{g}
  {
     {\IfNoValueTF{#1}
        {P}
        {P{(#1)}}%
     }
     }

\newcommand{\statev}{x}

\newcommand{\neighbor}[1]{N^{(#1)}}

\newcommand{\nactive}[1]{A^{(#1)}}

\newcommand{\statea}{i}
\newcommand{\stateb}{j} 
\chapter{Introduction and Motivation}
\label{chapter:intro}

Research in social networks involves the interplay of complex networks (dynamics of random graphs) and social analysis (stemming from the areas of economics and sociology). There are seminal books in this area
including~\cite{J08,V07}. In comparison,
this monograph deals with {\em sensing and  decision-making} in social networks.
The proliferation of  social media such as  real-time microblogging services (Twitter\footnote{On US Presidential election day in 2012, there were 15 thousand tweets per second resulting in 500 million tweets in the day. Twitter can be considered as a real-time sensor.}), online reputation and rating systems~(Yelp) together with app-enabled smartphones, facilitate real time sensing of social activities, social patterns,  and  behavior.

Sensing and decision making in social networks is an area that has witnessed remarkable progress in the last decade in electrical engineering, computer science, economics, finance, and sociology.
It is the aim of this monograph to survey some important topics in this area and present highly stylized examples that are readily accessible to a  signal processing, control/systems theory, and applied mathematics audience. Indeed, the main tools used in
this monograph are dynamic programming, Bayesian estimation (filtering), stochastic approximation (adaptive filtering) and  their convergence analysis (weak convergence and mean square analysis),
game-theoretic learning, and
graph theory.  There has been much recent activity in the signal processing community in the area of social networks.
``How global behavior emerges from simple local behavior of boundedly rational agents''
has been an underlying theme of  an NSF/UCLA workshop in 2010, special sessions
at  ICASSP 2011 and 2012 and the ICASSP 2011 expert summary in~\cite{ZKS11}. Also, the recent special issues~\cite{SBT13,ZCK13} deal with signal processing of social networks.


\section{Motivation}\label{intro:mot}

{\em Social sensing}~\cite{AA11,BEH06,CEL08,EP06} is defined as a process where physical sensors present in mobile devices such as GPS are used to infer social relationships and human activities. In this monograph, we work at a higher level of abstraction. We use the term {\em social sensor} or {\em  human-based sensor} to denote an agent that provides information about its environment (state of nature) on a social network after interaction with other agents. Examples of such social sensors include Twitter posts, Facebook status updates, and ratings on online reputation systems like Yelp and Tripadvisor. Such social sensors go beyond physical sensors for social sensing~\cite{RMZ11}. For example, user opinions/ratings (such as the quality of a restaurant) are available on Tripadvisor but are difficult to measure via  physical sensors. Similarly, future situations revealed by the Facebook status of a user  are impossible to predict using  physical sensors.

Statistical inference using social sensors is relevant in a variety of applications including localizing special events for targeted advertising~\cite{CCL10,LS10}, marketing~\cite{TBB10}, localization of natural disasters~\cite{SOM10}, and predicting sentiment of investors in financial markets~\cite{BMZ11,PL08}. It is demonstrated in~\cite{AH10} that models built from the rate of tweets for particular products can outperform  market-based predictors.
%
However, social sensors  present unique challenges from a statistical estimation point of view. First, social sensors  interact with and influence other social sensors. For example, ratings posted on online reputation systems strongly influence the behaviour of  individuals\footnote{It is reported in~\cite{IMS11} that 81\% of hotel managers  regularly check Tripadvisor reviews. It is reported in~\cite{Luc11} that a one-star increase in the Yelp rating maps to 5-9 \% revenue increase. \label{foot}}.
Such interactive sensing can result in  non-standard information patterns due to correlations introduced by the structure of the underlying social network. Second, due to privacy reasons and time constraints, social sensors typically do not reveal  raw observations of the underlying state of nature.
Instead, they reveal their decisions (ratings, recommendations, votes) which can be viewed as a low resolution (quantized)   function of their raw measurements and interactions with other social sensors.

As is apparent from the above discussion,  
there is strong motivation to construct mathematical models that capture the dynamics of interactive sensing involving social sensors. Such models facilitate understanding the dynamics of information flow in social networks and, therefore, the design of algorithms that can exploit these dynamics to estimate the underlying state of nature.
 In this monograph, {\em social learning}~\cite{Ban92,BHW92,Cha04}, {\em game-theoretic learning}~\cite{ FL98,HMB13}, and {\em stochastic approximation}~\cite{KY03,YZ10} serve as  useful mathematical abstractions for modelling  the interaction of social sensors.

\section{Main Results and Organization}\label{intro:results}

\begin{figure}[t]
\begin{center}
\includegraphics[width=0.95\textwidth]{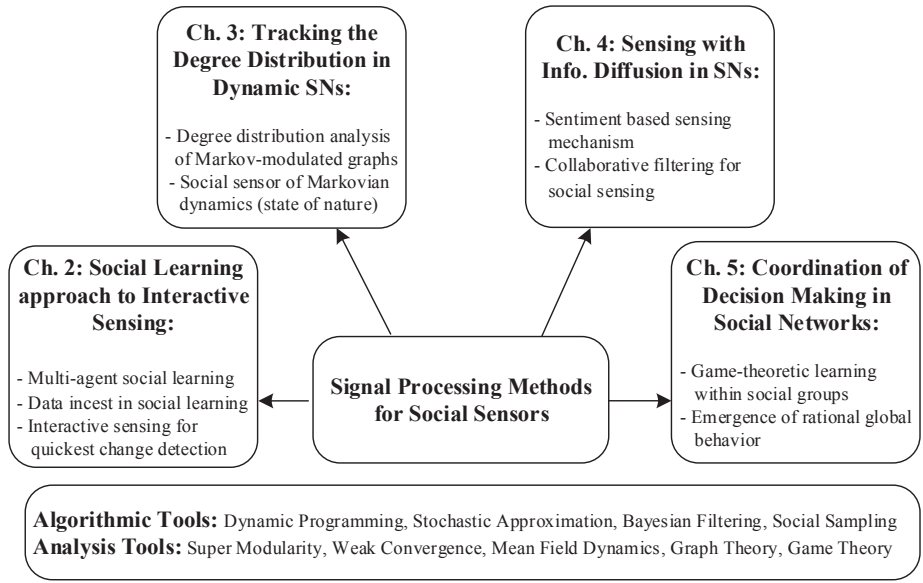}
\end{center}
\caption{Main results and organization of the monograph.}
\label{Fig:organization}
\end{figure}
As can be seen from Figure~\ref{Fig:organization}, this monograph
is organized into four chapters (excluding this introductory chapter) that
provide a survey, tutorial development, and discussion of four highly stylized examples: social learning for interactive sensing; tracking the degree distribution of social networks; sensing and information diffusion; and coordination of decision-making via game-theoretic learning.
Each of the four chapters is motivated by practical examples, and comprises of a literature survey together with careful problem formulation and mathematical analysis. The examples and associated analysis are readily accessible to a signal processing, control/systems theory, and applied mathematics audience.

In terms of information patterns, Chapter~\ref{Chapter:SL} considers Bayesian estimation and sequential decision making with sequential information flow and then information flow over small directed acyclic graphs.
In comparison, Chapter~\ref{Chapter:tracking} considers stochastic approximation algorithms for large random graphs that evolve
with time. Chapter~\ref{Chapter:diffusion} considers the asymptotics of  large graphs with fixed degree distribution but where the state of individual
node in the graph evolve over time---this models information diffusion. The mean field analysis in Chapter~\ref{Chapter:diffusion} results in a stochastic approximation type recursion, and the estimation problems are Bayesian (nonlinear filtering). Finally, Chapter~\ref{chapter:noncooperative} deals with learning in
non-cooperative repeated games comprising networks  of arbitrary size---the algorithms are of the stochastic approximation type.
In all these cases, sensors  interact with and influence other  sensors. It is the understanding of this interaction of local and global
behaviors in the context of social networks that constitutes the unifying theme of this monograph.

Below we give a brief synopsis of these four chapters. 

\subsubsection*{1. Social Learning Approach to Interactive Sensing}  Chapter~\ref{Chapter:SL}
 presents models and algorithms for interactive sensing in social networks where individuals act as sensors and  the information  exchange between individuals
is exploited to optimize sensing. Social learning is used as a mathematical formalism
  to model the interaction
between individuals that aim to estimate an underlying state of nature.

Social learning in multi-agent systems seeks to answer the following question:
\begin{quote} {\em How do decisions made by agents affect decisions made by subsequent agents?}  \end{quote}
In social learning, each agent chooses its action by optimizing its local utility function. Subsequent agents then use their private observations together with the decisions of previous agents to estimate (learn) the underlying state of nature. The setup is fundamentally different to classical signal processing in which sensors use noisy observations to compute estimates.

In the last decade, social learning has been used widely in economics, marketing, political science, and sociology to model the behavior of financial markets, crowds, social groups, and social networks; see~\cite{ADLO08,AO11,Ban92,BHW92,Cha04,LADO07} and numerous references therein. Related models have been studied in the context of sequential decision making in information theory~\cite{CH70,HC70} and statistical signal processing~\cite{CSL13,KP13} in the electrical engineering literature.

Social learning models for interactive sensing can predict unusual behavior. Indeed, a key result in social learning of an underlying random variable is that rational agents eventually herd~\cite{BHW92}; that is, they eventually end up choosing the same action irrespective of their private observations. As a result, the actions contain no information about the private observations and so the Bayesian estimate of the underlying random variable freezes. For a multi-agent sensing system, such behavior can be undesirable, particularly if individuals herd and make incorrect decisions.

In this context, the following questions are addressed in Chapter~\ref{Chapter:SL}:
 How can self-interested agents  that interact via social learning achieve a trade-off between
 individual privacy and reputation of the social group?
 How can  protocols be designed to prevent data incest in  online reputation blogs where individuals make recommendations?
How can sensing by  individuals that interact with each other be used by a global decision  maker
to detect changes in the underlying state of nature?
Chapter~\ref{Chapter:SL} presents an overview,  insights and discussion of social learning models in  the context of data incest propagation, change detection, and coordination of decision making.

  Several examples in social networks motivate Chapter~\ref{Chapter:SL}. Design of protocols to prevent data incest are motivated  by the design of fair online reputation systems
such as Yelp or Tripadvisor. In Online reputation systems, which  maintain logs of votes (actions)  by agents, social learning takes place with information exchange over a loopy graph  (where the agents form the vertices of the graph).
Due to the loops in the information exchange graph,
{\em data incest} (misinformation) can propagate: 
Suppose an agent wrote  a poor rating of a restaurant on a social media site.  Another agent is influenced by this rating, visits the restaurant, and then also gives  a poor rating on the social media site.
  The first agent visits the social media site  and notices that another agent has also given the restaurant a poor rating---this double confirms her rating and she enters another poor rating.
In a fair reputation system, such ``double counting'' or data incest should have been prevented by making the first agent  aware that the rating of the second agent was influenced by her own rating.

 As an example of change detection,
 consider measurement of the adoption of a  new  product using a micro-blogging platform like Twitter. The adoption of the technology diffuses through the market but its effects can only be observed through the tweets of select members of the population. These selected members act as  sensors for the parameter of interest.
 Suppose the state of nature  suddenly  changes due to  a sudden market shock or presence of a new competitor.
Based on the local actions of the multi-agent system that is performing social learning, a global decision maker (such as a market monitor or technology manufacturer) needs to decide  whether or not to declare if a change has occurred.
How can the global decision maker achieve such change detection to minimize a cost function comprised of false alarm rate and delay penalty? The local and global decision makers
interact, since
the local decisions determine the posterior distribution of subsequent agents which determines the global decision (stop or continue) which determines subsequent  local decisions.

\subsubsection*{2. Tracking Degree Distribution of Social Networks}
Chapter~\ref{Chapter:tracking}  considers dynamical random graphs.
The degree of a node in a network (also known as the connectivity) is the number of connections the node has in that network. The most important measure that characterizes the structure of a network (specially when the size of the network is large and the connections---adjacency matrix of the underlying graph---are not given) is the \textit{degree distribution} of the network. Chapter~\ref{Chapter:tracking} considers a Markov-modulated duplication-deletion random graph where, at each time instant, one node can either join or leave the network with probabilities that evolve according to the realization of a finite state Markov chain (state of nature). This chapter deals with the following questions: \begin{quote} {\em How can one estimate the state of nature using noisy observations of nodes' degrees in a social network?} and {\em How good are these estimates?}\end{quote}
Chapter~\ref{Chapter:tracking} comprises of two results.  First, motivated by social network applications, we analyze the asymptotic behavior of the degree distribution of the Markov-modulated random graph. From this degree distribution analysis, we can study the connectivity of the network, the size and the existence of a large connected component, the delay in searching such graphs, etc.~\cite{EGA04,J08,NWS02,N02}. Second, a stochastic approximation algorithm is presented to track the empirical degree distribution as it evolves over time. We further show
that the stationary degree distribution of Markov-modulated duplication-deletion random graphs depends on the dynamics of such graphs and, thus, on the state of nature. This means that, by tracking the empirical degree distribution, the social network can be viewed as a social sensor to track the state of nature. The tracking performance of the algorithm is analyzed in terms of mean square error. A functional central limit theorem is further presented for the asymptotic tracking error.

An important associated problem discussed in Chapter~\ref{Chapter:tracking} is how to actually construct random graphs via
simulation algorithms. In particular, for large social networks,  only the degree sequence is
available, and not the adjacency matrix. (The degree sequence is a non-increasing sequence of vertex degrees.)
Does a simple graph exist that realizes a particular degree sequence? How can all graphs that realize a degree sequence be constructed?
 Chapter~\ref{Chapter:tracking} presents a discussion of these issues.

\subsubsection*{3. Sensing and Information Diffusion in Social Networks}
Chapter~\ref{Chapter:diffusion} considers the following questions:
\begin{quote} {\em How does a behavior diffuse over a social network comprising of a population of interacting agents?} and {\em How can an underlying stochastic state be estimated based on sampling the population?}\end{quote}
As described in \cite{Pin08}, there is a wide range of social phenomena such as diffusion of technological innovations, cultural fads, and economic conventions~\cite{Cha04}, where
individual decisions are influenced by the decisions of others. Chapter~\ref{Chapter:diffusion}  considers two extensions of the widely used Susceptible-Infected-Susceptible (SIS) models  for  diffusion of  information in social networks~\cite{Pin06,Pin08,J08,PV01,V07}.
First,  the  states of individual nodes evolve  over time as a probabilistic function of the states of their neighbors {\em and} an underlying
target process. The underlying target  process can be viewed as the market conditions or competing  technologies that evolve with time and affect the information diffusion.
Second, the  nodes in the social network are sampled randomly to determine their state. Chapter~\ref{Chapter:diffusion} reviews  recent methods for sampling social networks such as  social sampling and respondent-driven sampling.
As the adoption of the new technology diffuses through the network, its effect is observed via sentiment  (such as tweets) of these selected members of the population.
 These selected nodes act as social sensors.
In signal processing terms, the underlying target  process can be viewed as a signal, and the social network can be viewed as a sensor.
The key difference compared to classical signal processing is that the social network (sensor) has dynamics due to the information diffusion.
Our aim is to estimate the underlying target state and the state probabilities of the nodes by sampling  measurements at nodes in the social network.
In a Bayesian estimation context, this is equivalent to a  filtering problem involving estimation of the state of a prohibitively large-scale Markov chain in noise. The key idea is to use {\em mean field dynamics} as
an approximation (with provable bounds)
for the information diffusion and, thereby, obtain a tractable model.

\subsubsection*{4. Coordination of Decisions as Non-cooperative Game-Theoretic Learning}

Chapter~\ref{chapter:noncooperative} studies game-theoretic learning in the context of social networks.
Game theory has traditionally been used in economics and social sciences with a focus on fully rational interactions where strong assumptions are made on the
information patterns available to individual agents. In comparison,  social sensors are agents with partial information and it is the dynamic
interactions among such agents that is of interest. This, together with the interdependence of agents' choices, motivates the need for game-theoretic learning models for agents interacting in social networks.

Chapter~\ref{chapter:noncooperative} deals with the question:
\begin{quote}
{\em When individuals are self-interested and possess limited sensing and communication capabilities,
can  a network
of such individuals
achieve sophisticated global behavior?}
\end{quote}
We  discuss a  non-cooperative game-theoretic learning approach for adaptive decision making in social networks. This can be viewed as non-Bayesian social learning. The aim is to ensure that all agents eventually choose actions from a common polytope of randomized strategies---namely, the set of correlated equilibria~\cite{A74,Aum87} of a non-cooperative game. The game-theoretic concept of equilibrium describes a condition of global coordination where all decision makers are content with the social welfare realized as the consequence of their chosen strategies.

We consider two examples of information exchange  among individuals. The first example comprises of fully social agents that can communicate with every other agent in the network. This provides a simple framework to present the ``regret-matching''~\cite{HM00,HMB13} decision making procedure that ensures convergence of the global behavior of the network to the correlated equilibria set. In the second example, we confine the information flow to social groups---each individual can only speak with her neighbors. Accordingly, the regret-matching procedure is revised to adapt to this more practical social network model. Finally, we consider the case of homogeneous social groups, where individuals share and are aware of sharing the same incentives. The regret-matching procedures is then adapted to exploit this valuable piece of information available to individuals within each social group. The final result in this chapter considers the scenario where the non-cooperative game model evolves with time according to the sample path of a finite-state Markov chain. It is shown that, if the speed of the Markovian jumps and the learning rate of the regret-matching algorithm match, the global behavior emerging from a network of individuals following the presented algorithms
properly tracks the time-varying set of correlated equilibria.

One of the main ideas in this chapter is that the limit system that represents the game-theoretic learning algorithms constitutes a differential inclusion. Differential inclusions are generalization of ordinary differential equations (ODEs)\footnote{A generic differential inclusion is of the form $dX/dt \in \mathcal{F}(X,t)$, where $\mathcal{F}(X,t)$ specifies a family of trajectories rather than a single trajectory as in the ordinary differential equations $dX/dt = F(X,t)$. See~\S\ref{subsec:proof-sketch-game} and Appendix~\ref{sec:theorem-proofs-game} for more details.}, and arise naturally in game-theoretic learning, since the strategies according to which others play are unknown. This is highly non-standard in the analysis of stochastic approximation algorithms in which the limit system is usually an ODE.

Chapter~\ref{chapter:noncooperative} ends with an example that shows how the presented algorithms can be applied in a social sensing application. We consider the problem of estimating the true value of an unknown parameter via a network of sensors. There have been a lot of recent works that study diffusion of information over graphs linking a multitude of agents; see~\cite{Say14,S13} and numerous references therein. We particulary focus on diffusion least mean square~(LMS) algorithms~\cite{LS08}: each sensor decides whether to activate, and if activates, (i) it will exchange estimate with neighbors and fuse the collected data; (ii) it will use the fused data and local measurements to refine its estimate via an LMS-type adaptive filter. Using a game-theoretic formulation, an energy-aware activation mechanism is devised that, taking into account the spatial-temporal correlation of sensors' measurements, prescribes sensors when to activate.
We first show that, as the step-size in the diffusion LMS approaches zero, the analysis falls under the unifying classical stochastic approximation theme of this chapter and, therefore, can be done using the well-known ODE method~\cite{KY03}. It is then shown that the proposed algorithm ensures the estimate at each sensor converges to the true parameter, yet the global activation behavior along the way tracks the set of correlated equilibria of the underlying activation control game.

\subsubsection*{5. Appendices} The two appendices at the end of this monograph present, respectively, a mean-square error analysis and weak convergence analysis
of two different types of stochastic approximation algorithms used to track time-varying behavior in social networks. These analysis are crucial in allowing us to predict the asymptotic dynamics
of such algorithms. The chapters provide sufficient intuition behind the theorems and the reader can skip the appendices without loss of continuity.
Appendix~\ref{ap:mu} generalizes the asymptotic analysis of duplication deletion random graphs in~\cite{CL06} to the case of Markov-modulated
graphs. It uses the concept of perturbed Lyapunov functions.
The weak convergence analysis presented in Appendix~\ref{sec:theorem-proofs-game} generalizes the convergence analysis
 provided in the seminal papers~\cite{HM00,HM01a} to the case where the game-theoretic learning algorithm can track a time-varying correlated equilibrium.
The convergence analysis in both appendices are presented in a tutorial fashion and are readily accessible to researchers in adaptive filtering,
stochastic optimization, and game theory.

\subsubsection*{Out-of-Scope Topics} Other important problems that have been extensively studied in the literature, but are outside the scope of this monograph include: consensus formation~\cite[Chapter~8]{J08},~\cite{KB08,TJ10}, metrics for measuring networks (other than degree distribution)~\cite[Chapter~2]{J08},~\cite{WKV13}, small world~\cite{Kle00,Wat99,WS98}, cooperative models of network formation~\cite[Chapter~1]{DW05},~\cite[Chapter~12]{J08},~\cite{SP09}, behavior dynamics in peer-to-peer media-sharing social networks~\cite{HK07,ZLL11}, and privacy and security modeling~\cite{Lan07,LYM03}. The interested reader is referred to the above cited works and references therein for extensive treatment of the topics.

\section{Perspective}\label{intro:prospective}

The social learning and game-theoretic learning formalisms considered in this monograph
 can be used either as descriptive tools, to predict the outcome of complex interactions amongst agents in sensing, or as  prescriptive tools, to design social networks and sensing
 systems around given interaction rules.
Information aggregation, misinformation propagation
and privacy are important issues in  sensing using social sensors. 
In this monograph, we  treat these issues in a highly stylized manner so as to provide easy accessibility to an electrical engineering audience.
The underlying  tools
used in this monograph are widely used by  the electrical engineering research community in the areas of signal processing, control, information theory and network communications.
The fundamental theory of network science is well-documented in seminal books such as~\cite{EK10,J08} and involves the interplay of random graphs and game theory.

In Bayesian estimation,
the twin effects of social learning (information aggregation with interaction amongst agents) and data incest
(misinformation propagation) lead to non-standard information patterns in estimating the underlying state of nature.
 Herding occurs when the public belief overrides the private observations and, thus, actions of agents are independent of their private observations.
Data incest results in bias in the public belief as a consequence of the unintentional re-use of identical actions in the formation of public belief in social learning---the information gathered by each agent is mistakenly considered to be independent. This results in overconfidence and bias in estimates of the state of nature.

Tracking a time-varying parameter that evolves according to a finite-state Markov chain (state of nature) is a problem of much interest in signal processing~\cite{BMP12,EKNS05,YK05b}. In social networks, sometimes the parameter under study (state of nature) cannot be sensed by pervasive sensors, e.g., the level of happiness in a community, the tendency of individuals to expand their networks, the strength of social links between individuals, etc. In such cases, social sensors can do much better than pervasive sensors. A social network with a large number of individuals can be viewed as an interactive sensing tool to obtain information about individuals or state of nature; this is a social sensor. Motivated by social network applications, a social sensor based framework is presented in Chapter~\ref{Chapter:tracking} to track the degree distribution of  Markov-modulated dynamic networks whose dynamics evolve over time according to a finite-state Markov chain.

Privacy issues impose important constraints on  social sensors. Typically, individuals are not willing to disclose private observations. Optimizing interactive  sensing with privacy constraints is an important problem. Privacy and trust pose conflicting requirements on human-based sensing:
 Privacy requirements  result in  noisier measurements or lower resolution actions, while maintaining a high degree
 of trust (reputation) requires accurate measurements.  Utility functions, noisy private measurements, and quantized actions are essential
 ingredients of the  social and game-theoretic  learning models presented in this monograph
that facilitate
modelling this trade-off between reputation and privacy.

In social sensor systems, the behavior is driven by the actions of a large number of autonomous individuals, who are usually self-interested and optimize their respective objectives. Often, these individuals form social contacts (i.e. links) by choice, rather than by chance. Further, there are always social and economic incentives associated with forming such social contacts based on the information obtained about the state of the nature or contribution to the diffusion of information across the network. The social network analysis using the common graph-theoretic techniques, however, fails to capture the behavior of such self-interested individuals and the dynamics of their interaction. This motivates the use of game-theoretic methods.
Game-theoretic learning explains how coordination in the decisions of such self-interested individuals might arise as a consequence of a long-run process of adaptation and learning in an interactive environment~\cite{FL08}. Interestingly enough, while each individual has limitations in sensing and communication, the coordinated behavior amongst individuals can lead to the manifestation of sophisticated behavior at the network level.

The literature in the areas of social learning, sensing, and networking is extensive.  In each of the following chapters, we provide  a brief review of relevant works together with references to experimental data.
 The book~\cite{Cha04} contains a complete treatment of social learning models with several remarkable insights.
For further references, we refer the reader to~\cite{Kri08,Kri11,Kri12,KMY08,MKZ09}.
In~\cite{Har05}, a  nice description is given of  how, if individual agents deploy simple heuristics, the global
system behavior can achieve
``rational'' behavior. The related problem of achieving
{\it coherence} (i.e., agents eventually choosing the same action or the same decision policy) among disparate sensors of decision agents without
cooperation has also witnessed intense research; see~\cite{PKP09} and~\cite{WKP11}.
 Non-Bayesian social learning models are also studied in~\cite{EF93,EF95}.

There is also a growing literature dealing with the interplay of technological networks and social networks~\cite{CCP13}. For example, social networks overlaid on technological networks account for a significant fraction of Internet use. Indeed, as discussed in~\cite{CCP13}, three key aspects of that cut across social and technological networks are the emergence of global coordination through local actions, resource sharing models and the wisdom of crowds (diversity and efficiency gains). These themes  are addressed in the current paper in the context of social learning.


\chapter{Social Learning Approach to Interactive Sensing }
\chaptermark{Social Learning Approach}
\label{Chapter:SL}

\section{Introduction}
This chapter comprises of three parts\footnote{The  chapter is an extended version of the paper~\cite{KP14}.}:

{\em 1. Survey and Review of Social Learning}:
Section~\ref{sec:classicalsocial} presents a  brief description of the classical social learning model.
We use social learning as the mathematical basis for modelling interaction of social sensors.
A key result in social learning is that rational agents eventually herd, that is, they pick the same
 action irrespective of their private observation and social learning stops.  To delay the effect of herding,  and thereby enhance social learning,  Chamley~\cite{Cha04}  (see also~\cite{SS97} for related work)
 has proposed a novel  constrained optimal social learning protocol. We describe how self-interested agents performing social learning can achieve  useful behavior in terms of
 optimizing a social welfare function. Such problems are motivated by privacy
 issues in sensing. If an agent reveals less information
  in its decisions,
 it maintains its privacy; on the other hand, as part of a social group, it has an incentive to optimize a social welfare
 function that helps estimate the state of nature.
 We review this protocol which is formulated as a  sequential stopping time problem.
 We show
  that the  constrained optimal social learning proposed by Chamley~\cite{Cha04}
  has a  threshold switching curve  in the space of public belief states.
Thus, the global decision to stop   can be implemented efficiently in a social learning model.

{\em 2. Data Incest in Online Reputation Systems}:
Section~\ref{sec:incest} deals with the question:
How can data incest (misinformation propagation) be prevented in online reputation blogs where social sensors  make recommendations?

In the classical social learning model,  each agent  acts once in a pre-determined order.
However, in online reputation systems  such as Yelp or Tripadvisor, which maintain logs of votes (actions)  by agents, social learning takes place with information exchange over a loopy graph  (where the agents form the vertices of the graph).
Due to the loops in the information exchange graph,
{\em data incest} (mis-information) can propagate: 
Suppose an agent wrote  a poor rating of a restaurant on a social media site.  Another agent is influenced by this rating, visits the restaurant, and then also gives  a poor rating on the social media site.
  The first agent visits the social media site  and notices that another agent has also given the restaurant a poor rating---this double confirms her rating and she enters another poor rating.

In a fair reputation system, such ``double counting'' or data incest should have been prevented by making the first agent  aware that the rating of the second agent was influenced by her own rating. Data incest results in a bias in the estimate of the state of nature. How can automated protocols be designed to prevent data incest, and thereby maintain a fair
 online reputation system? Section~\ref{sec:incest} describes how the administrator of a social network can maintain an unbiased (fair) reputation system.

3.   {\em  Interaction of Local and  Global Decision Makers for Change Detection}: Section~\ref{sec:socialc} deals with the question: In sensing, where individual agents perform social learning to estimate an underlying state of nature, how can changes in the  state of nature be detected?
Such sensing problems arise in a variety of applications such as financial trading, where individuals react to financial shocks~\cite{AS08}; marketing and advertising~\cite{Pin06,Pin08}, where consumers react to a new product; and localization of natural disasters (earthquakes and typhoons)~\cite{SOM10}.

Consider measurement of the adoption of a  new  product using a micro-blogging platform like Twitter. The
new technology diffuses through the market but its effects can only be observed through the tweets of select members of the population.
 Suppose the state of nature  suddenly  changes due to  a sudden market shock or presence of a new competitor.
Based on the local actions of the multi-agent system that is performing social learning, a global decision maker (such as a market monitor or technology manufacturer) needs to decide  whether or not to declare if a change has occurred.
How can the global decision maker achieve such change detection to minimize a cost function comprised of false alarm rate and delay penalty? The local and global decision makers
interact, since
the local decisions determine the posterior distribution of subsequent agents which determines the global decision (stop or continue), which in turn determines subsequent  local decisions.
 We show that this social learning based change detection problem leads to unusual behavior:
The optimal decision policy of the stopping time problem has multiple thresholds.    This  is unusual since,
 if it is optimal to declare that a change has occurred based on the posterior probability of change, it may not be optimal to declare a change when the posterior probability of change is higher!

\section{Multi-Agent Social Learning} \index{social learning}
   \label{sec:classicalsocial}

This section starts with a brief description of the classical social learning model. We review Chamley's novel constrained optimal social learning protocol, which is formulated as a  sequential stopping time problem, and delays the effect of herding~\cite{Cha04,SS97}.

\subsection{Motivation: What is social learning?}\label{sec:herd}

We start with a brief description of the `vanilla' social learning model\footnote{In typical formulations of social learning, the underlying state is assumed to be a random variable and not a Markov chain. Our description below is given in terms of a Markov chain since we wish to highlight the unusual structure of the social learning filter \index{nonlinear filtering!social learning filter} to a signal processing reader who is familiar with basic ideas in Bayesian filtering.
Also, we are interested in change detection problems in which the change time distribution can be modelled as the absorption time of a Markov chain.}. In social learning~\cite{Cha04}, agents estimate the underlying state of nature not only from their local measurements, but also from the actions of previous agents. (These previous actions  were taken by agents in response to their local measurements; therefore, these actions convey information about the underlying state). As we will describe below, the state estimation update in social learning has a drastically different structure compared to the standard optimal filtering recursion and can result in unusual behavior.

Consider a countable  number of agents performing social learning  to estimate the state of an underlying finite-state Markov chain $x$. Let $\X = \{1,2,\ldots,X\}$ denote a finite state space, $\tp$ the transition matrix and $\pi_0$ the initial distribution of the Markov chain.

Each agent acts once  in a predetermined sequential order indexed by $k=1,2,\ldots$. The index $k$ can also be viewed
as the discrete time instant when agent $k$ acts. A multi-agent system seeks to  estimate $x_k$.  Assume at the beginning of iteration $k$, all agents have access to the public belief $\pi_{k-1}$ defined in  Step \emph{(iv)} below. The social learning protocol proceeds as follows~\cite{BHW92,Cha04}:
\begin{enumerate}
 \item[\emph{(i)}] {\em Private Observation}: At time $k$, agent $k$ records a private observation $y_k\in \Y$ from the observation distribution $B_{iy} = \P(y|x=i)$, $i \in \X$. Throughout this chapter, we assume that $\Y = \{1,2,\ldots,Y\}$ is finite.
\item[\emph{(ii)}] {\em Private Belief}:  Using the public belief $\pi_{k-1} $ available at time $k-1$ (defined in Step (iv) below), agent $k$ updates its private posterior belief $\priv_k(i) =  \P(x_k = i| a_1,\ldots,a_{k-1},y_k)$ as the following Bayesian update (this is the classical \index{nonlinear filtering!hidden Markov model filter} Hidden Markov Model filter~\cite{EM02}):
\begin{align}
\label{eq:hmm}
\priv_k &=
\frac{B_{y_k} \tp^\p \pi}{ \mathbf{1}_X^\p B_y \tp^\p \pi}, \quad  
B_{y_k} = \text{diag}\left(\P(y_k|x=i),i\in \X\right) .
\end{align}
Here, $\mathbf{1}_X$ denotes the $X$-dimensional vector of ones, $\eta_k$ is an $X$-dimensional probability mass function (pmf), and $\tp^\p$ denotes transpose of the matrix $\tp$.
\item[\emph{(iii)}]  {\em Myopic Action}: Agent $k$ takes  action $a_k\in \ASL = \{1,2,\ldots, A\}$ to  minimize its expected cost
\begin{multline}
  a_k =  \argmin_{a \in \ASL} \E\{c(x,a)|a_1,\ldots,a_{k-1},y_k\} =\argmin_{a\in \ASL} \{c_a^\p\priv_k\}.    \label{eq:myopic}
\end{multline}
Here $\ca = (c(i,a), i \in \X)$ denotes an $X$-dimensional cost vector, and $c(i,a)$ denotes the cost  incurred when the underlying state is $i$ and the  agent chooses action $a$.\\
  Agent $k$ then broadcasts its  action $a_k$ to subsequent agents.
\item[\emph{(iv)}] {\em Social Learning Filter}: \index{nonlinear filtering!social learning filter}
Given the action $a_k$ of agent $k$ and the public belief $\pi_{k-1}$, each  subsequent agent $k' > k$
 computes the public belief $\pi_k$ according to the following ``social learning  filter'':\ \beq \pi_k = \Ts(\pi_{k-1},a_k), \text{ where }\; \Ts(\pi,a) =
 \frac{\Bs_a \tp ^\p\pi}{\sigs(\pi,a)} \label{eq:piupdate}\eeq
 and
$\sigs(\pi,a) = \mathbf{1}_X^\p \Bs_a \tp^\p \pi$ is the normalization factor of the Bayesian update.
In (\ref{eq:piupdate}),  the public belief $\pi_k(i)  = \P(x_k = i|a_1,\ldots a_k)$, and $\Bs_a  = \text{diag}(\P(a|x=i,\pi),i\in \X )$ has elements
\begin{equation} \label{eq:aprob}
\begin{split}
 & \P(a_k = a|x_k=i,\pi_{k-1}=\pi) = \sum_{y\in \Y} \P(a|y,\pi)\P(y|x_k=i) \\
 &   \P(a_k=a|y,\pi) = \begin{cases}  1, \text{ if }  c_a^\p B_y \tp^\p \pi \leq c_{\ta}^\p B_y \tp^\p\pi, \; \ta \in \ASL, \\
 0, \text{ otherwise. }  \end{cases}
\end{split}
\end{equation}
\end{enumerate}

The derivation of the social learning filter~(\ref{eq:piupdate}) is given in the discussion below.


\subsection{Discussion}  Let us pause to give some intuition about  the above social learning protocol.

{\em 1. Information Exchange Structure}: The information exchange in the above social learning protocol is sequential.
Agents send their hard decisions (actions) to subsequent agents.
Further, we have assumed that each agent acts once. Another way of viewing the  social learning  protocol
is that there are finitely  many agents that act repeatedly in some pre-defined order. If each agent chooses its local decision using the current public belief, the setting will be identical to the social learning setup. We also refer the reader to~\cite{AO11} for several recent results in social learning
over several types of network adjacency matrices.

{\em 2. Filtering with Hard Decisions}: Social learning can be viewed as agents making {\em hard} decision estimates at each time and sending these estimates to subsequent agents.
In conventional Bayesian state estimation, a {\em soft} decision is made, namely, the posterior distribution (or equivalently, observation) that is sent
to subsequent agents. For example, if $\ASL = \X$ and
the costs are chosen as $c_a = - e_a$ where $e_a$ denotes the unit indicator with $1$ in the $a$-th position, then
$\argmin_a c_a^\p \pi  = \argmax_a \pi(a)$, i.e.,  the maximum aposteriori probability~(MAP) state estimate. For this example, social learning is equivalent
to agents sending the hard MAP estimates to subsequent agents.

 Note that rather than sending a hard decision estimate, if
each agent chooses its action $a_k = y_k$ (that is, agents send their private observations),
the right-hand side of~(\ref{eq:aprob}) becomes $\sum_{y\in \Y} I(y = y_k)  \P(y|x_k=i) = \P(y_k|x_k=i)$ and, thus,
 the problem becomes a standard Bayesian filtering problem.

 It is also important to note  that the filtered distributions obtained via social learning are not commutative in the actions. That is
 $P(x|a_1=a,a_2 = \bar{a}) \neq P(x|a_1 = \bar{a}, a_2 = a)$. In comparison, for Bayesian estimation of random variable $x$ with conditionally independent  observations,
  $P(x|y_1=y,y_2 = \bar{y}) \neq P(x|y_1 = \bar{y}, y_2 = y)$.

 {\em 3. Dependence of Observation Likelihood on Prior}: The most unusual feature of the above protocol (to a signal processing audience) is
 the social learning filter~(\ref{eq:piupdate}).  In  standard state estimation via a Bayesian filter, the observation likelihood given the state is completely parameterized
 by the observation noise distribution and is functionally independent of the current prior distribution.
 In the social learning filter, \index{nonlinear filtering!social learning filter}
the likelihood of the action given the state (which is denoted by $\Bs_a$) is
an explicit function of the prior $\pi$!
Not only does the action likelihood depend on the prior, but it is also a discontinuous
function, due to the presence of the $\argmin$ in~(\ref{eq:myopic}).

{\em 4. Derivation of Social Learning Filter}: \index{nonlinear filtering!social learning filter}
The derivation of the social learning filter~(\ref{eq:piupdate})  is as follows:  Define the posterior as
 $\belief_k(j) = \P(x_k = j|a_1,\ldots, a_k)$. Then,
 \begin{align}
 &\belief_k(j) =  \frac{1}{\sigma(\pi_{k-1},a_k)} \P(a_k|x_k=j,a_1,\ldots,a_{k-1}) \nonumber\\ & \quad\times \sum_i \P(x_k=j|x_{k-1}=i)
 \P(x_{k-1}=i|a_1,\ldots,a_{k-1}) \nonumber \\
&= \frac{1}{\sigma(\pi_{k-1},a_k)}  \sum_y \P(a_k| y_k=y, a_1,\ldots,a_{k-1})\nonumber \\ &\quad\times \P(y_k=y|x_k=j)  \sum_i \P(x_k=j|x_{k-1}=i) \belief_{k-1}(i)\nonumber\\
&= \frac{1}{\sigma(\pi_{k-1},a_k)} \sum_y \P(a_k|y_k=y,\pi_{k-1})\P(y_k=y|x_k = j) \sum_i \tp_{ij} \belief_{k-1}(i)
\end{align}
where the normalization term is
\begin{align} \sigma(\pi_{k-1},a_k)= \sum_j\sum_y &\P(a_k|y_k=y,\pi_{k-1}) \nonumber\\ &\times\P(y_k=y|x_k = j) \sum_i \tp_{ij} \belief_{k-1}(i). \end{align}

  The above social learning protocol and social learning filter(\ref{eq:piupdate}) result in interesting dynamics in state estimation and decision making.
We will illustrate two interesting consequences that are  unusual to an electrical engineering audience:
\begin{itemize}
\item Rational Agents form herds and information cascades and blindly follow previous agents.
This is
 discussed in \S\ref{sec:cascade} below.

\item Making global decisions on change detection in a multi-agent system performing social learning results in multi-threshold behavior.
This is discussed in \S\ref{sec:socialc} below.
\end{itemize}



\subsection{Rational Agents form Information Cascades} \label{sec:cascade}
The first consequence of the unusual nature of  the social learning filter~(\ref{eq:piupdate}) is that social learning can result in  multiple
rational
agents
taking the same action independently of their observations.

Throughout this subsection, we  assume that  $x$ is a finite state random variable (instead of a Markov chain) with prior distribution $\pi_0$.
%
We start with the following definitions; see also \cite{Cha04}:
\begin{itemize}
\item  An individual agent $k$ {\em herds} \index{herding}  on the public belief $\pi_{k-1}$  if it chooses its action  $a_k = a(\pi_{k-1},y_k)$ in  (\ref{eq:myopic}) independently of its observation
$y_k$.
 \item  A {\em herd of agents} takes place at time $\bar{k}$, if the actions of all agents after time $\bar{k}$ are identical, i.e., $a_k = a_{\bar{k}}$ for all
time  $k > \bar{k}$.
\item An {\em information cascade} \index{information cascade}  occurs at time $\bar{k}$, if the public beliefs of all agents after time $\bar{k}$ are identical, i.e.
$\pi_k = \pi_{\bar{k}}$ for all $k > \bar{k}$.
\end{itemize}
Note that if an information cascade occurs, then since the public belief freezes, social learning ceases.
Also, from the above definitions, it is clear that an information cascade implies a herd of agents, but the reverse is not true; see~\S\ref{sec:numerical} for an example.

 The following
result, which is well known in the economics literature~\cite{BHW92,Cha04}, states that if agents follow the above social learning protocol, then   after some finite time $\bar{k}$, an
information cascade occurs\footnote{A nice analogy is provided in~\cite{Cha04}.
 If I see someone walking down the street with an umbrella, I assume (based on rationality) that he has checked the weather forecast and is carrying
 an umbrella since it might rain. Therefore, I also take an umbrella. Now, there are two people walking down the street
 carrying umbrellas. A third person
 sees two people with umbrellas and, based on the same inference logic, also takes an umbrella.  Even though each  individual is  rational,
 such herding behavior  might be irrational  since
 the first person who took the umbrella may not have checked the weather forecast.\\
 Another example is that of patrons who decide to choose a restaurant. Despite their menu preferences, each patron chooses the restaurant with the most  customers. Therefore, all patrons eventually herd to one restaurant. The paper~\cite{TBB10} quotes the following anecdote on user influence in a social network which can be interpreted as herding: ``... when a popular blogger left his blogging site for a two-week vacation, the site's visitor tally fell, and content produced by three invited substitute bloggers could not stem the decline.''}.
  The proof follows via an elementary application of the martingale convergence
 theorem.

\begin{theorem}[\cite{BHW92}]
\label{thm:herd} The social learning protocol described in~\S\ref{sec:herd}  leads to an information cascade  in finite time
with probability~1. That is, there
exists a finite time $\bar{k}$ after which social learning ceases, i.e., public belief $\pi_{k+1} = \pi_k$, $k \geq \bar{k}$, and all agents choose the same action, i.e., $a_{k+1} = a_k$, $k\geq \bar{k}$.
\end{theorem}

Instead of reproducing the proof, let us give some insight as to why Theorem~\ref{thm:herd} holds. It can be shown using martingale methods
that, at some finite time $k=\bar{k}$, the agent's probability $\P(a_k|y_k,\pi_{k-1})$ becomes  independent of the private observation $y_k$. Then, clearly
from~(\ref{eq:aprob}), $\P(a_k=a|x_k=i,\pi_{k-1}) =   \P(a_k=a|\pi)$. Substituting this into the social learning filter~(\ref{eq:piupdate}), we see
that $\pi_{k} = \pi_{k-1}$.  Thus, after some finite time $\bar{k}$, the social learning filter hits a fixed point and social learning stops.
As a result, all subsequent agents $k> \bar{k}$ completely disregard their private observations and take the same action $a_{\bar{k}}$,
thereby forming
 an information cascade (and therefore a herd).

\subsection[Individual Privacy vs Group Reputation]{Constrained Interactive  Sensing: Individual Privacy vs Group Reputation}
The above social learning protocol can be interpreted as follows.
Agents seek to estimate an underlying state of nature; however, they reveal their actions by maximizing their privacy according to the optimization~(\ref{eq:myopic}).
This leads to an information cascade and social learning stops.
In other words,
agents are interested in optimizing their own costs (such as maximizing privacy) and ignore the information benefits their action provides
to others.

\subsubsection{Partially Observed Markov Decision Process Formulation} \index{Partially Observed Markov Decision Process (POMDP)}
We now describe an optimized social learning procedure
that
 delays herding\footnote{In the restaurant problem, an obvious approach to prevent herding is as follows. If a restaurant knew that patrons choose the restaurant with the most  customers,
then the restaurant  could deliberately pay actors to sit in the restaurant so that it appears popular, thereby attracting customers. The methodology in this section, where herding is delayed by benevolent agents,  is a different approach.}.
 This approach is motivated by the following question:
How can agents assist social learning by  choosing their actions to trade off individual privacy (local costs) with
 optimizing the reputation\footnote{The papers~\cite{Mui02,GGG09} contain lucid  descriptions of
quantitative models for trust, reputation and privacy.}  of the entire social group? (See~\cite{Cha04} for an excellent discussion.) 

Suppose
agents seek to maximize the reputation of their social group
by  minimizing the following social welfare cost involving all agents in the social group (compared to
the myopic objective~(\ref{eq:myopic})  used in standard social learning):
\beq \label{eq:pomdp}
J_\mu(\pizero) = \Ep \lbr\sum_{k=1}^\infty \discount^{k-1} c_{a(\pi_{k-1},y_k,\mu(\pi_{k-1}))}^\p  \eta_k
 \rbr.  \eeq
In~(\ref{eq:pomdp}), 
$a(\pi,y,\mu(\pi)))$ denotes the decision rule that agents use to choose their actions as will be explained below. 
Also, $\discount\in [0,1)$ is an economic discount factor, and $\pi_0$ denotes the initial probability (prior)  of the state $x$. Finally,
 $\Pp$ and $\Ep$ denote the probability measure and expectation
of the evolution of the observations and underlying state, respectively, which are strategy dependent.

The key attribute of~(\ref{eq:pomdp})  is that each agent $k$ chooses its  action according to the
privacy constrained rule
\beq a_k = a(\pi_{k-1},y_k,\mu(\pi_{k-1})).
\label{eq:akp}
\eeq
Here,
the policy  $$\mu: \pi_{k-1} \rightarrow \{1,2\ldots, L\}$$
maps the available public belief to the set of $L$ privacy values.
The higher the privacy value, the less the agent reveals through its action.
This is  in contrast  to standard social
learning~(\ref{eq:myopic}) in which the action chosen is $a(\pi,y)$, namely, a myopic function of the private observation and public
belief.

The above formulation can be interpreted as follows:  Individual agents seek to maximize
their privacy according to social learning~(\ref{eq:akp}), but also seek to maximize the reputation of their entire
social group~(\ref{eq:pomdp}).

Determining the  policy $\mu^*$  that
minimizes~(\ref{eq:pomdp}), and thereby maximizes the social group reputation, is equivalent to solving a   stochastic control problem that is
called a  partially observed Markov decision process~(POMDP) problem~\cite{Cas98,Kri11}.
A POMDP comprises of a noisy observed Markov chain and the dynamics of the posterior distribution (belief state) is controlled by a policy ($\mu$ in our case).

\subsubsection{Structure of Privacy Constrained Sensing Policy}
 In general, POMDPs   \index{Partially Observed Markov Decision Process (POMDP)}  are computationally intractable to solve \cite{PT87} and, therefore, one cannot say anything useful about the structure\footnote{Characterizing
 the structure of the optimal policy of a POMDP is a difficult problem. We refer the reader
 to \cite{Lov87,Lov87a,Rie91,KD07,Kri11} for sufficient conditions (based on supermodularity \index{supermodularity} that yield a monotone optimal policy for a POMDP in terms of the monotone likelihood ratio stochastic order.} of the optimal policy $\mu^*$.
 However, useful insight can  be obtained by considering  the following extreme case of the
  above problem.  Suppose there are two privacy values  and each agent $k$ chooses action
  \begin{equation}
  a_k = \begin{cases} y_k,  & \text{ if } \mu(\pi_k) = 1 \text{ (no privacy)}, \\
  \argmin_a c_a^\p \pi_{k-1}, & \text{ if }  \mu(\pi_k) = 2 \text{ (full privacy}).
   						   \end{cases}
  \end{equation}
  That is, an agent either reveals its raw observation (no privacy) or chooses its action by completely neglecting its observation (full privacy).
 Once an  agent chooses the full privacy option, then all subsequent agents  choose exactly
 the same option and therefore herd---this follows since each agent's
  action reveals
  nothing about the underlying state of nature. Therefore, for this extreme example, determining the optimal policy $\mu^*(\pi)$ is equivalent to solving
  a stopping time problem:
Determine  the earliest time for agents to herd (maintain full privacy) subject to maximizing the
social group reputation.

  For  such a quickest herding  stopping time problem, one can say a lot about the structure of $\mu^*(\pi)$.
   Suppose the sensing system
wishes to determine if the state of nature is a specific target state (say state 1).
Then,~\cite{Kri11} shows that, under reasonable conditions
on the observation distribution and supermodular conditions \index{supermodularity} on the costs (\cite{MR07} discusses
supermodularity of influence in social networks),
 the dynamic programming recursion has a supermodular structure\footnote{The seminal book on supermodularity is \cite{Top98}, also see \cite{Ami05} for a lucid
 tutorial presentation.}
 (see also~\cite{Kri12,Kri13,KD07,Lov87,Rie91} for related results).
This implies that the optimal policy $\mu^*$ has the following structure:
There exists a threshold curve that partitions the  belief space such that, when the belief state is on one
side of the curve, it is optimal for agents to reveal full observations; if the belief state is on the other side of the curve,
then it is optimal to herd. Moreover, the target state~1 belongs to the region in which  it is optimal to herd\footnote{In standard POMDPs where agents do not perform social learning, it is well known~\cite{Lov87a} that the  set of beliefs
for which it is optimal to stop is convex.
Such convexity of the herding set does not hold in the current problem. But it is shown in~\cite{Kri11}
 that the set of beliefs for which it is optimal
to herd constitute a  connected set and so does the set of beliefs for which it is optimal to reveal full observations.}.
This threshold structure of the optimal policy means that if individuals deploy the simple heuristic
of
\begin{quote}
{\em``Choose increased privacy when belief is close to target state,''} \end{quote} then the group behavior
is sophisticated---herding is delayed
 and accurate estimates of the state of nature can be obtained.

 \section{Data Incest in Online Reputation Systems} \label{sec:incest} \index{data incest}
\index{online reputation}

This section generalizes the previous section by considering social learning \index{social learning} in a social network.
How can multiple social sensors interacting over a social network  estimate an underlying state of nature?  The state could be the position coordinates of an event~\cite{SOM10} or the quality of a social parameter such as quality of a restaurant or political party.

\begin{figure}[t]
\begin{center}
\includegraphics[width=.7\textwidth]{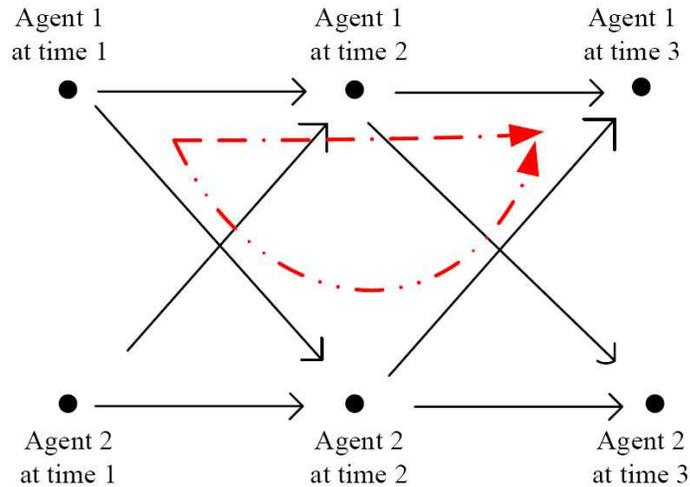}
\end{center}
\caption{Example of the information flow
in a social network with two agents over three event epochs. The arrows represent exchange of information.}
\label{fig:sample}
\end{figure}

The motivation for this section can be understood in terms of the following  sensing example.
 Consider the following interactions in a multi-agent social network where  agents seek to estimate an underlying state of nature. Each agent visits a restaurant based on reviews on an online reputation website. The agent then obtains  a private measurement of the state  (e.g., the quality of food in a restaurant) in noise. After that, he reviews the restaurant on the same online reputation website.  The information exchange in the social network is modeled by a directed graph.
 As mentioned in Chapter~\ref{chapter:intro}, data incest~\cite{KH13} arises due to the loops in the information exchange graph.
 This is illustrated in Figure~\ref{fig:sample}: Agents 1 and 2 exchange beliefs (or actions).
  The fact
 that there exist two distinct paths between Agent~1 at time~1 and Agent~1 at time~3 (depicted in red in Figure~\ref{fig:sample}) implies that
 the information of Agent~1 at time~1 is double counted, thereby leading to a data incest event.


How can data incest be removed so that agents obtain a fair (unbiased) estimate of the underlying state?
The methodology of this section  can be interpreted in terms of the recent {\em Time} article~\cite{Tut13}, which provides interesting rules for online reputation systems. These include: (i) review the reviewers, and  (ii) censor fake (malicious) reviewers. The data incest removal algorithm proposed in this section can be viewed as  ``reviewing the reviews'' of other agents to see if they are associated with data incest or not.

The rest of this section is organized as follows:
\begin{enumerate}
\item  Section~\ref{sec:social} describes the social learning model that is used to mimic the behavior of agents in online reputation systems.
The information exchange between agents in the social network is  formulated on a family of time-dependent directed acyclic graphs.
\item In~\S\ref{sec:repute}, a fair reputation protocol is presented and the criterion for achieving a fair rating is defined.
\item  Section~\ref{sec:removal} presents an incest removal algorithm so that the online reputation system
achieves a fair rating. A necessary and sufficient condition is given on the graph
that represents exchange of information between agents so as to achieve fair ratings.
\end{enumerate}

\paragraph*{Related works}  Collaborative recommendation systems are reviewed and studied in~\cite{AT05,KSJ09}. In~\cite{KT12},
a model of Bayesian social learning is considered in which  agents receive private information about the state of nature and observe actions of their neighbors in a tree-based network.
 Another type of misinformation caused by influential agents (agents who heavily affect  actions of other agents in social networks) is investigated in~\cite{AO11}. Misinformation in the context of this section is motivated by sensor networks where the term ``data incest'' is used~\cite{BK07}.   Data incest also arises in  belief propagation~(BP) algorithms~\cite{MWJ99,Pea86}, which are used in computer vision and error-correcting coding theory.
BP algorithms require passing local messages over the graph (Bayesian network) at each iteration. 
For graphical models with loops, BP algorithms are only approximate due to the over-counting of local messages~\cite{YFW05}, which is similar to data incest in  social learning.  With the algorithms presented in this section, data incest can be mitigated from Bayesian social learning over non-tree graphs that satisfy a topological constraint.
The closest work to the current section is~\cite{KH13}. However,  in~\cite{KH13}, data incest is considered in a network where agents exchange their private belief states---that is, no social learning is considered.  Simpler versions of this information exchange process and estimation were investigated in~\cite{Aum76,BV82,GP82}.


\subsection{Information Exchange Graph}\label{sec:social}
 Consider an online reputation system  comprised of social  sensors $\{1,2,\ldots,S\}$ that aim to estimate an underlying state of nature (a random variable).
 Let $x \in
 \X = \{1,2,\ldots,X\}$
 represent the state of nature (such as the quality of a hotel) with known prior distribution $\pi_0$. Let $k = 1,2,3,\ldots$ represent epochs at which events occur. These events involve taking observations, evaluating beliefs and choosing actions as  described below. The index $k$ marks the historical order of events, and not necessarily absolute time. However, for simplicity, we refer to $k$ as ``time''.

 To model the information exchange in the social network, we will use a family of directed acyclic graphs.
 It is convenient also to reduce the coordinates of time $k$ and agent $s$  to a single integer index $n$ as follows:
\begin{equation} \label{reindexing_scheme}
 n \ole s+ S(k-1), \quad
s \in \{1,\ldots, S\}, \; k = 1,2,3,\ldots.
\end{equation}
We  refer to $n$ as a ``node'' of a time-dependent information flow graph $G_n$ that we define below.

\subsubsection{Some Graph Theoretic Definitions}
Let \begin{equation}\label{eq:defG} G_{n} = (V_{n}, E_{n}), \quad n  = 1,2,\ldots \end{equation} denote a sequence of time-dependent graphs of information flow in the social network until and including time $k$ where $n = s + S(k-1)$. Each vertex in $V_{n}$ represents an agent $s'$ in the social network at time $k'$, and each edge $(n',n'')$ in $E_{n}\subseteq V_{n} \times V_{n}$ shows that the information (action) of node $n'$ (agent $s'$ at time $k'$) reaches node $n''$ (agent $s''$ at time $k''$).  It is clear that the communication graph $G_n$ is  a sub-graph of $G_{n+1}$. This means that the diffusion of actions  can be modelled via a family of time-dependent directed acyclic  graphs\footnote{Directed acyclic  graphs are directed graph with no directed cycles.}.

The algorithms below will involve specific columns of the adjacency matrix and transitive closure matrix of the graph $G_n$. The adjacency matrix \index{matrix!adjacency matrix} $A_n$ of $G_n$ is an $n\times n$ matrix with elements $A_n(i,j)$  given by
\beq \label{eq:adjacencymatrix}
A_n (i,j)=\begin{cases}
1, &\textrm{ if } (v_j,v_i)\in E \\
0, &\textrm{ otherwise}
\end{cases}\;, \quad A_n(i,i)=0.
\eeq
The transitive closure matrix \index{matrix!transitive closure matrix}   $T_n$ is the  $n\times n$ matrix
\begin{equation} T_n =  \textmd{sgn}((\mathbf{I}_n-A_n)^{-1}) \label{eq:tc} \end{equation}
where for any matrix $M$, the matrix $\text{sgn}(M)$ has elements
\begin{equation}
\text{sgn}(M)(i,j) = \begin{cases} 0, & \text{ if } M(i,j)=0, \\
1,  & \text{ if } M(i,j) \neq 0. \end{cases}
\end{equation}
Note that $A_n(i,j) = 1$ if there is a single-hop path between nodes $i$ and $j$. In comparison,
$T_n(i,j) = 1$ if there exists a (possibly multi-hop) path between node $i$ and $j$.

The information reaching node $n$ depends on the information flow graph  $G_n$.
The following two sets will be used  to specify the incest removal algorithms below:
\begin{align}
\history_n &= \left\{m : A_n(m,n) = 1 \right\},  \label{eq:history}  \\
\full_n &= \left\{m : T_n(m,n) = 1 \right\} . \label{eq:full}
\end{align}
Here, $\history_n$ denotes the set of previous nodes $m$ that communicate with node $n$ in a single-hop.
In comparison, $\full_n$
  denotes the set of previous nodes $m$ whose information eventually arrive at node $n$. Thus, $\full_n$  contains all possible  multi-hop connections by which information from a node $m$
 eventually reaches node $n$. 

Note that classical social learning of Sec.\ref{sec:herd}, is a special case with adjacency matrix
$A_n(i,j) = 1$ for $j=i+1$ and $A_n(i,j) = 0$ elsewhere.

\begin{example}
\label{example-1}
To illustrate the above notation
consider a social network consisting of $S=2 $ agents with the following information flow graph for three  time points $k=1,2,3$.
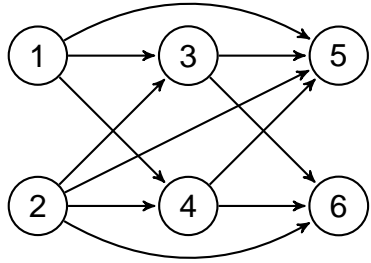
\begin{figure}[t]
\centering
\begin{tikzpicture}[->,>=stealth',shorten >=1pt,auto,node distance=2cm,
  thick,main node/.style={circle,fill=white!12,draw,font=\sffamily}]

  \node[main node] (1) {1};
  \node[main node] (2) [below  of=1] {2};
  \node[main node] (3) [right of=1] {3};
  \node[main node] (4) [below  of=3] {4};
  \node[main node] (5) [right of=3] {5};
  \node[main node] (6) [right of=4] {6};

  \path[every node/.style={font=\sffamily\small}]
    (1) edge node  {} (4)
        edge [bend left] node[left] {} (5)
        edge node {} (3)
    (2) edge node {} (3)
        edge node {} (4)
        edge  node {} (5)
        edge [bend right] node[left] {} (6)
    (3) edge node  {} (5)
        edge node  {} (6)
    (4) edge node  {} (5)
        edge node {} (6);
\end{tikzpicture}
\caption{Example of an  information flow network with two agents ($S=2$), namely, $s\in \{1,2\}$ and time instants $k=1,2,3$.  Circles represent the nodes indexed by $n= s + S(k-1)$
in the social network, and each edge depicts a communication link between two nodes.}
\label{sample}
\end{figure}
%
  Figure~\ref{sample} depicts the nodes  $n =1,2,\ldots,6$, where $n= s + 2(k-1)$.

  Note that, in this example, each node remembers all its previous actions as is apparent from Figure~\ref{fig:sample}.
  The information flow is characterized by
the family of directed acyclic graphs  $\{G_1,G_2,G_3,G_4,G_5,G_6\}$  with adjacency matrices  
\begin{displaymath}
\begin{array}{l}
A_1 = \begin{bmatrix}
0
\end{bmatrix},
A_2 = \begin{bmatrix}
0 & 0 \\
0 & 0
\end{bmatrix},
A_3 = \begin{bmatrix}
0 & 0 & 1 \\
0 & 0 & 1 \\
0 & 0 & 0
\end{bmatrix},\\
A_4 = \begin{bmatrix}
0 & 0 & 1 & 1 \\
0 & 0 & 1 & 1 \\
0 & 0 & 0 & 0 \\
0 & 0 & 0 & 0
\end{bmatrix},
A_5 = \begin{bmatrix}
0 & 0 & 1 & 1 & 1 \\
0 & 0 & 1 & 1 & 1 \\
0 & 0 & 0 & 0 & 1 \\
0 & 0 & 0 & 0 & 1 \\
0 & 0 & 0 & 0 & 0
\end{bmatrix}.
\end{array}
\end{displaymath}
Since nodes 1 and 2 do not communicate,
clearly $A_1$ and $A_2$ are zero matrices. Nodes 1 and  3 communicate as do nodes 2 and  3, hence,  $A_3$ has two
ones, etc.
Finally from~(\ref{eq:history}) and~(\ref{eq:full}),
\begin{displaymath}
\history_5=\{1,2,3,4\}, \quad \full_5 = \{1,2,3,4\}
\end{displaymath}
where $\history_5$ denotes all single-hop links to node~5, whereas $\full_5$ denotes all multi-hop links to node~5.

Note that $A_n$ is always the upper left  $n\times n$ submatrix of $A_{n+1}$. Also
 due to causality with respect to the time index $k$, the adjacency matrices are  always upper triangular.
\end{example}

\subsection{Fair Online Reputation System} \label{sec:repute}

The aim is to provide each node $n$ with an unbiased estimate
\beq \tbelief_{n-}(i) = P(x = i | \{a_m, m \in \full_n\})   \label{eq:aims}\eeq
subject to the following {\em social influence constraint}: \index{social influence constraint}
There exists a fusion algorithm $\mathcal{A}$ such that
  \beq  \tbelief_{n-} = \mathcal{A} (\pi_m, m \in \history_n).   \label{eq:si}\eeq

We  call $\tbelief_{n-}$ in (\ref{eq:aims})  the {\em true}  or {\em fair online rating} available to \index{online reputation}
  node $n$. Recall that $\full_n$, defined in (\ref{eq:full}), denotes all  information (multi-hop links) available  to node $n$. By definition, $\tbelief_{n-}$ is incest free since it is the desired conditional probability that we want.

The procedure summarized in Protocol~\ref{protocol:socialcons} aims to evaluate a fair reputation that uses social learning over a social network by eliminating incest.

\begin{algorithm}\floatname{algorithm}{Protocol}
(i) {\em  Information from Social Network}: 
\begin{enumerate}
\item {\em Recommendation from friends}: Node $n$ receives past actions  $\{a_m, m \in \history_n\}$ from previous nodes  $m \in \history_n$ in the social network. $\history_n$ is defined in~(\ref{eq:history}).
\item  {\em Automated Recommender System}: For these past actions $\{a_m, m \in \history_n\}$,   the network administrator
has already computed the public beliefs $(\pi_m, m \in \history_n)$ using Step (v) below.\\
The automated recommender system    fuses  public beliefs
$(\pi_m, m \in \history_n)$, into  the single   recommendation  belief  $\belief_{n-}$ as
\beq  \belief_{n-} = \mathcal{A} (\pi_m, m \in \history_n).  \label{eq:fuse}\eeq
The  fusion algorithm $\mathcal{A}$ will be designed below.
\end{enumerate}
(ii) {\em  Observation}: Node $n$  records   private observation  $y_n$ from  distribution $B_{iy} = \P(y|x=i)$, $i \in \X$.\\
(iii) {\em Private Belief}:  Node $n$ then  uses $y_n$ and public belief $\belief_{n-}$ to update its private belief via Bayes formula as
\beq   \priv_n =
\frac{B_{y_n}  \belief_{n-}}{ \mathbf{1}_X^\p B_y  \belief_{n-}}  \label{eq:privi}. \eeq
\\
(iv) {\em Myopic Action}: Node $n$ takes action
\begin{equation}a_n = \argmin_a c_a^\p \eta_n\end{equation} and inputs its action to the online reputation system.\\
(v) {\em Public Belief Update by Network Administrator}: Based on action $a_n$,
the network administrator (automated algorithm)  computes the public belief $\pi_n$ using  the social learning
filter (\ref{eq:piupdate}) with $P = I$.
\caption{Incest Removal for Social Learning  in Online Reputation Systems}\label{protocol:socialcons}
\end{algorithm}

If algorithm $\mathcal{A}$ is designed so that $\belief_{n-}(i)$ satisfies~(\ref{eq:aims}), then the computation~(\ref{eq:privi}) and Step~(v) yield
\begin{equation}   \begin{array}{l}\eta_n(i) =  \P(x = i | \{ a_m, m \in \full_n\},  y_n ),  \\
\pi_n(i)  = \P(x = i | \{a_m, m \in \full_n\},  a_n ),\end{array}\quad i \in \X \end{equation}
which are, respectively,   the correct private belief for node $n$ and the correct after-action public belief.

\subsubsection {Discussion of Protocol~\ref{protocol:socialcons}}
\indent
{\em (i) Data Incest}: It is important to note that without careful design of  algorithm~$\mathcal{A}$, \index{data incest}
due to  loops in the dependencies of actions on previous actions, the public rating $\belief_{n-}$ computed using~(\ref{eq:fuse}) can be substantially different from the fair online rating $\tbelief_{n-}$ of~(\ref{eq:aims}).
As a result,  $\eta_n $ computed via~(\ref{eq:privi}) will not be the correct private belief and
  incest will propagate in the network.  In other words,  $\eta_n$, $\belief_{n-}$, and $\belief_n$ are defined purely in terms of their computational expressions in Protocol~\ref{protocol:socialcons}---at this stage,
  they are not necessarily the desired  conditional probabilities unless algorithm $\mathcal{A}$ is designed to remove incest.

  Note that, instead of~(\ref{eq:fuse}), node $n$ could  naively (and incorrectly)  assume that the public beliefs $\belief_m, m\in \history_n$, that it received are
  independent. It would then
  fuse  these public beliefs as
  \beq  \belief_{n-} = \frac{\prod_{m\in \history_n} \belief_m }{\mathbf{1}_X^\p \prod_{m\in \history_n} \belief_m}. \label{eq:dataincest}\eeq
  This, of course, would result in data incest.

{\em (ii) How much does an individual remember?}: The above protocol has the flexibility of modelling  cases where
either each node remembers some (or all) of its past actions or none of its past actions. This facilitates modelling cases in which
people forget most of the past except for specific highlights.

{\em (iii) Automated Recommender System}: Steps (i) and (v)  of Protocol~\ref{protocol:socialcons} can be combined into an automated recommender system  that maps previous actions of agents
to a single recommendation (rating) $\belief_{n-}$ of~(\ref{eq:fuse}). This recommender system can operate completely opaquely to the actual user (node $n$). Node $n$ simply
uses the automated rating $\belief_{n-}$ as the current best available rating from the reputation system.

{\em (iv) Social Influence. Informational Message vs Social Message}: In Protocol~\ref{protocol:socialcons}, it is important that each individual $n$ deploys
Algorithm~$\mathcal{A}$  to fuse  the  beliefs $\{\belief_m, m \in \history_n\}$; otherwise, incest can propagate.
Here,  $\history_n$ can be viewed as the  ``social message'', i.e., personal friends of node $n$ since they directly communicate to node $n$, while the associated beliefs can be viewed as the
``informational message''.
The social message  from personal friends exerts a  large social influence---it provides significant incentive (peer pressure) for individual $n$ to comply with
Protocol~\ref{protocol:socialcons}, and thereby prevent incest.
 Indeed, a remarkable recent study described in~\cite{BFJ12} shows that  social messages (votes)
from known friends has significantly more influence on an individual than the information in the messages themselves. This study includes comparison of information messages and social messages on Facebook and their direct
effect on voting behavior.
 To quote~\cite{BFJ12},
\begin{quote}{\em ``The effect of social transmission on real-world voting
was greater than the direct effect of the messages themselves\ldots''}\end{quote}

{\em (v) Agent Reputation}:
The  cost function minimization in Step (iv) can be interpreted in terms of the reputation of agents in online reputation systems. If an agent continues to write bad reviews for high quality
restaurants on Yelp, his reputation declines among the users. Consequently, other people ignore reviews of that (low-reputation) agent in evaluating their opinion about the social unit under study (restaurant). Therefore, agents  minimize the penalty of writing inaccurate reviews (or equivalently increase their reputations) by choosing proper actions.

{\em (vi) Think and act}: Steps~(ii),~(iii),~(iv), and~(v)  of Protocol~\ref{protocol:socialcons} constitute standard social learning as described in~\S\ref{sec:herd}.
The key difference with standard social learning is  Step (i) that is performed by the network administrator.
Agents  receive public beliefs from the social network with arbitrary random delays.
These delays reflect the time an agent takes between reading the publicly available reputation and making its decision. It is typical behavior of people to read  published ratings multiple times and then think for an arbitrary amount of time before acting.

 \subsection{Incest Removal Algorithm in  Online Reputation System} \label{sec:removal}
 Below we design algorithm $\mathcal{A}$ in Protocol~\ref{protocol:socialcons} so that it yields the  fair  public rating $\tbelief_{n-}$ of~(\ref{eq:aims}).

\subsubsection{Fair Rating Algorithm}
 It is convenient to work with the logarithm of the un-normalized belief\footnote{The un-normalized belief proportional to $\belief_n(i)$ is the numerator of the social learning filter~(\ref{eq:piupdate}).
The corresponding un-normalized fair rating corresponding to $\tbelief_{n-}(i) $ is the joint distribution $\P(x=i, \{a_m, m \in \full_n\})$.
By taking logarithm of the un-normalized belief, Bayes formula merely becomes the sum of the log likelihood and log prior. This allows
 us to devise a data incest  removal algorithm based on linear combinations of the log beliefs.}.  Accordingly, define
\begin{equation} \lbelief_n(i) \propto \log \belief_n(i), \quad \lbelief_{n-}(i) \propto  \log \belief_{n-}(i), \quad i \in \X.\end{equation}

 The following theorem shows that the logarithm of the fair rating $\tbelief_{n-}$ defined in (\ref{eq:aims}) can be obtained as linear weighted combination of the logarithms of previous public beliefs.

  \begin{theorem}[Fair Rating Algorithm] \label{thm:socialincestfilter} Consider the online reputation system running  Protocol~\ref{protocol:socialcons}.
Suppose the  following algorithm
$\mathcal{A}(\lbelief_m, m \in \history_n)$ is
implemented in~(\ref{eq:fuse})  of  Protocol~\ref{protocol:socialcons} by the network administrator:
\begin{align}\label{eq:socialconstraintestimate}
\lbelief_{n-}(i)  =   w_n^\p \, \lbelief_{1:n-1}(i), 
 \;\;\text{ where } \;\;  w_n =  T_{n-1}^{-1}  t_n.
\end{align}
Then,  $\lbelief_{n-}(i) \propto \log \tbelief_{n-}(i)$. That is, algorithm~$\mathcal{A}$ computes the fair  rating $\log \tbelief_{n-}(i)$ defined in~(\ref{eq:aims}). \\
In (\ref{eq:socialconstraintestimate}),  $w_n$ is an  $n-1$ dimensional weight vector.
Recall  that
$t_n$ denotes the first $n-1$ elements of the  $n$th column of transitive closure matrix $T_n$.
\end{theorem}

Theorem~\ref{thm:socialincestfilter}  says that  the fair rating  $\tbelief_{n-}$ can be expressed as a linear function of the action log-likelihoods
 in terms of the transitive closure matrix $T_n$ of the information flow graph $G_n$. This is
intuitive since $\tbelief_{n-}$ can be viewed as the sum of information collected by the nodes such that there are paths between all these nodes and $n$.

\subsubsection{Achievability of Fair Rating by Protocol~\ref{protocol:socialcons}}
We are not quite done!
\begin{itemize}
\item On the one hand,  algorithm~$\mathcal{A}$  at node $n$ specified by~(\ref{eq:fuse}) has access only to beliefs
$\lbelief_m, m \in \history_n$---equivalently,  it  has access only to beliefs from previous nodes specified by $A_n(:,n)$, which denotes the  last column of the adjacency matrix $A_n$.
\item
On the other hand,  to provide incest free estimates, algorithm~$\mathcal{A}$ specified in~(\ref{eq:socialconstraintestimate})  requires  all previous beliefs $l_{1:n-1}(i)$ that are specified by the non-zero elements of the vector $w_n$.
\end{itemize}
The only way to reconcile the above points is  to  ensure  that  $A_n(j,n) = 0$ implies $w_n(j) = 0$ for $j=1,\ldots, n-1$. This condition means that the single hop
past estimates $\lbelief_m, m \in \history_n$,
available at node $n$ according to~(\ref{eq:fuse}) in Protocol~\ref{protocol:socialcons}  provide all the information that is required to compute
$w_n^\p \, \lbelief_{1:n-1}$ in~(\ref{eq:socialconstraintestimate}).
  This is essentially a condition on the information flow graph $G_n$. We formalize this condition in
the following theorem.

\begin{theorem}[Achievability of Fair Rating]\label{thm:sufficient}
Consider the fair rating algorithm specified by~(\ref{eq:socialconstraintestimate}). For Protocol~\ref{protocol:socialcons}  with available information $(\belief_m, m \in \history_n)$, to achieve the estimates $\lbelief_{n-}$ of algorithm~(\ref{eq:socialconstraintestimate}),
a necessary and sufficient condition on the information flow graph $G_n$ is \beq \label{constraintnetwork}
A_n(j,n)=0   \implies w_n(j)= 0.
\eeq
Therefore, for Protocol~\ref{protocol:socialcons} to generate incest free estimates for nodes $n=1,2,\ldots$, condition  (\ref{constraintnetwork}) needs to hold for each $n$.
(Recall
  $w_n$ is specified in~(\ref{eq:socialconstraintestimate}).)
\end{theorem}

Note that the constraint~(\ref{constraintnetwork}) is purely in terms of the adjacency matrix $A_n$ since the transitive closure matrix~(\ref{eq:tc}) is a function of the adjacency matrix. Therefore, Algorithm~(\ref{eq:socialconstraintestimate}), together with condition~(\ref{constraintnetwork}), ensures that incest free estimates are generated
by Protocol~\ref{protocol:socialcons}.

\begin{example}
Let us continue with Example~\ref{example-1}, where we already specified the adjacency matrices of the graphs
$G_1$, $G_2$, $G_3$, $G_4$, and $G_5$.
Using~(\ref{eq:tc}), the transitive closure matrices $T_n$  \index{matrix!transitive closure matrix} obtained from the adjacency matrices are given by:
\begin{displaymath}
\begin{array}{l}
T_1 = \begin{bmatrix}
1
\end{bmatrix},
T_2 = \begin{bmatrix}
1 & 0 \\
0 & 1
\end{bmatrix},
T_3 = \begin{bmatrix}
1 & 0 & 1 \\
0 & 1 & 1  \\
0 & 0 & 1
\end{bmatrix}, \\
T_4 = \begin{bmatrix}
1 & 0 & 1 & 1  \\
0 & 1 & 1 & 1 \\
0 & 0 & 1 & 0 \\
0 & 0 & 0 & 1
\end{bmatrix},
T_5 = \begin{bmatrix}
1 & 0 & 1 & 1 & 1 \\
0 & 1 & 1 & 1 & 1 \\
0 & 0 & 1 & 0 & 1 \\
0 & 0 & 0 & 1 & 1 \\
0 & 0 & 0 & 0 & 1
\end{bmatrix}.
\end{array}
\end{displaymath}
Note that $T_n(i,j) $ is non-zero only for $i\geq j$ due to causality---information sent by an agent can only arrive at another agent at a later time instant. The weight vectors are then
obtained from~(\ref{eq:socialconstraintestimate}) as
\begin{displaymath}
\begin{array}{l}
w_2 = \begin{bmatrix}0\end{bmatrix},\\
w_3 = \begin{bmatrix}1 & 1\end{bmatrix}^\p,\\
w_4 = \begin{bmatrix}1 & 1 & 0\end{bmatrix}^\p,\\
w_5 = \begin{bmatrix}-1 & -1 & 1 & 1\end{bmatrix}^\p.\\
\end{array}
\end{displaymath}
Let us examine these weight vectors. $w_2$ means that node $2$ does not use the estimate from node $1$. This formula is consistent with the constrained information flow because estimate from node $1$ is not available to node $2$; see Figure~\ref{sample}.
$w_3$ means that node $3$ uses estimates from node $1$ and $2$. $w_4$ means
that  node $4$  uses estimates only from node $1$ and node $2$ since the estimate from node $3$ is not available at node~$4$. As shown in Figure~\ref{sample}, the mis-information propagation occurs at node~$5$. The vector $w_5$ says that node~$5$ adds estimates from nodes $3$ and $4$ and removes estimates from nodes $1$ and $2$ to avoid double counting of these estimates that are already integrated into estimates from node $3$ and $4$. Indeed, using the algorithm~(\ref{eq:socialconstraintestimate}), incest is completely prevented in this example.

Let us now illustrate an example in which exact incest removal is impossible.
Consider the information flow graph of  Figure~\ref{sample}, but with the edge between nodes $2$ and $5$  deleted. Then, $A_5(2,5) = 0$, while $w_5(2) \neq 0$; therefore, the condition~(\ref{constraintnetwork}) does not hold. Hence, exact incest  removal is not possible for this case.
\end{example}

\subsection{Summary}
In this section,
we have outlined a controlled sensing problem over a social network in which the administrator controls (removes) data incest,
and thereby maintains an unbiased (fair) online reputation system.  The state of nature could be geographical coordinates of an event (in a target localization problem) or quality of a social unit (in an online reputation system).  As discussed above, data incest arises  due to the recursive nature of Bayesian estimation and non-determinism in the timing of
the  sensing by individuals.
Details of proofs, extensions and further numerical studies are presented in~\cite{HK13,KH13}.

How useful  are Bayesian social learning models?
Humans often make {\em monotone} decisions - the more favorable the private  observation,
the higher the recommendation. In addition, humans  typically  convert numerical attributes to ordinal scales before making a decision. For example,
it does not make a difference if the cost of a meal at a restaurant is \$200 or \$205; an individual would classify this cost as ``high".
Also credit rating agencies use ordinal symbols such as AAA, AA, A.
 It is shown in \cite{Kri12} that if the cost $c(x,a)$ satisfies a single crossing condition\footnote{The single crossing condition can be viewed as an ordinal generalization
 of  supermodularity \cite{Ami05}. Supermodularity is a sufficient condition for a single crossing condition to hold.} and the observation likelihoods satisfy a totally positive condition, then
 the recommendation $a_n$ made by  node $n$  is monotone increasing in its observation
$y_n$ and ordinal. In this sense, even if an agent does not exactly follow a Bayesian social learning model, its monotone ordinal behavior implies that such
a model is a useful idealization. \index{ordinal decisions}

\section{Interactive Sensing for  Quickest  Change  Detection}  \label{sec:socialc}

In this section, we
consider interacting social sensors in the context of detecting a change in the underlying state of nature.
Suppose a multi-agent system  performs social learning and makes local decisions as described in~\S\ref{sec:classicalsocial}. Given the public beliefs from the social learning protocol, how can quickest  change detection be achieved?
In other words, how can a global decision maker use the local decisions from individual agents to decide when a change has occurred?
It is shown below that making a global decision (change or no change) based on local decisions of individual agents has an unusual structure resulting in a non-convex stopping set.

A typical application of such social sensors  arises in the  measurement of the adoption of a new product using a micro-blogging platform like Twitter.
The adoption of the technology diffuses through the market but its effects can only be observed through the tweets of select individuals of the population.
 These selected individuals act as  sensors for estimating the diffusion. They interact and learn from the decisions (tweeted sentiments) of  other members and,
 therefore, perform
 social learning.  Suppose the state of nature suddenly changes due to a sudden market shock or presence of a new competitor.
 The goal for a market analyst or product manufacturer is to detect this change as quickly as possible by minimizing a cost function that involves the sum of the  false alarm and decision delay.

 \paragraph*{Related works}
 The papers~\cite{Pin06,Pin08} model diffusion in networks over a random graph with arbitrary degree distribution. The resulting diffusion is approximated using deterministic dynamics via a mean-field approach
~\cite{BW03}. In the seminal paper~\cite{EP06}, a sensing system for complex social systems is presented with data collected
 from cell phones.  This data is used
  to recognize social patterns, identify socially significant locations, and infer relationships.
 In~\cite{SOM10}, people using a microblogging service such as Twitter are considered as  sensors.
 A particle filtering  algorithm is then used to estimate the  centre of
 earthquakes and trajectories of typhoons. As pointed out in~\cite{SOM10}, an important characteristic of microblogging services such as Twitter is that they
 provide  real-time  sensing---Twitter users tweet several times a day, whereas  standard blog users  update information  once every several days.

Apart from the above applications in real-time  sensing,  change detection in social learning
also arises in mathematical finance models.
For example, in agent based models  for the microstructure of asset prices in high frequency trading in financial systems~\cite{AS08}, the
 state denotes the underlying
asset value that changes at a random time $\tau^0$.
Agents observe local individual decisions of previous agents via an order book, combine these observed decisions with their noisy private signals about the asset, selfishly optimize their expected local utilities, and then make their own individual decisions (whether to buy, sell or do nothing).
  The market evolves through the orders of trading agents.
 Given this order book information, the goal of the market maker (global decision maker) is to achieve quickest change point detection of when a shock occurs to the value of the asset~\cite{KA12}.

\subsection{Classical Quickest Detection}
The classical Bayesian quickest time detection problem~\cite{PH08} is as follows:
An underlying discrete-time state process $x$ jump-changes at a geometrically distributed random time $\tau^0$.
Consider a sequence of discrete time random measurements $\{y_k,k \geq 1\}$ such that,
 conditioned on the event $\{\tau^0 = t\}$, $y_k$, $k \leq t$,  are independent and identically distributed (i.i.d.) random variables with distribution
$B_1$ and $y_k, k >t$, are i.i.d. random variables with distribution $B_2$.
The quickest  detection problem involves detecting the change time $\tau^0$ with minimal cost. That is,
at each time $k=1,2,\ldots$, a decision $u_k \in \{ 1 \text{ (stop and announce change)}, 2 \text{ (continue)} \}$ needs to be made to optimize a tradeoff
between false alarm frequency and linear delay penalty.

To formalize this setup, let
\begin{equation}
\tp = \begin{bmatrix} 1 & 0 \\ 1-\tp_{22} & \tp_{22} \end{bmatrix}
\end{equation}
denote the transition matrix of  a two state Markov chain $x$
in which state 1 is absorbing. Then, it is easily seen that the geometrically distributed change time $\tau^0$ is equivalent to  the time
at which the Markov chain enters state 1. That is,
\begin{displaymath}
\tau^0 = \min \{k: x_k = 1\}, \;\text{ and }\; \E\{\tau^0\} = 1/(1-\tp_{22}).
\end{displaymath}
Let $\tau$ be the time at which the decision  $u_k = 1$ (announce change) is
made. The goal of quickest time detection is to minimize the  Kolmogorov--Shiryaev
criterion for detection of  a disorder~\cite{Shi63}:
\beq J_\mu(\pizero) =   d \,\Ep\left\{(\tau - \tau^0)^+\right\} +  f \Pp\left(\tau < \tau^0\right) .
\label{eq:ksd} \eeq
Here, $x^+ = x$ if $x\geq  0$, and $0$ otherwise. The non-negative constants $d$ and $f$ denote  the delay  and false alarm penalties, respectively.
Therefore, waiting too long to announce a change incurs a delay penalty $d$ at each time instant after the system has changed, while declaring
a change before it happens incurs a false alarm penalty $f$.
In~(\ref{eq:ksd}), $\mu$ denotes the  strategy of  the decision maker.  $\Pp$ and $\Ep$ are the probability measure and expectation
of the evolution of the observations and Markov state which are strategy dependent. Finally, $\pi_0$ denotes the initial distribution of the Markov chain $x$.

In  classical quickest detection, the  decision policy $\mu$  is a function of the two-dimensional
belief  state (posterior probability mass function)  $\pi_k(i) = \P(x_k = i | y_1,\ldots,y_k,u_1,\ldots,u_{k-1})$,
$i=1,2$, with  $\pi_k(1)+\pi_k(2) =1$.
It thus suffices to consider one element, say $\pi_k(2)$,  of this
probability mass function. Classical quickest  change detection (see for example~\cite{PH08}) says that the policy $\mu^*(\pi)$, which optimizes~(\ref{eq:ksd}), has the following  threshold structure:
There exists a threshold point $\pi^* \in [0,1]$ such that
\beq \mu^*(\pi_k) = \begin{cases} 2,  \text{ (continue) } & \text{ if }
\pi_k(2) \geq \pi^*, \\  1,  \text{ (announce change)  } &  \text{ if } \pi_k(2) < \pi^*.
\end{cases} \label{eq:onedim}
\eeq

\subsection{Multi-agent Quickest Detection Problem}
With the above classical formulation in mind, consider now the following multi-agent quickest change detection problem:
Suppose that a multi-agent system  performs
social learning  to estimate an underlying  state according to the social learning protocol of~\S\ref{sec:herd}.
 That is,
each agent acts once in a predetermined sequential order indexed by $k=1,2,\ldots$ (Equivalently, as pointed out in the discussion
in~\S\ref{sec:herd},  a finite number of agents act repeatedly in some pre-defined order and each
 agent chooses its local decision using the current public belief.)
  Given these local decisions (or equivalently the public belief),  the goal of the global decision maker is to minimize the quickest detection objective~(\ref{eq:ksd}).
The problem now is a
 non-trivial generalization of classical quickest detection. The posterior $\pi$ is now the public belief given by the social learning filter~(\ref{eq:piupdate})
 instead of a standard Bayesian filter. \index{nonlinear filtering!social learning filter}
There is now  interaction between  the local and global decision makers. The local decision $a_k$ from the social learning protocol determines the public belief state $\pi_k$ via the social learning filter~(\ref{eq:piupdate}), which determines
the global decision (stop or continue), which determines the local decision at the next time instant, and so on.

The global decision maker's policy $\mu^*:\pi \rightarrow \{1,2\}$ that optimizes the quickest detection objective~(\ref{eq:ksd}) and
the cost $J_{\mu^*}(\pi_0)$ of this optimal policy are the solution
of
 ``Bellman's dynamic programming  equation'':

{\small
\begin{equation} \label{eq:dp_alg}
\begin{split}
\mu^*(\pi)&= \argmin\lbr f \pi(2), \; d(1-\pi(2)) + \sum_{a \in \ASL}  V\left( T(\pi ,a) \right) \sigp(\pi,a)\rbr, \\
 V(\pi) &= \min \lbr  f \pi(2), \; d(1-\pi(2))  +  \sum_{a \in \ASL}  V\left( T(\pi,a) \right) \sigp(\pi,a)\rbr,\\
 J_{\mu^*}(\pi_0) &= V(\pi_0).
\end{split}
\end{equation}
}

\noindent
Here, $T(\pi,a)$  and $\sigma(\pi,a)$ are given by the social learning filter~(\ref{eq:piupdate})---recall   that $a$ denotes the local decision.
$V(\pi)$ is called the ``value function''---it is the cost incurred by the optimal policy when the initial belief state (prior) is $\pi$.
The above problem is more complex than a standard partially observed Markov decision process  (POMDP) \index{Partially Observed Markov Decision Process (POMDP)}  since the belief state update $T(\pi,a)$ now involves
the social learning filter \index{nonlinear filtering!social learning filter}  instead of the standard hidden Markov model filter. \index{nonlinear filtering!hidden Markov model filter}
In particular, as we will illustrate in the numerical example below, unlike in POMDPs, $V(\pi)$ is no longer concave. Also
the optimal policy $\mu^*(\pi)$ has a very different structure compared to the classical quickest detection.

\begin{example}
\label{sec:numerical}
We  now illustrate the unusual multi-threshold property of the global decision maker's optimal policy $\mu^*(\pi)$ in multi-agent quickest detection
with social learning.

Consider the social learning model of~\S\ref{sec:herd}  with the following parameters:
The underlying state is a 2-state Markov chain $x$ with state space $\X= \{1,2\}$ and transition probability matrix
\begin{displaymath}
\tp =  \begin{bmatrix} 1 &  0 \\  0.05 & 0.95 \end{bmatrix}.
\end{displaymath}
Therefore, the change time $\tau^0$  (i.e., the time the Markov chain jumps  from state 2 into absorbing state 1) is geometrically distributed with $E\{\tau^0\} = 1/0.05 = 20$.

{\em Social Learning Parameters}: Individual agents observe the Markov chain $x$ in noise with the observation symbol set $\Y=\{1,2\}$.  Suppose the observation
likelihood matrix with elements  $B_{iy} = \P(y_k=y | x_k = i)$ is
\begin{displaymath}
B = \begin{bmatrix}  0.9  & 0.1  \\  0.1 & 0.9 \end{bmatrix}.
\end{displaymath}
 Agents can choose their local actions $a$ from the action set $\ASL=\{1 ,2 \}$.
The state-dependent cost matrix of these actions is
\begin{displaymath}
c= (c(i,a), i\in X, a\in \ASL) = \begin{bmatrix}
4.57 & 5.57 \\ 2.57 & 0
 \end{bmatrix}.
 \end{displaymath}
 Agents perform social learning with the above parameters.
  The intervals $[0,\pi_1^*]$ and $[\pi_2^*,1] $ in Figure~\ref{fig:redgreen}a are regions where the optimal local actions taken by agents are independent of their observations.
For $\pi(2) \in  [\pi_2^*,1] $, the optimal local action is 2 and,  for $\pi(2) \in [0,\pi_1^*]$, the optimal local action is 1.
Therefore, individual agents herd for belief states in these intervals (see the definition in~\S\ref{sec:cascade})
and the local actions
 do not yield any information about the underlying state.
Moreover, the interval  $[0,\pi_1^*]$ depicts a region where all agents herd\footnote{Note that even if the agent $k$ herds so that its  action $a_k$
provides no information about its private observation $y_k$, the public belief still evolves according
to the predictor $\pi_{k+1} = \tp^\p \pi_k$. Therefore, an information cascade does not occur in this example.} (again see  the definition in~\S\ref{sec:cascade}),  meaning that once the belief state is in this region, it remains so
indefinitely and all agents choose the same local action~1.

{\em Global Decision Making}: Based on the local actions of the agents performing social learning, the global decision maker needs to perform quickest  change detection.
The global decision maker uses the
  delay penalty $d=1.05$ and false alarm penalty  $f=3$ in the objective function (\ref{eq:ksd}).
The optimal policy $\mu^*(\pi)$ of the global decision maker where $\pi = [1-\pi(2), \pi(2)]^\p$ is plotted versus $\pi(2)$  in Figure~\ref{fig:redgreen}a.
Note $\pi(2) = 1$ means that no change has occurred with certainty, while $\pi(2) = 0$ means a change has occurred with certainty.
The policy $\mu^*(\pi)$
was computed by constructing
a uniform grid of 1000 points for $\pi(2)\in [0,1]$ and then implementing the dynamic programming equation
(\ref{eq:dp_alg}) via a fixed point value  iteration algorithm
for 200 iterations.
The horizontal axis $\pi(2)$ is the posterior probability
of no change.
The vertical axis denotes the optimal decision:   $u=1$ denotes stop and declare change, while
 $u=2$ denotes continue.

\begin{figure}[t]
\begin{center}
\mbox{\subfloat[Optimal global decision policy $\mu^*(\pi)$]
{\includegraphics[scale=0.32]{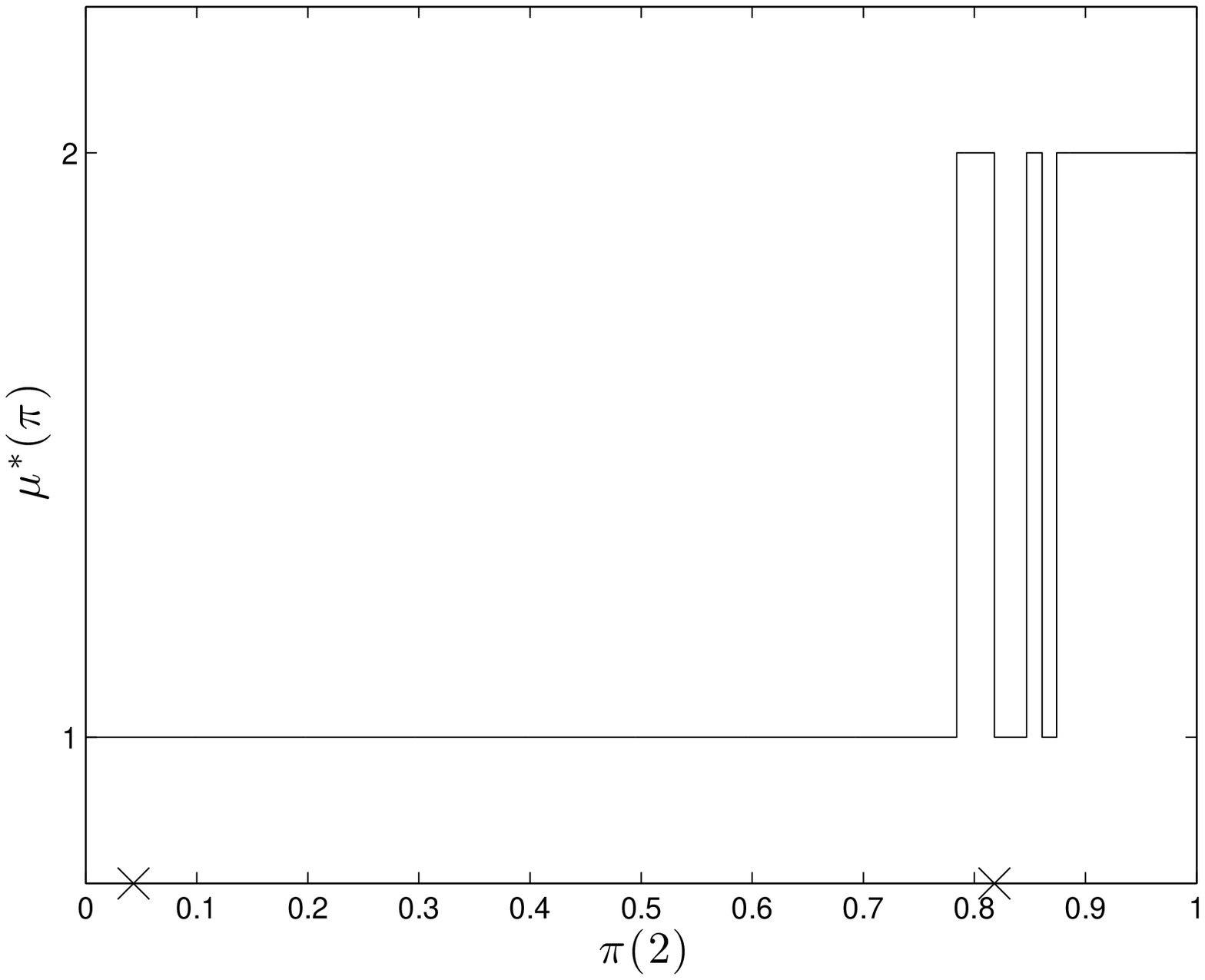}} \quad
\subfloat[Value function for  global decision policy]
{\includegraphics[scale=0.32]{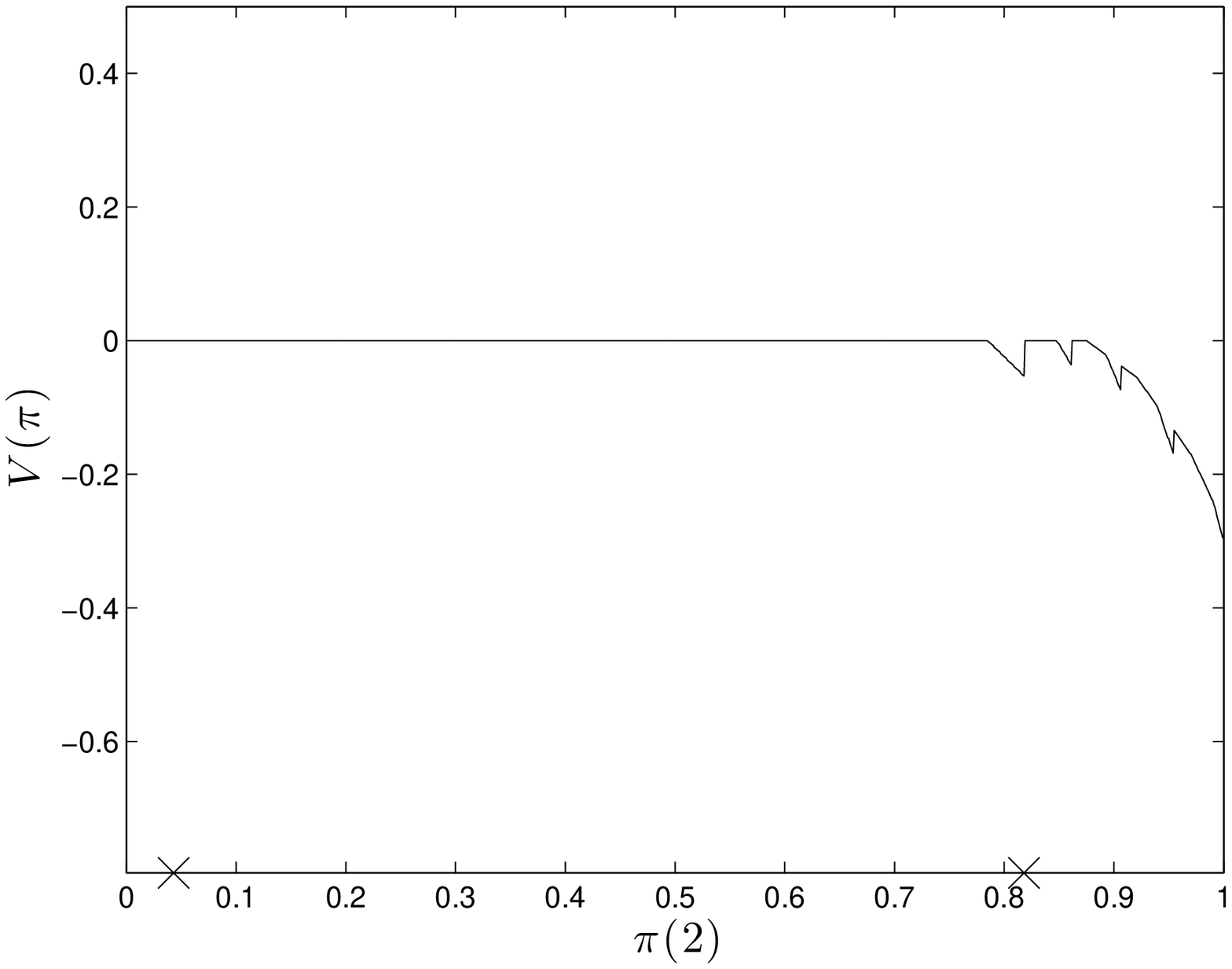}}   }
\end{center}
\caption{Optimal global decision policy for social learning based quickest time change detection
for geometric distributed  change time.  The parameters are specified in~\S\ref{sec:numerical}.  The optimal policy $\mu^*(\pi)\in \{1 \text{ (announce change) }, 2 \text{ (continue) }\}$ is characterized by a triple threshold---that is, it switches from 1 to 2 three times as the posterior $\pi(2)$ increases.
As explained in the text, for $\pi(2) \in [0,\pi_1^*]$, all agents herd, while for $\pi(2) \in [\pi_2^*,1]$ individual agents herd.
}
\label{fig:redgreen}
\end{figure}

 The most remarkable feature of Figure~\ref{fig:redgreen}a is  the multi-threshold behavior of  the
global decision maker's optimal policy $\mu^*(\pi)$. Recall $\pi(2)$ depicts the posterior probability of no change.
 Consider the region where  $\mu^*(\pi)  = 2$ and sandwiched between
 two regions where $\mu^*(\pi) = 1$.  Then, as  $\pi(2)$ (posterior probability of  no change) increases, the optimal policy
  switches from $\mu^*(\pi) = 2$ to $\mu^*(\pi) = 1$. In other words, the optimal global decision policy ``changes its mind''---it switches from no change to  change as  the posterior probability of a change decreases!
 Thus, the global decision (stop or continue) is a non-monotone function of the posterior probability obtained from local decisions.

Figure~\ref{fig:redgreen}b shows the associated value function obtained via stochastic dynamic programming~(\ref{eq:dp_alg}). Recall that the value function
$V(\pi)$  is the cost
incurred by the optimal policy with initial belief state $\pi$.
Unlike standard sequential detection problems, in which the value function is concave, the figure shows
that the value function is non-concave and
discontinuous.
To summarize, Figure~\ref{fig:redgreen}  shows
that  social learning based quickest detection results in fundamentally different decision policies compared to classical quickest time detection (which has a single threshold). Thus, making global decisions (stop or continue) based on local decisions (from social learning) is non-trivial.
In~\cite{Kri12}, a detailed analysis of the problem is given together with the characterization of this multi-threshold behavior. More general phase-distributed change times are further considered in~\cite{Kri12}.
\end{example}

\section{Closing Remarks}

In this chapter, we used social learning as a model for interactive sensing with social sensors.
We summarize here some  extensions of the social learning framework that are relevant to interactive sensing.
\paragraph{Wisdom of Crowds}

Surowiecki's  book~\cite{Sur05}  is an excellent popular piece that   \index{wisdom of crowds}
explains the wisdom-of-crowds hypothesis. The wisdom-of-crowds hypothesis predicts that the independent judgments of a crowd of individuals (as measured by any form of central tendency) will be relatively accurate, even when most of the individuals in the crowd are ignorant and error prone. The book also
studies situations (such as rational bubbles) in which  crowds are not wiser than individuals.
\begin{quote}{\em Collect enough people on a street corner staring at the sky, and everyone who walks past will look up.}\end{quote} Such herding behavior is typical in social learning.

\paragraph{In which order should agents act?}
In the social learning protocol, we assumed that the agents act sequentially in a pre-defined order.
However, in many social networking applications, it is important to optimize the order in which agents  act. For example,
consider an online review site where individual reviewers with different reputations  make their reviews publicly available.
If a reviewer with high reputation publishes its review first, this review will unduly affect the decision of a reviewer with lower reputation.
In other words, if the most senior agent ``speaks'' first, it will unduly affect the decisions of more junior  agents. This could lead to an increase in bias of the underlying state estimate\footnote{To quote a recent paper from Haas School of Business, U.C. Berkeley~\cite{AK09}: ``In  94\% of cases, groups (of people) used the first answer provided as their final answer... Groups tended to commit to the first answer provided by
any group member.''  People with dominant personalities tend to speak first and most forcefully ``even when they actually lack competence''.}.
On the other hand, if the most junior agent is polled first, since its variance is large, several agents will need to be polled in order
to reduce the variance. We refer the reader to~\cite{OS01} for an interesting description of who should speak first in a public debate\footnote{As described
in~\cite{OS01}, seniority is considered in the rules of debate and voting in the U.S.\ Supreme Court. ``In the past, a vote was taken after the newest
justice to the Court spoke. With the justices voting in order of ascending seniority largely, it was said, to avoid the pressure from long-term members
of the Court on their junior colleagues.''}.
It turns out that, for two agents, the seniority rule is always optimal for any prior---that is, the  senior agent speaks first followed
by the junior agent; see~\cite{OS01} for the proof. However, for more than two agents, the optimal order
depends on the prior and the observations in general.

\paragraph{Global Games for Coordinating Sensing}  In the classical Bayesian  social learning model of~\S\ref{sec:classicalsocial}, agents act sequentially in time. The global games model that has been studied
in economics during the last two decades, considers multiple agents that act simultaneously by predicting the behavior
of other agents.
The theory of global games was first introduced in~\cite{CD93} as a tool for refining equilibria in economic game theory;
see~\cite{MS00} for an excellent exposition.
Global games  \index{game!global game}
are an ideal method for decentralized coordination amongst agents. They
have  been used to model speculative currency attacks and regime change in social systems; see~\cite{AHP07,KLM07,MS00}.

 The most widely studied form of a global game is a one-shot Bayesian game which proceeds as follows: Consider a  continuum of agents  in which each agent $i$ obtains noisy measurements $y^i$
of  an underlying state of nature $x$. Assume all agents have the same
observation likelihood  density $p(y|x)$; however, the individual measurements obtained by agents
are statistically independent of those obtained by other agents. Based on its observation $y^i$, each agent
takes an action $a^i \in \{1, 2\}$ to optimize its expected utility $\E\{U(a^i, \alpha) | y^i\}$, where
$\alpha \in [0,1]$ denotes the fraction of all agents that take action 2.  Typically, the utility $U(1,\alpha)$ is set to zero.

For example,  suppose $x$  (state of nature) denotes the quality of a social group and  $y^i$ denotes the measurement of this quality by agent $i$.  The action $a^i = 1$  means that agent $i$ decides not to  join the social group, while  $a^i = 2$ means that agent $i$ joins the group.
The utility function $U(a^i=2,\alpha)$ for joining the social group depends on $\alpha$, where $\alpha$ is the fraction of people that decide to join the  group.
In~\cite{KLM07}, the utility function is chosen as follows:
If $\alpha \approx 1$, i.e.,  too many people join the group, then  the utility to each agent is small since the group is too congested and agents do not receive sufficient individual service.
On the other hand, if $\alpha \approx 0$, i.e.,  too few people join the group, then the utility  is also small since there is not enough social interaction.

Since each agent is  rational, it  uses its  observation $y^i$ to predict $\alpha$, i.e., the fraction of  other agents
that choose action 2. The main question is then: What is the optimal strategy for each agent $i$ to maximize its expected utility?

It has been shown that, for
a variety of
  measurement noise models (observation likelihoods $p(y|x)$) and  utility functions $U$, the
 symmetric Bayesian
Nash equilibrium of the global game is unique and
has a threshold structure in the observation. This means that, given its observation $y^i$, it is optimal for each agent $i$ to choose
 its actions as follows:
 \begin{equation} \label{eq:thresbne}
  a^i = \begin{cases} 1, &  \textmd{if } y^i < y^* \\ 2,  &  \textmd{if } y^i \geq y^* \end{cases}
 \end{equation}
where the threshold $y^*$  depends on the prior, noise distribution, and utility function.

In the above example of joining a social group, this result means that, if agent $i$ receives a   measurement $y^i$ of the quality of the group and $y^i$  exceeds a threshold $y^*$, then it should join.
This is yet another example of simple local behavior (act according to a threshold strategy) resulting in global sophisticated behavior (Bayesian Nash equilibrium).
As can be seen, global games provide a decentralized way of achieving coordination amongst  social sensors.
In~\cite{AHP07}, the above one-shot Bayesian game is generalized to a dynamic (multi-stage)  game operating over a possibly infinite horizon. Such games facilitate modelling the dynamics of how people join, interact, and leave
social groups.

The papers~\cite{Kri08,Kri09} use global games to model networks of sensors and cognitive radios.
In~\cite{KLM07}, it has been shown that the above threshold structure~(\ref{eq:thresbne}) for the Bayesian Nash equilibrium breaks
down if the utility function $U(2,\alpha)$  decreases too rapidly due to congestion. The equilibrium structure becomes much more complex
and  can be described by the following  quotation~\cite{KLM07}:
\begin{quote} {\em ``Nobody goes there anymore. It's too crowded''} -- Yogi Berra \end{quote}

\chapter{Tracking Degree Distribution in Dynamic Social Networks}\label{Chapter:tracking}
\chaptermark{Tracking Degree Distribution}
\section{Introduction}

Social networks can be viewed as complex sensors that provide information about interacting individuals and an underlying state of nature\footnote{For example, real-time event detection from Twitter posts is investigated in~\cite{SOM10} or the early detection of contagious outbreaks via social networks is studied in~\cite{CFH10}.}. In this chapter, we consider a dynamic social network  where  at each time instant one node can join or leave the network. The probabilities of joining or leaving evolve according to the realization of a finite state Markov chain that represents the state of nature. This chapter presents two results.  First, motivated by social network applications, the asymptotic behavior of the degree distribution of the Markov-modulated random graph is analyzed. Second, using noisy observations of nodes' connectivity, a ``\textit{social sensor}'' \index{social sensor} is designed for tracking the underlying state of nature as it evolves over time.

\subsection{Motivation}
 \textbf{Why analyze the degree distribution?} The  degree distribution\index{degree distribution}  yields useful information about the connectivity  of the random graph~\cite{AB02,KRR01,N03a}. 
 The degree distribution can further be used to investigate the diffusion of information or disease through social networks \cite{Pin08,V07}. The existence of a ``giant component''\footnote{A giant component is a connected component with size $O(n)$, where $n$ is the total number of vertices in the graph.  If the average degree of a random graph is strictly greater than one, then there exists a unique giant component with probability one~\cite{CL06}, and the size of this component can be computed from the expected degree sequence.} \index{giant component} in complex networks can be studied using the degree distribution. The size and existence of a giant component has important implications in social networks in terms of modeling information propagation  and spread of human disease~\cite{EGA04,N02,NWS02}. The degree distribution is also used to analyze the ``\textit{searchability}'' \index{searchability} of a network. The ``\textit{search}'' problem  arises when a specific node in a network faces a problem (request) whose solution is at other node, namely, destination (e.g., delivering a letter to a specific person, or  finding a web page with specific information)~\cite{AA05,V07}. The searchability of a social network~\cite{V07} is  the average number of nodes that need to be accessed  to reach the destination. Degree distribution is also used to investigate the  robustness and  vulnerability of a network in terms of the network response to attacks on its nodes or links~\cite{CNS00,HKY02}. The papers~\cite{WK12,WKV13} further use degree-dependent tools for classification of social networks.

\textbf{Social sensors for tracking a Markovian target:} Tracking a time-varying parameter that evolves according to a finite-state Markov chain has several applications in target tracking~\cite{EKNS05}, change detection~\cite{BMP12}, multi-user detection in wireless systems~\cite{YK05b}, and economics~\cite{KR98}. In this chapter, we consider a dynamic social network where the interactions between nodes evolve over time according to a Markov process that undergoes infrequent jumps (the state of nature). An example of such social networks is the friendship network among residents of a city, where the dynamics of the network change in the event of a large festival. (Further examples are provided in Chapter~\ref{Chapter:diffusion}.) In this chapter, we introduce Markov-modulated random graphs to mimic social networks where the interactions among nodes evolve over time due to the underlying dynamics (state of nature). These social networks
can be used as a \textit{social sensor} for tracking the underlying state of nature. That is, using noisy measurements of the degree distribution of the network,
the jumps in the underlying state of nature can be tracked.

\subsection{Main Results}

\begin{figure}[t]
\begin{center}
\includegraphics[width=.95\textwidth]{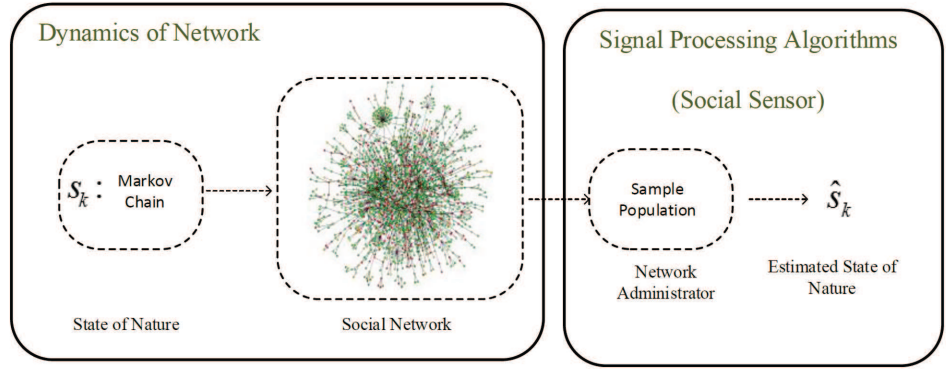}
\end{center}
\caption{Tracking the underlying state of nature using a Markov-modulated random graph as a social sensor.}
\label{Fig:SocialSensor}
\end{figure}

 Markov-modulated random graphs are introduced in~\S\ref{sec:dynamic-graph}. We then provide a degree distribution analysis for such graph
in~\S\ref{CH1:Result1} that allows us to determine the relation between the structure of the network (in terms of connectivity) and the underlying state of nature. Indeed, it will be shown in~\S\ref{CH1:Result1} that there exists a unique stationary degree distribution for the Markov-modulated graph for each state of the underlying Markov chain. It thus suffices to estimate the degree distribution in order to track the underlying state of nature. Next, a stochastic approximation algorithm is proposed to track the empirical degree distribution of the Markov-modulated random graph.
In particular, we address the following two questions in~\S\ref{CH1:Result2}:
\begin{itemize}
\item \textit{How can a social sensor estimate (track) the empirical degree distribution using a stochastic approximation algorithm with no knowledge of the Markovian dynamics?}
\item \textit{How accurate are the estimates generated by the stochastic approximation algorithm when the random graph evolves according to the duplication-deletion model with Markovian switching?}
\end{itemize}
By tracking the degree distribution of a Markov-modulated random graph, we can design a social sensor to track the underlying state of nature \index{state of nature} using the noisy measurements of nodes' connectivity; see Figure~\ref{Fig:SocialSensor}.

\subsection{Related Works}
For the background and fundamentals on social and economic networks, we refer to~\cite{J08}. The reader is also referred to~\cite{KY03} for a comprehensive development of stochastic approximation algorithms. Adaptive algorithms and stochastic approximations are used in a wide range of applications such as system identification, control theory, adaptive filtering, state estimation, and wireless communications~\cite{MMB12,Mou98,DM05}. Tracking capability of regime switching stochastic approximation algorithms is further investigated in~\cite{YKI04} in terms of the mean square error. Different applications of social sensors in detection and estimation are investigated in~\cite{SOM10,CFH10,AMO13}. The differences between social sensors, social sensing, and pervasive sensors along with challenges and open areas in social sensors are further presented in~\cite{RMZ11}.

Here, the related literature on dynamic social networks is reviewed briefly. The book~\cite{D07} provides a detailed exposition of random graphs. The dynamics of random graphs are investigated in the mathematics literature, for example, see~\cite{CL06,ER60,LHN05} and the reference therein.
In~\cite{PSS03}, a duplication model is proposed where at each time step a new node joins the network. However, the dynamics of this model do not evolve over time.  In~\cite{CL06}, it is shown that the degree distribution of such networks satisfies a \index{power law} \textit{power law}. In random graphs which satisfy the power law, the number of nodes with an specific degree depends on a parameter called ``power law exponent''~\cite{JMB01,W01}. A generalized Markov graph model for dynamic social networks along with its application in social network synthesis and classification  is also presented in~\cite{WKV13}.

\subsection*{Experimental Studies:}
\textit{1) Degree distribution analysis:} The degree distribution analysis of real-world networks has attracted much attentions recently,~\cite{AHK07,C12,EMB02,GN02,NWS02,N01,KW06}. A large network dataset collection can be found in~\cite{SNAP}, which includes datasets from social networks, web graphs, road networks, internet networks, citation networks, collaboration networks, and communication networks. The paper~\cite{N01} investigates the structure of scientific collaboration networks in terms of  degree distribution, existence of giant component, and the average degree of separation. In the scientific collaboration networks, two scientists are connected if they have co-authored a paper. Another example is the degree distribution of the affiliation network of actors.

\textit{2) Social Networks as Social Sensors:} With a large number of rational agents, social networks can be viewed as  social sensors \index{social sensor} for extracting information about the world or people. For example, the paper~\cite{SOM10} presents a social sensor (based on Twitter) for real-time event detection of earthquakes in Japan, namely, the target event. They perform semantic analysis on tweets (which are related to the target event) in order to detect the occurrence and the location of the target event (earthquake). Another example is the use of the social network as a sensor for early detection of contagious outbreaks~\cite{CFH10}. Using the fact that central individuals in a social network are likely to be infected sooner than others, a social sensor is designed for the early detection of contagious outbreaks in~\cite{CFH10}. The performance of this social sensor was verified during a flu outbreak at Harvard College in 2009---the experimental studies  showed that this social sensor provides a significant additional time to react to epidemics in the society.

\section{Markov-Modulated Random Graphs}
\label{sec:dynamic-graph}
To mimic dynamic social networks, we consider
Markov-modulated random graphs of the duplication-deletion type. Let $\tim=0,1,2,\ldots$ denote discrete time. Denote by $\mc_\tim$ a discrete-time Markov chain \index{Markov chain} with state space
 \begin{equation}\label{eq:mcstates}
\mathcal{\sizemc} = \{1,2,...,\sizemc\},
\end{equation}
evolving according to the $\sizemc \times \sizemc$ transition probability matrix
\beq \label{eq:A}
\A^{\emc}  = \identity + \emc \generatormatrix,
\eeq
 and initial probability distribution $\initialprob$.
 Here, $\identity$ is an $\sizemc\times \sizemc$ identity matrix, $\emc$ is a small positive real number, and $\generatormatrix$ is an irreducible generator of a continues-time Markov chain satisfying
\begin{equation}
q_{ij} >0,\; \textmd{ for } \; i\neq j, \;\textmd{ and }\; Q\mathbf{1} = \mathbf{0}
\end{equation}
where $\mathbf{1}$ and $\mathbf{0}$ represent column vectors of ones and zeros, respectively. The transition probability matrix  $\A^\emc$ is therefore close to identity matrix. Here and henceforth, we refer to such a Markov chain  $\mc_\tim$ as a ``slow'' Markov chain.

 A Markov-modulated duplication-deletion random graph is parameterized by the 7-tuple $(\sizemc,\A^{\emc},\initialprob,\pdupstep,\pdup,\pdel,\graph_0)$. Here, $\pdup $  and $\pdel$ are $\sizemc$-dimensional vectors with elements $\pdup(i)$ and $\pdel(i)  \in [0,1]$, $i=1,\ldots,\sizemc$, where $\pdup(i)$ denotes the connection probability, and $\pdel(i)$ denotes the deletion probability. Also, $\pdupstep \in [0,1]$ denotes the probability of duplication step, and $\graph_0$ denotes the  initial graph at time~$0$. In general, $\graph_0$ can be any finite simple connected graph. For simplicity, we assume that $\graph_0$  is a simple connected graph with size $\sizegraph_0$.

 The \index{graph!duplication-deletion graph}duplication-deletion random graph\footnote{For convenience in the analysis, assume that a node generated in the duplication step cannot be eliminated in the deletion step immediately after its generation---that is, the new node, which is generated in the vertex-duplication step of Step~2, remains unchanged in Step~3. Also, the nodes whose degrees are changed in the edge-deletion part of Step~3, remain unchanged in the duplication part of Step~3 at that time instant. To prevent the isolated nodes, assume that the neighbor of a node with degree one cannot be eliminated in the deletion step. The duplication step in Step 2
 ensures that the graph size does not decrease.} is constructed via Algorithm~\ref{alg:duplication}.

\begin{algorithm}
 At time $\tim$, given the graph $\graph_\tim$ and Markov chain state
$\mc_\tim$,  simulate the following events: \\
\textbf{Step 1: Duplication step}: With probability
$\pdupstep$ implement the following steps:
\begin{itemize}
\item Choose node $\parentnode$ from graph $\graph_\tim$ randomly with uniform
distribution.
\item  \textit{Vertex-duplication}: Generate a new node $\newnode$.
\item \textit{Edge-duplication}:
\begin{itemize}
\item Connect node $\parentnode$  to node $\newnode$. (A new edge between $\parentnode$ and $\newnode$ is
added to the graph.)
\item Connect each neighbor of node $\parentnode$  with probability
$\pdup(\mc_\tim)$  to  node $\newnode$. These connection events are statistically
independent.
\end{itemize}
\end{itemize}

\textbf{Step 2: Deletion Step}: With probability
$\pdel(\mc_\tim)$ implement the following steps: \begin{itemize}
 \item \textit{Edge-deletion}: Choose node $\delnode$ randomly from
$\graph_\tim$ with uniform distribution. Delete node $\delnode$ along with the connected edges in graph $\graph_\tim$.
 \item \textit{Duplication Step}:  Implement Step 1.

 \end{itemize}
 \textbf{Step 3}: Denote the resulting graph by $\graph_{\tim+1}$.\\Generate $\mc_{\tim+1}$ (Markov chain) using transition
matrix $\A^{\emc}$.\\
 \textbf{Step 4: Network Manager's Diagnostics}: The network manager
computes the estimates of the expected degree distribution.\\
 Set $\tim \rightarrow \tim+1$ and go to Step 1. \caption{Markov-modulated Duplication-deletion Graph parameterized by
$(\sizemc,\A^{\emc},\initialprob,\pdupstep,\pdup,\pdel,\graph_0)$}
 \label{alg:duplication}
\end{algorithm}

The \index{graph!Markov-modulated random graph} Markov-modulated random graph generated by Algorithm~\ref{alg:duplication} mimics social networks where the interactions between nodes  evolve over time due to the underlying dynamics (state of nature) such as seasonal variations (e.g., the high school friendship social network  evolving over time with different winter/summer dynamics). In such cases, the connection/deletion probabilities $\pdup,\pdel$ depend on the state of nature and evolve with time. Algorithm~\ref{alg:duplication} models these time variations as a finite state Markov chain $\mc_\tim$ with transition matrix $\A^{\emc}$.

\subsection*{Discussion:}
 The connection/deletion probabilities $\pdup,\pdel$ can be determined by the solution of a utility maximization problem. Let $\utilityfunc^{\rm join} : [0,1] \times \mathcal{\sizemc} \rightarrow \rr$  denote a utility function that gives payoff to an individual who considers to expand his neighbors in ``Edge-duplication step'' of Algorithm~\ref{alg:duplication}  as a function of $(\pdup,\mc)$. Similarly, let $\utilityfunc^{\rm leave} : [0,1] \times \mathcal{\sizemc} \rightarrow \rr$  denote a utility function that pays off to an individual who considers to leave the network in ``Deletion step'' of Algorithm~\ref{alg:duplication} as a function of $(\pdel,\mc)$. With the above utility functions, the probabilities of connection/deletion when the state of nature is $\mc$ can be viewed as the solutions of the following maximization problems:
\begin{equation}
\begin{split}
\pdup(\mc) &= \argmaxx_{\pdup}\lbr\utilityfunc^{\rm join}(\pdup,\mc)\rbr,\\
\pdel(\mc) &= \argmaxx_{\pdel}\lbr\utilityfunc^{\rm leave}(\pdel,\mc)\rbr.
\end{split}
\end{equation}
These utility functions can be interpreted in terms of mutual benefits and \index{privacy}privacy concerns. One example could be $\utilityfunc^{\rm join}(\pdup,\mc) = \ben^{\rm join}(p,\mc) - \costmc$, where $\ben^{\rm join}(\pdup,\mc)$ is the benefit one obtains by expanding his network with probability $\pdup$ when the underlying state of nature is $\mc$, and $\costmc$ is the cost incurred by sacrificing his ``privacy''. In this example, when an individual decides to leave the network, the utility he obtains will be $\utilityfunc^{\rm leave}(\pdel,\mc) = \ben^{\rm leave}(q,\costmc) - \costleave(\mc)$, where $\ben^{\rm leave}(\pdel,\costmc)$ is the benefit he earns by preserving  privacy  and $\costleave(\mc)$ is the benefit he loses by leaving the network when the underlying state of nature is $\mc$.

\section[{Degree Distribution Analysis of Dynamic Random Graphs}]{Degree Distribution Analysis of \\ Markov-modulated Random Graphs%
\sectionmark{Degree Distribution Analysis}}
\sectionmark{Degree Distribution Analysis}
\label{CH1:Result1}

This section presents degree distribution \index{degree distribution} analysis of the \textit{fixed size Markov-modulated duplication-deletion random graph} generated according to Algorithm~\ref{alg:duplication}.
As we will show shortly, the expected degree distribution of such  a graph depends on the underlying dynamics of the model that follow the state of nature. Therefore, the expected degree distribution of the graph generated can be used to track the state of nature. Therefore, the entire social network forms a social sensor. Before proceeding, let us introduce some notation. 

\paragraph{Notation.}
At each time $\tim$, let $\sizegraph_\tim$ denote the number of nodes of graph $\graph_\tim$. Also, let $f_\tim(i)$ denote the number of vertices of graph $\graph_\tim$ with degree $i$. Clearly, $\sum_{i\geq 1} f_\tim(i) = \sizegraph_\tim$. Define the ``empirical vertex degree distribution'' as
\beq \label{eq:sample}
g_\tim(i) = \frac{f_\tim(i)}{\sizegraph_\tim}, \quad \text{for } 1\leq i\leq \sizegraph_\tim.\eeq
 Note that $g_\tim(i)$  can be viewed as a probability mass function since $g_\tim(i) \geq 0$ and $\sum_i g_\tim(i) = 1$.
 Let  $\bg_\tim = \E\{g_\tim\}$ denote the ``expected vertex degree distribution,'' where $\g_\tim$ is the empirical degree distribution defined in~(\ref{eq:sample}).

Consider the sequence of  finite duplication-deletion random graphs~$\{\graph_\tim\}$, generated by Algorithm \ref{alg:duplication} with $\pdupstep = 0$. Clearly, the number of vertices in the graph
satisfies $\sizegraph_\tim  = \sizegraph_0$ for $ \tim = 1,2,\ldots$, i.e., size of the graph is fixed. The following assumption characterizes the Markovian dynamics of the state of the nature.
\begin{Assumption}
\label{assumption-maz}
The slow Markov chain  $\mc_\tim$ evolves according to the transition matrix \index{matrix!transition matrix}  $\A^{\emc} = \identity + \emc \generatormatrix$,  where $\emc$ is a small positive real number and  $\generatormatrix = [\generatorelement_{ij}]$ is an irreducible\footnote{The irreducibility assumption implies that there exists a unique
stationary distribution $\pi \in \mathbb{R}^{\sizemc
\times 1}$ for this Markov chain such that
$\pi' = \pi' \A^{\emc}$.
} generator matrix \index{matrix!generator matrix}satisfying
\begin{displaymath}
\generatorelement_{ij} \geq 0, \;\textmd{ if } \;i\neq j,\; \textmd{ and }\;  \sum_{j=1}^{\sizemc}\generatorelement_{ij} = 0,\;\forall i.
\end{displaymath}
The initial distribution  $\initialprob$ is further independent of $\emc$.
\end{Assumption}

Theorem~\ref{theo:mu} below asserts that the expected degree distribution \index{degree distribution!expected degree distribution} of the fixed size Markov-modulated duplication-deletion random graph satisfies a recursive equation. Using this recursive equation, one can solve for the expected degree distribution.
\begin{theorem}\label{theo:mu}
Consider the  fixed size Markov-modulated duplication-deletion random graph generated according to Algorithm~\ref{alg:duplication},
where $\A^\emc = \identity
+ \emc \generatormatrix$ and $\pdupstep =0$. Let $\bg^{\mc}_\tim =
\E\{\g_\tim|\mc_\tim = \mc\}$ denote the expected degree distribution of nodes when the state of the underlying Markov chain is $\mc_\tim = \mc$. Then,
$\bg^{\mc}_\tim$ satisfies the
following recursion
\beq
\label{eq:true1}
\bg^{\mc}_{\tim+1} = (\identity + \frac{1}{\sizegraph_0}\transition'(\mc))\bg^{\mc}_{\tim}
\eeq
where  $C^\prime$ denotes transpose of a matrix $C$, and $\transition(\mc)$
is a generator matrix\footnote{That is, each row adds to zero and each non-diagonal element of $\transition(\mc)$  is positive.} with elements $(\text{for $1\leq i,j\leq \sizegraph_0$}$):
\beq \label{eq:L}
      \transitionelem_{ji}(\mc) = \left\{\begin{array}{ll}
       0, & j < i-1, \\
      \pdel(\mc) \pdup(\mc)^{i-1} + \pdel(\mc) \big(1 + \pdup(\mc)(i -
1)\big), &j = i - 1, \\
       i \pdel(\mc)\pdup(\mc)^{i-1} (1 - \pdup(\mc)) -\pdel(\mc)\big(i + 2 +
\pdup(\mc)i \big),  &j = i,  \\
      \pdel(\mc) {{i+1}\choose{i-1}}\pdup(\mc)^{i-1}(1-\pdup(\mc))^{2} +
\pdel(\mc)(i+1), & j = i + 1, \\
     \pdel(\mc)  {{j}\choose{i-1}}\pdup(\mc)^{i-1}(1-\pdup(\mc))^{j-i+1}, &
j > i + 1.
          \end{array}\right.
       \eeq
\begin{proof}
The proof is presented in Appendix~\ref{ap:mu}.
\end{proof}
\end{theorem}\vspace{3mm}

Theorem~\ref{theo:mu} shows that evolution of the expected degree distribution in a fixed size Markov-modulated duplication-deletion random graph satisfies~(\ref{eq:true1}). One can rewrite~(\ref{eq:true1}) as
\beq \bg_{\tim+1}^\mc
= \ttrue^{\prime}(\mc)\bg_{\tim}^\mc \eeq where  $\label{eq:B}
\ttrue(\mc) = \identity + \frac{1}{\sizegraph_0} \transition(\mc).$
Since $\transition(\mc)$ is a generator matrix, $\ttrue(\mc)$ can
be considered as the transition matrix of a slow Markov chain. It is also straightforward to show that for each $\mc \in \mathcal{\sizemc}$, $\ttrue(\mc)$ is irreducible and aperiodic.
Hence, for each state of the Markov chain $\mc \in \mathcal{\sizemc}$, there exists a unique stationary distribution $\bg(\mc)$ such that \beq \label{eq:gbar} \bg(\mc) = \ttrue^{\prime}(\mc) \bg(\mc).\eeq
 Note that the underlying Markov chain $\lbr\mc_\tim\rbr$ depends on the small parameter $\emc$. The main idea is that, although $\mc_\tim$ is time-varying but it is piecewise constant (since $\emc$ is small parameter)---it changes slowly over time. Further, in light of~(\ref{eq:true1}), the evolution of $\bg_\tim^\mc$ depends on $\frac{1}{\sizegraph_0}$. Our assumption throughout this chapter is that $\emc \ll \frac{1}{\sizegraph_0}$. Therefore, the evolution of $\bg_\tim^\mc$ is faster than the evolution of $\mc_\tim$. That is, $\bg_\tim^\mc$ reaches its stationary distribution $\bg(\mc)$ before the state of $\mc_\tim$ changes. From~(\ref{eq:gbar}), the expected degree distribution of the fixed size Markov-modulated duplication-deletion random graph can be uniquely computed for each state of the underlying Markov chain $\mc_\tim = \mc$. This allows us to track the state of nature $\mc_\tim$ via estimating the expected degree distribution as will be shown in~\S\ref{CH1:Result2}.

\section[{Power Law Analysis of Infinite Duplication-deletion Graph}]{Case Study: Power Law Analysis of Infinite Graph%
\sectionmark{Power Law Analysis}}
\sectionmark{Power Law Analysis}
So far in this chapter, a degree distribution analysis is provided for the fixed size Markov-modulated random graph generated according to Algorithm~\ref{alg:duplication} with $\pdupstep = 0$. Motivated by applications
in social networks, this section extends this analysis to the \textit{infinite duplication-deletion random graphs without Markovian dynamics}. Here, we investigate the random graph generated according to Algorithm~\ref{alg:duplication} with $\pdupstep = 1$, and when there are no Markovian dynamics ($\sizemc = 1$). Since $\pdupstep = 1$ for  $\tim \geq 0$, $\graph_{\tim+1}$ has an extra vertex as compared to  $\graph_\tim$. In particular, since $\graph_0$ is an empty set, $\graph_\tim$ has
$\tim$ nodes, i.e., $\sizegraph_\tim = \tim$. In the rest of this section, employing the same approach as in the proof of Theorem~\ref{theo:mu}, it will be shown that the infinite duplication-deletion random graph without Markovian dynamics
satisfies a power law\index{power law}. An expression is further derived for the power law exponent\index{power law!power law exponent}.

Let us first define the power law:
\begin{definition}[Power Law]
\label{def1}
Consider the infinite duplication-deletion random graph without Markovian dynamics
 generated according to
Algorithm~\ref{alg:duplication}.
Let $\PLn_k$ denote the number of nodes of degree
$k$ in a random graph $\graph_\tim$. Then, $\graph_\tim$ satisfies a
power law distribution if $\PLn_k$ is proportional to $k^{-\beta}$ for a
fixed $\beta > 1$, i.e.,
\begin{displaymath}
\log \PLn_k = \alpha -\beta \log k
\end{displaymath}
where $\alpha$ is a constant. $\beta$ is called the {\em power law exponent}.
\end{definition}
 The power law distribution is satisfied in many networks such as WWW-graphs, peer-to-peer networks, phone call graphs, co-authorship graph and various massive online social networks (e.g. Yahoo, MSN, Facebook)~\cite{BR99,BRS01,CF03,FFF99,KKR99,SHH88,S01}. The power law exponent describes asymptotic degree distribution of networks from which characteristics of networks such as maximum degree, searchability, existence of giant component, and diameter of the graph can be investigated \cite{CL06,V07}. The following theorem
 states  that the graph generated according to Algorithm~\ref{alg:duplication} with $\pdupstep = 1$ and $\sizemc = 1$ satisfies a power law.
\begin{theorem}
\label{theo:pl}
    Consider the infinite random graph with Markovian dynamics $\graph_\tim$ obtained by Algorithm \ref{alg:duplication} with 7-tuple $(1,1,1,1,\pdup,\pdel,\graph_0)$ with the expected degree distribution $\bg_\tim$. As $\tim \rightarrow \infty$, $G_\tim$ satisfies  a power law. That is, \begin{equation}\lim_{n\to\infty} \log \bg_\tim(i) = \alpha -\beta \log i \end{equation} where the power law exponent, $\beta$, can be computed from
        \begin{align}\label{eq:pl}
(1+\pdel)( \pdup^{\beta - 1 }+\pdup\beta - \pdup ) =1 + \beta \pdel.
\end{align}
Here, $\pdup$ and $\pdel$ are the probabilities defined in duplication and deletion steps, respectively.
\begin{proof}
The detailed proof is provided in~\cite{HKY13}. Here, we only present an outline of the proof that comprises of two steps: (i) finding the power law exponent, and (ii) showing that the degree distribution converges to a power law with the computed compnent as $\tim \rightarrow \infty$. To find the power law exponent, we derive a recursive equation for the number of nodes with degree $i+1$ at time $\tim + 1 $, denoted by $\degree_{\tim+1}(i+1)$, in terms of the degrees of nodes in graph $\graph_\tim$. Then, rearranging this recursive equation yields an equation for the power law exponent. To prove that the degree
distribution satisfies a power law, we define a new parameter $h_\tim(i) = \frac{1}{\tim}\sum_{k=1}^{i}\E\{\degree_{\tim}(k)\}$ and show that $\lim_{\tim \rightarrow \infty} h_\tim(i) = \sum_{k=1}^iCk^{-\beta}$, where $\beta$ is the power law exponent computed in the first step.
\end{proof}
\end{theorem}
Theorem~\ref{theo:pl} asserts that the infinite duplication-deletion random graph without Markovian dynamics generated by Algorithm~\ref{alg:duplication} satisfies a power law and provides an expression for the power law exponent. The significance of this theorem is that it ensures, with use of one single parameter (the power law exponent), we can describe the degree distribution of large numbers of nodes in graphs that model social networks. The above result slightly extends~\cite{CLD03,PSS03}, where only a duplication model is considered. However,  a graph generated by any arbitrary pure duplication step may not satisfy the power law\footnote{Bebek et al. in \cite{BBC06} provide conditions on the dynamics of the duplication process such that the resulting graph satisfies a power law.}.
Theorem~\ref{theo:pl} allows us to explore characteristics (such as searchability, diffusion, and existence/size of the giant component) of large networks which can be modeled with the infinite duplication-deletion random graphs.

\begin{figure}[t]
\begin{center}
\includegraphics[width=\textwidth]{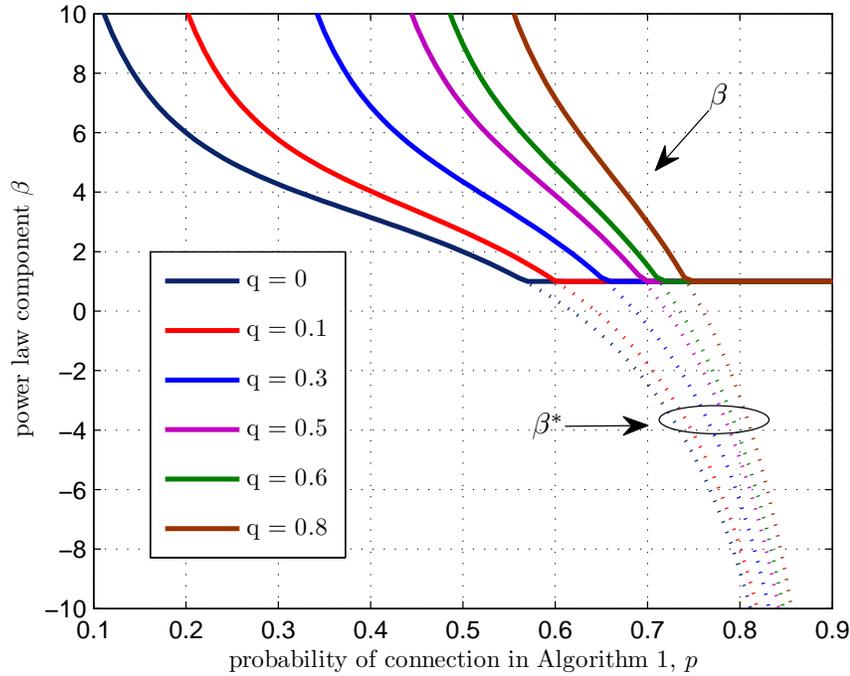}
\end{center}
\caption{The power law exponent for the non-Markovian random graph
generated according to Algorithm~\ref{alg:duplication} obtained by (\ref{eq:pl}) for
different values of $\pdup$ and $\pdel$  in Algorithm~\ref{alg:duplication}. }
\label{beta}
\end{figure}

\begin{remark}[Power Law Exponent]\index{power law!power law exponent}
Let $\beta^*$ denote the solution of (\ref{eq:pl}). Then, the power law exponent is defined as  $\beta =\max\{1,\beta^*\}$. Figure~\ref{beta} shows the power law exponent and $\beta^*$  versus $\pdup$ for different values of probability of deletion $\pdel$. As can be seen in Figure~\ref{beta}, the power law exponent is increasing in $\pdel$ and decreasing in $\pdup$.
\end{remark}

\section{Social Network as a Social Sensor}\label{CH1:Result2}
In~\S\ref{CH1:Result1}, a degree distribution analysis has been provided for the fixed size Markov-modulated duplication-deletion random graph generated by Algorithm~\ref{alg:duplication}.   
It was further shown in~\S\ref{CH1:Result1} that there exists a unique  stationary degree distribution for each state of the underlying Markov chain $\lbr\mc_\tim\rbr$ which represents state of nature. In this section, we assume that the empirical degree distribution of the graph, $\g_\tim$, is observed in noise. Motivated by social sensor applications, the aim here is to track the unknown state of nature---without knowing the dynamics of the graph---using noisy observations of degree distribution. Since the expected degree distribution of the Markov-modulated duplication-deletion random graph depends on the underlying Markov chain, the state of nature can be estimated via tracking the expected degree distribution; see~(\ref{eq:L}).  Here, the social network  is used to track the underlying state of nature. Therefore, the entire social network forms a social sensor\index{social sensor}.

The degree distribution of the network generated according to Algorithm~\ref{alg:duplication}
is measured in noise as follows
\beq \hat{\degree}_\tim = \degree_\tim + \noisee_\tim.   \eeq
Here, at each time $\tim$, the elements $\noisee_\tim(i)$  of the noise
vector are integer-valued  zero mean random variables, and
$\sum_{i\geq1} \noisee_\tim(i) = 0$. The zero sum assumption ensures that
$\hat{\degree}_\tim$ is a valid empirical distribution. The observation process can be viewed as counting the number of nodes with specific degree by the administrator of the network. In terms of the empirical vertex distribution, one can rewrite this
measurement process as
$$\obs_\tim(i) = \frac{\hat{\degree}_\tim(i)}{\sum_{i\geq 0}
\hat{\degree}_\tim(i)} =  \frac{\hat{\degree}_\tim(i)}{\sizegraph_0} =
\g_\tim(i) + \frac{1}{\sizegraph_0}\noisee_\tim(i). $$
That is, the vertex distribution $\g_\tim$ of the graph
$\graph_\tim$
is measured in noise:
\beq \obs_\tim = \g_\tim + \noise_\tim  \eeq where $\noise_\tim = \frac{\noisee_\tim}{\sizegraph_0}$. Recall that  $\sizegraph_\tim = \sizegraph_0$ when $\pdupstep = 0$. The normalized noisy observations $\obs_\tim$ are used to estimate the empirical probability mass function of the degree of each node. To estimate  a time-varying probability mass function (pmf), the following stochastic approximation algorithm with (small positive) constant step-size $\esa$
is used
\begin{equation}\label{eq15}
\hg_{\tim+1} = \hg_\tim +\esa\left(\obs_\tim - \hg_\tim\right).
\end{equation}
 Note that the stochastic approximation algorithm (\ref{eq15}) does not
assume any knowledge of the Markov-modulated dynamics of the graph (state of nature).  The
Markov chain assumption for the random graph dynamics is only used in our
convergence and tracking analysis. By means of the stochastic approximation (\ref{eq15}), the social sensor can track the expected degree distribution and, consequently, the underlying state of nature.\\

\begin{example} Consider the scenario that the underlying Markov chain (state of nature) has $\sizemc =2$ states with slow transition matrix $\A^\emc = \identity + \emc \generatormatrix$.  
Assume that the size of the graph $\sizegraph_0$ is sufficiently large such that the evolution of $\bg_\tim^\mc$ is faster than the evolution
of $\mc_\tim$. This means that the expected degree distribution $\bg_\tim^\mc$ reaches its stationary distribution ($\bg(\mc)$) before the state of $\mc_\tim$ changes. The stationary distribution in this example, which can be computed from~(\ref{eq:true1}), can be either $\bg(1)$ (if the state of nature is $\mc = 1$) or $\bg(2)$ (if the state of nature is $\mc = 2$). Assume that there exist a network administrator who has access to noisy measurements of  nodes' degrees $\obs_\tim$ (e.g., by counting the number of nodes with specific degrees). Then, by use of the stochastic approximation algorithm~(\ref{eq15}), the network administrator is able to precisely track the expected degree distribution. From the estimated degree distribution, the network administrator can estimate the current state of nature. In this example, the social network is used to estimate the unknown state of nature from noisy observations recorded from the network, therefore, can be viewed as a social sensor. In the next subsection, we show that the distance from the expected degree distribution and the estimates obtained by\index{stochastic approximation} stochastic approximation algorithm~(\ref{eq15}) is bounded.
\end{example}

 \subsection{Tracking Error of the Stochastic Approximation Algorithm}\label{subsec:bound}
 The goal here is to analyze how well the algorithm tracks the empirical degree distribution of the graph (and consequently the state of nature). To this end, we analyze the asymptotic behavior of the estimated degree distribution.
Define the tracking error as $\tg_\tim = \hg_\tim -\bg(\mc_\tim)$.
Theorem~\ref{theo2} below shows that the difference between the sample path and the expected probability mass function is small---implying that the stochastic approximation algorithm can successfully track the Markov-modulated node distribution given the noisy measurements. We again emphasize that no knowledge of the Markov chain parameters are required in the algorithm. It also finds the order of this difference in terms of $\esa$ and $\emc$.

\begin{theorem}
\label{theo2}
Consider the random graph $(\sizemc, \A^{\emc},\initialprob,\pdup,\pdel,\pdupstep,
\graph_0)$. Suppose\footnote{In this chapter, we assume that $\emc = O(\esa)$. Therefore, $\emc^2 = o(\esa)$.} that $\emc^2 = o(\esa)$.
Then, for sufficiently large $\tim$, the tracking error of the stochastic
approximation algorithm~(\ref{eq15}) satisfies
\begin{equation}
\mathbf{E}|\tg_\tim|^2 = O\left(\esa+\emc +\frac{\emc^2}{\esa}\right).
\end{equation}
\begin{proof}
The proof uses the perturbed Lyapunov function method and is provided in Appendix~\ref{ap:bound}.
\end{proof}
\end{theorem}
As a corollary of  Theorem \ref{theo2}, we obtain the following mean square error convergence\index{mean square error} result.
\begin{corollary} Under the conditions of Theorem \ref{theo2}, if $\emc = O(\esa)$,

$$ \mathbf{E} |\tg_\tim|^2 = O(\esa).$$
Therefore,
$$\limsup_{\e\to 0} \mathbf{E} |\tg_\tim|^2 = 0.$$
\end{corollary}

\subsection{Limit System Characterization}\label{subsec:ode}
The following theorem asserts that the sequence of estimates generated by the stochastic approximation algorithm (\ref{eq15}) follows the dynamics of a Markov-modulated ordinary differential equation\index{ordinary differential equation} (ODE).

Before proceeding with the main theorem below, let us recall a definition.
\begin{definition}[Weak Convergence]\index{weak convergence}
\label{def:weak-convergence}
Let $Z_k$ and $Z$ be ${\mathbb R}^r$-valued random vectors. We say $Z_k$ \emph{converges weakly} to $Z$ ($Z_k \Rightarrow Z$) if for any bounded and continuous function $f(\cdot)$, $Ef(Z_k)\to Ef(Z)$ as $k\to \infty$.
\end{definition}

Weak convergence is a generalization of convergence in distribution to a function space\footnote{We refer the interested reader to~\cite[Chapter~7]{KY03} for further details on weak convergence and related matters. Appendix \ref{ap:conv-pf} contains a brief outline.}.

\begin{theorem}\label{theo3}
Consider the Markov-modulated random graph generated according to Algorithm~\ref{alg:duplication}, and the sequence of estimates $\{\hg_\tim\}$, generated by the stochastic approximation algorithm~(\ref{eq15}). Suppose Assumption~\ref{assumption-maz} holds and $\emc = O(\esa)$. Define the continuous-time interpolated process
\begin{equation}
\hg^\esa(\ct) = \hg_\tim, \  \mc^\esa(\ct) = \mc_\tim
 \ \text{ for }\ct\in[\tim\esa,(\tim+1)\esa).
\end{equation}
 Then, as $\esa\rightarrow 0$,
$(\hg^\esa(\cdot),\mc^\esa(\cdot))$ converges weakly to
$(\hg(\cdot),\mc(\cdot))$, where $\mc(\cdot)$ is a continuous-time Markov chain
with generator $\generatormatrix$, $\hg(\cdot)$ satisfies the Markov-modulated ODE
 \begin{equation}
 \label{diff2}
 \frac{d\hg(\ct)}{d\ct} = -\hg(\ct) + \bg(\mc(\ct)), \quad\hg(0) = \hg_0
 \end{equation}
 and $\bg(\mc)$ is defined in (\ref{eq:gbar}).

\end{theorem}

The above theorem asserts that the limit system associated with the stochastic approximation algorithm~(\ref{eq15}) is a  Markovian switched ODE~(\ref{diff2}).  As mentioned in~\S\ref{CH1:Result1}, this is unusual since typically in the averaging theory analysis of stochastic approximation algorithms, convergence occurs to a deterministic ODE. The intuition behind this somewhat unusual result is that the Markov chain evolves on the same time-scale as the stochastic approximation algorithm.
 If the Markov chain evolved on a faster time-scale, then the
limiting dynamics would indeed be a deterministic ODE weighed by the
stationary distribution of the Markov chain. If the Markov chain evolved
slower than the dynamics of the stochastic approximation algorithm, then
the asymptotic behavior would also be a deterministic ODE with the Markov
chain being a constant.

\subsection{Scaled Tracking Error}\label{subsec:er}
Next, we study the behavior of the scaled tracking error\index{tracking error}
between the estimates generated by the stochastic approximation algorithm~(\ref{eq15}) and the expected degree distribution.  The following theorem states that the tracking error should also satisfy a switching diffusion equation and provides a functional central limit theorem for this scaled tracking error. Let $\ser_k = \frac{\hg_k -\mathbf{E}\{\bg(\mc_k)\}}{\sqrt{\esa}}$ denote the scaled tracking error.

\begin{theorem}\label{theo4}
Suppose Assumption~\ref{assumption-maz} holds.
Define
$\ser^\esa(t) = \ser_k$ for $t \in [k\esa, (k+1)\esa)$.
Then, $(\nu^\e\cd,\theta^\e\cd)$ converges
weakly to $(\nu\cd,\theta\cd)$ such that
$\nu\cd$ is the solution of the following Markovian switched diffusion
process
\begin{equation}\label{de}
\ser(t) = -\int_0^t \ser(s) ds + \int_0^t
\cov^{\frac{1}{2}}(\mc(\tau))d\omega(\tau).
\end{equation}
Here, $\omega(\cdot)$ is an $\rr^{\sizegraph_0}$-dimensional standard Brownian motion.
The covariance matrix
$\cov(\mc)$ in~(\ref{de}) can be explicitly computed as
\beq\label{eq:cov3}
\cov(\mc) = Z(\mc)'D(\mc) + D(\mc)Z(\mc) - D(\mc) - \bg(\mc)\bg'(\mc).
\eeq
Here, $D(\mc) = \diag(\bg(\mc,1),\ldots,\bg(\mc, \s))$ and $Z(\mc) =
\left(\identity - \ttrue(\mc) +\mathbf{1}\bg'(\mc)\right)^{-1}$, where
$\ttrue(\mc_\tim)$ and $\bg(\mc)$ are defined in (\ref{eq:B})
and~(\ref{eq:gbar}), respectively.
\end{theorem}

For general switching processes, we refer to \cite{YZ10}. In fact, more
complex continuous-state dependent switching rather than Markovian switching\index{Markov chain!Markovian switching} are considered there.
 Equation~(\ref{eq:cov3}) reveals that the covariance matrix\index{matrix!covariance matrix} of the tracking
error depends on $\ttrue(\mc)$ and $\bg(\mc)$ and, consequently,
on the parameters $\pdup$ and $\pdel$ of the random graph. Recall from \S\ref{CH1:Result1} that
$\ttrue(\mc)$ is the transition matrix of the Markov chain which
models the evolution of the expected degree distribution in Markov
modulated random graphs and can be computed from Theorem~\ref{theo:mu}.
The covariance of the tracking error, which can be explicitly computed from (\ref{eq:cov3}), is useful for computational purposes.

\section{A Note on Degree-based Graph Construction} \label{sec:graphsimulate}

 \index{graph!graph construction}The first step in numerical studies of social networks is the graphical modeling of such networks.  A graph can be uniquely determined by the adjacency matrix\index{matrix!adjacency matrix} (also known as the connectivity matrix) of the graph. However, in the graphical modeling of social networks (specially when the size of the network is relatively large), the only available information is the degree sequence\index{degree sequence} of nodes, and not the adjacency matrix of the network.

 \begin{definition}
 The degree sequence, denoted by $\degreeseq$,  is a non-increasing sequence comprising of the vertex degrees of the graph vertices.
 \end{definition}

The degree sequence, in general, does not specify the graph uniquely; there can be a large number of graphs that realize a given degree sequence. It is straightforward  to show that not all integer sequences represent a true degree sequence of a graph. For example, sequence $\degreeseq = \{2,1,1\}$ represents a tree with two edges, but $\degreeseq = \{3,2,1\}$ cannot be realized as the degree sequence of a simple graph. Motivated by social network applications, this section addresses the following two questions given a degree sequence $\degreeseq$:
\begin{itemize}
\item \it{Existence Problem: Is there any simple graph that realizes $\degreeseq$?}
\item \it{Construction Problem: How can we construct all simple graphs that realize a true degree sequence $\degreeseq$? }
\end{itemize}
There are two well-known results that address the existence problem: (i) the Erd\"{o}s-Gallai theorem~\cite{EG60} and the Havel-Hakimi theorem~\cite{H55, H62}. These theorems provide necessary and sufficient conditions for a sequence of non-negative integers to be a true degree sequence of a simple graph. Here, we recall these results without proofs.
\begin{theorem}[Erd\"{o}s-Gallai\index{Erd\"{o}s-Gallai theorem}, \cite{EG60}]
Let $\degreeGC_1 \geq \degreeGC_2 \geq \cdots \geq \degreeGC_n > 0$ be integers. Then, the degree sequence $\degreeseq = \{\degreeGC_1,\cdots, \degreeGC_n\}$ is graphical if and only if
\begin{enumerate}
\item $\sum_{i=1}^\GCnode\degreeGC_i$ is even;
\item for all $ 1 \leq k < \GCnode$:
\end{enumerate}
\vspace{-0.4cm}
\beq
\label{eq:GC1}
\sum_{i=1}^k\degreeGC_i \leq k(k-1) + \sum_{i=k+1}^{\GCnode} \min{\{k,\degreeGC_i\}}.
\eeq
\end{theorem}
It is shown in \cite{TV03} that there is no need to check (\ref{eq:GC1}) for all $1 \leq k \leq \GCnode -1$; it suffices to check (\ref{eq:GC1}) for  $1 \leq k \leq s$, where $s$ is chosen such that $\degreeGC_s \geq s$ and $\degreeGC_{s+1} < s +1$. Note that, in degree-based graph construction, we only care about nodes of degree greater than zero; zero-degree nodes are isolated nodes which can be added to the graph consisting of nodes of degree greater than zero.

The Havel-Hakimi theorem\index{Havel-Hakimi theorem} also provides necessary and sufficient conditions for a degree sequence to be graphical.
It also gives a  greedy algorithm to construct a graph from a given graphical degree sequence.

\begin{theorem}[Havel-Hakimi, \cite{H55,H62}]
Let $\degreeGC_1 \geq \degreeGC_2 \geq \cdots \geq \degreeGC_\GCnode > 0$ be integers. Then, the degree sequence $\degreeseq = \{\degreeGC_1,\cdots, \degreeGC_\GCnode\}$ is graphical if and only if the degree sequence $\degreeseq' =\{\degreeGC_2-1, \degreeGC_3-1,\cdots, \degreeGC_{\degreeGC_1+1}-1,  \degreeGC_{\degreeGC_1+2}, \cdots,  \degreeGC_{\GCnode}\}$ is graphical.
\end{theorem}

In the following, we provide algorithms to construct a simple graph from a true degree sequence. In the construction problem, the degree sequence is treated as a collection of \textit{half-edges}\index{half-edge}; a node with degree $\degreeGC_i$ has $\degreeGC_i$ half-edges. One end of these half-edges are fixed at node $i$, but the other ends are free. An edge between node $i$ and node $j$ is formed by connecting a half-edge from node $i$ to a half-edge from node $j$. The aim is to connect all these half edges such that no free half-edge is left. The Havel-Hakimi theorem provides a recursive procedure to construct a graph from a graphical degree sequence. This procedure is presented in Algorithm~\ref{alg:GC}
\begin{algorithm}[t]
Given a graphical sequence $\degreeGC_1 \geq \degreeGC_2 \geq \cdots \geq \degreeGC_\GCnode > 0$:\\
Start from $\GCnodeinitial = 1$
\begin{itemize}
\item [(i)] Initialize $\GCnodefinal = \GCnode$.
\item[ (ii)]  Connect (one half-edge of) node $\GCnodeinitial$ to (a half-edge of) node $\GCnodefinal$
\item[ (iii)] Check that the resulting degree sequence is graphical \begin{itemize}
\item \textbf{if Yes:}
\begin{enumerate}
\item Let $\GCnodefinal = \GCnodefinal - 1$.
\item Repeat (i).
\end{enumerate}
\item \textbf{if No:}
\begin{enumerate}
\item Save the connection between node $\GCnodeinitial$ and node $\GCnodefinal$
\item If node $\GCnodeinitial$ has any half-edges left, let $\GCnodefinal = \GCnodefinal-1$ and repeat~(i)
\end{enumerate}
\end{itemize}
\item[(iv)] If $\GCnodeinitial < \GCnode $, then, $\GCnodeinitial \leftarrow \GCnodeinitial + 1$  and repeat~(i).
\end{itemize}
\caption{Creating a sample graph from a given degree sequence}
 \label{alg:GC}
\end{algorithm}

Using Algorithm~\ref{alg:GC}, one can  sample from graphical realizations of a given degree sequence. In this algorithm, each vertex  is first connected to nodes with lower degrees. Therefore, Algorithm~\ref{alg:GC} generates graphs where high-degree nodes tend to connect to low-degree nodes; the resulting graph has assortative property~\cite{KTE09,N02}. To overcome this problem, one way is to perform \textit{edge swapping} repeatedly such that the final graph looses its assortative property. In the edge swapping method, two edges (for example (1,2) and (3,4)) can be swapped (to (1,4) and (2,3)) without changing the degree sequence. Edge swapping method is also used to generate all samples from a given degree sequence; one sample is generated via Algorithm~\ref{alg:GC} and then, by use of Markov chain Monte-Carlo algorithm based on edge swapping \cite{T81}, other samples from the graphical realizations of the degree sequence are obtained.

In \cite{KTE09} a swap-free algorithm is proposed to generate all graphical realizations of a true degree sequence. Before proceeding to Algorithm~\ref{alg:GCC}, we first provide definitions which will be used in this algorithm.
\begin{definition}
Let $\degreeseq = \{\degreeGC_1,\cdots, \degreeGC_\GCnode\}$ be a degree sequence of a simple graph and $\GCneighbor(\GCnodeinitial)$ be the set of adjacent nodes of node $i$. Then, the degree sequence reduced by $\GCneighbor(\GCnodeinitial)$ is denoted by $\degreeseq|_{\GCneighbor(\GCnodeinitial)} = \{\degreeGC_1|_{\GCneighbor(\GCnodeinitial)},\cdots, \degreeGC_n|_{\GCneighbor(\GCnodeinitial)}\}$ with elements defined as follows
\begin{align}
\degreeGC_k|_{\GCneighbor(\GCnodeinitial)}=\Bigg\{\begin{array}{l} \degreeGC_k - 1, \quad \text{if $k \in \GCneighbor(\GCnodeinitial)$}, \\0, \quad\hspace{.9cm}\text{if $k =\GCnodeinitial$},\\\degreeGC_k, \hspace{.8cm}\quad \text{otherwise.}
\end{array}
\end{align}
\end{definition}
\begin{definition} \label{def:cr}
Let $(a_1,a_2,\ldots,a_n)$ and $(b_1,b_2,\ldots,b_n)$ be two sequences. Then, $(a_1,a_2,\ldots,a_n) <_{CR} (b_1,b_2,\ldots,b_n)$ if and only if there exists an index $m$ such that $1 \leq m \leq n$ and $a_m < b_m$ and $a_i = b_i$ for all $m < i \leq n$.
\end{definition}

\begin{algorithm}[htbp]
Given a graphical sequence $\degreeGC_1 \geq \degreeGC_2 \geq \cdots \geq \degreeGC_\GCnode > 0$\\
Start from $\GCnodeinitial = 1$\\
\textbf{Step 1: Find neighbors with highes index of node $\GCnodeinitial$} \\
The aim is to find $A_R(\GCnodeinitial)$:\vspace{-3mm}
\begin{itemize}
\item [(i)] Initialize $\GCnodefinal = \GCnode$.
\item[(ii)]  Connect node $\GCnodeinitial$ to node $\GCnodefinal$
\item[(iii)] Check that the resulting degree sequence is graphical \begin{itemize}
\item \textbf{if Yes:}
\begin{enumerate}
\item Let $\GCnodefinal = \GCnodefinal - 1$
\item Repeat (i).
\end{enumerate}
\item \textbf{if No:}
\begin{enumerate}
\item Save the connection between node $\GCnodeinitial$ and node $\GCnodefinal$
\item If node $\GCnodeinitial$ has any stubs left, let $\GCnodefinal = \GCnodefinal-1$ and repeat (i)
\end{enumerate}
\end{itemize}
\end{itemize}
\vspace{-2mm}
\textbf{Step 2: Find all possible neighbors of node $\GCnodeinitial$}\\
With $<_{CR}$ defined in (\ref{def:cr}),
the aim is to find \\$\mathcal{A}(\GCnodeinitial) = \{ \GCneighbor(\GCnodeinitial) = \{v_1, \cdots, v_{\degreeGC_\GCnodeinitial}\}; \GCneighbor(\GCnodeinitial) <_{CR} A_R(\GCnodeinitial)$ and $\degreeseq|_{\GCneighbor(\GCnodeinitial)}$ is graphical$\}$
\begin{itemize}
\item[(i)] Find all sets of nodes who are colexicographically smaller than $A_R(1)$ (prospective neighbor sets).
\item[(ii)] Connect node $\GCnodeinitial$ to those neighbors and check if the resulting degree sequence is graphical.
\end{itemize}
\textbf{Step 3:} For every $\GCneighbor(\GCnodeinitial) \in \mathcal{A}(\GCnodeinitial)$:
\begin{itemize}
\item  Connect node $\GCnodeinitial$ to $\GCneighbor(\GCnodeinitial)$
\item Discard node $\GCnodeinitial$ and compute the reduced degree sequence $\degreeseq|_{\GCneighbor(\GCnodeinitial)}$
\item Create all graphs from degree sequence $\degreeseq|_{\GCneighbor(\GCnodeinitial)}$ using this algorithm
\end{itemize}
\caption{Constructing all graphs from a graphical degree sequence \cite{KTE09}}
 \label{alg:GCC}
\end{algorithm}

Let $\degreeseq$ be a non-increasing graphical degree sequence. In order to construct the graph, we need to find all possible neighbors $\GCneighbor(i)$ (``allowed set'') of each node  $i$ such that if we connect this node to its allowed set, then the resulting reduced degree sequence $\degreeseq|_{\GCneighbor(i)}$ is also graphical, i.e.,  the graphicality is preserved. Algorithm~\ref{alg:GCC} provides a systematic way (swap-free) to generate all graphical realizations of a true degree sequence (by means of finding all possible neighbors of each node).

\section{Numerical Examples}\label{sec:num}
In this section, numerical examples are given to illustrate the results
from~\S\ref{CH1:Result1} and~\S\ref{CH1:Result2}.
The main conclusions are:
\begin{enumerate}
\item The infinite duplication-deletion random graph without Markovian dynamics generated by Algorithm~\ref{alg:duplication} satisfies
a power law as stated in Theorem~\ref{theo:pl}. This is illustrated in Example~\ref{ex1} below.
\item The degree distribution of the fixed size duplication-deletion random graph generated by Algorithm~\ref{alg:duplication} can be computed from Theorem~\ref{theo:mu}. This is shown in Example~\ref{ex2} below.
\item The estimates obtained by stochastic approximation algorithm~(\ref{eq15}) follow the expected probability distribution precisely
without information about the Markovian dynamics. This is illustrated in
Example~\ref{ex3} below.
\end{enumerate}

\begin{example}{\label{ex1}} Consider an infinite duplication-deletion random graph without Markovian dynamics generated by Algorithm~\ref{alg:duplication} with $\pdup = 0.5$ and $\pdel = 0.1$. Theorem~\ref{theo:pl} implies that the degree sequence of the resulting graph satisfies a power law with exponent computed using (\ref{eq:pl}). Figure~\ref{sim2} shows the number of nodes with specific degree on a logarithmic scale for both horizontal and vertical axes. It can be inferred from the linearity in  Figure~\ref{sim2} (excluding the nodes with very small degree), that the resulting graph from duplication-deletion process satisfies a power law. As can be seen in Figure~\ref{sim2}, the power law is a better approximation for the middle points compared to the both ends.
\begin{figure}[t]
\begin{center}
\includegraphics[width=0.9\textwidth]{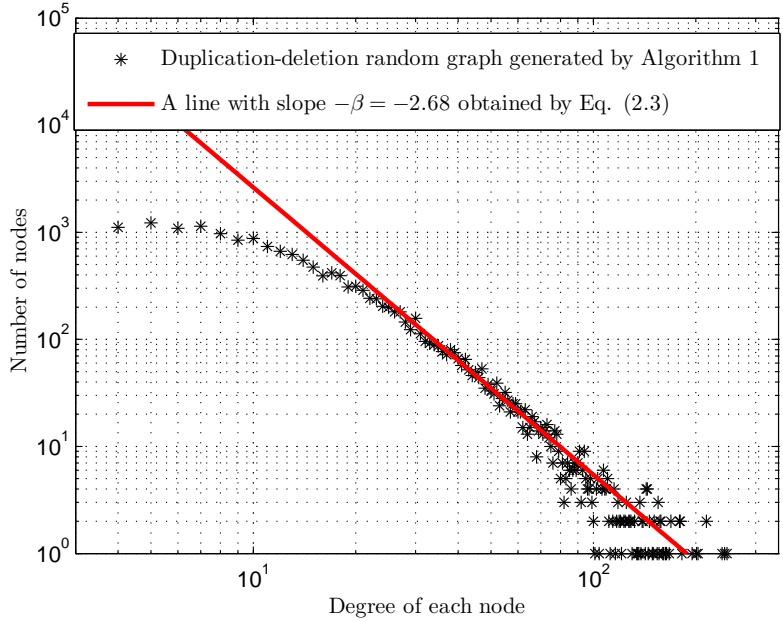}
\end{center}
\caption{The degree distribution
of the duplication-deletion random graph satisfies a power law. The
parameters are specified in Example~\ref{ex1} of \S\ref{sec:num}. }
\label{sim2}
\end{figure}
\end{example}
\begin{example}{\label{ex2}} Consider the fixed size duplication-deletion random graph obtained by Algorithm~\ref{alg:duplication} with $\pdupstep = 0$, $\sizegraph_0 = 10$, $p = 0.4$, and $q = 0.1$. (We consider no Markovian dynamics here to illustrate Theorem~\ref{theo:mu}.) Figure~\ref{mc} depicts the degree distribution of the fixed size duplication-deletion random graph obtained by Theorem~\ref{theo:mu}. As can be seen in Figure~\ref{mc}, the computed degree distribution is close to that obtained by simulation.

\begin{figure}[p]
\begin{center}
\includegraphics[width=0.73\textwidth]{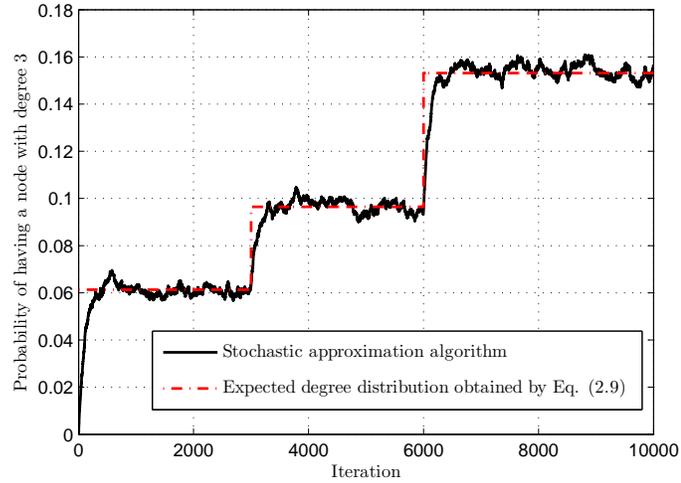}
\end{center}
\caption{The estimated degree distribution obtained by the stochastic approximation algorithm
(\ref{eq15}) and the expected degree distribution computed from (\ref{eq:gbar}).}
\label{samplepath}
\end{figure}
\begin{figure}[p]
\begin{center}
\includegraphics[width=0.73\textwidth]{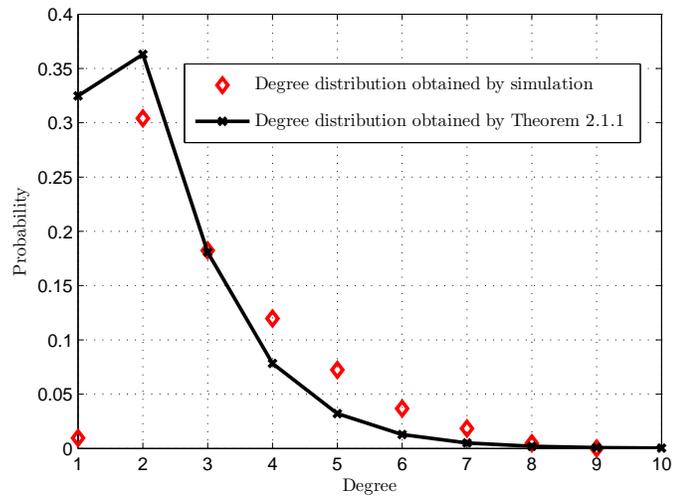}
\end{center}
\caption{The degree distribution
of the fixed size Markov-modulated duplication-deletion random graph. The parameters are specified in Example~\ref{ex2} of \S\ref{sec:num}.}
\label{mc}
\end{figure}


\end{example}
\begin{example}{\label{ex3}} Consider the fixed size Markov-modulated duplication-deletion random graph
generated by Algorithm \ref{alg:duplication} with $\pdupstep = 0$ and $\sizegraph_0 = 500$. Assume that the underlying Markov chain has three states,
$\sizemc=3$. We choose the following values for probabilities of connection and
deletion: state (1): $p = q = 0.05$, state (2): $p = 0.2$ and $q = 0.1$,
and state (3): $p = 0.4$, $q = 0.15$.  The sample path of the Markov chain jumps at time $\tim=3000$ from state (1) to state (2) and at time $\tim=6000$ from state (2) to state (3). As the state of
the Markov chain changes, the expected degree distribution, $\bg(\mc)$,
obtained by (\ref{eq:gbar}) evolves over time. Only one element of
the expected degree distribution vector is shown in
Figure~\ref{samplepath} via a dotted line. The estimated probability mass
function, $\hg_\tim$, obtained by the stochastic approximation algorithm (\ref{eq15}) is plotted in
Figure~\ref{samplepath} using a solid line. The figure shows that the
estimates obtained by the stochastic approximation algorithm (\ref{eq15})
follow the expected degree distribution obtained by (\ref{eq:gbar})
precisely without any information about the Markovian dynamics.\end{example}

\section{Closing Remarks}

\subsection*{Summary}The interaction between nodes in dynamic social networks is not always fixed and may evolve over time. An example of such time-varying dynamics is the seasonal variations in friendship among college students. The Markov-modulated random graph generated by Algorithm~\ref{alg:duplication} mimics such networks where the dynamics (the connection/deletion probabilities $\pdup,\pdel$) depend on the state of nature and evolve over time. Algorithm~\ref{alg:duplication} models these time variations as a finite state Markov chain $\lbr\mc_\tim\rbr$.
This model forms our basis to analyze social networks.

We analyzed Markov-modulated duplication-deletion random graphs in terms of degree distribution. When the size of graph is fixed ($\pdupstep = 0$) and  $\emc$ is small,
 the expected degree distribution of the Markov-modulated duplication-deletion random graph can be uniquely computed from~(\ref{eq:true1}) for each state of the underlying Markov chain. This result allows us to express the structure of network (degree distribution) in terms of the dynamics of the model.

We also showed that, when the size of the graph is fixed and there is no Markovian dynamics ($\sizemc =1, \pdupstep = 1$), the random graph generated according to Algorithm~\ref{alg:duplication} satisfies a power law with exponent computed from~(\ref{eq:pl}). The importance of this result is that a single parameter (power law exponent) characterizes the structure of a possibly very large dynamic network.

We further used a stochastic approximation algorithm to estimate the empirical degree distribution of random graphs.
 The stochastic approximation algorithm~(\ref{eq15}) does not assume any knowledge of the Markov-modulated dynamics of the graph (state of nature). Since the expected degree distribution can be uniquely computed, a social sensor can be designed based on (\ref{eq15}) to track the state of nature using the noisy observations of nodes' degrees. Theorem~\ref{theo2}  showed that the tracking error of the stochastic approximation algorithm is small and is in order of $O(\esa)$.

Finally, a discussion on graph construction was provided for simulation purposes.

\subsection*{Extensions} Here, we discuss some possible extensions of the framework used in this chapter and avenues for future research which are relevant to social sensors.

In addition to degree distribution, there are some other measures that characterize large networks such as diameter, average path length, clustering, and centrality~\cite{J08}; for example, \cite{WKV13} uses degree distribution and clustering coefficient to classify social networks. An extension to this work is to use stochastic approximation algorithms and graph theoretic tools employed in this chapter for tracking other characteristics of large networks.

In this chapter, we used probabilistic sampling to obtain measurements from degree distribution. That is, some nodes are randomly chosen and enquired about the number of their neighbors. Another extension to this work is to employ Respondent-driven sampling (RDS)~\cite{Hec97,Hec02,Lee09} as an approach for sampling from hidden networks in the society.
To quote~\cite{Hec97},  \begin{quote}{\em ``hidden populations have two characteristics: first, no sampling frame exists so that the size and boundaries of the population are unknown and  second, there exist strong privacy concerns, because membership involves stigmatized or illegal behavior, leading  individuals to  refuse to cooperate or provide unreliable answers to protect their privacy.''}\end{quote} An example of hidden populations is the network of  active injection drug users. Algorithm~\ref{alg:duplication} mimics these hidden populations as well: A person usually becomes active injection drug user through another active drug user. Then, he expands his network by connecting to other drug users in this population.
An extension to this work is to employ RDS\footnote{RDS is described in  Chapter~\ref{Chapter:diffusion} of this monograph.} to record observations about nodes' connectivity in hidden populations in order to track the structure (degree distribution) of such populations.

\chapter[Sensing with Information Diffusion in Social Networks]{Sensing with Information Diffusion in Complex Social Networks}
\chaptermark{Sensing with Information Diffusion}
\label{Chapter:diffusion}

\section{Introduction}
This chapter considers how a behavior diffuses over a social network comprising of a  population of interacting agents and how an underlying stochastic state can be estimated based on sampling
the population. As described in~\cite{Pin08}, there is a wide range of social phenomena such as diffusion of technological innovations, cultural fads, and economic conventions~\cite{Cha04} where
individual decisions are influenced by the decisions of others. A large body of research on social networks has been devoted to the diffusion of information (e.g., ideas, behaviors, trends) \cite{Gra78}, and particularly on finding a set of target nodes so as to maximize the spread of a given product \cite{MR07,Che09}.

Consider a social network where the states of individual nodes evolve  over time as a probabilistic function of the states of their neighbors and an underlying
target process.
The evolution in the state of agents in the network can be viewed as diffusion of information in the network.
Such  Susceptible-Infected-Susceptible (SIS) models  for  diffusion of  information in social networks 
has been extensively studied in~\cite{J08,Pin06,Pin08,PV01,V07} to  model, for example, the adoption of a new technology in a consumer market.

In this chapter we consider two  extensions of the SIS model:
First,  the  states of individual nodes evolve  over time as a probabilistic function of the states of their neighbors {\em and} an underlying
target process. The underlying target  process can be viewed as the market conditions or competing  technologies that evolve with time and affect the information diffusion.
Second, the  nodes in the social network are sampled randomly to determine their state. (We will review two recent methods for sampling social networks, namely, social sampling and respondent-driven sampling.)
As the adoption of the new technology diffuses through the network, its effect is observed via sentiments  (such as tweets) of these selected members of the population.  \index{sentiment}
 These selected nodes act as social sensors.
In signal processing terms, the underlying target  process can be viewed as a signal, and the social network can be viewed as a sensor.
The key difference compared to classical signal processing is that the social network (sensor) has dynamics due to the information diffusion.

\subsection{Aim}
Our aim is to estimate the underlying target state and the state probabilities of the nodes by sampling  measurements at nodes in the social network.
In a Bayesian estimation context, this is equivalent to a  filtering problem involving estimation of the state of a prohibitively large scale Markov chain in noise. The key idea is to use {\em mean field dynamics} as  \index{mean field dynamics}
an approximation (with provable bounds)
for the information diffusion and, thereby, obtain a tractable model.
Such mean field dynamics\footnote{Recently, there has also been substantial interest in mean field games~\cite{HCM12}. These are outside the scope of the current monograph.} have been studied in~\cite{BW03} and applied to social networks in~\cite{Pin06,Pin08,V07}.
For an excellent recent  exposition of interacting particle systems comprising of agents each with a finite state space, see~\cite{Ald13}, where the more apt term
``Finite Markov Information Exchange (FMIE) process'' is used.

\subsection{Motivation}

A typical application of such social sensors  arises in the  measurement of the adoption of a new product using a micro-blogging platform like Twitter.
The adoption of the technology diffuses through the market but its effects can only be observed through the tweets of select individuals of the population.
 These selected individuals act as  sensors for estimating the
 diffusion. They interact and learn from the decisions (tweeted
 sentiments) of  other members.
(This is similar to social learning.)
  Suppose the state of nature changes  suddenly  due to a sudden market shock or presence of a new competitor.
 The goal for a market analyst or product manufacturer is to estimate
 the target state so as to detect the market shock or new competitor.
This is a Bayesian filtering problem\footnote{A more general approach (which
we will not elaborate on)  is to
formulate the problem as a
 Bayesian quickest detection  problem (or more generally a stopping
 time problem) that seeks to  minimize a cost
 function that involves the sum of the  false alarm and decision
 delay. Bayesian filtering is of course an integral part of such
 change detection.}.

To quote from Wikipedia:  \index{sentiment}
``Sentiment analysis refers to the use of natural language processing, text analysis and computational linguistics to identify and extract subjective information in source materials.''
The following excerpt from \cite{Ben11} illustrates the increasing importance of social sentiment with the growth of social networking:
\begin{itemize}
\item ``53\% of people on Twitter recommend companies and/or products in their Tweets, with 48\% of them delivering on their intention to buy the product.
(ROI Research for Performance, June 2010)
\item The average consumer mentions specific brands over 90 times per week in conversations with friends, family, and co-workers.  (Keller Fay, WOMMA, 2010)
\item Consumer reviews are significantly more trusted---nearly 12 times more---than descriptions that come from manufacturers, according to a survey of US mom Internet users by online video review site EXPO.
(eMarketer, February 2010)
\item In a study conducted by social networking site myYearbook, 81\% of respondents said they had received advice from friends and followers relating to a product purchase through a social site; 74\% of those who received such advice found it to be influential in their decision. (ClickZ, January 2010)
\end{itemize}

 As another example,  \cite{SOM10} considers each  Twitter user as a sensor and uses a particle filtering  algorithm to estimate the  centre of
 earthquakes and trajectories of typhoons. As pointed out in \cite{SOM10}, an important characteristic of microblogging services such as Twitter is that they
 provide  real-time  sensing---Twitter users tweet several times a day, whereas  standard blog users  update information  once every several days.

\subsection{Experimental Studies}
Here, we mention papers that  investigate the diffusion of information in real-world social networks such as Facebook, Twitter, and Blogs. Motivated by marketing applications, \cite{SRM09} studies the diffusion (contagion) behaviour in Facebook\footnote{In US in 2010, \$1.7 billion was spent on advertising through social media. The share of Facebook is 53\% in this market~\cite{RCG12}.}. Using data on around 260000 Facebook \textit{pages}\footnote{Facebook pages usually advertise  products, services, bands, celebrities, etc.}, \cite{SRM09} analyzes how information diffuses on Facebook. The topological and temporal characteristics of information diffusion on weblogs are also studied in~\cite{LMF07}. Twitter with around 200 million active users and 400 million tweets per day, has become a powerful information sharing tool~\cite{KLP10,TW}.

\section{Social Network Model}

A social network is modelled as a  graph with $N$ vertices:
\begin{equation}
\network = (\Vertexset,\edgeset),  \text{ where } \Vertexset= \vertexset, \text{ and } \edgeset  \subseteq \Vertexset \times \Vertexset.
\end{equation}
Here, $\Vertexset$ denotes the finite set of vertices, and $\edgeset $ denotes the set of edges.
 In social networks, it is customary to use the terminology {\em network}, {\em nodes} and {\em links} for {\em graph}, {\em vertices} and {\em edges}, respectively.

We use the notation $(\nodem,\noden)$ to refer to a link between node $\nodem$ and $\noden$. The network may be undirected in which case $ (\nodem,\noden) \in \edgeset$ implies $(\noden,\nodem) \in \edgeset$. In undirected graphs, to simplify notation, we use the notation $\nodem,\noden$ to denote the undirected link between node $\noden$ and $\nodem$. If the graph is directed, then $(\nodem,\noden) \in \edgeset$ does not imply that $(\noden,\nodem) \in \edgeset$. We will assume that self loops (reflexive links) of the form $i,i$  are excluded from $\edgeset$.

An important parameter  of a social network $\network = (\Vertexset,\edgeset)$ is the connectivity of its nodes.
Let $\nbhood{\nodem}$ and $\degreediff{\nodem}$ denote the neighbourhood set
and   degree (or connectivity)  of a node $\nodem \in \Vertexset$, respectively. That is,  
with $|\cdot|$ denoting cardinality,
\begin{equation}
\nbhood{\nodem} =  \{\noden \in \Vertexset :  \nodem,\noden \in \edgeset \},
\quad \degreediff{\nodem} =  \big| \nbhood{\nodem}\big|.
\end{equation}
For convenience, we assume that the maximum degree of the network is uniformly bounded by some fixed integer $\degmax$.

Let $\vertexnum{\deg}$ denote the number of nodes with degree $\deg$, and let the degree distribution
 $ \degdist{\deg}$ specify the fraction of nodes with degree $\deg$. That is, for $\deg=0,1,\ldots,\degmax$,
$$
\vertexnum{\deg} =  \sum_{\nodem \in \Vertexset}  \indicator\lbr\degreediff{\nodem} = \deg\rbr , \quad
\degdist{\deg} =  \frac{ \vertexnum{\deg} }{\vertexnum} .
 $$
Here, $\indicator\lbr\cdot\rbr$ denotes the indicator function. Note that
 $\sum_\deg \degdist{\deg}=1$. The degree distribution  can be viewed as the probability that a node selected  randomly with uniform distribution on $\Vertexset$ has a connectivity $\deg$.

 Random graphs generated to have a degree distribution $\degdist$ that is Poisson were formulated by Erd\"{o}s and Renyi~\cite{ER59}.
 Several recent works show that large scale social networks are characterized by connectivity distributions that are different to Poisson distributions.
 For example, the internet, www have a power law connectivity distribution $\degdist(\deg) \propto \deg^{-\gamma}$, where $\gamma$ ranges between 2 and 3.
Such scale free networks are studied in~\cite{BR99}. In the rest of this chapter, we assume that the degree distribution of the social network is arbitrary but known---allowing an arbitrary degree distribution facilities modelling complex networks.

\subsection{Target Dynamics}
Let $\dtime = 0,1,\ldots$ denote discrete time.
Assume the target process $\target$ is a finite state Markov chain with transition probability
\begin{equation}
\tptarget{\tstate}{\tstatep} = \P\left(\target{\dtime+1}=\tstatep  |  \target{\dtime} = \tstate \right).
\end{equation}
In the example of technology diffusion, the target process can denote the availability of competition or market forces that determine whether a node
adopts the technology.  In the model below, the target state will affect the  probability that an agent adopts the new technology.

\subsection{Diffusion of Information in Social Network}
The model we present below for the diffusion of information in the social network is called the Susceptible-Infected-Susceptible (SIS) model \cite{PV01,V07}.
The diffusion of information  is modelled by the time evolution of the state of individual nodes in the network.
Let  $\state{\dtime}{\nodem} \in \{0,1\}$ denote the state at time $\dtime$ of each node $\nodem$ in the social network.
Here, $\state{\dtime}{\nodem}= 0$ if the agent at time $\dtime$ is susceptible and $\state{\dtime}{\nodem} = 1$  if the agent is infected.
At time $\dtime$, the state vector of the $\vertexnum$ nodes is
\begin{equation}
\state{\dtime} = \lb\state{\dtime}{1} , \ldots, \state{\dtime}{\vertexnum}\rb^\p \in \{0,1\}^{\vertexnum}.
\end{equation}

Assume that the process $\statev$  evolves as a discrete time Markov process with transition
law depending on the target state $\target$.
If node $\nodem$ has degree $\degreediff{\nodem}=\deg$, then the probability of node $\nodem$ switching from state $\statea$ to $\stateb$ is
\begin{equation}
\begin{split}
&\P\left(\state{\dtime+1}{\nodem} =\stateb | \state{\dtime}{\nodem}= \statea, \state{\dtime}{i-},\target{\dtime}=\tstate\right) \\
&\hspace{0.7cm}= \tpdiff{\statea\stateb}{\deg}{\nactive{\nodem}_\dtime}{\tstate},\; \statea,\stateb \in \{0,1\}.
\label{eq:tp}
\end{split}
\end{equation}
Here, $\nactive{\nodem}_\dtime$ denotes the number of infected neighbors of node $\nodem$ at time $\dtime$. That is,
\begin{equation}
\nactive{\nodem}_\dtime = \sum_{\noden \in \neighbor{\nodem} } I\lbr\noden:  \state{\dtime}{\nodem} = 1\rbr.
\end{equation}
In words, the transition probability of an  agent depends on its degree distribution and the number of active neighbors.

With the above probabilistic model, we are interested in modelling the evolution of infected agents over time.
Let   $\infectdist{\dtime}{\deg}$  denote the fraction of infected nodes at each time $\dtime$ with degree $\deg$.
We call $\infectdist$ as the {\em infected   node distribution}.
So
\begin{equation}
\infectdist{\dtime}{\deg} =  \frac{1}{\vertexnum(\deg)} \sum_{\nodem \in \Vertexset} \indicator\lbr\degreediff{\nodem} = \deg,  \state{\dtime}{\nodem} = 1\rbr  , \quad \deg=0,1,\ldots,\degmax
\end{equation}
We assume that the infection spreads according to the following dynamics:
\begin{enumerate}
\item
At each time instant $\dtime$, a single agent, denoted by $\nodem$,  amongst the $\vertexnum$ agents is chosen uniformly.
%
Therefore, the probability that the chosen agent $\nodem$ is infected and of degree $\deg$  is $\infectdist{\dtime}{\deg}\, \degdist{\deg}$. The probability that the chosen agent $\nodem$ is  susceptible and of degree $\deg$ is $(1-\infectdist{\dtime}{\deg})\, \degdist{\deg}$.
\item
 Depending on whether its state $\state{\dtime}{\nodem}$ is infected
or susceptible, the
 state of agent $\nodem$ evolves according
to the transition probabilities  specified in~(\ref{eq:tp}).
\end{enumerate}

With the Markov chain transition dynamics of individual agents specified above, it is clear that the infected  distribution
$\infectdist{\dtime} =
 \big(\infectdist{\dtime}{1},\ldots, \infectdist{\dtime}{\degmax}\big)$  is an $\prod_{\deg=1}^{\degmax} \vertexnum{\deg}$ state Markov  chain. Indeed,
 given $\infectdist{\dtime}{\deg}$, due to the infection dynamics specified above
\begin{equation}
\infectdist{\dtime+1}{\deg} \in \left\{ \infectdist{\dtime}{\deg} - \frac{1}{\vertexnum{\deg} }, \;
 \infectdist{\dtime}{\deg} + \frac{1}{\vertexnum{\deg} }
\right\}.
\end{equation}
Our aim below is to specify the transition probabilities of the Markov chain $\infectdist$.
Let us start with the following statistic that forms a convenient parametrization of the transition probabilities. Given the infected node distribution $\infectdist{\dtime}$ at time $\dtime$,
define $\pa(\infectdist{\dtime})$ as the probability that at time $\dtime$ a uniformly sampled   link in the network points to an infected node. We call $\pa(\infectdist{\dtime})$ as the
{\em infected
link probability}. Clearly 
\begin{align}
\pa(\infectdist{\dtime})
&= \frac{ \sum_{\deg=1}^{\degmax}\text{(\# of links from infected node of degree $\deg$)}}
{\sum_{\deg=1}^{\degmax}\text{(\# of links  of degree $\deg$)}}  \nonumber\\
&=
 \frac{ \sum_{\deg=1}^{\degmax} \deg \, \degdist{\deg}\, \infectdist{\dtime}{\deg} } {\sum_{\deg}^{\degmax} \deg \, \degdist{\deg} }.  \label{eq:infectasymp}
\end{align}

In terms of the infected link probability $\pa$, we can now specify  the scaled transition probabilities\footnote{The transition probabilities are scaled by the degree distribution $\degdist{\deg}$ for notational convenience. Indeed, since $\vertexnum{\deg} = \vertexnum \degdist{\deg}$,
by using these scaled probabilities we can express the dynamics of the process $\infectdist$ in terms of the same-step size $1/\vertexnum$ as described
in Theorem~\ref{thm:minc}.
Throughout this chapter, we
assume that  the degree distribution $\degdist(\deg)$, $\deg \in \{1,2,\ldots,\degmax\}$, is uniformly bounded away from zero.  That is, $\min_\deg \degdist{\deg} > \epsilon$ for some positive
constant $\epsilon$.
} of the process
$\infectdist$:
\begin{align}
 \atp{01}{\deg}{\pa{\dtime}}{\tstate} 
&\ole \frac{1}{\degdist{\deg}}\,
\P\left(\infectdist{\dtime+1}{\deg} =  \infectdist{\dtime}{\deg} + \frac{1}{\vertexnum{\deg} } \Big\vert  \target{\dtime}=\tstate \right) \nonumber \\
&= (1-\infectdist{\dtime}{\deg})  \,  \sum_{\aindex = 0}^\deg \tpdiff{01}{\deg}{\aindex}{\tstate} \,  \P
(\text{$\aindex$ out of $l$ neighbors infected})
\nonumber\\
& = (1-\infectdist{\dtime}{\deg})  \,  \sum_{\aindex = 0}^\deg \tpdiff{01}{\deg}{\aindex}{\tstate} \binom{\deg}{\aindex}
\pa{\dtime}{\aindex}  (1 - \pa{\dtime})^{\deg -\aindex}
\end{align}

\begin{align}
 \atp{10}{\deg}{\pa{\dtime}}{\tstate}  &\ole \frac{1}{\degdist{\deg}}\,
\P\left(\infectdist{\dtime+1}{\deg} =  \infectdist{\dtime}{\deg} - \frac{1}{\vertexnum{\deg} } \Big\vert \target{\dtime} = \tstate\right) \nonumber \\ &=
\infectdist{\dtime}{\deg} \,\sum_{\aindex = 0}^\deg \tpdiff{10}{\deg}{\aindex}{\tstate}\,  \binom{\deg}{\aindex}
\pa{\dtime}{\aindex}  (1 - \pa{\dtime})^{\deg -\aindex}.
\end{align}
In the above, the notation  $\pa{\dtime}$ is the short form for $\pa(\infectdist{\dtime})$.
The  transition probabilities  $\atp{01}$ and $\atp{10}$ defined above model the diffusion of information about the target state $\target$ over the social network.
We have the following martingale representation theorem for the evolution of Markov process $\infectdist$.

Let $\sigmaf_\dtime$ denote the sigma algebra generated by $\{\infectdist_{0},\dots,\infectdist_{\dtime}, \target{0},\ldots\target{\dtime} \}$.

\begin{theorem} For $\deg=1,2,\ldots, \degmax$, the infected distributions evolve as

{\small
\begin{equation} \label{eq:minc}
\infectdist{\dtime+1}{\deg} = \infectdist{\dtime}{\deg} + \frac{1}{\vertexnum}
\left[
  \atp{01}{\deg}{\pa(\infectdist{\dtime})}{\target{\dtime}} -
   \atp{10}{\deg}{\pa(\infectdist{\dtime})}{\target{\dtime}}
    + \minc_{\dtime+1}  \right]
\end{equation}
}

\noindent
where $\minc$ is a martingale increment process, that is $\E\{\minc_{\dtime+1}| \sigmaf_\dtime \} = 0$. Recall $\target$ is the finite state Markov chain that models  the target process. \label{thm:minc}
\end{theorem}

The above theorem is a well-known martingale representation of a Markov chain \cite{EAM95}---it says that a discrete time Markov process can be obtained by discrete
time filtering of a martingale increment process.
The theorem implies that the infected distribution dynamics resemble what is commonly called a stochastic approximation (adaptive filtering) algorithm in statistical signal
processing: the new estimate is the old estimate plus a noisy update (the ``noise'' being a martingale increment)  that is weighed by a small step size $1/\vertexnum$ when $\vertexnum$ is large. Subsequently, we will exploit the structure in Theorem \ref{thm:minc} to devise a mean field dynamics  \index{mean field dynamics} model which has a state of dimension~$\degmax$.
This is to be compared with the intractable state dimension $\prod_{\deg=1}^{\degmax} \vertexnum{\deg}$ of the Markov chain $\infectdist$.

\begin{remark}[Data Incest]  \index{data incest}
In the above susceptible-infected-susceptible (SIS) model, the following sequence of events that is similar to a data incest event (discussed in \S\ref{sec:incest}) is possible:
 Node $\nodem$  infects its neighbor node $\noden$; then, node $\nodem$ switches in state from infected to susceptible; then, node $\noden$ re-infects node $\nodem$.
Therefore, misinformation can propagate in the network according to such loops.
An alternative model is the susceptible-infected-recovered (SIR) model where each node has three states: susceptible, infected, and recovered. If the recovered state
is made absorbing, then such incest events can be eliminated.
\end{remark}

\begin{example}
We discuss  examples of transition probabilities $\tpdiff{\statea\stateb}{\deg}{\nactive{\nodem}_\dtime}{\tstate}$ defined in~(\ref{eq:tp})  for information diffusion. These examples are provided in~\cite{Pin06} and deal with
how the use of a new technology/product spreads in a social network. Let the state  of agent $\nodem$ be $\state{\dtime}{\nodem}= 0$ if the agent at time $\dtime$ has not adopted the new technology, and $\state{\dtime}{\nodem} = 1$  if the agent
has adopted the new technology.
As mentioned earlier,
the target state $\target{\dtime}$ determines the availability of a competing technology/product or, alternatively,  the available market for the product.
\end{example}

In deciding whether to adopt the new technology, each agent $\nodem$ considers the following costs and rewards:
\begin{enumerate}
\item The cost node $\nodem$ pays for adopting the new technology is $\cost{\nodem}$. Assume that, at time $\dtime$, the costs $\cost{\nodem}$, $m=1,\ldots,\vertexnum$ are independent and identically distributed random variables with cumulative distribution function $\costpdf$. The fact that this distribution depends on the target state $\target{\dtime}$ (competing technology that evolves as a Markov chain) permits correlation of the costs $\cost$ over time.
\item If an agent $\nodem$ has $ \nactive{\nodem}_\dtime$ neighbors at time $\dtime$, then it obtains a benefit of $\reward \nactive{\nodem}_\dtime$ for adopting the technology, where $\reward$ is a positive real number.
\end{enumerate}
The agent is a myopic optimizer  and, hence, chooses to adopt the technology only if $\cost{\nodem} >  \reward  \nactive{\nodem}_\dtime$.
Therefore,  the transition probabilities  in (\ref{eq:tp}) are
\beq
\tpdiff{01}{\deg}{\nactive{\nodem}_\dtime}{\target{\dtime}} = \P\left( \cost{\nodem} >  \reward  \nactive{\nodem}_\dtime\right) =1- \costpdf\left( \reward  \nactive{\nodem}_\dtime\right).
\label{eq:ex1}
\eeq
Assume that if the product deteriorates/fails, the agent can no longer use the product (and will then need to reconsider  the possibility of adopting it).
If the product fails with probability $\failprob$, then
\beq \label{eq:failprob}
 \tpdiff{10}{\deg}{\nactive{\nodem}_\dtime}{\target{\dtime}} = \failprob. \eeq

Notice that the transition probabilities in~(\ref{eq:ex1}) do not depend on the node's connectivity $\deg$.  Constructing cases where the transition probabilities
depend on $\deg$ is straightforward. For example, suppose a node picks a single neighbor uniformly from its $\deg$ neighbors and then receives
a benefit of $\reward$ if this randomly chosen neighbour has adopted the product. The probability of choosing an active neighbor from $  \nactive{\nodem}_\dtime$ active
neighbors, given a total of $\deg$ neighbors, is clearly $\nactive{\nodem}_\dtime/ \deg$.
Then, assuming the agent acts as a myopic optimizer, it will adopt the product if  $ \P\big( \cost{\nodem} >  \reward \nactive{\nodem}_\dtime/ \deg\big) $. Therefore,
\begin{equation}
\tpdiff{01}{\deg}{\nactive{\nodem}_\dtime}{\target{\dtime}}  =1- \costpdf\left( \frac{\reward  \nactive{\nodem}_\dtime}{\deg}\right).
\end{equation}

\section{Sentiment-Based Sensing Mechanism}  \index{sentiment}
\label{sec:sensing}

We now describe  the sensing mechanism used to measure the active link probability distribution $\infectdist{\dtime}$ and target state in the above social network.

If  node  $\nodem$  has   state $\state{\dtime}{\nodem} = \state$ at time $\dtime$, and the target state is $\target{\dtime} = \target$, then node $\nodem$ communicates  with message
$\nodeobs{\dtime}{\nodem}$ where
\begin{equation} \nodeobs{\dtime}{\nodem} \sim   \oprob{\state}{\nodeobs}, \quad
\text{ where }   \quad
\nodeobs\in \{0,1,\ldots,\nodeobsdim\}. \end{equation}
That is, the message $\nodeobs{\dtime}$ is generated according to the  conditional probabilities
$ \oprob{\state}{\obs} = \P \big(\nodeobs{\dtime}{\nodem} = \nodeobs  \vert  \state{\dtime}{\nodem} =\state\big) $.
These elements correspond to the number of
tweets or sentiment of node $\nodem$ about the product based on its current state.
It is assumed that $\nodeobs{\dtime}{\nodem}$ is statistically independent of  $\nodeobs{\dtime}{\noden}$, $\noden \neq \nodem$.

An important question regarding sensing in a social network is:
 How can one construct a small but representative sample of a social network with a large number of nodes? Leskovec \& Faloutsos in \cite{LF06} study and compare several scale-down and back-in-time sampling procedures.
The simplest possible sampling scheme for a population is uniform sampling.  We also briefly describe {\em social sampling} and  {\em respondent-driven sampling} which
are recent methods that have become increasingly popular.

\subsection{Uniform Sampling}
Consider the following sampling-based measurement strategy. At each period $\dtime$,   $\sample{\deg}$ individuals are
sampled\footnote{For large population sizes $\vertexnum$, sampling with and without replacement are equivalent.}  independently and uniformly  from
the population $\vertexnum{\deg}$ comprising of agents with connectivity  degree $\deg$.
That is, a uniform distributed i.i.d. sequence of nodes, denoted by$\{\nodem{\seq}, \seq=1: \sample{\deg}\}$, is generated from the population  $\vertexnum{\deg}$.
The messages  $\nodeobs{\dtime}{\nodem{\seq}}$
of these  $\sample{\deg}$  individuals  are recorded.
From these independent samples, the empirical sentiment distribution $\sentiment{\dtime}{\deg}$ of degree $\deg$ nodes at each time $\dtime $ is obtained as
\beq \sentiment{\dtime}{\deg}{\nodeobs} = \frac{1}{\sample{\deg} }  \sum_{
\seq=1}^{\sample{\deg}} I\lbr\nodeobs{\dtime}{\nodem{\seq}}=\nodeobs\rbr,
\quad \nodeobs =1 , \ldots, \nodeobsdim.  \label{eq:sentiment} \eeq
At each time $\dtime$, the empirical sentiment distribution $\sentiment{\dtime}$ can be viewed as noisy observations of the infected distribution $\infectdist{\dtime}$ and target state process
$\target{\dtime}$.

\subsection{Non-Uniform Social Sampling} \index{social sampling}
Social sampling is an extensive area of research; see~\cite{DKS12} for recent results. In social  sampling,  participants in a poll respond with a summary of their friend's
responses. This leads to a reduction in the number of samples required.
If the average degree of nodes in the network is $\deg$, then  the savings in the number of samples is by a factor of $\deg$, since a
randomly chosen node summarizes the results form $\deg$ of its friends. However, the variance and bias of the estimate
depend strongly on the social network structure\footnote{In~\cite{DKS12}, a nice intuition is provided in terms of intent polling and expectation polling. In intent polling,
individual are sampled and asked who they intend to vote for. In expectation polling,  individuals are sampled and  asked who they think would win the election.
For a given sample size, one would believe that expectation poling is more accurate than intent polling
since in expectation polling, an individual would typically consider its own intent together with the intents of its friends.}.
In~\cite{DKS12}, a social sampling method is introduced and analyzed where nodes of degree $\deg$ are sampled with probability proportional to $1/\deg$.  This is intuitive since weighing neighbors' values by the reciprocal of the degree undoes the bias introduced by large degree nodes.
It  then illustrates  this social sampling method and variants on the {\sc LiveJournal} network (livejournal.com)  comprising of more than 5 million nodes and 160 million
 directed edges.

\subsection{MCMC Based Respondent-Driven Sampling (RDS)} \index{respondent-driven sampling}
Respondent-driven sampling~(RDS) was   introduced by Heckathorn~\cite{Hec97,Hec02,Lee09} as an approach for sampling from hidden populations
in social networks and  has gained
enormous popularity in recent years.
There are more than 120 RDS studies worldwide involving sex workers and  injection drug users~\cite{MJK08}.
%
As mentioned in~\cite{GS09}, the U.S. Centers for Disease Control and Prevention~(CDC) recently selected RDS for a 25-city study of injection drug users that is part of the National HIV Behavioral Surveillance System~\cite{LACH07}. 

RDS  is a variant of the well known method of snowball sampling where current sample members recruit future sample members. The RDS procedure is as follows:  A small number of people in the target
population serve as seeds. After participating in the study, the seeds recruit other people they know through the social network in the target population. The sampling continues according to this procedure  with current sample members recruiting
the next wave of sample members until the desired sampling size is reached. Typically,  monetary compensations are provided for participating in the data collection and recruitment.

RDS can be viewed as a form of Markov Chain Monte Carlo~(MCMC) sampling (see~\cite{GS09} for an excellent exposition).
 Let $\{\nodem{\seq},\seq = 1:\sample{\deg}\}$  be the realization of an aperiodic irreducible Markov chain with
state space  $\vertexnum{\deg}$ comprising of nodes
of degree $\deg$. This Markov chain models the individuals of degree $\deg$ that are snowball sampled, namely, the first individual $\nodem{1}$ is sampled and then recruits the second
individual $\nodem{2}$ to be sampled, who then recruits $\nodem{3}$ and so on.
Instead  of the independent sample estimator~(\ref{eq:sentiment}),
an asymptotically unbiased MCMC estimate is then generated as
\beq \frac{ \sum_{\seq = 1}^{\sample{\deg}}  \frac{I(\nodeobs{\dtime}{\nodem{\seq}}=\nodeobs)}{\steady{\nodem{\seq}}} }{ \sum_{\seq=1}^{\sample{\deg}}  \frac{1}
{\steady{\nodem{\seq}}}
}
\label{eq:mcmcrds}
\eeq
where  $\steady(\nodem)$, $\nodem \in \vertexnum{\deg}$, denotes the stationary distribution of the Markov chain.  For example, a reversible Markov chain  with
prescribed stationary distribution  is straightforwardly generated by the Metropolis Hastings algorithm.

In RDS, the transition matrix  and, hence, the stationary distribution $\steady$
in  the estimator~(\ref{eq:mcmcrds})
 is specified as follows: Assume that  edges between any two nodes $\nodem$ and $\noden$ have symmetric weights $\weight{\nodem}{\noden}$
(i.e.,
 $\weight{\nodem}{\noden} = \weight{\noden}{\nodem}$, equivalently, the network is undirected). In RDS,
node  $\nodem$ recruits node $\noden$ with transition probability
  $\weight{\nodem}{\noden}/ \sum_{\noden} \weight{\nodem}{\noden}$. Then, it can be easily seen that
the stationary distribution is
$\pi(\nodem) = \sum_{\noden \in \Vertexset} \weight{\nodem}{\noden}/ \sum_{\nodem \in \Vertexset, \noden \in \Vertexset} \weight{\nodem}{\noden}$. Using this stationary
distribution, along with the above transition probabilities for sampling agents in~(\ref{eq:mcmcrds}), yields the RDS algorithm.

It is well known that a Markov chain over a non-bipartite connected undirected network $\network$ is aperiodic. Then, the initial seed for the RDS
algorithm can be picked arbitrarily, and the above estimator is an asymptotically unbiased estimator.

Note the difference between RDS and social sampling: RDS uses the network to recruit the next respondent, whereas social sampling seeks
to reduce the number of samples by using people's knowledge of their friends' (neighbors') opinions.

\subsection{Extrinsic Observations}
In addition to the above sentiment based observation process, often extrinsic measurements of the target state are available. For example, if the target state
denotes the price/availability of products offered by competitors or the current market, then economic indicators yield noisy measurements.
Let $\targetobs{\dtime}$ denote noisy measurements of the target state $\target{\dtime}$ at time $\dtime$.  These observations are obtained as
\beq  \targetobs{\dtime} \sim  \targetoprob{\target}{\targetobs} \ole \P(\targetobs{\dtime}=\targetobs| \target{\dtime} = \target), \text{ where }  \targetobs \in \{1,2,\ldots, \targetobsdim\}.  \label{eq:extrinsic}
\eeq

\begin{remark}
The reader may be familiar with the DARPA network challenge in 2009 where the locations of 10 red balloons in the continental US
were to be determined using social networking. In this case, the winning MIT Red Balloon Challenge Team used a recruitment based sampling method.
The strategy can also be viewed as a variant of the Query Incentive Network model of~\cite{KR05}.
\end{remark}

\section{Mean Field Dynamics for Social Sensing} \index{graph!mean field dynamics}
The aim here is to
construct statistical signal processing algorithms for
computing the minimum mean square error estimate of the infected distribution $\infectdist{\dtime}$ and
target state $\target{\dtime}$  given the sequence of sentiment observations $\sentiment{1:\dtime} =  \{\sentiment{1},\ldots,\sentiment{\dtime}\}$.
This problem
 is intractable due to the dimension
$\prod_{\deg=1}^{\degmax} \vertexnum{\deg}$
of the state $\infectdist{\dtime}$ when the number of agents $\vertexnum$ is  large.  Fortunately, the {\em  mean field dynamics}  approximation of the information
diffusion has a state dimension of $\degmax$ and will be shown to be an excellent approximation as the number of agents $\vertexnum$ becomes large.  This is the subject of the current section.

Depending on the time scale on which the target state $\target$ evolves, there are two  possible asymptotic limits for the mean field dynamics for information
diffusion in the social network:
\begin{enumerate}
\item  Diffusion matched to target dynamics;
\item  Diffusion faster than target dynamics.
\end{enumerate}
We consider both these cases below and formulate estimation of the states $\infectdist{\dtime},\target{\dtime}$,  given the   sequence of sentiment observations $\sentiment{1:\dtime}$, as a nonlinear  filtering problem.
The main consequence is that the resulting mean field dynamics filtering problem can be solved straightforwardly  via the use of sequential Markov chain Monte-Carlo methods. This is in contrast to the original filtering problem which is intractable as $\vertexnum \rightarrow \infty$.


\section[Information Diffusion Matched to Target State Dynamics]{Mean Field Dynamics: Information Diffusion\\ Matched to Target State Dynamics}
\sectionmark{Mean Field Dynamics}

The main assumption here is that the diffusion of information flow in the social network, namely $\infectdist$, evolves  on the same time-scale
as the target state $\target$. This is made explicit in the following assumption.
\begin{Assumption}\label{asm:1}
The target state $\target{\dtime}$ has transition probability matrix $\tptarget = I + \frac{1}{\vertexnum} Q$, where
$Q$ is a generator matrix, i.e., $q_{ij}\geq0$ for $i\neq j$ and $Q\mathbf{1} = \mathbf{0}$, where $\mathbf{1}$ and $\mathbf{0}$ are column vectors of ones and zeros, respectively.
\end{Assumption}

\subsection{Mean Field Dynamics}
\index{mean field dynamics}
The mean field dynamics state
that as the number of agents $\vertexnum$ grows to infinity, the dynamics of the infected distribution  $\infectdist$, described by (\ref{eq:minc}),
in the social
network evolves according to an ordinary differential equation that is modulated by a continuous-time Markov chain which depends
on the target state evolution $\target$.
More specifically, under~Assumption~\ref{asm:1}, the mean field dynamics for the system~(\ref{eq:minc}) are as follows:
For $\deg=1,2,\ldots, \degmax$,
\begin{align}  \frac{d \infectdist{\ctime}{\deg} }{ d \ctime}  &=  \atp{01}{\deg}{\pa(\infectdist{\ctime})}{\target{\ctime}} -
   \atp{10}{\deg}{\pa(\infectdist{\ctime})}{\target{\ctime}} \label{eq:system} \\
%
\atp{01}{\deg}{\pa}{\target{\ctime}} & = (1-\infectdist{\ctime}{\deg})  \,  \sum_{\aindex = 0}^\deg \tpdiff{01}{\deg}{\aindex}{\target{\ctime}}
\binom{\deg}{\aindex}
\pa^{\aindex}  (1 - \pa)^{\deg-\aindex}  \nonumber\\
 \atp{10}{\deg}{\pa}{\target{\ctime}} &= \infectdist{\ctime}{\deg}  \,  \sum_{\aindex = 0}^\deg \tpdiff{10}{\deg}{\aindex}{\target{\ctime}}
\binom{\deg}{\aindex}
\pa^{\aindex}  (1 - \pa)^{\deg-\aindex} \nonumber  \\
%
\pa(\infectdist{\ctime}) & = \frac{ \sum_{\deg=1}^{\degmax} \deg \, \degdist{\deg}\, \infectdist{\ctime}{\deg} } {\sum_{\deg}^{\degmax} \deg \, \degdist{\deg} }
\nonumber  
 \end{align}
 where $\target{\ctime}$ is a continuous time Markov chain with generator~$Q$. 

That the above mean field dynamics follow from~(\ref{eq:minc}) is intuitive. Such averaging results are well known in the adaptive filtering community where they are deployed
to analyze the convergence of adaptive filters. The difference here is that the limit mean field dynamics are not deterministic but Markov modulated. Moreover, the
mean field dynamics here constitute a model for information diffusion, rather than the asymptotic behavior of an adaptive filtering algorithm.
As mentioned earlier, from an engineering point of view, the mean field dynamics yield a tractable model for estimation.

To formalize the  convergence of the discrete time dynamics~(\ref{eq:minc}) of the infected distribution to the mean field dynamics~(\ref{eq:system}) as the number
of agents $\vertexnum\rightarrow \infty$,
 we first need to re-write the discrete time process $\infectdist{\dtime}$ in~(\ref{eq:minc})
as a continuous time process. This is straightforwardly done by constructing a piecewise linear interpolation between the discrete-time points---in classical
signal processing, this constitutes what is commonly called a first-order hold interpolation. Define the sampling period as $1/\vertexnum$ and then
define the continuous time linear interpolated process\footnote{One can also consider a piecewise constant interpolation as widely used in~\cite{KY03} for weak convergence analysis of stochastic approximation algorithms; then, the appropriate
function space is the space of C\`{a}dl\`{a}g functions, which is usually denoted $D[0,\finaltime]$.}
 associated with $\infectdist{\dtime}$ as
\begin{equation}\tinfectdist{\ctime}{\deg} =  \infectdist{\dtime}{\deg} + \frac{\ctime - \dtime/ \vertexnum}{1/\vertexnum} \left(
\infectdist{\dtime+1}{\deg} - \infectdist{\dtime}{\deg}
\right)
\end{equation}
for continuous time $\ctime \in \big[\frac{\dtime}{\vertexnum}, \frac{(\dtime+1 )}{\vertexnum}\big)$, $\dtime = 0,1,\ldots$.

We then have the following exponential bound  result for the error $\left\| \tinfectdist{\ctime} - \infectdist{\ctime} \right\|_\infty$;  recall
$\infectdist{\ctime}$ is given the mean field dynamics (\ref{eq:system}) while  the actual system's infected distribution is  $\tinfectdist{\ctime} $:

\begin{theorem}  \label{thm:mftmatched}
Suppose Assumption~\ref{asm:1} holds. Then, for $\vertexnum$ sufficiently large,
\begin{equation}
\P\lbr \max_{0 \leq \ctime \leq \finaltime} \left\| \tinfectdist{\ctime} - \infectdist{\ctime} \right\|_\infty \geq \epsilon\rbr
\leq  C_1  \exp(-C_2 \epsilon^2 \vertexnum)
\end{equation}
where $C_1$ and $C_2$ are positive constants and $\finaltime$ is any finite time horizon.
\end{theorem}

The proof of the above theorem follows from~\cite[Lemma 1]{BW03}.
The exponential bound follows from an application of the Azuma-Hoeffding inequality.
The above theorem provides an exponential bound (in terms of the number of agents $\vertexnum$)  for the probability of  deviation of the sample path of the infected distribution from the mean field dynamics for any finite time interval~ $\finaltime$.

{\em Remark}:
For the signal processing reader more familiar with discrete-time averaging theory, given the discrete time system (\ref{eq:minc}), the discrete-time  mean field dynamics are:
\beq
 \minfectdist{\dtime+1}{\deg}   =  \minfectdist{\dtime}{\deg} +  \frac{1}{\vertexnum} \left[ \atp{01}{\deg}{\pa(\minfectdist{\dtime})}{\target{\dtime}} -
   \atp{10}{\deg}{\pa(\minfectdist{\dtime})}{\target{\dtime}}\right]  \label{eq:dsystem} .\eeq
  Then the following is the discrete time equivalent of Theorem \ref{thm:mftmatched}: For a discrete time horizon of $\finaltime$ points,  the deviation between the mean field dynamics $\minfectdist{\dtime}$ in (\ref{eq:dsystem}) and actual infected distribution in  $\infectdist{\dtime} $ (\ref{eq:minc}) satisfies
\beq
\P\lbr \max_{0 \leq \dtime \leq \finaltime} \left\| \minfectdist{\dtime} - \infectdist{\dtime} \right\|_\infty \geq \epsilon\rbr
\leq  C_1  \exp(-C_2 \epsilon^2 \vertexnum)
\end{equation}
providing $\finaltime = O(\vertexnum)$.

\subsubsection{Some Perspective}
The stochastic approximation and adaptive filtering literature~\cite{BMP90,KY03} has several averaging analysis methods for
recursions of the form~(\ref{eq:minc}).  The well studied mean square error analysis~\cite{BMP90,KY03} computes bounds on $ \E\| \tinfectdist{\ctime} - \infectdist{\ctime} \|^2$ instead of the maximum
deviation in Theorem \ref{thm:mftmatched}. A mean square error analysis of estimating a Markov modulated empirical distribution is given in~\cite{YKI04}. Such mean
square analysis assume a finite but small step size $1/\vertexnum$ in~(\ref{eq:minc}).
Another powerful class of analysis methods involves weak convergence. \index{weak convergence} These seek to prove that
$ \lim_{\vertexnum \rightarrow \infty}  \P \left\{ \sup_{\ctime \leq \finaltime} \| \tinfectdist{\ctime} - \infectdist{\ctime} \| \geq \epsilon \right\} = 0 $.
Weak convergence methods are powerful enough to tackle Markovian noise, whereas in Theorem~\ref{thm:mftmatched} the noise term is a much simpler martingale increment.

On the other hand, unlike weak convergence results, Theorem~\ref{thm:mftmatched} offers an exponential bound. This is useful in  the important problem
of determining the  asymptotic behavior as $\finaltime \rightarrow \infty$ and then taking the limit  $\vertexnum \rightarrow \infty$. (In comparison, the
 above analysis  characterizes the behavior as $\vertexnum \rightarrow \infty$ for finite time $\finaltime$.)
 In general it is not true that the limits
${\vertexnum \rightarrow \infty}$ and
$\finaltime \rightarrow \infty$ can be interchanged. Weak convergence proofs for the asymptotic behavior in $\finaltime$ require construction of perturbed Lyapunov functions
\cite{KY03}.
In comparison, 
as explained in~\cite{BW03}, since  the exponential bound $\sum_{N} \exp(- C N)$ is summable, the Borel-Cantelli
lemma applies. This facilities concluding important results on the exit times of the process $\tinfectdist{\ctime} $. In particular, given an open  set $U \in \reals^{\degmax}$, define
the exit time as the random variable
\beq \tau^{\vertexnum}(U) = \inf \left\{\ctime \geq 0: \tinfectdist{\ctime} \notin U \right\}. \eeq
Let $\gamma^+(\infectdist)$ denote the closure of the set of states visited by the mean field trajectory $\infectdist{\ctime}$, $\ctime \geq 0$.
Then, from Theorem \ref{thm:mftmatched},  it can be shown that \cite[Proposition 1]{BW03}
\begin{equation} \P\left\{ \lim_{\vertexnum\rightarrow \infty} \tau^{\vertexnum}(U) = \infty\right\}  = 1.\end{equation}
In words, the exit time of the infected process  from any neighbourhood of the mean field dynamics trajectory is probabilistically very large. Indeed, Bena\"{\i}m \& Weibull in~\cite{BW03} go on to
conclude that attractors of the mean field dynamics are good predictors of the stochastic process when the initial conditions are close.

\subsection{Sentiment-Based Observations} \index{sentiment}
Next, consider the sensing mechanism with observations defined in~\S\ref{sec:sensing}. Assuming the sample size is sufficiently large,
it follows from the central limit theorem (for i.i.d. processes, in the case of independent sampling, and for Markov processes, in the case of MCMC sampling)
that

\beq
\sentiment{\dtime}{\deg}{\obsdiff} \approx  \obsm{\obsdiff} \, \infectdist{\dtime}{\deg}  +  \feedforward{\obsdiff} + \onoise{\dtime}{\deg}{\obsdiff}
\eeq
where
\begin{align}
\feedforward{\obsdiff} &= \oprob{0}{\obsdiff} \degdist{\deg},  \nonumber\\
\obsm{\obsdiff} &= (\oprob{1}{\obsdiff} - \oprob{0}{\obsdiff}) \degdist{\deg}, \nonumber\\
 \onoise{\dtime}{\deg}{\obsdiff} &\sim \normal(0, \var{\infectdist{\dtime}{\deg}}{\obsdiff}). \label{eq:cltonoise}   \end{align}
 For i.i.d. sampling, the variance is  easily evaluated as
 \begin{equation}
 \begin{split}
 \var{\infectdist{\dtime}{\deg}}{\obsdiff} = \frac{1}{\sample{\deg}} &\left( \obsm{\obsdiff} \, \infectdist{\dtime}{\deg}  + \feedforward{\obsdiff} \right)\\
 &\times \left( 1-  \obsm{\obsdiff} \, \infectdist{\dtime}{\deg}  -  \feedforward{\obsdiff} \right).
\end{split}
\end{equation}
 For MCMC sampling, the variance is evaluated by approximating an infinite series (alternatively, excellent bounds exist~\cite{Bre99}).

We can then define the following observation process
\beq
\sentiment{\dtime}{\obsdiff} \ole \frac{\sum_{\deg=1}^{\degmax} \deg \, \degdist{\deg} \sentiment{\dtime}{\deg}{\obsdiff}}{\sum_{\deg=1}^{\degmax} \deg\, \degdist{\deg} }.
\eeq
The key point is that $\sentiment{\dtime}{\obsdiff}$ is a linear combination of $\sentiment{\dtime}{\deg}{\obsdiff}$. It then follows from~(\ref{eq:infectasymp}) that
\beq \label{eq:samplesentiment}
\sentiment{\dtime}{\obsdiff} =  \pa{\dtime} +  \onoise{\dtime}{\obsdiff} \eeq
where
\begin{align}  \onoise{\dtime}{\obsdiff} & \sim \normal(0, \var{\obsdiff}), \nonumber \\
 \var{\obsdiff} &=   \frac{\sum_{\deg=1}^{\degmax} \deg \, \degdist{\deg} \var{\infectdist{\dtime}{\deg}}{\obsdiff} }{\sum_{\deg=1}^{\degmax} \deg\, \degdist{\deg} }. \nonumber
\end{align}
\subsection{Bayesian Filtering Problem}  \index{nonlinear filtering}
Given the extrinsic and sentiment observations described above, how can the infected degree distribution $\infectdist_\dtime$ and target state $\target_\dtime$ be estimated at each time instant?
Below we discuss estimation of $\infectdist_\dtime$ and $\target_\dtime$  as a Bayesian filtering problem.

 The partially observed state space model with continuous time dynamics~(\ref{eq:system}) and
discrete time observations~(\ref{eq:samplesentiment}) and~(\ref{eq:extrinsic}), constitutes a system with continuous time dynamics and discrete time observations.
Alternatively, the partially observed state space model with discrete time dynamics~(\ref{eq:dsystem}) and observation
equation~(\ref{eq:samplesentiment}), and~(\ref{eq:extrinsic}) constitutes a system with discrete time dynamics and discrete time observations.
In either case, computing the conditional mean estimate of $\target{\ctime}, \infectdist{\ctime}$ in the continuous time case  or $\target{\dtime}, \infectdist{\dtime}$ in the discrete time case, given the observation sequence
 $(\sentiment{1:\dtime}, \targetobs{1:\dtime})$
 is a Bayesian filtering problem. In fact, filtering of such jump Markov linear systems have been studied extensively in the signal processing literature~\cite{DGK01,LK99}
 and can be solved via the use of sequential Markov chain Monte-Carlo methods.

For large $\vertexnum$ in the mean field dynamics (\ref{eq:dsystem}), it is reasonable to expect that a linearized approximation  in $\minfectdist$ to the dynamics is accurate.
Denote the resulting linearized model as
   \begin{align}
 \minfectdist{\dtime+1} =  \bar{F}(\target{\dtime}) \minfectdist{\dtime}.
 \end{align}
 In the special case when only extrinsic observations $\targetobs{\dtime}$ of $\target{\dtime}$ are available (recall the extrinsic observations are defined in (\ref{eq:extrinsic})),
computing the filtered estimate $\E\{ \pa{\dtime}| \targetobs{1:\dtime}\}$ can be done  via a finite-dimensional  filtering algorithm. Indeed, this model coincides with
the so-called image-based tracking model that has widely been studied in signal processing~\cite{DB96,KE97b,KE97a,SH89}.

\section[Mean Field Dynamics]{Mean Field Dynamics: Fast Information Diffusion and Slow Target State Dynamics}

The  assumption here is that the diffusion of information flow, namely $\infectdist$, in the social network occurs on a faster time-scale
compared to the evolution of the target state $\target$.  That is, the transition matrix  $\tptarget$ of the Markovian target $\target$ is
almost identity and the target state displays piecewise constant behavior with infrequent jumps from one state to another.
This is made explicit in the following assumption.
\begin{Assumption}\label{asm:2}
The target state $\target{\dtime}$ has transition probability matrix $\tptarget = I + \frac{1}{\vertexnum^2} Q$ where
$Q$ is a generator matrix.
\end{Assumption}
Considering the diffusion dynamics~(\ref{eq:minc}) together with~Assumption~\ref{asm:2}, it is clear that we have a two-time-scale system, where the target
process $\target$ evolves on a slower time scale as compared with the information diffusion in the network.
For this two-time-scale case, we have the following mean field dynamics for the infected distribution: \index{mean field dynamics}

Suppose Assumption~\ref{asm:2} holds.
Then, for $\deg=1,2,\ldots, \degmax$,
\begin{align}  \frac{d \infectdist{\ctime}{\deg} }{ d \ctime} & =  \atp{01}{\deg}{\pa(\infectdist{\ctime})}{\target{\stime}} -
   \atp{10}{\deg}{\pa(\infectdist{\ctime})}{\target{\stime}},\;  0 \leq \ctime \leq \infty. \label{eq:slowsystem} \end{align}
   In the above equation,
 $\target{\stime}$ is a constant on the time-scale $\ctime$; it evolves as a continuous time Markov chain with generator $Q$ on the slow time scale $\stime$.

A similar exponential probability bound to Theorem~\ref{thm:mftmatched} can be proved for the above mean field dynamics system comprised of
 $\degmax$ Markov modulated differential equations~(\ref{eq:slowsystem}).
Also, in analogy to (\ref{eq:dsystem}), for large $\vertexnum$, the discrete time mean field dynamics read:
 \beq
 \minfectdist{\dtime+1}{\deg}   =  \minfectdist{\dtime}{\deg} +  \frac{1}{\vertexnum} \left[ \atp{01}{\deg}{\pa(\minfectdist{\dtime})}{\target{\sdtime}} -
   \atp{10}{\deg}{\pa(\minfectdist{\dtime})}{\target{\sdtime}}\right]  \label{eq:slowdisc} .\eeq
Here, $\dtime$ indexes  a fast time scale and $\sdtime$ indexes a slow time scale.
In particular,  the slow process $\target{\sdtime}$ is a constant on the fast time scale $\dtime$; it evolves as a discrete time Markov chain with transition matrix  $\tptarget = I + \frac{1}{\vertexnum^2} Q$ on the slow time scale $\sdtime$.

 The sentiment based observation process is modelled as in~(\ref{eq:samplesentiment}), and the extrinsic observation process $\targetobs_{\sdtime}$ on the slow time-scale $\sdtime$ is modelled as in~(\ref{eq:extrinsic}).

\subsection{Bayesian Filtering Problem}  \index{nonlinear filtering}
In the case of fast information diffusion and slow target dynamics,   estimating the slow target state $\target$ and fast infected distribution $\infectdist$ are decoupled due to the
two-time-scale nature of the problem. In particular, at the slow time-scale $\sdtime$, the  conditional mean estimate  $\E\{\target_{\sdtime}|   \targetobs_{1:\sdtime}\}$
 is obtained using a Hidden Markov model filter
 \index{nonlinear filtering!hidden Markov model filter}
  (or Wonham filter in continuous time)~\cite{EAM95}.  Then, estimating the state $\infectdist{\dtime}$ given
 noisy observations  $\sentiment{\dtime}{\obsdiff}$ becomes a nonlinear regression problem since there is no state noise process driving (\ref{eq:dsystem}).
Alternatively, a stochastic approximation algorithm can be used to track the Markov chain; see \cite{YK05b} for details.

\subsection{Numerical Example}
Below we illustrate the mean field dynamics model for the case of fast information diffusion and slow target dynamics.
In this example, we simulate the diffusion of information through a network comprising of $ N=100$ nodes (with maximum degree $\degmax = 17$) in two different scenarios as described below. We further assume that at time $\dtime = 0 $, $5\%$ of nodes are infected. The mean field dynamics model is investigated in terms of the infected link probability~(\ref{eq:infectasymp}). The infected link probability for the case of fast information diffusion and slow target dynamics can be computed by
 \begin{equation} \label{eq:linf} \pa(\minfectdist{\dtime})  = \frac{ \sum_{\deg=1}^{\degmax} \deg \, \degdist{\deg}\, \minfectdist{\dtime}{\deg} } {\sum_{\deg}^{\degmax} \deg \, \degdist{\deg} }.\end{equation} Here, $\minfectdist{\dtime}$ is obtained via discrete-time mean field dynamics~(\ref{eq:slowdisc}).

{\em Scenario~1:}
Assume each agent is a myopic optimizer  and, hence, chooses to adopt the technology only if $\cost{\nodem} >   \nactive{\nodem}_\dtime$; $\reward = 1$. At time $k$, the costs $\cost{\nodem}$, $m = 1,2,\ldots,100$, are i.i.d. random variables simulated from uniform distribution $U[0,C(\target{\dtime})]$. Therefore, the transition probabilities  in (\ref{eq:tp}) are
\begin{equation}
\tpdiff{01}{\deg}{\nactive{\nodem}_\dtime}{\target{\dtime}} = \P\left( \cost{\nodem} >    \nactive{\nodem}_\dtime\right) =\left\{\begin{array}{l} \frac{\nactive{\nodem}_\dtime)}{C(\target{\dtime})}, \quad \nactive{\nodem}_\dtime \leq C(\target{\dtime}), \\ 1, \quad \hspace{.7cm} \nactive{\nodem}_\dtime > C(\target{\dtime}).  \end{array}\right.\nn
\end{equation}

\begin{figure}[t]
\centering
\subfloat[]{
  \includegraphics[width=0.5\linewidth]{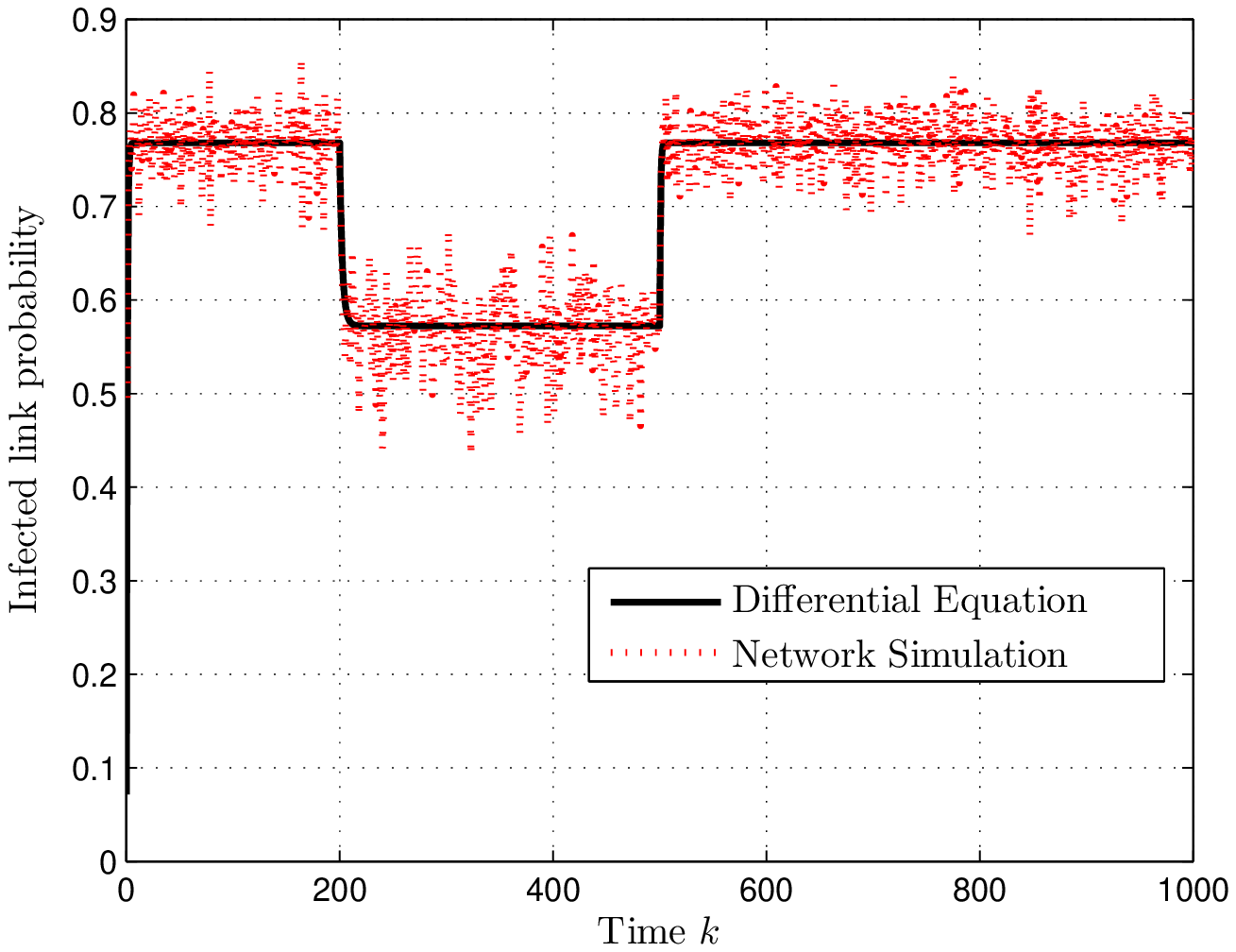}
  \label{fig:diffusion_a}
}%
\subfloat[]{
  \includegraphics[width=0.5\linewidth]{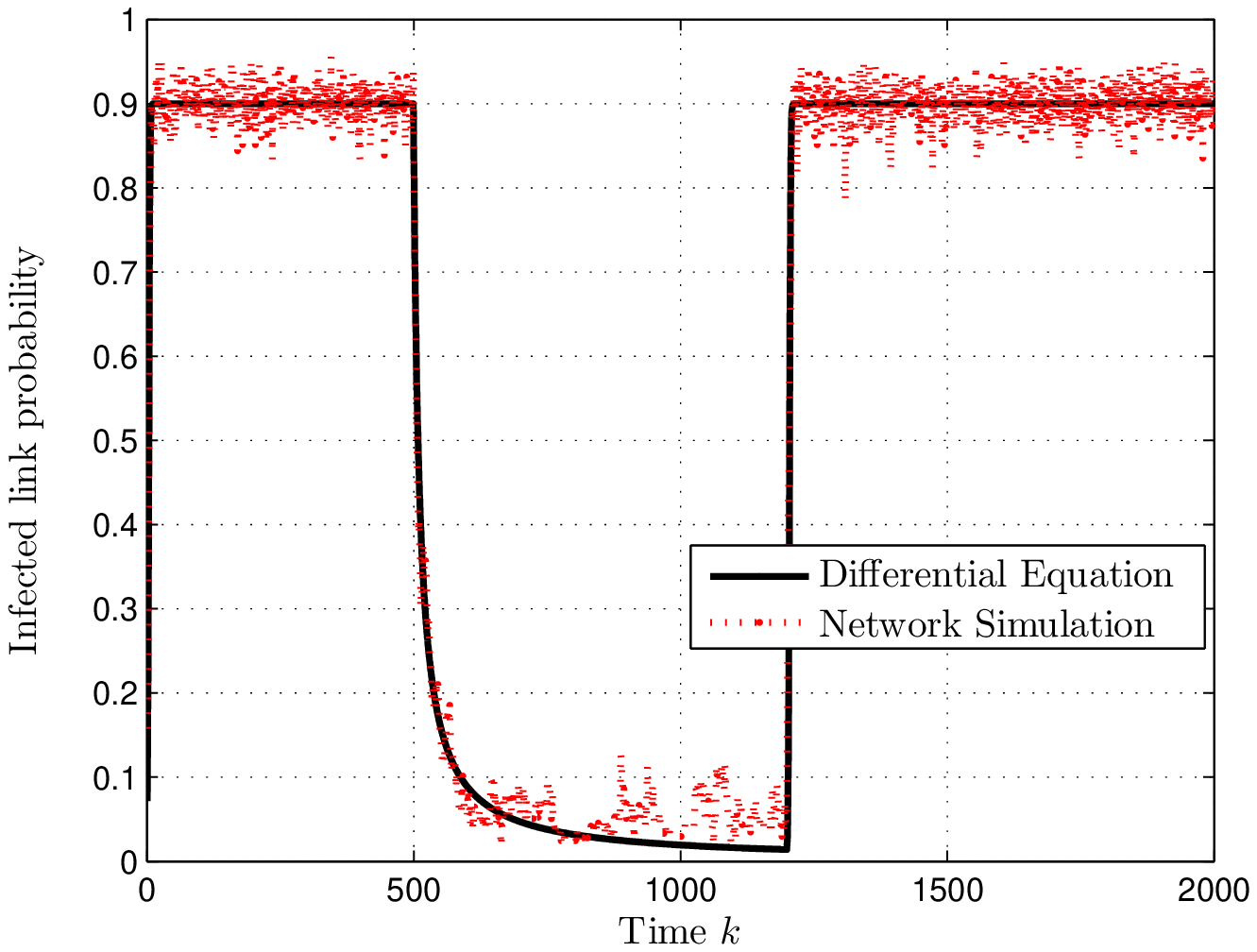}
  \label{fig:diffusion_k}
}
\caption{The infected link probability obtained from network simulation compared to the one obtained from the mean field dynamics model~(\ref{eq:linf}) in two different scenarios: (a) The transition probabilities in (\ref{eq:ex1}) depend only on the  number of infected neighbors $A^(m)_k$ (the parameters are defined in Scenario~1). (b) The transition probabilities in (\ref{eq:ex1}) depend on the node's connectivity $\deg$ and the number of infected neighbors $\nactive{\nodem}_\dtime$ (the parameters are defined in Scenario~2).}
\label{fig:diffusion}
\end{figure}

The probability that a product fails is  $\failprob =0.3$, i.e.,  $$\tpdiff{10}{\deg}{\nactive{\nodem}_\dtime}{\target{\dtime}} = 0.3.$$ The infected link probabilities obtained from network simulation~(\ref{eq:infectasymp}) and from the discrete-time mean field dynamics model~(\ref{eq:linf}) are illustrated in Figure~\ref{fig:diffusion_a}. To illustrate that the infected link probability computed from (\ref{eq:linf}) follows the true one (obtained by network simulation), we assume that the value of $C$ jumps from $1$ to $10$ at time $\dtime = 200$, and from $10$ to $1$ at time $\dtime = 500$. As can be seen  in Figure~\ref{fig:diffusion_a}, the mean field dynamics provide an excellent approximation to the true infected distribution.

{\em Scenario~2:}
The transition probabilities in~(\ref{eq:tp}) depend on the node's connectivity $\deg$.
These probabilities are
\begin{equation}
\tpdiff{01}{\deg}{\nactive{\nodem}_\dtime}{\target{\dtime}}
=\left\{\begin{array}{l} \frac{\nactive{\nodem}_\dtime)}{\deg C(\target{\dtime})}, \quad \nactive{\nodem}_\dtime \leq \deg C(\target{\dtime}), \\ 1, \quad \hspace{.7cm} \nactive{\nodem}_\dtime > \deg C(\target{\dtime}).  \end{array}\right.
\end{equation}
and
\begin{equation}\tpdiff{10}{\deg}{\nactive{\nodem}_\dtime}{\target{\dtime}} = 0.3.
\end{equation}
The infected link probability obtained from network simulation (\ref{eq:infectasymp}) and from the discrete-time mean field dynamics model~(\ref{eq:linf}) are illustrated in Figure~\ref{fig:diffusion_k}. Similar to Scenario~1, here, we assume that the value of $C$ jumps from $1$ to $10$ at time $\dtime = 500$, and from $10$ to $1$ at time $\dtime = 1200$ to show that the infected link probability computed from (\ref{eq:linf}) follows the one obtained from network simulation. Again the mean field dynamics provide an excellent approximation
to the actual infected distribution.

\section{Closing Remarks}

This chapter has shown how the dynamics of the spread of information in a graph can be asymptotically modelled via Markov modulated differential equations.
These mean field dynamics serve
as a tractable model for filtering of underlying states of the network. As an immediate extension, one can consider sequential change detection and, more generally, control
of the mean field dynamics.

Below we outline two extensions.

\subsubsection*{Reactive Information Diffusion}
A key difference between social sensors and conventional sensors in statistical signal processing is that social sensors are reactive:  A social sensor uses  additional information gained to modify
its behavior.  Consider the case where the sentiment-based observation process is made available in a public blog. Then, these observations will affect the transition dynamics of the agents and, therefore,
the mean field dynamics.   \index{sentiment}


\subsubsection*{How Does Connectivity Affect Mean Field Equilibrium?}
The papers~\cite{Pin06,Pin08} examine the structure of  fixed points of the mean field differential equation~(\ref{eq:system}) when the underlying target process $\target$ is not present (equivalently, $\target$ is a one state process).  They consider the case where the agent transition probabilities are parameterized by $\tpdiff{01}{\deg}{\aindex}  = \mu F(\deg,\aindex)$ and $\tpdiff{10} = \failprob$; see~(\ref{eq:failprob}).
Then, defining $\lambda = \mu/\failprob$,
 they study how the following two thresholds behave with
the degree distribution and diffusion mechanism:
\begin{enumerate}
\item {\em Critical threshold $\lambda_c$:} This is defined as the  minimum value of $\lambda$ for which there exists a fixed point of~(\ref{eq:system}) with positive fraction of infected agents, i.e.,
$\infectdist{\infty}{\deg} > 0$ for some $\deg$ and, for $\lambda \leq \lambda_c$,
such a fixed point does not exist.
\item  {\em Diffusion threshold $\lambda_d$:} Suppose the initial condition $\infectdist{0}$ for the infected distribution is infinitesimally small. Then, $\lambda_d$ is the minimum value of $\lambda$ for which
$\infectdist{\infty}{\deg} > 0$ for some $\deg$, and such that, for $\lambda \leq \lambda_d$, $\infectdist{\infty}{\deg}  = 0$  for all $\deg$.
\end{enumerate}
Determining how these thresholds vary with degree distribution and diffusion mechanism is very useful for understanding the long term behavior of agents in the social network.



\chapter{Non-Cooperative Game-Theoretic Learning}
\label{chapter:noncooperative}

\section{Introduction}
\label{sec:introduction}

Many social and economic situations involve interactive decision making with possibly diverging interests. Often there is an advantage to having individuals coordinate on their decisions~\cite{ZJV13}. For instance, a person may choose the same cellphone carrier as the majority of family and friends to take advantage of the free talk times. Social networks spread information, thereby facilitate coordination of such self-interested units by speeding up the resolution of uncertainties. For instance, individuals form friendship groups within which members are able to observe others' decisions.


This chapter examines how global coordination can be obtained among a group of decision makers
when each decision maker has limited awareness of the outside world and is only capable of `simple' local behavior. Such coordination can potentially lead to higher social welfare to each individual decision maker, and hence equilibrates the social network. We focus on no (or low) {\em internal regret} algorithms as a strategy of play in game-theoretic learning. The internal regret\footnote{In comparison, the external regret compares the performance of a strategy selecting actions to the performance of the best of those actions in hindsight.}
compares the loss of a strategy to the loss of a modified strategy, which consistently replaces one action by another---for example, ``every time you bought Windows, you should have bought Apple instead.'' We refer the reader to~\cite{BM07} for an excellent discussion of internal and external regret-based algorithms.

Game theory \index{game} has traditionally been used in economics and social sciences with a focus on fully rational interactions where strong assumptions are made on the information patterns available to individual agents. In comparison, social sensors are agents with partial information and it is the dynamic interactions between agents that are of interest. This, together with the interdependence of decision-makers' choices, motivates the need for game-theoretic learning models for agents interacting in social networks.

Learning dynamics in games can be typically classified into Bayesian learning~\cite{ADLO08,C03}, adaptive learning~\cite{HM01b}, \index{learning!adaptive learning}and evolutionary dynamics~\cite{HS03,LHN05}. This chapter focuses on adaptive learning where individuals deploy simple rule-of-thumb strategies. The aim is to determine if such simple individual behaviour can result in sophisticated global behaviour. We finish this chapter with an example to show how the presented regret-based rule-of-thumb strategies can be employed to devise an energy-aware sensing mechanism for parameter estimation va diffusion least mean squares.

\subsection{Literature}
\label{sec:literature}
The theory of learning in games formalizes the idea that equilibrium arises as a result of players learning from experience. It further examines how and what kind of equilibrium might arise as a consequence of a long-run process of adaptation and learning in an interactive environment~\cite{FL08}. Due to space limitations, this chapter
focuses on adaptive learning models \index{learning!adaptive learning} in which players try to maximize their own payoffs while simultaneously learning about others' play\footnote{The evolutionary game theory also provides a rich framework for modeling the dynamics of adaptive opponent strategies for large population of players. The interested reader is referred to~\cite{HS03,S10} for recent surveys.}. The question is then when self-interested learning and adaptation will result in the emergence of equilibrium behavior. The literature on game-theoretic learning has relied on the idea that learning rules should strike a balance between performance and complexity~\cite{FL08}. That is, simple learning rules are expected to perform well in simple environments, whereas larger and more complex environments call for more sophisticated learning mechanisms.

The simplest setting is one in which players' decisions can be observed by all other players at the end of each round. Fictitious play~\cite{FL98} is a simple stylized model of learning in non-cooperative repeated games with such communication model. The player behaves as if she is Bayesian. That is, she believes that the opponents' play corresponds to draws from an unknown stationary distribution and simply best responds to her belief about such a distribution\footnote{Obviously, if all players make decisions according to the fictitious play, the actual environment is not stationary, hence, players have the wrong model of the environment.}. Another simple model, namely, stochastic fictitious play~\cite{FL98}, forms beliefs as in fictitious play, however, chooses actions according to a stochastic best-response function. This new model brings about two advantages: (i) it avoids the discontinuity inherent in fictitious play, where a small change in the belief can lead to an abrupt change in behavior; (ii) it is ``Hannan-consistent''~\cite{H57}: its time average payoff is at least as good as maximizing against the time-average of opponents' play, which is not true for fictitious play.

Regret-matching~\cite{HM00,HM01a} as a strategy of play in long-run interactions has been long known to guarantee convergence to the correlated equilibria set~\cite{Aum87}. The regret-based adaptive procedure in~\cite{HM00} assumes a complete connectivity graph\footnote{A complete graph is a simple undirected graph in which every pair of distinct vertices is connected by a unique edge.} for information exchange among players, whereas the regret-based reinforcement learning algorithm in~\cite{HM01a} \index{learning!reinforcement learning} assumes a set of isolated players who rely only on their realized payoffs. This chapter is inspired by the work of Hart \& Mas-Colell in~\cite{HM00,HM01a} and focuses on learning algorithms in scenarios  where players form social groups and disclose information of their decisions only within their social groups.

The concept of regret, well-known in the decision theory, has also been introduced to the realm of random sampling~\cite{D91,DG93}. These methods are of particular interest in the multi-armed bandit literature, which is concerned with optimizing the cumulative objective function values realized over a period of time~\cite{ACF02,ACFS95}, and the
pure exploration problem~\cite{ABM10}, which involves finding the best arm after a given number of arm pulls. In such problems, the regret value of a candidate arm measures the worst-case consequence that might possibly result from selecting another arm. Such a regret is sought to be minimized via devising random sampling schemes that establish a proper balance between exploration and exploitation.

\subsection{Experimental Studies}
\label{sec:exper-study-game}
Parallel to the growing interest in social networks, a new game-theoretic paradigm has been developed to incorporate some aspects of social networks such as information exchange and influence structure into a game formalism. One such example is the so-called graphical games, where each player's influence is restricted to his immediate neighbors~\cite{K07,KLS01}. Early experimental studies on these games indicate significant network effects. For instance, Kearns et al. present in~\cite{KSM06} the results of experiments in which human subjects solved a version of the vertex coloring problem\footnote{The vertex coloring problem is an assignment of labels (colors) to the vertices of a graph such that no two adjacent vertices share the same label. The model has applications in scheduling, register allocation, pattern matching, and community identification in social networks~\cite{TBK07,MT10}.} that was converted into a graphical game. Each participant had control of a single vertex in the network. One task was to coordinate (indirectly) with their neighbors so that they all had the same color.

Griffin \& Squicciarini in~\cite{GS12} tackle the problem of information release in social media via a game-theoretic formulation. They analyze the dynamics of social identity verification protocols and, based on some real-world data, develop a deception model for online users. The proposed model captures a user's willingness to release, withhold or lie about information depending on the behavior of the user's circle of friends. The empirical study infers a relationship between the qualitative structure of the game equilibria and the automorphism group of the social network.

The correlated equilibrium arguably provides a natural way to capture conformity to social norms~\cite{CW12}. It can be interpreted as a mediator instructing people to take actions according to some commonly known probability distribution. Such a mediator can be thought of as a social norm that assigns roles to individuals in a society~\cite{XV12}. If it is in the interests of each individual to assume the role assigned to him by the norm, then the probability distribution over roles is a correlated equilibrium~\cite{Aum87,DM96,F86}. The Fact that actions are conditioned on signals or roles indicates how conformity can lead to coordinated actions within social groups~\cite{JJ91,S80}. Norms of behavior are important in various real-life behavioral situations such as public good provision\footnote{The public good game is a standard of experimental economics. In the basic game, players secretly choose how many of their private tokens to put into a public pot and keep the rest. The tokens in this pot are multiplied by a factor (greater than one) and this ``public good'' payoff is evenly divided among players.}, resource allocation, and the assignment of property rights; see~\cite{BK07a} for an extensive treatment of public good games in social networks.

The works~\cite{CS07,DF10} further report results from an experiment that explores the empirical validity of correlated equilibrium. It is found in~\cite{DF10} that when the private recommendations from the mediator are not available to the players, the global behavior is characterized well by mixed-strategy Nash equilibrium. Their main finding, however, was that players follow recommendations from the third party only if those recommendations are drawn from a correlated equilibrium that is ``payoff-enhancing'' relative to the available Nash equilibria.

\section[Non-Cooperative Games and Correlated Equilibrium]{Non-Cooperative Games\\ and Correlated Equilibrium}
\sectionmark{Non-Cooperative Games}
\label{sec:strategic-form-game}
This section starts with introducing the standard representation of non-cooperative games in~\S\ref{sec:strategic-form}. We then proceed to introduce and elaborate on two prominent solution concepts of such games in~\S\ref{sec:CE}.

\subsection{Strategic Form Non-Cooperative Games}
\label{sec:strategic-form}
\index{game!non-cooperative game}
The initial step in performing a non-cooperative game-theoretic analysis is to frame the situation in terms of all conceivable actions of agents and their associated payoffs. The standard representation of a non-cooperative game~$\game$, known as \emph{normal form} or \emph{strategic form} game~\cite{M86,OR94}, comprises three elements:
\begin{equation}
\label{eq:game}
\game = \left(\plyrset,\big(\actset^\plyrind\big)_{\plyrind\in\plyrset},\big(\utilityk\big)_{\plyrind\in\plyrset}\right).
\end{equation}
Each element is described as follows:

\textit{(i) Player Set:} $\plyrset = \lbr 1,\cdots,\fplyr\rbr$. Essentially, a player models an entity that is entitled to making decisions, and whose decisions affect others' decisions. Players may be people, organizations, firms, etc., and are indexed by $\plyrind\in\plyrset$.

\textit{(ii) Action Set:} $\actset^\plyrind = \lbr 1,\cdots,\fact\rbr$, that denotes the actions, also referred to as \emph{pure strategies}, available to player $\plyrind$ at each decision point. A generic action taken by player $\plyrind$ is denoted by $\actk$, where $\actk\in\actset^\plyrind$. The actions of players may range from deciding to establish or abolish links with other agents~\cite{BG00} to choosing among different products/technologies~\cite{ZW03}.

Subsequently, a generic element of the \emph{action profile} of all players is denoted by $\actprof = \big(\act^1,\cdots,\act^\fplyr\big)$, and belongs to the set $\actset = \actset^1\times\cdots\times\actset^\fplyr$, where $\times$ denotes the Cartesian product. Following the common notation in game theory, one can rearrange the action profile as $\actprof = \big(\actk,\actprof^{-\plyrind}\big)$, where $ \actprof^{-\plyrind} = \big( \act^1,\cdots, \act^{\plyrind-1}, \act^{\plyrind+1}, \newline\cdots,\act^\fplyr\big)$ denotes the action profile of all players excluding player $\plyrind$.

\textit{(iii) Payoff Function:} $\utilityk:\actset\to\mathbb{R}$ is bounded, and gives payoff to player $\plyrind$ as a function of the action profile $\actprof\in\actset$ taken by all players. The interpretation of such a payoff is the sum of rewards and costs associated with the chosen action as the outcome of the interaction. The payoff function can be quite general. For instance, it could reflect reputation or privacy, using the models in~\cite{GGG09,Mui02}, or benefits and costs associated with maintaining links in a social network, using the model in~\cite{BG00,ZV13b}. It could also reflect benefits of consumption and the costs of production, download, and upload in content production and sharing over peer-to-peer networks~\cite{GLML01,PV10}, or the capacity available to users in communication networks~\cite{JLL12,LZ09}.

In contrast to the pure strategy that specifies a possible decision or action, a \emph{mixed strategy} \index{mixed strategy} is a probability distribution that assigns to each available action in the action set a likelihood of being selected. More precisely, a mixed strategy for player $\plyrind$ is of the form
\begin{equation}
\label{eq:mixed-strategy}
\textstyle
\mixedstrat^\plyrind = \big(\mixedstratind^\plyrind(1),\cdots,\mixedstratind^\plyrind(\fact)\big),\quad 0\leq \mixedstratind^\plyrind(i)\leq 1,\;\sum_{i\in\actset^\plyrind} \mixedstratind^\plyrind(i) = 1.
\end{equation}
A mixed strategy can be interpreted as how frequently each action is to be played in a repeated game. Note that, even if the set of pure strategies is finite, there are infinitely many mixed strategies.
%
The set of all mixed strategies available to player $\plyrind$ forms an $\fact$-simplex, denoted by $\Delta\actset^\plyrind$.  An individual
uses a mixed strategy only when: (i) he/she is indifferent between several pure strategies; (ii) keeping the opponent guessing is desirable, i.e., the opponent can benefit from knowing the next move; (iii) the player needs to experiment in order to learn his/her optimal strategy. The significance of mixed strategies becomes more clear when we talk about existence of Nash equilibrium in~\S\ref{sec:CE}.

In framing the game-theoretic analysis, another important issue is the information available to individuals at each decision point. This essentially has bearing on the timing of interactions and is ascertained via answering the following questions:

\emph{(i) Is the interaction repeated?} \index{game!repeated game}Repeated games refer to a situation where the same base game, referred to as \emph{stage game}, is repeatedly played at successive decision points. For instance, suppliers and buyers make deals repeatedly, nations engage in ongoing trades, bosses try to motivate workers on an ongoing basis, etc. Repeated interaction allows for socially beneficial outcomes, essentially substituting for the ability to make binding agreements. It further allows learning as the outcome of each interaction conveys information about actions and preferences of other players.

\emph{(ii) Do players make decisions simultaneously or sequentially?} \index{game!simultaneous move game} Simultaneous move games arise when players make decisions simultaneously, without being aware of the choices and preferences of others. For instance, two firms independently decide whether to develop and market a new product, or individuals independently make a choice as which social club to join. In contrast, moving after another player in sequential games gives the advantage of knowing and accounting for the previous players' actions prior to taking an action.

\emph{(iii) What does each player realize/observe once the interaction is concluded?} Embodied in the answer to this question is the social knowledge of each player. With the prevalence of social networks, individuals may obtain information about the preferences and choices of (some of) other decision makers in addition to the payoff that they realize as the outcome of making choices. A particular choice made by a self-interested rational player is understood to be a best response to his limited perception of the outside world. Observing such choices, other players could adjust their strategies so as to maximize their payoffs via decreasing others.

In this chapter, we concentrate on \emph{simultaneous move} non-cooperative games that are \emph{infinitely repeated}. Three different models will be studied for the information flow and social knowledge of players. The first model assumes \emph{social} agents who observe the decisions of all agents playing the same game. The second model considers \emph{social groups}, within which agents share their decisions. Hence, each agent is only aware of successive decisions of a subset of the agents who play the same game. The third model assumes formation of \emph{homogeneous social groups}, where agents share identical interests and exchange their beliefs.

\subsection{Solution Concepts: Correlated vs. Nash Equilibrium}
\label{sec:CE}
Having framed the interactions as a non-cooperative game, one can then make predictions about the behavior of decision makers. Strategic dominance is the most powerful analytical tool to make behavioral predictions. A \emph{dominant strategy} is one that ensures the highest payoff for an agent irrespective of the action profile of other players. A more precise definition is provided below.

\begin{definition}[Dominant Strategy]
An action $\act^\plyrind_{_d}\in\actset^\plyrind$ is a dominant strategy for player $\plyrind$ if for all other actions $\act^\plyrind\in\actset^\plyrind-\big\lbrace\act^\plyrind_{_d}\big\rbrace$:
\begin{equation}
\label{eq:dominant}
\utilityk\big(\act^\plyrind_{_d},\actprof^{-\plyrind}\big) > \utilityk\big(\act^\plyrind,\actprof^{-\plyrind}\big),\quad\forall\actprof^{-\plyrind} \in\actset^{-\plyrind}.
\end{equation}
\end{definition}
Dominant strategies make game-theoretic analysis relatively easy. They are also powerful from the decision makers perspective since no prediction of others' behavior is required. However, such strategies do not always exist, hence, one needs to resort to equilibrium notions in order to make behavioral predictions.

The Nash equilibrium~\cite{N51} and its refinements are without doubt the most well-known game-theoretic equilibrium notion in both  economics and engineering. One can classify Nash equilibria into two types: (i) \emph{Pure strategy Nash equilibria}, where all agents are playing pure strategies, and (ii) \emph{Mixed strategy Nash equilibria}, where at least one agent  is playing a mixed strategy. The underlying assumption in Nash equilibrium is that players act independently. That is, the probability distribution on the action profiles is a product measure. More precisely,
\begin{equation}
\label{eq:independence}
\mixedstratind\big(i_1,\cdots,i_\fplyr\big) \ole \mathbb{P}\left(\act^1 = i_1,\cdots,\act^\fplyr = i_\fplyr\right) = \prod_{k=1}^{\fplyr} \mixedstratind^\plyrind(i_\plyrind)
\end{equation}
where $\mixedstratind^\plyrind(i_\plyrind)$ denotes the probability of agent $\plyrind$ playing action $i_\plyrind$.

A Nash equilibrium is a strategy profile with the property that no single player can benefit by deviating unilaterally to another strategy; see below for precise definitions.
\begin{definition}[Nash Equilibrium]
An action profile $\actprof = \big(\actk,\actprof^{-\plyrind}\big)$ is a pure strategy Nash equilibrium \index{Nash equilibrium} if
\begin{equation}
\label{eq:nash}
\utilityk\big(\actk,\actprof^{-\plyrind}\big) \geq \utilityk\big(\underline{\act}^\plyrind,\actprof^{-\plyrind}\big),\quad\forall \underline{\act}^\plyrind\in\actset^\plyrind,\;\forall\plyrind\in\plyrset.
\end{equation}
A profile of mixed strategies $\mixedstrat = \big(\mixedstrat^1,\cdots,\mixedstrat^\fplyr\big)$ forms a mixed strategy Nash equilibrium if for all ${\underline{\act}}^\plyrind\in\actset^\plyrind$ and $\plyrind\in\plyrset$:
\begin{equation}
\label{eq:nash-mixed}
\sum_{\actk,\actprof^{-\plyrind}}\mixedstratind\big(\actk,\actprof^{-\plyrind}\big) \utilityk\big(\actk,\actprof^{-\plyrind}\big) \geq \sum_{\actprof^{-\plyrind}}\mixedstratind\big(\underline{\act}^k,\actprof^{-\plyrind}\big) \utilityk\big(\underline{\act}^\plyrind,\actprof^{-\plyrind}\big).
\end{equation}
\end{definition}

John F. Nash proved in his famous paper~\cite{N51} that every game with a finite set of players and actions has at least one mixed strategy Nash equilibrium.
However, as asserted by Robert J. Aumann\footnote{Robert J. Aumann was awarded the Nobel Memorial Prize in Economics in 2005 for his work on conflict and cooperation through game-theoretic analysis. He is the first to conduct a full-fledged formal analysis of the so-called infinitely repeated games.}
in the following extract from~\cite{Aum87}, ``Nash equilibrium does make sense if one starts by assuming that, for some specified reason, each player knows which strategies the other players are using.'' Evidently, this assumption is rather restrictive and, more importantly, is rarely true in any strategic interactive situation. He adds:
\begin{quote} {\em
``Far from being inconsistent with the Bayesian view of the world, the notion of equilibrium is an unavoidable consequence of that view. It turns out, though, that the appropriate equilibrium notion is not the ordinary mixed strategy equilibrium of Nash (1951), but the more general notion of correlated equilibrium.''} -- Robert J. Aumann
\end{quote}
This, indeed, is the very reason why correlated equilibrium~\cite{Aum87} best suits and is central to the analysis of strategic decision-making in social networks in this chapter.

Correlated equilibrium \index{correlated equilibrium} is a generalization of the well-known Nash equilibrium and describes a condition of competitive optimality among decision makers. It can most naturally be viewed as imposing equilibrium conditions on the joint action profile of agents (rather than on individual actions as in Nash equilibrium). An intuitive interpretation of correlated equilibrium is coordination in decision-making as described by the following example: Consider a traffic game where two drivers meet at an intersection and have to decide whether to ``go'' or ``wait''. If they simultaneously decide to go, they will crash and both incur large losses. If one waits and the other one goes, the one that passes the intersection receives positive utility, whereas the one who waits receive zero utility. Finally, if they both decide to wait, they both will receive small negative utilities due to wasting their time. The natural solution is the traffic light. In game-theoretic terminology, the traffic light serves as a fair randomizing device that recommends actions to the agents---signals one of the cars to go and the other one to wait; it is fair in the sense that at different times it makes different recommendations.

In light of the above example, suppose that a mediator is observing a non-cooperative game being repeatedly played among a number of agents. The mediator, at each round of the game, gives private recommendations as what action to choose to each agent. The recommendations are correlated as the mediator draws them from a joint probability distribution on the action profile of all agents; however, each agent is only given recommendations about
her own action. Each agent can freely interpret the recommendations and decide if to follow. A correlated equilibrium results if neither of agents wants to deviate from the provided recommendation.
That is, in correlated equilibrium, agents' decisions are coordinated as if there exists a global coordinating device that all agents trust to follow.

Let us now formally define the set of correlated equilibrium for the game~$\game$, defined in \S\ref{sec:strategic-form}.
\begin{definition}[Correlated Equilibrium]
\label{def:CE}
Let $\mixedstrat$ denote a probability distribution on the space of action profiles $\actset$, i.e.,
\begin{equation}
\textstyle
0\leq \mixedstratind(\actprof) \leq 1, \forall \actprof\in\actset,\; \textmd{and}\; \sum_{\actprof} \mixedstratind(\actprof) = 1.
\end{equation}
The set of correlated $\epsilon$-equilibria \index{correlated equilibrium} is the convex polytope
{\small
\begin{equation}
\label{eq:CE_defn}
\hspace{-0.1cm}\CEe = \lbr \mixedstrat:
\sum_{\actprof^{-\plyrind}} \mixedstratind^\plyrind\big(i, \actprof^{-\plyrind}\big)\left[U^{\plyrind}\big(j, \actprof^{-\plyrind}\big) - U^\plyrind\big(i, \actprof^{-\plyrind}\big)\right] \leq \epsilon, \forall i,j\in\actset^\plyrind, \plyrind\in\plyrset \rbr
\end{equation}
}

\vspace{-0.3cm}
\noindent
where $\mixedstratind^\plyrind\big(i, \actprof^{-\plyrind}\big)$ denotes the probability of player $\plyrind$ playing action $i$ and the rest playing $\actprof^{-\plyrind}$. If $\epsilon = 0$ in~(\ref{eq:CE_defn}), the convex polytope represents the set of correlated equilibria and is denoted by $\CE$.
\end{definition}

There is much to be said about correlated equilibrium; see Aumann~\cite{A74,Aum87} for rationality arguments. Some advantages that make it ever more appealing include:
\begin{enumerate}
    \item \emph{Realistic:} Correlated equilibrium is realistic in multi-agent learning. Indeed, Hart and Mas-Colell observe in~\cite{HM01a} that for most simple adaptive procedures, ``\ldots there is a natural coordination device: the common history, observed by all players. It is thus reasonable to expect that, at the end, independence among players will not obtain;''

    \item \emph{Structural Simplicity:} The correlated equilibria set constitutes a compact convex polyhedron, whereas the Nash equilibria are isolated points at the extrema of this set~\cite{NCH04};

    \item \emph{Computational Simplicity:} Computing correlated equilibrium only requires solving a linear feasibility problem (linear program with null objective function) that can be done in polynomial time, whereas computing Nash equilibrium requires finding fixed points;

    \item \emph{Payoff Gains:} The coordination among agents in the correlated equilibrium can lead to potentially higher payoffs than if agents take their actions independently (as required by Nash equilibrium)~\cite{Aum87};

    \item \emph{Learning:} There is no natural process that is known to converge to a Nash equilibrium in a general non-cooperative game that is not essentially equivalent to exhaustive search. There are, however, natural processes that do converge to correlated equilibria (the so-called law of conservation of coordination~\cite{HM03}), e.g., \index{regret-matching procedure} regret-matching~\cite{HM00}.
\end{enumerate}

\begin{figure}[!t]
\centerline{\includegraphics[width=0.8\textwidth]{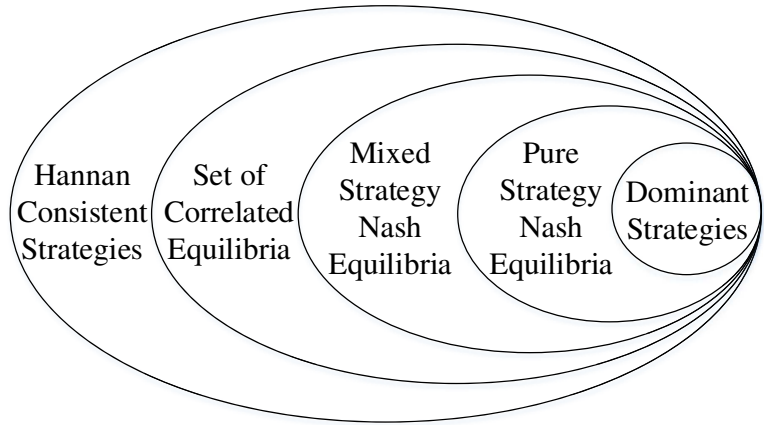}}
\caption{Equilibrium notions in non-cooperative games. Enlarging the equilibria set weakens the behaviorial sophistication on the player's part to distributively reach equilibrium through repeated plays of the game.}
\label{fig:strategy-sets}
\end{figure}

Existence of a centralized coordinating device neglects the distributed essence of social networks. Limited information at each agent about the strategies of others further complicates the process of computing correlated equilibria. In fact, even if agents could compute correlated equilibria, they would need a mechanism that facilitates coordinating on the same equilibrium state in the presence of multiple equilibria---each describing, for instance, a stable coordinated behavior of manufacturers on targeting influential nodes in the competitive diffusion process~\cite{TAM12}. This highlights the significance of adaptive learning algorithms \index{learning!adaptive learning} that, through repeated interactive play and simple strategy adjustments by agents, ensure reaching correlated equilibrium. The most well-known of such algorithms, fictitious play, was first introduced in 1951~\cite{R51}, and is extensively treated in~\cite{FL98}. It, however, requires monitoring the behavior of all other agents that contradicts the information exchange structure in social networks. The focus of this chapter is on the more recent regret-matching learning algorithms~\cite{BHS06,C04,HM00,HM01a}. We use tools from stochastic approximation~\cite{KY03}, to adapt these algorithms to the information exchange structure in social networks, and Markov-switched systems~\cite{YZ98,YZ10}, to track time-varying equilibria under evolution of the environment and the social network.

Figure~\ref{fig:strategy-sets} illustrates how the various notions of equilibrium are related in terms of the relative size and inclusion in other equilibria sets. As discussed earlier in this subsection, dominant strategies and pure strategy Nash equilibria do not always exist---the game of ``Matching Pennies'' being a simple
example. Every finite game, however, has at least one mixed strategy Nash equilibrium. Therefore, the ``nonexistence critique'' does not apply to any notion that generalizes the mixed strategy Nash equilibrium in Figure~\ref{fig:strategy-sets}. A Hannan consistent strategy (also known as ``universally consistent'' strategies~\cite{FL95}) is one that ensures, no matter what other players do, the player's average payoff is asymptotically no worse than if she were to play any \emph{constant} strategy for in all previous periods. Hannan consistent strategies guarantee no asymptotic external regrets and lead to the so-called ``coarse correlated equilibrium''~\cite{MV78} notion that generalizes the Aumann's correlated equilibrium.
%


\section{Local Adaptation and Learning}
\label{sec:local-behavior-game}

\begin{figure}[!t]
\centerline{\includegraphics[width=1.1\textwidth]{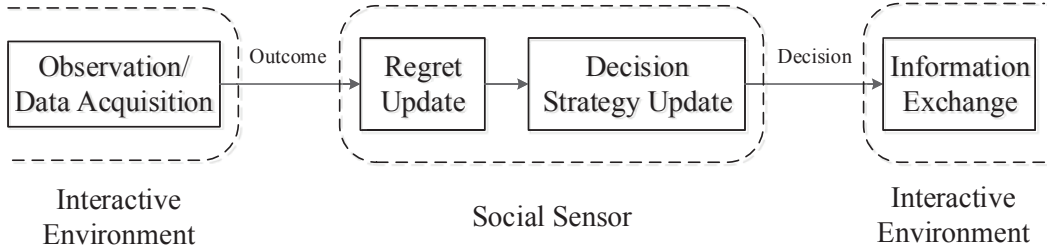}}
\caption{Regret-based local adaptation and learning. Individuals update regrets after observing the outcome of the previous decisions and, accordingly, make next decisions. These new decisions will then be shared within social groups.}
\label{fig:social-sensor-game}
\end{figure}

This section presents an adaptive decision-making procedure that, if locally followed by every individual decision maker, leads to rational global behavior. The global behavior is meant to capture the decision strategies of all individuals taking part in the decision-making process, and will be made clear in the next section. This procedure is viewed as particularly simple and intuitive as no sophisticated updating,  prediction, or fully rational behavior is essential on the part of agents. The procedure can be simply described as follows: At each period, an agent may either continue with the same action as the previous period, or switch to other actions, with probabilities that are proportional to a ``regret measure''~\cite{HM00}; see Figure~\ref{fig:social-sensor-game}. Willingness to proceed with the previous decision mimics the ``inertia'' that is existent in human's decision-making process. In what follows, we present three algorithms, in the ascending level of sophistication, that adapt this simple procedure to various social network architectures.

The communication among agents (hence, the level of ``social knowledge'') can be captured by a connectivity graph \index{graph!connectivity graph}.
Below, a formal definition is given:

\begin{definition}[Connectivity Graph]
\label{def:connectivity-graph}
It is an undirected graph $\gamegraph = (\gamevertex,\gameedge)$, where $\gamevertex = \plyrset$ is the set of agents, and
\begin{equation}
\plyrind, l \in \gameedge \Leftrightarrow \plyrind\; \textmd{knows}\; \act^{l}_\dtimee\;\textmd{and}\; l\;\textmd{knows}\;\actk_\dtimee\; \textmd{at the end of period}\; k.\nonumber
\end{equation}
\end{definition}
Accordingly, the neighborhood of each agent $\plyrind$ is defined as:
\begin{equation}
\label{eq:neighbors}
\begin{array}{ll}
\textmd{Open Neighborhood:} & \neighborhoodk \ole \lbr l\in\plyrset ; (\plyrind,l)\in\gameedge\rbr\\
\textmd{Closed Neighborhood:} & \cneighborhoodk \ole \neighborhoodk \cup \lbr \plyrind\rbr
\end{array}
\end{equation}
The communication protocol which allows agents to exchange information only with neighbors is referred to as \emph{neighborhood monitoring}~\cite{NKY13a}. This generalizes the perfect monitoring assumption, standard in the game theory literature~\cite{HM01b}, where agents share decisions with all other agents at the end of each period\footnote{Note that sharing actions differs from exchanging strategies according to which decisions are made; the latter is more cumbersome and less realistic, however, conveys information about the payoff functions of opponents, hence, can lead to faster coordination.}.

\subsection{Case I: Fully Social Players}
\label{sec:social-game}

For simplicity of exposition, we start with the adaptive learning procedure \index{learning!adaptive learning} for the scenario with the highest level of social knowledge: \emph{perfect monitoring}. That is, once a decision is made by an agent, it is observable by all other agents\footnote{In certain situations, the aggregate decisions of all agents affects individual agent's strategy update, e.g., proportion of neighbors adopting a new technology or proportion of neighboring sensors that have chosen to activate. Therefore, one can assume that this aggregate information is available to each agent (instead of individual decisions) at the end of each period.}. More precisely,
\begin{equation}
\label{eq:fully-social}
\textmd{The graph $\gamegraph$ is complete and } \cneighborhoodk = \plyrset.
\end{equation}
Although this is a rather impractical assumption in the emerging large-scale social networks, it sheds light and provides insight on the grounds of the decision-making procedure and how it can be deployed in environments with more complex connectivity architecture.

Suppose the game $\game$, defined in \S\ref{sec:strategic-form}, is being repeatedly played over time: $k = 1,2,\ldots$. At time $k$, each agent $\plyrind$ makes a decision $\actk_\dtimee\in\actset^\plyrind$ according to a probability distribution $\mixedstrat^\plyrind_\dtimee$, which is defined via the following ``regret-matching'' \index{regret-matching procedure} procedure~\cite{HM00}:

Let $I_{\lbr\cdot\rbr}$ denote the indicator operator and $0<\stepsize<1$ be a small parameter. Define the matrix $\regmatplyr_\dtimee$, where each element $r^n_\dtimee(i,j)$ records
\begin{equation}
\label{eq:regret-def}
r^n_\dtimee(i,j) = \stepsize \sum_{\tau \leq k} (1-\stepsize)^{k-\tau} \lb\utilityk\big(j,\actprof^{-\plyrind}_\tau\big) - \utilityk \big(\actprof_\tau\big)\rb \indicatoritau.
\end{equation}
%
The above expression for $r^n_\dtimee(i,j)$ has a clear interpretation as a measure of the
``regret'' that agent $\plyrind$ would experience for not having played action $j$, every time action $i$ was picked in the past. The strategy $\mixedstrat^\plyrind_\dtimee = \big(\mixedstratind^\plyrind_\dtimee(1),\cdots,\mixedstratind^\plyrind_{\dtimee}(\fact)\big)$ is then defined as follows
\begin{equation}
\label{eq:strategy-MC}
\begin{split}
\mixedstratind^{\plyrind}_\dtimee(i) &\ole \mathbb{P}\big(\act^\plyrind_\dtimee = i\big| \act^\plyrind_{\dtimee-1} = j\big)\\
&= \left\{
\begin{array}{ll}
\frac{1}{\inertia^\plyrind} \big|r^\plyrind_\dtime(j,i)\big|^{+}, & i\neq j\\
1-\sum_{j\neq i} p^{\plyrind}_\dtimee(j), & i = j
\end{array}
\right.
\end{split}
\end{equation}
where $\left|x\right|^{+} = \max\lbrace0,x\rbrace$, and $\inertia^\plyrind >0$ is a large enough number so
that $\mixedstrat^\plyrind_\dtimee$ is a valid probability mass function\footnote{It suffices to let $\inertia^\plyrind > \fact\left|\utilityk_{\textmd{max}} - \utilityk_{\textmd{min}} \right|$, where $\utilityk_{\textmd{max}}$ and $\utilityk_{\textmd{min}}$ represent the upper and lower bounds on the payoff function $\utilityk\cd$ for agent $\plyrind$, respectively.}.

Intuitively speaking, the regret-matching procedure governs the decision-making process by propensities to depart from the current decision. It is natural to postulate that, if a decision maker decides to switch her action, it should be to actions that are perceived as being better. The regret-matching procedure assigns positive probabilities to all such better choices. In particular, the better an alternative action seems, the higher will be the probability of choosing it next time. More precisely, the probabilities of switching to different actions are proportional to their regrets relative to the current action.

The regret-matching \index{regret-matching procedure} procedure in the case of fully social players is summarized in Algorithm~\ref{alg:social-player}.

\begin{algorithm}[H]
Let $\oneK = [1,\cdots,1]^T$ denote an $N\times 1$ column vector of ones, $\utilityk_{\textmd{max}}$ and $\utilityk_{\textmd{min}}$ represent the upper and lower bounds on the payoff function $\utilityk\cd$ for agent $\plyrind$, respectively, and $I_{\lbr\cdot\rbr}$ denote the indicator function.\\

\textbf{Initialization}:  Set $\inertia^\plyrind > \fact\left|\utilityk_{\textmd{max}} - \utilityk_{\textmd{min}} \right|$, $\mixedstrat^\plyrind_0 = {1\over \fact}\cdot\mathbf{1}_{\fact}$, and $\regmatplyr_0 = \mathbf{0}$.\\


\textbf{Step 1: Action Selection}.
Choose $\act^\plyrind_\dtimee\sim\mixedstrat^\plyrind_\dtimee$, where
\begin{equation}
\mixedstratind^{\plyrind}_\dtimee(i) = \left\{
\begin{array}{ll}
\frac{1}{\inertia^\plyrind} \left|r^\plyrind_\dtimee\big(\act^\plyrind_{\dtimee-1},i\big)\right|^{+}, & i\neq \act^\plyrind_{\dtimee-1}\\
1-\sum_{j\neq i} p^{\plyrind}_\dtimee(j), & i = \act^\plyrind_{\dtimee-1}
\end{array}
\right.
\end{equation}
where $\left|x\right|^{+} = \max\lbrace0,x\rbrace$.

\textbf{Step 2: Information Exchange}. Share decisions $\act^\plyrind_\dtimee$ with others.

\textbf{Step 3: Regret Update}.
\begin{equation}
\label{eq:reg-update}
\regmatplyr_{\dtimenext} = \regmatplyr_\dtimee + \stepsize\lb B^\plyrind \left(\actprof_\dtimee\right) - \regmatplyr_\dtimee \rb
\end{equation}
where $B^\plyrind \left(\actprof_\dtimee\right) = \big[ b^{\plyrind}_{ij}\left(\actprof_\dtimee\right)\big]$ is an $\fact\times\fact$ matrix with elements
\begin{equation}
\label{eq:B-game}
b^{\plyrind}_{ij}\left(\actprof_\dtimee\right) = \lb \utilityk\big(j,\actprof^{-\plyrind}_\dtimee\big) - \utilityk\big(i,\actprof^{-\plyrind}_\dtimee\big)\rb\cdot\indicatori.\\
\end{equation}

\textbf{Step 4: Recursion}.
Set $k\leftarrow k+1$, and go Step 1.
 \caption{Local Adaptation and Learning for Social Players}
 \label{alg:social-player}
\end{algorithm}

\subsection*{Discussion and Intuition:}

\emph{1) Adaptive Behavior:} In~(\ref{eq:regret-def}), $\stepsize$ serves as a forgetting factor to foster adaptivity to the evolution of the non-cooperative game parameters or the social network architecture. That is, as agents repeatedly take actions, the effect of the old underlying parameters on their current decisions vanishes; see~\S\ref{sec:regime-switching-game} for a somewhat detailed treatment of regime switching non-cooperative games.

\emph{2) Inertia:} The choice of $\inertia^\plyrind$ guarantees that there is always a positive probability of playing the same action as the last period. Therefore, $\inertia^\plyrind$ can be viewed as an ``inertia'' parameter: A higher $\inertia^\plyrind$
yields switching with lower probabilities. It plays a significant role in breaking away from bad cycles. It is worth emphasizing that the speed of convergence to the correlated equilibria set is closely related to this inertia parameter.

\emph{3) Better-reply vs. Best-reply:} In light of the above discussion, the most distinctive feature of the regret-matching procedure, that differentiates it from other works such as~\cite{FV97,FL98,FL99a}, is that it implements a better-reply rather than a best-reply strategy\footnote{This has the additional effect of making the behavior
continuous, without need for approximations~\cite{HM00}.}. This inertia assigns positive probabilities to any actions that are just better. Indeed, the behavior of a regret-matching decision maker is very far from that of a rational decision maker that makes optimal decisions given his (more or less well-formed) beliefs about the environment. Instead, it resembles the model of a reflex-oriented individual that reinforces decisions with ``pleasurable'' consequences~\cite{HM01a}.


\emph{(4) Computational Complexity:} The computational burden (in terms of calculations per iteration) of the ``regret-matching'' procedure is small. It does not grow with the number of agents, hence, is scalable. At each iteration, each agent needs to execute two multiplications, two additions, one comparison and two table lookups (assuming random numbers are stored in a table) to calculate the next decision. Therefore, it is suitable for implementation in sensors with limited local computational capability.

\emph{(5) Decentralized Adaptive Filtering:} \index{stochastic approximation!adaptive filtering} As will be seen later in \S\ref{sec:main-results-game}, if every agent follows Algorithm~\ref{alg:social-player}, the global behavior converges to the set of correlated equilibria. Theoretically, finding a correlated equilibrium in a game is equivalent to solving a linear feasibility problem; see~(\ref{eq:CE_defn}). Algorithm~\ref{alg:social-player} is thus an adaptive filtering algorithm for solving this problem, which is quite interesting due to its decentralized nature.

\subsection{Case II: Social Groups}
\label{sec:social-cliques}
\index{social group}
As pointed out above, the ``perfect monitoring'' assumption is quite restrictive in the emerging large-scale social networks. A natural extension is to devise an adaptive learning procedure that relies on the ``neighborhood monitoring'' model for exchange of information among agents. That is, once a decision is made by agent $\plyrind$, it is only observable by her neighbors $\neighborhoodk$ on the connectivity graph $\gamegraph$; see Definition~\ref{def:connectivity-graph}. In fact, agents are oblivious to the existence of other agents except their neighbors.

Social groups are characterized by the coalition of agents that perform the same localized task and interact locally. However, not only does the payoff of each agent depend on her local interaction and contribution to carrying out the local tasks, but also it is affected by the inherent interaction with those outside her social group. In particular, the payoff that each agent receives as the outcome of her decision comprises of two terms: (i) \emph{Local payoff}, due to local interaction (e.g., performing tasks) within social groups; (ii) \emph{Global payoff}, due to the implicit strategic global interaction with agents outside social group. More precisely, the payoff function can be expressed as~\cite{NKY13a}
%
\begin{equation}
\label{eq:utility-neighb-monitor}
\utilityk\big(\actk,\actprof^{-\plyrind}\big) = \utilitykL\big(\actk,\actprof^{\neighborhoodk}\big) + \utilitykG\big(\actk,\actprof^{\nonneighbork}\big).
\end{equation}
Here, $\nonneighbork = \plyrset - \cneighborhoodk$ denotes the set of ``non-neighbors'' of agent $\plyrind$.

Now, suppose the game $\game$, defined in \S\ref{sec:strategic-form}, with payoff function~(\ref{eq:utility-neighb-monitor}) is being repeatedly played over time: $k = 1,2,\ldots$. At time $k$, each agent $\plyrind$ makes a decision $\actk_\dtimee$ and receives a payoff\footnote{Note that $\utilitykn\cd = \utilityk\big(\cdot,\actprof^{-\plyrind}_\dtimee\big)$, i.e., the time-dependency of $\utilitykn\cd$ corresponds to the time-variations of other agents' action profile $\actprof^{-\plyrind}_\dtimee$.} $\utilitykn\big(\actk_\dtimee\big)$. Assume that agents know their local payoff functions; hence, knowing the action profile of neighbors, they are able to evaluate their local stage payoffs. In contrast, even if agents know their global payoff functions, they could not directly compute global payoffs as they do not acquire the action profile of agents outside their social groups. However, they can compute their \emph{realized} global payoffs by
\begin{equation}
\label{eq:realized-global}
\utilitykGn\big(\actk_\dtimee\big) \ole \utilitykn\big(\actk_\dtimee\big) - \utilitykL\big(\actk,\actprof^{\neighborhoodk}_\dtimee\big).
\end{equation}

\begin{figure}[!thbp]
\centerline{\includegraphics[width=0.75\textwidth]{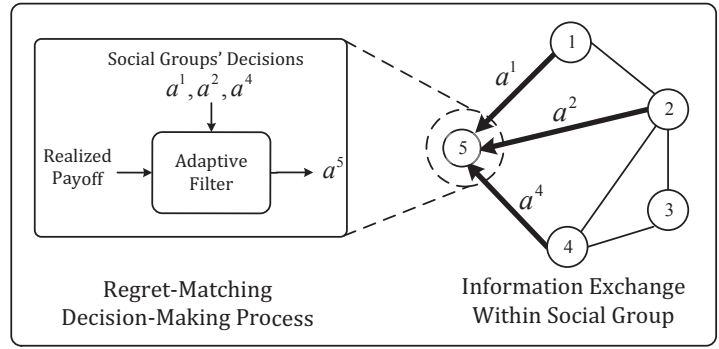}}
\caption{Game-theoretic learning within social groups. Circles represent individuals in the social network. Agent 5 exchanges decision with her social group members $N^5 = \lbr 1,2,4\rbr$. The exchanged decisions, together with agent 5's realized payoff, are then fed into the regret-matching adaptive filter to determine the next action for agent 5.}
\label{fig:game_learning}
\end{figure}

The ``revised regret-matching'' procedure, depicted in Figure~\ref{fig:game_learning}, is then as follows~\cite{NKY13a,HM01a}: The local regret matrix $\regmatplyrL_\dtimee$ is defined as before; that is, each element $r^{L,\plyrind}_\dtimee(j,i)$ records
\begin{equation}
\label{eq:local-regret}
r^{L,\plyrind}_\dtimee(j,i) = \stepsize \sum_{\tau\leq k} (1-\stepsize)^{k-\tau} \lb\utilitykL\big(j,\actprof^{\neighborhoodk}_\tau\big) - \utilitykL\big(\actk_\tau,\actprof^{\neighborhoodk}_\tau\big)\rb \indicatoritau.
\end{equation}
However, since agent $\plyrind$ knows neither her global payoff function nor the decisions $\actprof^{\nonneighbork}_\dtimee$ of those outside her social group, she is unable to perform the thought experiment to evaluate global payoffs for alternative actions as she does in~(\ref{eq:local-regret}). She thus employs an unbiased estimator $\regmatplyrG_\dtimee$ of the global regrets, which relies only on the realizations of global payoffs $\big\lbrace \utilitykGn(\actk_\dtimee)\big{\rbrace}$. More precisely, each element $r^{G,\plyrind}_\dtimee(j,i)$ can be expressed as follows:
{\small
\begin{align}
\label{eq:global-regret}
\begin{split}
r^{G,\plyrind}_\dtimee(j,i) = &\stepsize \sum_{\tau\leq k} (1-\stepsize)^{k-\tau} \\ &\times \lb \frac{\mixedstratind^\plyrind_\tau(i)}{\mixedstratind^\plyrind_\tau(j)}\utilitykGtau\big(\actk_\tau\big)\indicatorjtau - \utilitykGtau\big(\actk_\tau\big)\indicatoritau\rb.
\end{split}
\end{align}
}

\vspace{-0.2cm}
\noindent
The agent combines local and global regrets
\begin{equation}
\label{eq:combined-regrets}
\regmatplyr_\dtimee \ole \regmatplyrL_\dtimee+\regmatplyrG_\dtimee
\end{equation}
to update her strategy $\mixedstrat^{\plyrind}_\dtimee$ as follows: Let $0 < \explor < 1$, and recall $\fact$ represents the cardinality of the action set $\actset^\plyrind$. The play probabilities are then given by
\begin{equation}
\label{eq:strategy-MC-social-group}
\begin{split}
\mixedstratind^{\plyrind}_\dtimee(i) &= \mathbb{P}\big(\act^\plyrind_\dtimee = i\big| \act^\plyrind_{\dtimee-1} = j\big)\\
&= \left\{
\begin{array}{ll}
(1-\explor)\min\lbr \frac{1}{\inertia^\plyrind}\big|r^\plyrind_\dtimee(j,i)\big|^{+},\frac{1}{\fact}\rbr + \frac{\explor}{\fact}, & i\neq j\\
1-\sum_{j\neq i} p^{\plyrind}_\dtimee(j), & i = j
\end{array}
\right.
\end{split}
\end{equation}
where, as before, $\left|x\right|^{+} = \max\lbrace0,x\rbrace$, and $\inertia^\plyrind >0$ is a large enough number.

The local adaptation and learning algorithm within social groups is summarized in Algorithm~\ref{alg:social-clique}.

\begin{algorithm}[H]

\textbf{Initialization}:   Set $0 < \explor < 1$, $\inertia^\plyrind > \fact\left|\utilityk_{\textmd{max}} - \utilityk_{\textmd{min}} \right|$, $\mixedstrat^\plyrind_0 = {1\over \fact}\cdot\mathbf{1}_{\fact}$, and $\regmatplyrL_0 = \regmatplyrG_0 = \mathbf{0}$.\\


\textbf{Step 1: Action Selection}.
Choose $\act^\plyrind_\dtimee$ according to the mixed strategy $\mixedstrat^\plyrind_\dtimee$:
{\small
\begin{equation}
\begin{split}
\mixedstratind^{\plyrind}_\dtimee(i)
= \left\{
\begin{array}{ll}
(1-\explor)\min\lbr \frac{1}{\inertia^\plyrind}\left|r^{\plyrind}_\dtimee\left(\act^\plyrind_{\dtimee-1},i\right)\right|^{+},\frac{1}{\fact}\rbr + \frac{\explor}{\fact}, & i\neq \act^\plyrind_{\dtimee-1}\\
1-\sum_{j\neq i} p^{\plyrind}_\dtimee(j), & i = \act^\plyrind_{\dtimee-1}
\end{array}
\right.
\end{split}
\end{equation}
}

\vspace{-0.3cm}
\noindent
where $\left|x\right|^{+} = \max\lbrace0,x\rbrace$.

\textbf{Step 2: Information Exchange}. Share decisions $\act^\plyrind_\dtimee$ with neighbors $\neighborhoodk$.

\textbf{Step 3: Regret Update}.

(i) For $m\in\lbrace L, G\rbrace$:
\begin{equation}
\label{eq:reg-update-2}
\regmatplyrm_{\dtimenext} = \regmatplyrm_\dtimee + \stepsize\lb B^{m,\plyrind} \left(\actprof_\dtimee\right) - \regmatplyrm_\dtimee \rb
\end{equation}
where $B^{m,\plyrind} \left(\actprof_\dtimee\right) = \big[ b^{m,\plyrind}_{ij}\left(\actprof_\dtimee\right)\big]$ is an $\fact\times\fact$ matrix:
\begin{equation}
\label{eq:B-social-clique}
\begin{split}
b^{L,\plyrind}_{ij}\left(\actprof_\dtimee\right) &= \lb \utilitykL\big(j,\actprof^{-\plyrind}\big) - \utilitykL\big(i,\actprof^{-\plyrind}\big)\rb \indicatori\\
b^{G,\plyrind}_{ij}\left(\actprof_\dtimee\right) &= \bigg[ \frac{\mixedstratind^{\plyrind}_\dtimee(i)}{\mixedstratind^{\plyrind}_\dtimee(j)} \utilitykGn\big(\act^\plyrind_\dtimee\big) \indicatorj- \utilitykGn\big(\act^\plyrind_\dtimee\big) \indicatori\bigg]\\
\end{split}
\end{equation}
(ii) Combine local and global regrets:
\begin{equation}
\regmatplyr_\dtimee = \regmatplyrL_\dtimee+\regmatplyrG_\dtimee.
\end{equation}

\vspace{-0.1cm}
\textbf{Step 4: Recursion}.
Set $k\leftarrow k+1$, and go Step 1.
 \caption{Local Adaptation and Learning Within Social Groups}
 \label{alg:social-clique}
\end{algorithm}

\subsection*{Discussion and Intuition:}

Before proceeding, we should emphasize that the remarks on the ``regret-matching'' procedure~\cite{HM01a}, discussed in \S\ref{sec:social-game}, will continue to hold for the revised regret-matching algorithm in the presence of social groups.

\emph{(1) Global Regret Estimate:} Close scrutiny of~(\ref{eq:global-regret}) reveals that the second term is similar to the second term in~(\ref{eq:local-regret}), and simply evaluates the weighted average global payoffs realized in periods when action $i$ was picked. The first term, roughly speaking, estimates the weighted average global payoffs if action $j$ replaced action $i$ in exactly those periods when action $i$ was chosen\footnote{Strictly speaking, the limit sets of the process (\ref{eq:global-regret}) and the true global regret updates (if agents observed the action profile of agents outside their social group) coincide; see~\cite{NKY13a} for details. Alternatively, the conditional expectation of the difference between the global regret estimates and the true global regrets can be proved to be zero; see~\cite{HM01a} for details.}. To this end, it only uses realized payoffs of those periods when action $j$ was actually played. The normalization factor $\mixedstratind^\plyrind_\tau(i)/\mixedstratind^\plyrind_\tau(j)$, intuitively speaking, makes the lengths of the respective periods comparable~\cite{HM01a}.

\emph{(2) Play Probabilities:} The randomized strategy~(\ref{eq:strategy-MC-social-group}) is simply a weighted average of two probability vectors: The first term, with weight $1 - \explor$, is proportional to the positive part of the combined regrets $\big|\regmatplyrL_\dtimee+\regmatplyrG_\dtimee\big|^+$ in a manner similar to (\ref{eq:strategy-MC}). Taking the minimum with
$\frac{1}{\fact}$ guarantees that $\mixedstrat^\plyrind_\dtimee$ is a valid probability distribution, i.e., $\sum_{i} \mixedstratind^\plyrind_\dtimee(i) = 1$. The second term, with weight $\explor$, is just a uniform distribution over the action space $\actset^\plyrind$~\cite{NKY13a,HM01a}.

\emph{(3) Reinforcement Learning:} Assuming that no player is willing to form a social group, Algorithm~\ref{alg:social-clique} simply implements a reinforcement learning \index{learning!reinforcement learning} procedure that only requires the realized payoff to prescribe the next decision.
The probability of choosing a particular action $i$ at time $\dtimee+1$ gets ``reinforced'' by picking $i$ at time $\dtimee$, while the probabilities of the other actions $j \neq i$ decrease. In fact, the higher the payoff realized as the consequence of choosing $i$ at time $\dtimee$, the greater will be this reinforcement~\cite{HM01a}.

\emph{(4) Exploration vs. Exploitation:} The second term, with weight $\explor$, simply forces every action to be played with some minimal frequency (strictly speaking, with probability $\explor/\fact$). The ``exploration'' factor\footnote{As will be discussed later, larger $\explor$ will lead to the convergence of the global behavior to a larger $\CEdistance$-distance of the correlated equilibria set.} $\explor$ is essential to be able to estimate the contingent payoffs using only the realized payoffs; it can, as well, be interpreted as exogenous statistical ``noise.''

\subsection{Case III: Homogeneous Social Groups}
\label{sec:homogeneous-groups}
\index{social group!homogeneous social group} Next, we consider non-cooperative games in the presence of social groups with closer and more intense ties: Individuals in each social group have identical interests in the sense that they all share, and are aware of sharing, the same payoff function. This opens up opportunity for agents to collaborate within social groups to optimize their payoffs, while retaining their autonomy~\cite{NK13c}.

More precisely, agents form $\fcom$ non-overlapping communities $\communitys\subset\plyrset$, $\comind\in\comset = \lbr 1,\cdots,\fcom\rbr$, each with cardinality $J_s$ and identical payoff functions\footnote{Without loss of generality, agents belonging to the same community are assumed to have identical action sets.}, i.e.,
\begin{equation}
\utilityk\cd = U^{l}\cd \Leftrightarrow\; \plyrind, l\in\communitys\;\;\textmd{for some}\;\;\comind\in\comset.
\end{equation}
Accordingly, agents have to take actions from the same action set within each community, i.e.,
\begin{equation}
\actset^\plyrind = \actset^{l}\cd \Leftrightarrow\; \plyrind, l\in\communitys\;\;\textmd{for some}\;\;\comind\in\comset.
\end{equation}
For notational convenience, denote by $U^s\cd$ and $\actset^s$ the payoff function and action set, respectively, of community $\communitys$. Let further $A^s = \big|\actset^s\big|$, where $|\cdot|$ denotes the cardinality operator.

Within communities, agents form social groups. For notational convenience, we continue to use the connectivity graph $\gamegraph$ in Definition~\ref{def:connectivity-graph} to represent such ``homogeneous social groups.'' Individuals in a social group are more closely tied in the sense that they have identical interests and are willing to share the extra (and strategically of great value) piece of information that they share the same payoff function\footnote{Note that, although individuals within the same social group are mindful of sharing the same payoff function, they are unaware of the strategies according to which group members make their decisions.}. In particular,
\begin{equation}
\label{eq:clique-def}
(\plyrind,l)\in\gameedge \textmd{ in graph } G \Leftrightarrow \textmd{agents $\plyrind$ and $l$ both know}\; \utilityk\cd = U^{l}\cd.
\end{equation}
The social group for each agent $\plyrind$ is represented by the set of neighbors $\cneighborhoodk$ on the connectivity graph $\gamegraph$; see~(\ref{eq:neighbors}). Finally, for clarity of exposition and without loss of generality, we adopt the ``perfect monitoring''~\cite{HM01b} model for information exchange among agents; see~(\ref{eq:fully-social}). It is straightforward for the interested reader to extend the local adaptation and learning algorithm that follows to the case of ``neighborhood monitoring,'' which was discussed in~\S\ref{sec:social-cliques}.

\begin{example}
Consider the problem of multi-target tracking using a network of ZigBee-enabled sensors~\cite{KMY08}. Sensors close to each target share the same goal, i.e., to localize the same target, hence, have identical payoff functions. The entire network thus incorporates several communities of sensors with identical payoff functions, each localizing a particular target. The measurements collected by such communities of sensors are correlated both in time and space. Therefore, nodes can save battery power by choosing to sleep while their neighbors are active. Two advantages directly follow by letting such sensors become social and exchange information within the social groups that they form: (i) Sharing regrets will lead to faster coordination of their activation behavior; (ii) Sharing local estimates will lead to improved localization performance.
\end{example}

In what follows, inspired by the idea of diffusion least mean squares~\cite{LS08,S13}, we enforce cooperation among individuals in each social group via exchanging and fusing regret information.
Such diffusion of regret information within social groups is rewarding for all members as they all share the same payoff function. This provably leads to faster coordination among agents and enables them to respond in real time to changes in the game.

The regret matrix $\regmatplyr$ is defined as in \S\ref{sec:social-game}; see~(\ref{eq:regret-def}). The cooperative diffusion protocol is then implemented as follows: Agents exchange regret information within social groups and fuse the collected data via a linear combiner. More precisely, define the \emph{weight matrix} \index{matrix!weight matrix} for each community $\communitys$ as $\weightmat = \lb \weightkl\rb$ satisfying
\begin{equation}
\label{eq:weight-matrix}
\weightmat = I_{J_s} + \diffstep \Cmat,\quad 0 < \diffstep < 1
\end{equation}
where $\identityK$ denotes the $\fplyr\times\fplyr$ identity matrix, the matrix $\Cmat = \lb \Cmatkl\rb$ is symmetric, i.e., $[\Cmat]^\top = \Cmat$ where $'$ denotes the transpose operator, and satisfies
%
%
\begin{enumerate}
    \item[(i)] $|\Cmatkl|\leq 1$;
    \item[(ii)] $\Cmatkl > 0$ if and only if $(k,l)\in\gameedge$;
    \item[(iii)] $\Cmat\mathbf{1} = \mathbf{0}$, where $\mathbf{1}$ and $\mathbf{0}$ are column vectors of ones and zeros, respectively.
\end{enumerate}
Each row $\plyrind$ of the matrix $\weightmat$ simply gives the weights that agent $\plyrind$ assigns to the regret information collected from neighbors. These weights may depend on the  reputation of the group members~\cite{GBS08,N07,ZV13a}. The fused regret matrix for agent $\plyrind$ at time $k$, denoted by $\regmatplyrdiff_\dtimee$, is then computed via
\begin{equation}
\label{eq:diff-regret}
\regmatplyrdiff_\dtimee = \sum_{l\in\cneighborhoodk} \weightkl \regmatplyrl_\dtimee.
\end{equation}
Recall that all neighbors of agent $\cneighborhoodk$ belong to the same community as agent $\plyrind$.
Therefore, they all share the same payoff function. The fused regrets are then promptly fed back into the stochastic approximation algorithm that updates regrets:
\begin{equation}
\label{eq:reg-update-3}
\regmatplyr_{\dtimenext} = \regmatplyrdiff_\dtimee + \stepsize\lb B^\plyrind \left(\actprof_\dtimee\right) - \regmatplyrdiff_\dtimee \rb
\end{equation}
where $B^\plyrind \left(\actprof_\dtimee\right)$ is a $\fplyr\times\fplyr$  matrix with elements identical to~(\ref{eq:B-game}). Finally, agent $\plyrind$ picks her next action according to the same strategy $\mixedstrat^\plyrind$ as in~(\ref{eq:strategy-MC}). Simply put, agent $\plyrind$ follows the ``regret-matching'' procedure~\cite{HM01a}, summarized in Algorithm~\ref{alg:social-player}, with the exception that fused-regret matrix $\regmatplyrdiff_\dtimee$ replaces the regret matrix $\regmatplyr_\dtimee$.

The cooperative regret-matching procedure in the presence of homogeneous social groups is summarized in Algorithm~\ref{alg:homo-social-clique}.

\begin{algorithm}{}

\textbf{Initialization}:  Set $\inertia^\plyrind > \fact\left|\utilityk_{\textmd{max}} - \utilityk_{\textmd{min}} \right|$, $\mixedstrat^\plyrind_0 = {1\over \fact}\cdot\mathbf{1}_{\fact}$, and $\regmatplyr_0 = \mathbf{0}$.\\


\textbf{Step 1: Action Selection}.
Choose $\act^\plyrind_\dtimee$ according to the mixed strategy $\mixedstrat^\plyrind_\dtimee$:
\begin{equation}
\label{eq:strategy-MC-homog-social-group}
\begin{split}
\mixedstratind^{\plyrind}_\dtimee(i)
= \left\{
\begin{array}{ll}
\frac{1}{\inertia^\plyrind} \big|r^\plyrind_\dtimee\big(\act^\plyrind_{\dtimee-1},i\big)\big|^{+}, & i\neq \act^\plyrind_{\dtimee-1}\\
1-\sum_{j\neq i} p^{\plyrind}_\dtimee(j), & i = \act^\plyrind_{\dtimee-1}
\end{array}
\right.
\end{split}
\end{equation}
where $\left|x\right|^{+} = \max\lbrace0,x\rbrace$.

\textbf{Step 2: Information Exchange}. Share decisions $\act^\plyrind\dtimee$ with others.

\textbf{Step 3: Regret Update}.
\begin{equation}
\label{eq:reg-update-4}
\regmatplyr_{\dtimenext} = \regmatplyrdiff_\dtimee + \stepsize\lb B^\plyrind \left(\actprof_\dtimee\right) - \regmatplyrdiff_\dtimee \rb
\end{equation}
where $B^\plyrind \left(\actprof_\dtimee\right)$ is an $A^s \times A^s$ matrix identical to Algorithm~\ref{alg:social-player}.

\vspace{0.2cm}
\begin{center}
{\em (The first two steps implement the regret-matching procedure~\cite{HM00}.)}
\end{center}
\vspace{0.2cm}

\textbf{Step 4: Regret Fusion}.
\begin{equation}
\label{eq:linear-combiner-2}
\regmatplyrdiff_{\dtimenext} = \regmatplyrdiff_{\dtimee}+ \mu\lb D^\plyrind(\actprof_\dtimee) - \regmatplyrdiff_{\dtimee}\rb
\end{equation}
where $D^{\plyrind}(\actprof_\dtimee)=\big[d^{\plyrind}_{ij}(\actprof_\dtimee)\big]$ is an $A^s \times A^s$ matrix:
\begin{equation}
\label{eq:linear-combiner-3}
d^{\plyrind}_{ij}(\actprof_\dtimee) = \sum_{l\in\neighborkc} \weightkl \left[U^s(j,\actprof^{-l}_\dtimee) - U^s(i,\actprof^{-l}_\dtimee)\right] I\lbr \act^l_\dtimee = i\rbr.
\end{equation}
and $s$ denotes the index of the community to which agent $\plyrind$ belongs.

\vspace{0.2cm}
\begin{center}
{\em (This step implements the diffusion protocol~\cite{LS08,S13}.)}
\end{center}
\vspace{0.2cm}

\textbf{Step 5: Recursion}.
Set $k\leftarrow k+1$, and go Step 1.
 \caption{Adaptation and Learning Within Homogeneous Social Groups}
 \label{alg:homo-social-clique}
\end{algorithm}

\subsection*{Discussion and Intuition:}

Before proceeding, we should emphasize that the remarks on the ``regret-matching'' procedure~\cite{HM01a}, discussed in \S\ref{sec:social-game}, will continue to hold for the revised regret-matching algorithm in the presence of homogeneous social groups.

\emph{(1) Diffusion of Regrets:} Within each community of agents,
the social group that an agent forms may differ from that of his neighbor's. Therefore, the cooperative diffusion protocol~(\ref{eq:diff-regret}) helps fuse information across the community into agent~$\plyrind$.

\emph{(2) Data Fusion Timescale:}
The particular choice of the weight matrix $\weightmat$ allows to perform regret update and fusion of social group's information on the same timescale by choosing $\inertia = \stepsize$ in~(\ref{eq:weight-matrix}). This enables the agents to respond in real-time to the evolution of the game, e.g., changes in community memberships or payoffs of communities.

\section{Emergence of Rational Global Behavior}
\label{sec:main-results-game}
This section characterizes the global behavior emergent from individuals following the local adaption and learning algorithms in \S\ref{sec:local-behavior-game}. Section~\ref{subsec:global-behavior} provides a formal definition for the global behavior of agents. Section~\ref{subsec:CE-convergence} then entails the main theorem of this chapter that shows the global behavior manifested as the consequence of each agent individually following algorithms in~\S\ref{sec:local-behavior-game} converges to the correlated equilibria set. Finally, \S\ref{subsec:proof-sketch-game} lays out a less technically involved sketch of the convergence proofs that will further shed light on the dynamics of the ``regret-matching'' procedure and how it could be adapted to strategic decision making in social networks. Detailed proofs are relegated to Appendix~\ref{sec:theorem-proofs-game} for clarity of presentation.

\subsection{Global Behavior}
\label{subsec:global-behavior}
The global behavior $\globbehav_\dtimee$ at time $k$ is defined as the empirical frequency of joint play of all agents up to period $k$. Formally,
\begin{equation}
\label{eq:global-behavior}
\globbehav_\dtimee = \sum_{\tau\leq k} (1-\stepsize)^{k-\tau} \unitvec_{\actprof_\tau}
\end{equation}
where $\unitvec_{\actprof_\tau}$ denotes the unit vector with the element corresponding to the joint play $\actprof_\tau$ being equal to one. Further, $\stepsize$ serves as a forgetting factor to foster adaptivity to the evolution of game. That is, the effect of the old
game model on the decisions of individuals vanishes as agents repeatedly take actions. Given $\globbehav_\dtimee$, the average payoff accrued by each agent can be straightforwardly evaluated, hence the name global behavior. It is more convenient to define $\globbehav_\dtimee$ via the stochastic approximation recursion
\begin{equation}
\label{eq:global-behavior-SA}
\globbehav_\dtimee = \globbehav_{k-1} + \stepsize \lb \unitvec_{\actprof_\dtimee} - \globbehav_{k-1}\rb.
\end{equation}

The global behavior $\globbehav_\dtimee$ is a system ``diagnostic'' and is only used for the analysis of the emergent collective behavior of agents. That is, it does not need to be computed by individual agents.
In real-life application such as smart sensor networks, however, a network controller can monitor $\globbehav_\dtimee$ and use it to adjust agents' payoff functions to achieve the desired global behavior~\cite{MKZ09}.

\subsection{Main Result: Convergence to Correlated Equilibrium}
\label{subsec:CE-convergence}
In what follows, the main theorem of this chapter is presented that reveals both the local and the global behavior emerging from each agent individually following the algorithms presented in \S\ref{sec:local-behavior-game}. We use
$\regmatplyr_\dtimee$ and $\globbehav_\dtimee$ as indicative of agent's local and global experience, respectively.

The main theorem simply asserts that, if an agent individually follows any of the algorithms of \S\ref{sec:local-behavior-game} (depending on the particular game and connectivity graph model), she will experience regret of at most $\epsilon$ (for a small number $\epsilon$) after sufficient repeated plays of the game. Indeed, the number $\epsilon$ diminishes and asymptotically goes to zero in Algorithms~\ref{alg:social-player} and~\ref{alg:homo-social-clique}, where agents exhibit no exploration. This theorem further states that if now all agents start following any of algorithms of \S\ref{sec:local-behavior-game} independently, the global behavior converges to the correlated equilibria set. Differently put, agents can coordinate their strategies in a distributed fashion so that the distribution of their joint behavior belongs to the correlated equilibria set. From the game-theoretic point of view, we show non-fully rational local behavior of individuals, due to utilizing a ``better-reply'' rather than a ``best-reply'' strategy, can lead to the manifestation of globally sophisticated behavior.

Before proceeding with the theorem, recall that weak convergence is a generalization of convergence in distribution to a function space\footnote{We refer the interested reader to~\cite[Chapter~7]{KY03} for further details on weak convergence and related matters.}; see Definition~\ref{def:weak-convergence} in~\S\ref{subsec:ode}.


\begin{theorem}
\label{theorem:main-game}
Suppose the game $\game$, defined in \S\ref{sec:strategic-form}, is being repeatedly played. Define the continuous-time interpolated sequence of the global behavior iterates $\globbehav_\dtimee$ as follows:
\begin{equation}
\label{eq:global-interpol}
\globinterpolt = \globbehav_\dtimee
\quad \textmd{for}\quad t\in[k\stepsize,(k+1)\stepsize)
\end{equation}
where $\globbehav_\dtimee$ is defined in~(\ref{eq:global-behavior}). With a slight abuse of notation, denote by $\regmatplyr_\dtimee$ the regret matrices rearranged as a vector of length $(\fact)^2$, and define the continuous-time interpolated sequence
\begin{equation}
\label{eq:regret-interpol}
\regmatplyrinterpol = \regmatplyr_\dtimee\quad \textmd{for}\quad t\in[k\stepsize,(k+1)\stepsize)
\end{equation}
where $\regmatplyr_\dtimee$ is given in~(\ref{eq:regret-def}) and~(\ref{eq:combined-regrets}).
Let further $\mathbb{R}^-$ represent the negative orthant and $\|\cdot\|$ denote the Euclidean norm.
Then, the following results hold:

\vspace{0.2cm}
\noindent
\textbf{Result 1:} Consider ``Case I'' that examines fully social players; see \S\ref{sec:social-game}. If every agent follows the ``regret-matching'' procedure~\cite{HM00}, summarized in Algorithm~\ref{alg:social-player}, then as $\stepsize\to 0 $ and $t\to\infty$:

(i) $\regmatplyrinterpol$ converges weakly \index{weak convergence} to the negative orthant in the sense that
\begin{equation}
\textmd{dist}\big[ \regmatplyrinterpol,\mathbb{R}^-\big] = \inf_{\boldsymbol{r} \in \mathbb{R}^-}\big\| \regmatplyrinterpol - \boldsymbol{r}\big\|  \Rightarrow 0;
\end{equation}

(ii) $\globinterpolt$ converges weakly to the correlated equilibria set $\CE$ in the sense that
\begin{equation}
\label{eq:convergence-1}
\textmd{dist}\lb \globinterpolt,\CE\rb = \inf_{\boldsymbol{z} \in \CE}\left\| \globinterpolt - \boldsymbol{z}\right\| \Rightarrow 0.
\end{equation}

\vspace{0.2cm}
\noindent
\textbf{Result 2:} Consider ``Case II'' that examines formation of social groups; see \S\ref{sec:social-cliques}. For each $\epsilon$, there exists an upper bound $\hat{\explor}(\epsilon)$ on the exploration factor $\explor$ such that, if every agent follows Algorithm~\ref{alg:social-clique} with $\explor < \hat{\explor}(\epsilon)$, as $\stepsize\to 0 $ and $t\to\infty$:

(i) $\regmatplyrinterpol$ converges weakly \index{weak convergence} to $\epsilon$-distance of the negative orthant, i.e., $$\textmd{dist}\big[ \regmatplyrinterpol,\mathbb{R}^-\big] - \epsilon \Rightarrow 0;$$

(ii) $\globinterpolt$ converges weakly to the correlated $\epsilon$-equilibria set $\CEe$, i.e., $$\textmd{dist}\lb \globinterpolt,\CEe\rb \Rightarrow 0.$$

\vspace{0.2cm}
\noindent
\textbf{Result 3:} Consider ``Case III'' that examines formation of homogeneous social groups; see \S\ref{sec:homogeneous-groups}. If every agent follows
Algorithm~\ref{alg:homo-social-clique}, then as $\stepsize\to 0 $ and $t\to\infty$:

(i) $\regmatplyrinterpol$ converges weakly \index{weak convergence} to the negative orthant.

(ii) $\globinterpolt$ converges weakly to the correlated equilibria set $\CE$.
\end{theorem}

\noindent
\textbf{Discussion:}

\emph{1) Convergence to Set:} Note that Theorem~\ref{theorem:main-game} proves convergence of $\globbehav_\dtimee$ to the correlated equilibria set $\CE$, rather than a point in the set $\CE$. In fact, once the convergence occurs, $\globbehav_\dtimee$ can generally move around in the polytope $\CE$. The same argument holds for convergence of regrets $\regmatplyr_\dtimee$ to the negative orthant.

\emph{2) Time-invariant Games:} If the parameters of the game $\game$ (such as the payoff functions, action sets, etc.) do not evolve with time, one can replace the constant adaptation rate $\stepsize$ with the decreasing step-size $\stepsize_\dtimee = 1/k$ in Algorithms~\ref{alg:social-player}--\ref{alg:homo-social-clique}. This results in achieving stronger convergence results in the sense that one can now prove almost sure convergence to the correlated equilibria set. More precisely, there exists with probability one $K(\epsilon)>0$ such that for $k > K(\epsilon)$, one can find a correlated equilibrium joint distribution $\pi\in\CE$ at most at $\epsilon$-distance of $\globbehav_\dtimee$~\cite{HM00}.

\subsection{Sketch of Convergence Proof}
\label{subsec:proof-sketch-game}

Let us now proceed with a sketch of the proof of the main results summarized in Theorem~\ref{theorem:main-game}. For better readability, details for each step of the proof are postponed to Appendix~\ref{sec:theorem-proofs-game}. Certain technical details, that are out of the scope of this chapter, are also passed over; however, adequate citation will be provided for the interested reader. The convergence analysis is based on~\cite{BHS06} and is organized into three steps as follows: First, it will be shown that the limit systems for the discrete time iterates of the presented algorithms are differential inclusions\footnote{Differential inclusions \index{differential inclusion} are generalization of the concept of ordinary differential equation. A generic differential inclusion is of the form $dX/dt \in \mathcal{F}(X,t)$, where $\mathcal{F}(X,t)$ specifies a family of trajectories rather than a single trajectory as in the ordinary differential equations $dX/dt = F(X,t)$. See Appendix~\ref{sec:theorem-proofs-game} for a formal definition.}. Differential inclusions arise naturally in game-theoretic learning, since the strategies according to which others play are unknown. Next, using Lyapunov function methods, stability of the limit dynamical system is studied and its set of global attractors is characterized. Accordingly, asymptotic stability of the interpolated processes associated with the discrete-time iterates is proved. Up to this point, we have shown that each agent asymptotically experiences zero (or at most $\epsilon$) regret by following any of the algorithms presented in \S\ref{sec:local-behavior-game}. In the final step, it will be shown that the global behavior emergent from such limit individual dynamics in the game is attracted to the correlated equilibria set.

For the sake of simplicity of exposition, we shall focus in what follows on the proof for the ``regret-matching'' procedure (Algorithm~\ref{alg:social-player}). We then elaborate on how to modify each step of the proof so as to adapt it to the assumptions made in ``Case II'' and ``Case III'' in \S\ref{sec:local-behavior-game}.

Before proceeding with the proof, we shall study properties of the sequence of decisions $\lbr\actk_\dtimee\rbr$ made according to Algorithms~\ref{alg:social-player}--\ref{alg:homo-social-clique}. Careful consideration of the strategy $\mixedstrat^\plyrind_\dtimee$
in the ``regret-matching'' procedure indicates that the sequence of decisions $\lbr \actk_\dtimee\rbr$ simply forms a finite-state Markov chain. Standard results on Markov chains show that the transition probabilities in~(\ref{eq:strategy-MC}) admits (at least) one invariant measure. Let $\statdistk\big(\regmatplyr\big)$ denote such an invariant measure. Then, the following lemma holds.

\begin{lemma}
\label{lemma:stationary-distr}
The invariant measure $\statdistk\left(\regmatplyr\right)$ satisfies\footnote{Equation~(\ref{eq:stat-distr-prop}) together with $\statdistk\left(\regmatplyr\right) \geq 0$ and $\sum_{i}\statdistindi\left(\regmatplyr\right) = 1$ forms a linear feasibility problem (null objective function). Existence of $\statdistk\left(\regmatplyr\right)$ can alternatively be proved using strong duality; see~\cite[Sec.~5.8.2]{BV04} for details.}
\begin{equation}
\label{eq:stat-distr-prop}
\sum_{i\neq j} \statdistindi\left(\regmatplyr\right) \left| r^\plyrind(i,j)\right|^+ = \sum_{i\neq j} \statdistindj\left(\regmatplyr\right) \left| r^\plyrind(j,i)\right|^+.
\end{equation}
\begin{proof}
Letting $\statdistk\left(\regmatplyr\right)$ denote the invariant measure of transition probabilities~(\ref{eq:strategy-MC}), one can write
\begin{align}
\statdistindi\left(\regmatplyr\right) = \statdistindi\left(\regmatplyr\right) \lb 1 - \sum_{j\neq i} {\left|r^\plyrind(j,i)\right|^+\over \inertia^\plyrind}\rb + \sum_{j\neq i}  \statdistindj\cdot {\left|r^\plyrind(i,j)\right|^+\over \inertia^\plyrind}.\nonumber
\end{align}
Cancelling out $\statdistindi\big(\regmatplyr\big)$ on both sides
and rearranging the equality yields the desire result.
\end{proof}
\end{lemma}
We shall now proceed with the proof.

\subsection*{Step 1: Characterization of Limit Individual Behavior}
The first step characterizes the asymptotic behavior of individual agents as a differential inclusion when they follow algorithms of \S\ref{sec:local-behavior-game}. We borrow techniques from the theory of stochastic approximations~\cite{KY03} and  work with the piecewise-constant continuous-time interpolations of the discrete-time iterates of regrets. The regret matrix $\regmatplyr_\dtimee$ incorporates all information recorded in $\globbehav_\dtimee$, however, translated to each agent's language based on her payoff function. Since it is more convenient to work with $\regmatplyr_\dtimee$, we characterize the limiting process for $\regmatplyr_\dtimee$ and analyze its stability instead in the next step. Technical details are tedious and are only provided for ``Case I''. The interested reader is referred to~\cite{NKY13a},~\cite[Chapter 8]{KY03} for more details.

\emph{Case I:} Let $\mixedstrat^{-\plyrind}$ denote a probability measure over the joint action space $\actset^{-\plyrind}$ of all agents excluding agent $\plyrind$, and denote by $\simplexminus$ the simplex of all such measures. The expected payoff to agent $\plyrind$ given the joint strategy $\mixedstrat^{-\plyrind}$ is then defined as
\begin{equation}
\label{eq:expected-payoff-game}
\utilityk\big(\actk,\mixedstrat^{-\plyrind}\big) \ole \sum_{\actprof^{-\plyrind}} \mixedstratind^{-\plyrind}\big(\actprof^{-\plyrind}\big)\utilityk\big(\actk,\actprof^{-\plyrind}\big)
\end{equation}
where $\mixedstratind^{-\plyrind}\big(\actprof^{-\plyrind}\big)$ denotes the probability of all agents excluding agent $\plyrind$ jointly picking $\actprof^{-\plyrind}$.

\begin{theorem}
\label{theorem:limit-system-1}
As $\stepsize\to 0$, the interpolated process $\regmatplyrinterpolcdot$ converges weakly to $\regmatplyr(\cdot)$, that is a solution to the \index{differential inclusion} differential inclusion\footnote{Since the r.h.s. in~(\ref{eq:thrm-1-1}) is a compact convex set, the linear growth condition~(\ref{eq:linear_growth}) for the differential inclusion is trivially satisfied; see Definition~\ref{def:diff-inclusion} for detials.}
\begin{equation}
\label{eq:thrm-1-1}
{d\regmatplyr \over dt} \in \diffinclsocial - \regmatplyr
\end{equation}
where $\diffinclsocial$ is an $\fact\times\fact$ matrix with elements defined below:
\begin{equation}
\label{eq:thrm-1-2}
\diffinclsocialind \ole \lbr \big[\utilityk\big(j,\mixedstrat^{-\plyrind}\big) - \utilityk\big(i,\mixedstrat^{-\plyrind}\big)\big] \statdistindi\big(\regmatplyr\big) ; \mixedstrat^{-\plyrind}\in \simplexminus \rbr.
\end{equation}
\begin{proof}
See Appendix~\ref{sec:app_B_1}.
\end{proof}
\end{theorem}

{\em Discussion and Intuition:} The asymptotics of a stochastic approximation algorithm is typically captured by an ordinary differential equation (ODE). Here, although agents observe $\actprof^{-\plyrind}$, they are oblivious to the strategies $\mixedstrat^{-\plyrind}$ from which $\actprof^{-\plyrind}$ has been drawn. Different $\mixedstrat^{-\plyrind}\in\simplexminus$ form different trajectories of $\regmatplyr\ctimeet$. Therefore, $\regmatplyrinterpol$ converges weakly to the trajectories of a differential inclusion rather than an ODE \index{ordinary differential equation}.

\vspace{0.2cm}
\emph{Case II:} Let $\neibmixedstrat$ denote a probability measure over the joint action space of the open neighborhood of agent $\plyrind$,
and denote by $\simplexneighbor$ the simplex of all such measures. The expected local payoff to agent $\plyrind$ given the joint strategy $\neibmixedstrat$ can be defined similar to~(\ref{eq:expected-payoff-game}). It is straightforward to show that the invariant measure of the transition probabilities~(\ref{eq:strategy-MC-homog-social-group}) now takes the form
\begin{equation}
\label{eq:statdistr-groups}
\statdistindgroup = (1-\explor)\statdistk\big(\regmatplyr\big) + \frac{1}{\fact}\cdot\oneK
\end{equation}
where $\statdistk\big(\regmatplyr\big)$ is the stationary distribution satisfying~(\ref{eq:stat-distr-prop}); see Lemma~\ref{lemma:stationary-distr}. That is, $\statdistk\big(\regmatplyr\big)$ corresponds to the stationary distribution of the transitions probabilities~(\ref{eq:strategy-MC-homog-social-group}) when the exploration factor $\explor = 0$.

\begin{theorem}
\label{theorem:limit-system-2}
Define the continuous-time interpolated sequence
\begin{equation}
\begin{array}{c}
R^{L,n,\varepsilon}(t) = \regmatplyrL_\dtimee\\
R^{G,n,\varepsilon}(t) = \regmatplyrG_\dtimee
\end{array}
\quad \textmd{for}\quad t\in[\dtimee\stepsize,(\dtimee+1)\stepsize)
\end{equation}
where $\regmatplyrL_\dtimee$ and $\regmatplyrG_\dtimee$ are given in~\S\ref{sec:social-cliques}. Then, as $\stepsize\to 0$, the interpolated pair processes $\big(\regmatplyrLinterpolcdot,\regmatplyrGinterpolcdot\big)$ converges weakly to $\big(\regmatplyrL(\cdot),\regmatplyrG(\cdot)\big)$, that is a solution to the system of interconnected differential inclusion \index{differential inclusion}
\begin{equation}
\label{eq:thrm-2-1}
\left\{
\begin{array}{l}
{d\regmatplyrL \over dt} \in \diffinclsocialL - \regmatplyrL,\\
{d\regmatplyrG \over dt} \in \diffinclsocialG - \regmatplyrG.
\end{array}
\right.
\end{equation}
where $\diffinclsocialL$ and $\diffinclsocialG$ are $\fact\times\fact$ matrices with elements defined as follows:
\begin{equation}
\label{eq:thrm-2-2}
\begin{split}
&h^{L,n}_{ij} \ole \lbr \big[\utilitykL\big(j,\neibmixedstrat\big) - \utilitykL\big(i,\neibmixedstrat\big)\big] \statdistindgroupind ; \neibmixedstrat\in \simplexneighbor \rbr\\
&h^{G,n}_{ij} \ole \lb \utilitykGt(j) - \utilitykGt(i)\rb \statdistindgroupind
\end{split}
\end{equation}
where $\utilitykGt(\cdot)$ is the interpolated process of the global payoffs accrued from the game, $\statdistindgroup$ is given in~(\ref{eq:statdistr-groups}), and $\regmatplyr = \regmatplyrL + \regmatplyrG$.
\end{theorem}

{\em Discussion and Intuition:} The interconnection of the dynamics in~(\ref{eq:thrm-2-1}) is hidden in $\statdistindgroup$
since
it is a function of the pair $\big(\regmatplyrL,\regmatplyrG\big)$; see~(\ref{eq:combined-regrets}). Agents are oblivious to the strategies $\neibmixedstrat$ from which the decision $\actprof^{\neighborhoodk}$ has been made. Therefore, following the same argument as in ``Case I,'' $\regmatplyrLinterpolcdot$ converges weakly to the trajectories of a differential inclusion. The same argument holds for $\regmatplyrGinterpolcdot$ except that the limit dynamics follows an ODE. This is due to agents being oblivious to the facts: (i) non-neighbors exist, and (ii) global payoffs are affected by non-neighbors' decisions. However, they realize the time-dependency of $\utilitykGn\left(\cdot\right)$ as taking the same action at various times results in different payoffs.

\vspace{0.2cm}
\emph{Case III:} In light of the diffusion protocol, agents' successive decisions affect, not only their own future strategies, but also their social groups' policies. This suggests looking at the dynamics of the regret for the entire social group to account for such collaboration. Accordingly, define
\begin{equation}
\label{eq:network-regret}
\globalregret_\dtimee \ole \textmd{col}\left\{ R^{\plyrind_1}_\dtimee,\ldots,R^{\plyrind_{_{J_s}}}_\dtimee\right\}, \quad\forall n_j\in\communitys
\end{equation}
and the associated sequence of interpolated process
\begin{equation}
\label{eq:global-regret-interpol}
\globalregretinterpol\ctimeet = \globalregret_\dtimee\;\;\textmd{for}\;\; t\in[k\stepsize,(k+1)\stepsize).
\end{equation}
Let further $\otimes$ denote the Kronecker product.

\begin{theorem}
\label{theorem:limit-system-3}
As $\stepsize\rightarrow 0$, the interpolated process $\globalregretinterpol\cd$ converges weakly to $\globalregret\cd$ that is a solution of the differential inclusion \index{differential inclusion}
\begin{equation}
\label{eq:thrm-3-1}
{d\globalregret\over dt} \in \diffinclglobal^s\left(\globalregret\right) + (\Cmatkron-I) \globalregret
\end{equation}
where $\Cmatkron = \Cmat\otimes I_{A^s}$ (see (\ref{eq:weight-matrix})), $A^s$ denotes the cardinality of the action set of community $\communitys$, and
\begin{equation}
\label{eq:thrm-3-2}
\begin{array}{c}
\mathbf{h}^s_{ij}\left(\globalregret\right) \ole \left\{\left[U^s(\iota,\mixedstrat^{-\kappa}) - U^s(j,\mixedstrat^{-\kappa})\right]\psi_{\iota}^\kappa; \mixedstrat^{-\kappa}\in\Delta\actset^{-\kappa}\right\},\\
\iota = i\bmod A^s,\ \kappa = \left\lfloor {i\over A^s}\right\rfloor+1.
\end{array}
\end{equation}
Further,
$\statdistk$ represents the stationary distribution of the transition probabilities (\ref{eq:strategy-MC-homog-social-group}), which satisfies~(\ref{eq:stat-distr-prop}).
\end{theorem}

\subsection*{Step 2: Stability Analysis of Limit Dynamical System}
\begin{figure}
\centering
\subfloat[]{
  \includegraphics[width=0.4\linewidth]{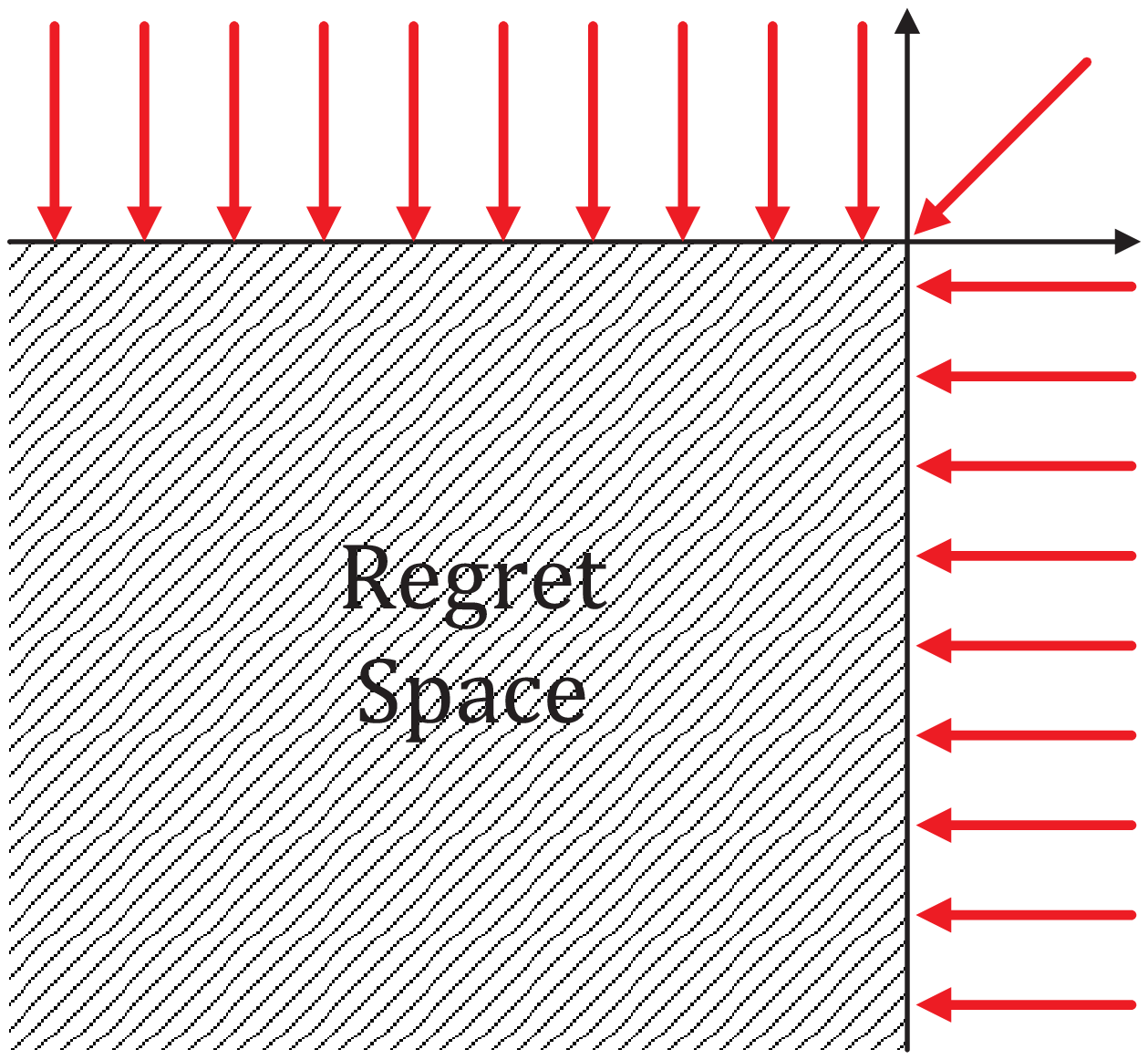}
  \label{fig:proofs-game-1}
}%
\subfloat[]{
  \includegraphics[width=0.4\linewidth]{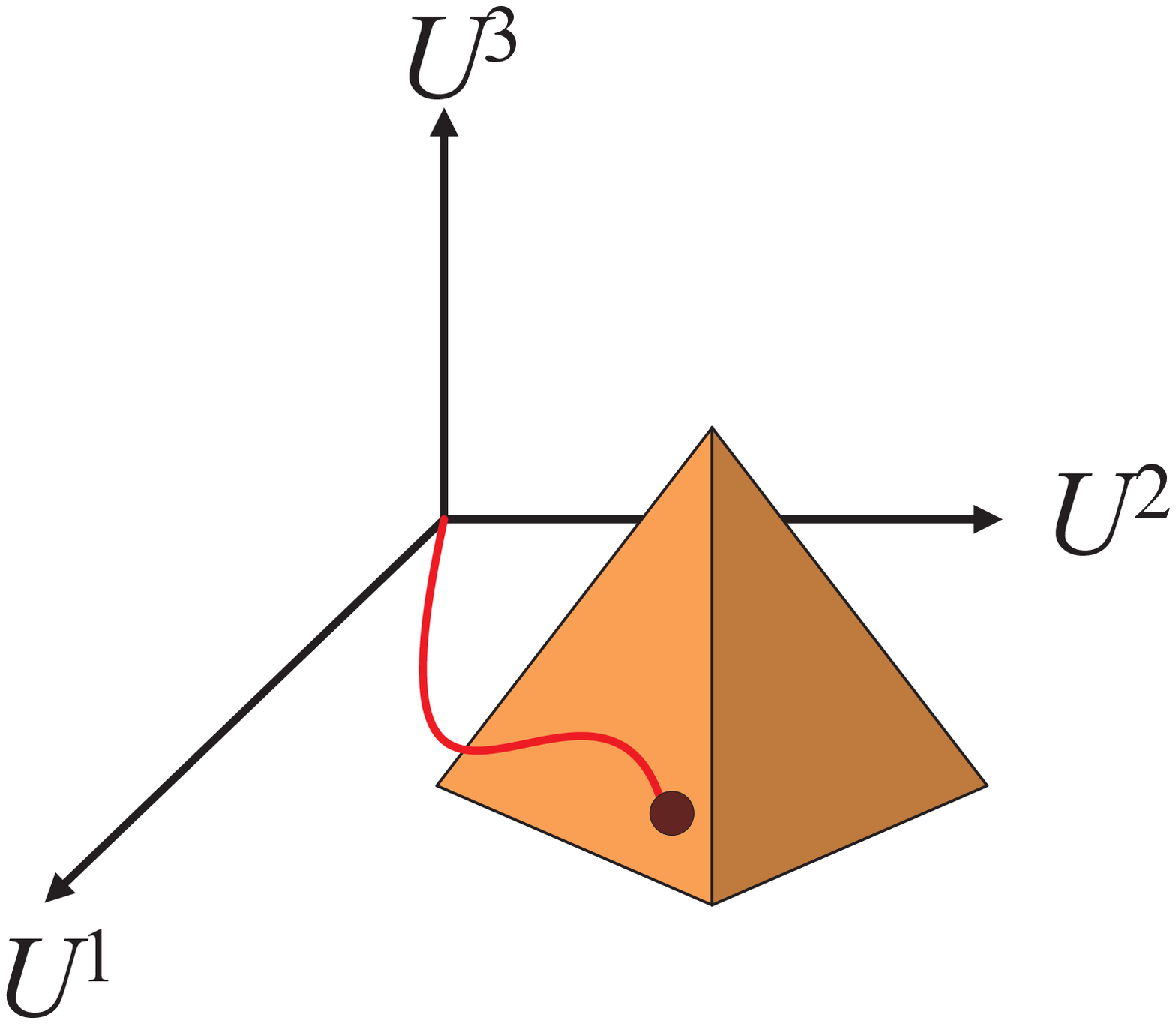}
  \label{fig:proofs-game-2}
}
\caption{(a) Asymptotic stability and convergence of regrets to the negative orthant (Corollary~\ref{corollary-game}). (b) A typical correlated equilibria polytope in a three-player game. The red line depicts the trajectory of average payoff accrued by each player (Theorem~\ref{theorem:CE-convergence}).}
\label{fig:proofs-game}
\end{figure}

This step examines stability of the limit systems that have been shown to represent the local dynamical behavior of agents in Step~1. The set of global attractors of these limit systems are then shown to comprise the negative orthant $\mathbb{R}^-$; see Figure~\ref{fig:proofs-game-1}. Therefore, agents asymptotically experience zero (or at most as small as $\epsilon$) regret via exhibiting the simple local behavior prescribed by Algorithms~\ref{alg:social-player}--\ref{alg:homo-social-clique}.

In what follows, we unfold the regret matrix and rearrange its elements as a vector. With a slight abuse of notation, we continue to use the same notation for this regret vector.
\begin{theorem}
\label{theorem:asymp-stability}
The limit systems presented in Theorems~\ref{theorem:limit-system-1}--\ref{theorem:limit-system-3} are globally asymptotically stable. In particular,
\begin{equation}
\label{eq:thrm-4-1}
\lim_{t\to\infty} \textmd{dist}\lb \regmatplyr\ctimeet,\mathbb{R}^-\rb = 0
\end{equation}
where $\textmd{dist}\lb\cdot,\cdot\rb$ denotes the usual distance function.
\begin{proof}
See Appendix~\ref{sec:theorem-proofs-game}.
\end{proof}
\end{theorem}

The following corollary then follows from Theorems~\ref{theorem:limit-system-1}--\ref{theorem:limit-system-3} and Theorem~\ref{theorem:asymp-stability}. It asserts that the result of Theorem~\ref{theorem:asymp-stability} also holds for the interpolated processes associated with the regrets---they also converge to the negative orthant; see Figure~\ref{fig:proofs-game-1}.
\begin{corollary}
\label{corollary-game}
Consider the interpolated process $\regmatplyrinterpolcdot$, defined in~(\ref{eq:regret-interpol}). Then,
\begin{equation}
\label{eq:corollary-1}
\lim_{t\to\infty} \lim_{\stepsize\to 0} \big|\regmatplyrinterpol\big|^+ = \mathbf{0}
\end{equation}
where $|\cdot|^+$ denotes the element-wise positive operator, and $\mathbf{0}$ represents the zero vector of the same size as $\regmatplyrinterpol$.
\end{corollary}

\subsection*{Step 3: Convergence to Correlated Equilibria Set}
In the final step of the proof we characterize the global behavior of agents whose local behavior ensures no (or small $\epsilon$) asymptotic regret; see Figure~\ref{fig:proofs-game}. It will be shown that the regret of individual agents converging to the negative orthant provides the necessary and sufficient condition for convergence of global behavior to the correlated equilibria set.
\begin{theorem}
\label{theorem:CE-convergence}
Recall the interpolated process for global behavior $\globinterpol\cd$, defined in~(\ref{eq:global-interpol}), and the interpolated regret process $\regmatplyrinterpolcdot$, defined in~(\ref{eq:regret-interpol}). Then,
\begin{equation}
\label{eq:thrm-5-1}
\globinterpol\cd \Rightarrow \CE\quad\textmd{\em iff}\quad \regmatplyrinterpolcdot\Rightarrow\mathbb{R}^-\;\textmd{for all}\; \plyrind\in\plyrset.
\end{equation}
\begin{proof}
The detailed proof is provided in Appendix~\ref{sec:theorem-proofs-game}.
\end{proof}
\end{theorem}
The above theorem, together with Corollary~\ref{corollary-game}, completes the proof.

\section{Markov Modulated Non-Cooperative Games}
\label{sec:regime-switching-game}
In this section, motivated by applications in cognitive networks~\cite{NKY13b,KMY08,MKZ09}, we introduce a general class of Markov modulated noncooperative games.
 We then comment on whether the results of \S\ref{sec:main-results-game} can be generalized to such settings. In particular, we examine if the simple local adaptation and learning algorithms,  presented in \S\ref{sec:local-behavior-game}, can lead to a global behavior that is agile enough to track the regime-switching polytope of correlated equilibria.

\index{game!regime-switching game}
Suppose the non-cooperative game model, formulated in \S\ref{sec:strategic-form}, evolves with time due to: (i) agents joining or leaving the game; (ii) changes in agents' incentives (payoffs); and (iii) the
connectivity graph $\gamegraph$ varying with time. Suppose further that all time-varying parameters are finite-state
and absorbed to a vector indexed by $\markovgame$. It is thus reasonable to model evolution of the game via a discrete-time finite-state Markov chain $\lbr\markovgame_\dtimee\rbr$. Without loss of generality, assume that the Markov chain has state-space $\statespacegame = \lbr 1,\ldots,\fstate\rbr$ and transition probability matrix \index{matrix!transition matrix} \index{Markov chain!Markovian switching}
\begin{equation}
\label{eq:Markov-chain-game}
P^{\markovspeed} = I_\fstate +\markovspeed \contmarkovmat.
\end{equation}
Here, $\markovspeed > 0$ is the small parameter that, roughly speaking, determines the speed of jumps of $\lbr\markovgame_\dtimee\rbr$, $I_\fstate$ denotes the $\fstate\times\fstate$ identity matrix, and $\contmarkovmat =\lb \contmarkovmatind\rb$ is the generator \index{matrix!generator matrix} of a continuous-time Markov chain satisfying
\begin{itemize}
    \item[(i)] $\contmarkovmat\oneM = \zeroM$;
    \item[(ii)] $\contmarkovmatind\geq0$ for $i\neq j$;
    \item[(iii)] $|\contmarkovmatind|\leq 1$ for all $i,j\in\statespacegame$.
\end{itemize}
Choosing $\markovspeed$ small enough ensures that $p_{ij}^\markovspeed = \delta_{ij}+\markovspeed \contmarkovmatind >0$ in the entries of~(\ref{eq:Markov-chain-game}), where $\delta_{ij}$ denotes the Kronecker $\delta$ function satisfying $\delta_{ij} = 1$ if $i=j$ and 0 otherwise.

For better exposition, let us assume that the payoff functions vary with the state of the Markov chain $\lbr\markovgame_\dtimee\rbr$ in the rest of this section. More precisely, the payoff function for each agent $\plyrind$ takes the form
\begin{equation}
\label{eq:payoff-switching}
\utilityk\big(\actk,\actprof^{-\plyrind},\markovgame\big): \actset^{\plyrind}\times\actset^{-\plyrind}\times\statespacegame \to \mathbb{R}.
\end{equation}
The main assumption here is that agents do not observe the Markov chain $\lbr\markovgame_\dtimee\rbr$, nor are they aware of its dynamics. Agents however may realize time-dependency of the payoff functions as taking the same action at different time instants may result in different payoffs. Put differently, it does not enter implementation of the local adaptation and learning algorithms. The Markov chain dynamics are  used in analyzing the tracking capability of the algorithms.

As the payoff functions jump change according to the Markov chain $\lbr\markovgame_\dtimee\rbr$, so does the correlated equilibria set as illustrated in Figure~\ref{fig:switching-game}. We thus need to modify Definition~\ref{def:CE} to incorporate $\markovgame$-dependency of the payoff functions as follows:
\begin{definition}
\label{eq:CE-switching}
Let $\mixedstrat$ denote a probability distribution on the space of action profiles $\actset$. The set of correlated equilibria of the regime-switching game \index{game!regime-switching game} is the convex polytope

\vspace{-0.1cm}
{\small
\begin{equation}
\label{eq:CE_defn_switching}
\CEe(\markovgame_\dtimee) = \lbr \mixedstrat:
\sum_{\actprof^{-\plyrind}} \mixedstratind^{\plyrind}\big(i, \actprof^{-\plyrind}\big)\left[U^{\plyrind}\big(j, \actprof^{-\plyrind},\markovgame_\dtimee\big) - U^\plyrind\big(i, \actprof^{-\plyrind},\markovgame_\dtimee\big)\right] \leq 0, \forall i,j,n \rbr\nonumber
\end{equation}
}
\begin{figure}
\centerline{\includegraphics[width=0.6\textwidth]{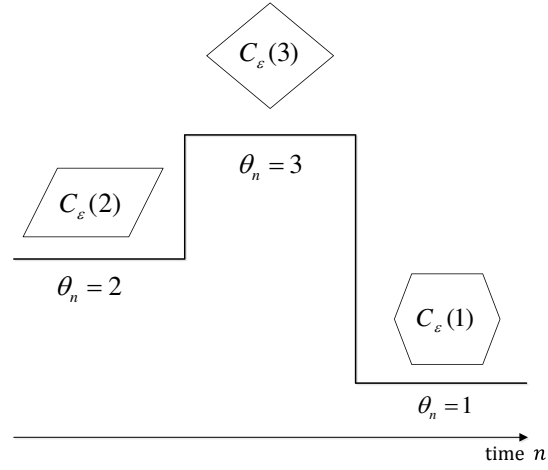}}
\caption{Tracking convex polytope of correlated equilibria $\CE(\markovgame_\dtimee)$ randomly evolving with time according to a Markov process $\lbr\markovgame_\dtimee\rbr$. Theorem~\ref{theorem:CE-convg-switch} asserts that such tracking is attainable by each agent individually following the regret-matching algorithms presented in \S\ref{sec:local-behavior-game}.}
\label{fig:switching-game}
\end{figure}

\vspace{-0.1cm}
\noindent
where $\mixedstratind^\plyrind\big(i, \actprof^{-\plyrind}\big)$ denotes the probability of agent $\plyrind$ playing action $i$ and the rest playing $\actprof^{-\plyrind}$.
\end{definition}

The following theorem then provides a condition that ensures the local adaptation and learning algorithms of \S\ref{sec:local-behavior-game} (with no further modifications) generate trajectories of global behavior $\globbehav_\dtimee$ that tracks the regime switching set of correlated equilibrium $\CE(\markovgame_\dtimee)$.

\begin{theorem}
\label{theorem:CE-convg-switch}
Consider the non-cooperative game $\game$ regime switching according to a finite-state Markov chain $\lbr \theta_\dtimee\rbr$ with transition probabilities given by~(\ref{eq:Markov-chain-game}). Suppose that the step-size of Algrithms~\ref{alg:social-player}--\ref{alg:homo-social-clique} is small, i.e., $0<\stepsize\ll 1$. Assume further that that
\begin{equation}
\label{eq:condition-Markov-chain}
\textmd{the process $\markovgame_\dtimee$ is ``slow'' in the sense that $\markovspeed = \stepsize$.}
\end{equation}
Recall the interpolated sequence of regrets $\regmatplyrinterpolcdot$, defined in~(\ref{eq:regret-interpol}), and interpolated global behavior $\globbehav^\stepsize\cd$, defined in~(\ref{eq:global-interpol}). If every agent follows Algorithms~\ref{alg:social-player}--\ref{alg:homo-social-clique}, as $\stepsize\to 0$ and $t\to\infty$\footnote{To be more precise, the following results hold for an $\epsilon$ neighborhood of the convergent sets in ``Case II''.}:

(i) $\regmatplyrinterpol$ converges weakly to the negative orthant
in the sense that
\begin{equation}
\textmd{dist}\big[ \regmatplyrinterpol,\mathbb{R}^-\big] = \inf_{\boldsymbol{r} \in \mathbb{R}^-}\big\| \regmatplyrinterpol - \boldsymbol{r}\big\|  \Rightarrow 0;
\end{equation}

(ii) $\globbehav^\stepsize\cd$ converges weakly to the regime-switching polytope of correlated equilibria $\CE(\markovgame\cd)$. More precisely,
\begin{equation}
\label{eq:T0_2}
\textmd{dist}\lb\globbehav^\stepsize\cd,\CE(\markovgame\cd)\rb=\inf_{\globbehav\in\CE(\markovgame\cd)} \left|\globbehav^\stepsize\cd-\globbehav\right| \Rightarrow 0.
\end{equation}
where $\markovgame\cd$ is a continuous-time Markov chain with generator $\contmarkovmat$; see~(\ref{eq:Markov-chain-game}).
\begin{proof}
We skip the detailed proof for the sake of brevity. Here, we only provide a descriptive sketch of the steps of the proof: First, by a combined use of updated treatment on stochastic approximation~\cite{KY03}, and Markov-switched systems~\cite{YZ98,YZ10}, we show that the limiting behavior converges weakly to the differential inclusions presented in Theorems~\ref{theorem:limit-system-1}--\ref{theorem:limit-system-3} modulated by a continuous-time Markov chain with generator \index{matrix!generator matrix} $\contmarkovmat$. Next, using the multiple Lyapunov function method~\cite{CL07b,CL07a},~\cite[Chapter~3]{L03}, we show that the limiting dynamical system is globally asymptotically stable almost surely. The interested reader is referred to~\cite{NKY13a} for the detailed proof in ``Case II.''
\end{proof}
\end{theorem}

\paragraph*{Discussion and Intuition:}
It is well known that: (i) If the underlying parameter $\markovgame_\dtimee$ change too drastically, there is no chance one can track the time-varying properties via an adaptive stochastic approximation algorithm. (Such a phenomenon is known as trackability~\cite{BMP90}.) b) If the parameters evolve on a slower time-scale as compared to the stochastic approximation algorithm, they remain constant in the fast time-scale and their variation can be asymptotically ignored. Condition~(\ref{eq:condition-Markov-chain}) simply ensures that the parameters of the game model evolve according to a Markov chain that evolves on the same timescale as Algorithms~\ref{alg:social-player}--\ref{alg:homo-social-clique}. The above theorem then asserts that, if the ``matching-speed'' condition holds, Algorithms~\ref{alg:social-player}--\ref{alg:homo-social-clique} are agile in tracking the jump changes in the correlated equilibria set. From the game-theoretic point of view, non-fully-rational behavior of individual agents (due to utilizing a ``better-reply'' rather than a ``best-reply'' strategy) leads to a sophisticated globally rational behavior that can adapt to time inhomogeneities in the game model.

\section{Example: Energy-Aware Sensing via Diffusion LMS}
\label{sec:diffusion-LMS}
This section provides an example of how the adaptation and learning algorithms of~\S\ref{sec:local-behavior-game} can be applied to achieve sophisticated global behavior in a social sensing application. We consider energy-aware activation control of sensors aiming to collaboratively estimate the true value of a common parameter of interest\footnote{The material in this section is based on the recent paper~\cite{NKY13b}.}. We focus on the diffusion least mean square~(LMS) algorithms~\cite{LS08,Say03,Say14}; however, the proposed framework can applied to any collaborative non-Bayesian estimation setting.

\subsection{Centralized Parameter Estimation Problem}
\label{sec:game-estimation}
Consider a set of $\fplyr$ sensors $\plyrset = \lbr 1,\ldots,\fplyr\rbr$ with the objective to estimate an unknown $M\times 1$ parameter vector $\boldsymbol{\psi}^o$  via recording noisy measurements
\begin{equation}
\label{eq:observation}
y_\dtimee^\plyrind = \boldsymbol{u}_\dtimee^\plyrind \boldsymbol{\psi}^o + v_\dtimee^\plyrind,\quad \plyrind = 1,\ldots,\fplyr.
\end{equation}
Here, $\lbr \boldsymbol{u}_\dtimee^\plyrind\rbr$ and $\lbrace v_\dtimee^\plyrind\rbrace$ denote the sequences of $1\times M$ random regressor vectors and zero-mean local measurement noises, respectively. Such linear models are well-suited to approximate input-output relations for many practical applications~\cite{Say03}. The sequence $\lbr \boldsymbol{u}_\dtime^\plyrind, v_\dtime^\plyrind\rbr$ can generally be correlated as long as the remote past and distant future are asymptotically independent; see~\cite[Section~II-C]{NKY13b} for details and examples of such processes. For simplicity of presentation, we assume here that the sequence $\lbr \boldsymbol{u}_\dtime^\plyrind, v_\dtime^\plyrind\rbr$ is temporally white and spatially independent such that~\cite{LS08}:
\begin{equation}
\begin{split}
&\E \lbr\lb\boldsymbol{u}_\dtimee^\plyrind\rb^\top \boldsymbol{u}_{\underline\dtimee}^l\rbr = R_d^\plyrind \cdot \delta_{\dtimee\underline\dtimee} \cdot \delta_{\plyrind l}\\
&\E \lbr v_\dtimee^\plyrind v_{\underline\dtimee}^l\rbr = \sigma^2_{v,\plyrind}\cdot \delta_{\dtimee \underline\dtimee} \cdot \delta_{\plyrind l}
\end{split}
\end{equation}
where $\boldsymbol{x}^\top$ denotes the transpose of vector $\boldsymbol{x}$, $R_d^\plyrind$ is positive-definite and $\delta_{ij}$ is the Kronecker delta function: $\delta_{ij} = 1$ if $i = j$, and 0 otherwise. The noise sequence $\lbrace v_\dtimee^\plyrind\rbrace$ is further uncorrelated with the regression data $\lbrace \boldsymbol{u}_{\underline\dtimee}^l\rbrace$ for all $\lbrace \dtimee, \underline\dtimee, \plyrind, l\rbrace$:
\begin{equation}
\E \lbr\lb \boldsymbol{u}_{\dtimee}^\plyrind\rb^\top v^l_{\underline\dtimee}\rbr = \mathbf{0}.
\end{equation}
The network of sensors then seeks to solve the least mean square parameter estimation problem
\begin{equation}
\label{eq:centralized_problem}
\min_{\boldsymbol{\psi}} \ \E\left\|\boldsymbol{Y}_\dtimee - \mathbf{u}_\dtimee\boldsymbol{\psi}\right\|^2
\end{equation}
where $\|\cdot\|$ denotes Euclidean norm and
\begin{displaymath}
\mathbf{u}_\dtimee \ole \textmd{col}\left( \boldsymbol{u}^1_\dtimee,\ldots,\boldsymbol{u}_\dtimee^\fplyr\right), \quad
\boldsymbol{Y}_\dtimee \ole \textmd{col}\left( y_\dtimee^1,\ldots,y_\dtimee^\fplyr\right).
\end{displaymath}

\subsection{Diffusion Least Mean Squares (LMS)}
We focus on the diffusion LMS algorithm~\cite{LS08} that adopts a peer-to-peer diffusion protocol and provides a distributed solution to the centralized parameter estimation problem~(\ref{eq:centralized_problem}). Sensors form social groups within which estimates are exchanged. Sensors then fuse the collected data and combine it with the local measurements to refine their estimates. Deploying such ``social sensors'' has been proved to improve the estimation performance, yet yield savings in computation, communication and energy expenditure~\cite{Say14,S13}.

Recall the connectivity graph $\gamegraph = (\gamevertex,\gameedge)$, from Definition~\ref{def:connectivity-graph}, and the closed neighborhood $\cneighborhoodk$ in~(\ref{eq:neighbors}). For simplicity of presentation, we assume the connectivity graph $\gamegraph$ is fixed and strongly connected, i.e., there exists a path between each pair of nodes\footnote{Intermittent sensor failures can be captured by a random graph model where the probability that two sensors are connected is simply the probability of successful communication times the indicator function that shows the two sensors are neighbors in the underlying fixed graph. In this case, mean connectedness of the random graph is sufficient, i.e., $\lambda_2(\bar{L})>0$, where $\bar{L} = \E L_n$, $L_n$ denotes the Laplacian of the random graph process $\lbr\gamegraph_n\rbr$, and $\lambda_2\cd$ denotes the second largest eigenvalue; see~\cite{KMR12} for details.}. Let $0<\mu\ll 1$ be a small positive step size
and define the weight matrix $W = [w_{ij}]$ for the diffusion protocol as follows:
\begin{equation}
\label{eq:C-1}
\begin{array}{c}
W = I_{K} + \mu Q,\\
|q_{ij}|\leq 1,\;\;\forall i,j, \;\;q_{ij} \geq 0, \;\;\textmd{for}\;\; i\neq j,\\
Q\mathbf{1} = \mathbf{0},\;\;\textmd{and} \;\; Q^\top = Q.
\end{array}
\end{equation}
The diffusion LMS \index{diffusion least mean square} then requires each sensor to employ a stochastic approximation algorithm of the form
\begin{equation}
\label{eq:LMS-non-activation}
\begin{split}
\textmd{(Data Assimilation)}\quad \boldsymbol{\psi}_{\dtimee+1}^\plyrind &= \boldsymbol{\phi}_{\dtimee}^\plyrind + \mu \lb\boldsymbol{u}^\plyrind_\dtimee\rb^\top\left[y_\dtimee^\plyrind - \boldsymbol{u}^\plyrind_\dtimee \boldsymbol{\phi}^\plyrind_{\dtimee}\right]\\
\textmd{(Data Fusion)}\qquad\;\boldsymbol{\phi}_{\dtimee+1}^\plyrind &= \sum_{l\in\cneighborhoodk} w_{\plyrind l}\boldsymbol{\psi}^l_{\dtimee+1}
\end{split}
\end{equation}
where the second step is the local fusion of estimates via a linear combiner. Note that, since each sensor has a different neighborhood, the fusion rule~(\ref{eq:LMS-non-activation}) helps fuse data across the network into node's estimate. This enables the network to respond in real-time to the temporal and spatial variations in the statistical profile of the data.

Define the interpolated process associated with each sensor's estimate of the true parameter as follows:
\begin{displaymath}
\boldsymbol{\psi}^{\plyrind,\mu}(\ctime) = \boldsymbol{\psi}_{\dtimee}^\plyrind\quad\textmd{for}\quad t\in[\dtimee\mu,(\dtimee+1)\mu).
\end{displaymath}
Using the well-known ODE method for convergence analysis of stochastic approximation algorithms, it is shown in~\cite{NKY13b} that, as $\stepsize\to 0$,
\begin{equation}
\label{eq:lms-convg}
\boldsymbol{\psi}^{\plyrind,\mu}(\ctime) \Rightarrow \boldsymbol{\psi}^o \quad\textmd{as}\quad t\to\infty.
\end{equation}
%

To provide some intuition on the particular choice of the weight matrix in~(\ref{eq:C-1}), consider the standard diffusion LMS algorithm in~\cite{LS08}: it uses the same updates as in~(\ref{eq:LMS-non-activation}), however, with weight matrix $\hat{W}$ that is only assumed to be stochastic, i.e., $\hat{W}\mathbf{1} = \mathbf{1}$. Define the network's global estimate of the true parameter as
\begin{displaymath}
\boldsymbol{\Psi}_{\dtimee} = \textmd{col}\left( \boldsymbol{\psi}^1_\dtimee,\ldots,\boldsymbol{\psi}^\fplyr_\dtimee\right).
\end{displaymath}
Let further
\begin{displaymath}
\boldsymbol{U}_\dtimee = \textmd{diag}\left(\boldsymbol{u}^1_\dtimee,\ldots,\boldsymbol{u}^\fplyr_\dtimee\right), \quad \hat{\mathbf{W}} = \hat{W}\otimes I_M
\end{displaymath}
where $\otimes$ represents the Kronecker product.
The global recursion for the diffusion LMS across the network can then be expressed as
\begin{equation}
\label{eq:diffusion-only}
\boldsymbol{\Psi}_{\dtimee+1} = \hat{\mathbf{W}}\boldsymbol{\Psi}_{\dtimee} + \mu I_{\fplyr M} \boldsymbol{U}_\dtimee^\top \lb\boldsymbol{Y}_\dtimee - \boldsymbol{U}_\dtimee \hat{\mathbf{W}} \boldsymbol{\Psi}_{\dtimee}\rb.
\end{equation}
\begin{remark}
A classical stochastic approximation algorithm \index{stochastic approximation} is of the form $\boldsymbol{\Psi}_{\dtimee+1} = \boldsymbol{\Psi}_{\dtimee} + \mu g(\boldsymbol{\Psi}_{\dtimee},\boldsymbol{v}_\dtimee)$, where $\boldsymbol{v}_\dtimee$ denotes a noise sequence and $0<\mu\ll 1$ denotes the step-size. Clearly, the global recursion for diffusion LMS~(\ref{eq:diffusion-only}) is not in such form because of the data fusion captured by the first term on the r.h.s in~(\ref{eq:diffusion-only}).
\end{remark}

The following theorem shows that the diffusion LMS~(\ref{eq:diffusion-only}) is ``equivalent'' to a classical stochastic approximation algorithm.
%
Below, we use the notation $\log(A)$ to denote the matrix logarithm of a matrix $A$. Equivalently, $B = \log(A)$ if $A = \exp(B)$ where $\exp(B)$ denotes the matrix exponential.
\begin{theorem}
\label{theorem:diffusion-LMS}
Let $\hat{\boldsymbol{W}}$ be a stochastic matrix. Then, as the sampling time $\Delta\to 0$, the discrete-time data fusion recursion~(\ref{eq:diffusion-only})
is equivalent to a standard stochastic approximation algorithm of the form
\begin{equation}
\label{eq:fusion-revised}
\begin{split}
\boldsymbol{\Psi}_{\dtimee+1} &= \boldsymbol{\Psi}_{\dtimee} + \mu I_{\fplyr M} \lb \tilde{\boldsymbol{W}}\boldsymbol{\Psi}_{\dtimee} + \boldsymbol{U}_\dtimee^\top \lb\boldsymbol{Y}_\dtimee - \boldsymbol{U}_\dtimee \hat{\mathbf{W}} \boldsymbol{\Psi}_{\dtimee}\rb\rb,\\ &\textmd{ where}\;\tilde{\boldsymbol{W}} = \ln\big(\hat{\boldsymbol{W}}\big)/\Delta,\;
\textmd{and } 0<\mu\ll 1
\end{split}
\end{equation}
 in the sense that they both form discretizations of the same ODE \index{ordinary differential equation}
\begin{displaymath}
\frac{d\boldsymbol{\Psi}}{dt} = \tilde{\boldsymbol{W}}\boldsymbol{\Psi}.
\end{displaymath}
%
\begin{proof}
Suppose $\Delta$ time units elapse between two successive iterations of~(\ref{eq:diffusion-only}). Then, the diffusion protocol (the first term on the r.h.s of~(\ref{eq:diffusion-only})) can be conceived as a discretization of the ODE $d\boldsymbol{\Psi}/dt = \tilde{\boldsymbol{W}}\boldsymbol{\Psi}$, where $\exp\big(\tilde{\boldsymbol{W}}\Delta\big) = \hat{\boldsymbol{W}}$.
Taking Taylor expansion yields
\begin{equation}
\label{eq:W-taylor}
\hat{\boldsymbol{W}} = I + \Delta \tilde{\boldsymbol{W}} + o(\Delta),\quad\textmd{where }\; o(\Delta)\to 0\;\textmd{ as }\; \Delta\to 0.
\end{equation}
Therefore,
as $\Delta\to 0$, the standard diffusion LMS~(\ref{eq:diffusion-only}) is equivalent to the standard stochastic approximation iterate~(\ref{eq:fusion-revised}).
\end{proof}
\end{theorem}

Further inspection of the matrix $\tilde{\boldsymbol{W}}$ in~(\ref{eq:fusion-revised}) shows that, since $\tilde{\boldsymbol{W}} = \newline lim_{\Delta\to 0} (\exp(\tilde{\boldsymbol{W}}\Delta) - I)/\Delta$,
\begin{equation}
\label{eq:Q-properties}
\tilde{\boldsymbol{W}}\mathbf{1} = \mathbf{0}, \;\textmd{ and }\; \tilde{w}_{ij} > 0 \;\textmd{ for }\; i\neq j.
\end{equation}
Now, comparing~(\ref{eq:Q-properties}) with~(\ref{eq:C-1}), together with Theorem~\ref{theorem:diffusion-LMS}, confirms the equivalence between the standard diffusion protocol~\cite{LS08} and the diffusion protocol constructed by using the weight matrix~(\ref{eq:C-1}). The advantages of the particular choice of the weight matrix~(\ref{eq:C-1}) in the adaptive filtering algorithm~(\ref{eq:LMS-non-activation}) are however threefold: (i) both data assimilation and fusion takes place on the same timescale; (ii) simpler derivation of the known results can be obtained by employing the powerful ordinary differential equation~(ODE) method~\cite{BMP90,KY03}; (iii) one can use weak convergence methods~\cite[Chapter 8]{KY03} to show how responsive the adaptive filtering algorithm is to the time-variations of the true parameter.

\subsection{Energy-Aware Activation Control of Diffusion LMS}
\index{diffusion least mean square!energy-aware diffusion LMS}
Can one make the above decentralized solution to the sensing problem even more efficient by allowing sensors to activate only when their contributions outweigh the activation costs? We equip the sensors with an activation mechanism that, taking into account the spatial-temporal correlation of their measurements, prescribes sensors to sleep when the the energy cost of acquiring new measurement outweighs its contribution to the estimation task. When a node is active, it
updates its estimate and performs fusion of its own estimate with those received from neighboring nodes, whereas inactive
nodes do not update their estimates. More precisely, each sensor runs local updates of the form
\begin{equation}
\boldsymbol{\psi}_{\dtimee+1}^\plyrind = \left\{
\begin{array}{ll}
\boldsymbol{\phi}_{\dtimee}^\plyrind + \mu \lb\boldsymbol{u}^\plyrind_\dtimee\rb^\top\left[y_\dtimee^\plyrind - \boldsymbol{u}^\plyrind_\dtimee \boldsymbol{\phi}^\plyrind_{\dtimee}\right],& \textmd{if }\; \actk_\dtimee = 1\; \textmd{ (Active)}\\
\boldsymbol{\psi}_{\dtimee}^\plyrind, & \textmd{if } \;\actk_\dtimee = 0 \;\textmd{ (Sleep)}
\end{array}
\right.
\end{equation}
where $\boldsymbol{\phi}_{\dtimee}^\plyrind$ is the fused local estimate, defined in~(\ref{eq:LMS-non-activation}). Due to the interdependence of sensors' activation behavior, a game-theoretic approach is a natural choice to model their interaction.

The problem of each sensor $\plyrind$ is then to successively pick action $\actk_\dtimee$ from the set $\actset^\plyrind = \lbrace 0~\textmd{(sleep)}, 1~\textmd{(activate)}\rbrace$ to strategically optimize a utility function. This utility function captures the trade-off between energy expenditure and the ``value'' of sensor's contribution. Let $\boldsymbol{\Psi} = \big[\boldsymbol{\psi}^1,\cdots,\boldsymbol{\psi}^\fplyr\big]$ denote the profile of estimates across the network. The utility function for each sensor $\plyrind$ can then be defined as follows:
%
\begin{equation}
\label{eq:utility-function-lms}
U^\plyrind\big(\actk,\actprof^{-\plyrind};\boldsymbol{\Psi}\big) = U_{L}^\plyrind\big(\actk,\actprof^{\neighborhoodk};\boldsymbol{\Psi}\big) + U_{G}^\plyrind\big(\actk,\actprof^{\nonneighbork}\big).
\end{equation}

The \emph{local} utility function $U^\plyrind_{L}(\cdot)$ captures the trade-off between the value of the measurements collected by sensor~$\plyrind$ and the energy costs associated with it.
If too many of sensor $\plyrind$'s neighbors activate simultaneously, excessive energy is consumed due to the spatial-temporal correlation of sensor measurements---that is, the data collected by sensor $\plyrind$ is less valuable.
On the other hand, if too few of sensor $\plyrind$'s neighbors activate, their fused estimates lack ``innovation.''
The \emph{local} utility of sensor $k$ is then given by
\begin{equation}
\label{eq:local-utility}
\begin{split}
U^\plyrind_{L}\big(\actk,\actprof^{\neighborhoodk};\boldsymbol{\Psi}\big) = \Big[&
K_{l,1}\left( 1- \exp\left(-\gamma_l\|\boldsymbol{\psi}^\plyrind - \boldsymbol{\phi}^\plyrind\|^2 s(\eta^k)\right)\right) \\&- K_{l,2}\big(E_{\textmd{Tx}}\big(\eta^\plyrind\big) + E_{\textmd{Act}}\big)\Big]I\lbr\actk = 1\rbr.
\end{split}
\end{equation}
where
\begin{displaymath}
\label{eq:local-details}
\eta^k\big(\actk,\actprof^{\neighborhoodk}\big) = \actk + \sum_{l\in\neighborhoodk} \textmd{a}^l.
\end{displaymath}
In addition, $K_{l,1}$ and $\gamma_l$ are the pricing parameters related to the ``reward'' associated with the data collected by sensor $\plyrind$, $K_{l,2}$ is the pricing parameter related to the energy costs associated with activation $E_{\textmd{Act}}$ and broadcasting measurements $E_{\textmd{Tx}}\left(\cdot\right)$. Finally, $s(\eta^\plyrind)$ denotes the probability of successful transmission\footnote{The probability of successful transmission is given by \begin{displaymath}s(\eta^\plyrind) = \sum_{m=0}^{B_{\textmd{max}}}p(\eta^\plyrind)(1-p(\eta^\plyrind))^m.\end{displaymath} Here, $B_{\textmd{max}}$ denotes the maximum number of back-offs, $p(\eta^\plyrind) = (1-q^\plyrind)^{\eta^\plyrind-1}$ denotes the probability that the channel is clear, and $q^\plyrind$ represents the probability that sensor $\plyrind$ is transmitting at a given time; see~\cite[p. 6099]{KMY08} for details in the unslotted CSMA/CA scheme.}.
A sensor is thus motivated to activate when majority of its neighbors are in the sleep mode and/or its estimate $\boldsymbol{\psi}^\plyrind$ is far from the local fused estimate $\boldsymbol{\phi}^\plyrind$.

The \emph{global} utility function $U^\plyrind_{G}(\cdot)$ concerns the connectivity of the network and diffusion of estimates.
Let $\mathcal{R}_r^\plyrind$ denote the set of sensors within radius $r$ from sensor $\plyrind$ excluding the neighborhood $\cneighborhoodk$. Define the number of active sensors in $\mathcal{R}_r^\plyrind$ by
\begin{displaymath}
\zeta^\plyrind\big(\actk,\actprof^{\nonneighbork}\big) = \actk + \sum_{l\in\mathcal{R}_r^\plyrind} a^l.
\end{displaymath}
The \emph{global} utility of sensor $\plyrind$ is then given by
\begin{equation}
\label{eq:global-utility}
U^\plyrind_{G}\big(\actk,\actprof^{\nonneighbork}\big) =
K_{g}\big(e^{-\gamma_g\zeta^\plyrind}\big)I\lbr\actk = 1\rbr
\end{equation}
where $K_g$ and $\gamma_g$ are the pricing parameters. Higher $\zeta^\plyrind$ lowers the global utility due to less importance of sensor $\plyrind$'s contribution to the diffusion of estimates across the network and keeping connectivity in $\mathcal{R}^k_r$.
Sensor $\plyrind$ is thus motivated to activate when majority of the nodes in its geographic region~$\mathcal{R}^\plyrind_r$ are in the sleep mode.


Each sensor $\plyrind$ realizes $\eta^\plyrind_\dtimee = \eta^\plyrind\big(\actk_\dtimee,\actprof_\dtimee^{\neighborhoodk}\big)$ as a consequence of receiving estimates $\lbrace \boldsymbol{\psi}^l_\dtimee\rbrace_{l\in\neighborhoodk}$ from neighbors, therefore, is able to evaluate its local utility at each period $\dtimee$. However, it does not observe the actions of non-neighbors, therefore, does not realize $\zeta_\dtimee^\plyrind = \zeta^\plyrind\big(\actk_\dtimee,\actprof^{\nonneighbork}_\dtime\big)$ and is unable to evaluate its global utility. The above game-theoretic model for activation control of sensors matches the setting of~\S\ref{sec:social-cliques}. Therefore, we employ Algorithm~\ref{alg:social-clique} with the above utility function to devise the activation mechanism. This mechanism is then embedded into the diffusion LMS algorithm such that the overall energy-aware diffusion LMS forms a two-timescale stochastic approximation algorithm: the fast timescale corresponds to the game-theoretic activation mechanism (step-size $\stepsize$) and the slow timescale is the diffusion LMS (step-size $\mu = \mathcal{O}(\stepsize^2)$).

The resulting energy-aware diffusion LMS algorithm is summarized in Algorithm~\ref{alg:diffusion-lms}.
\begin{algorithm}[H]
\caption{Energy-Aware Diffusion LMS \index{diffusion least mean square!energy-aware diffusion LMS}}
 \label{alg:diffusion-lms}
\textbf{Initialization}:  Set $0<\delta<1$, $\inertia^\plyrind > \fact\left|\utilityk_{\textmd{max}} - \utilityk_{\textmd{min}} \right|$, $\mixedstrat^\plyrind_0 = \big[\frac{1}{2},\frac{1}{2}\big]^\top$, and $R^{L,\plyrind}_0 = R^{G,\plyrind}_0 = \mathbf{0}_{2\times 2}$. Choose $0<\varepsilon,\rho\ll 1$ such that $\mu = \mathcal{O}(\stepsize^2)$, and $\boldsymbol{\psi}^\plyrind_0 = \boldsymbol{\phi}^\plyrind_0 = \mathbf{0}$.\\

\textbf{Step 1: Node Activation}.
Choose $\actk_\dtimee\in\lbr 1 \textmd{ (Activate)}, 0 \textmd{ (Sleep)}\rbr$  according to the randomized strategy $\mixedstrat^\plyrind_\dtimee$:
\begin{equation}
\begin{split}
\mixedstratind^{\plyrind}_\dtimee(i)
= \left\{
\begin{array}{ll}
(1-\explor)\min\lbr \frac{1}{\inertia^\plyrind}\left|r^{\plyrind}_\dtimee\big(\act^\plyrind_{\dtimee-1},i\big)\right|^{+},\frac{1}{2}\rbr + \frac{\explor}{2}, & \act^\plyrind_{\dtimee-1} \neq i\\
1-\sum_{j\neq i} p^{\plyrind}_\dtimee(j), & \act^\plyrind_{\dtimee-1} = i
\end{array}
\right.
\end{split}
\end{equation}
%

\textbf{Step 2: Diffusion LMS}.
\begin{equation}
\boldsymbol{\psi}_{\dtimee+1}^\plyrind = \left\{
\begin{array}{ll}
\boldsymbol{\phi}_{\dtimee}^\plyrind + \mu \lb\boldsymbol{u}^\plyrind_\dtimee\rb^\top\left[y_\dtimee^\plyrind - \boldsymbol{u}^\plyrind_\dtimee \boldsymbol{\phi}^\plyrind_{\dtimee}\right],& \textmd{if }\; \actk_\dtimee = 1\; \textmd{ (Active),}\\
\boldsymbol{\psi}_{\dtimee}^\plyrind, & \textmd{if } \;\actk_\dtimee = 0 \;\textmd{ (Sleep).}
\end{array}
\right.
\end{equation}

\textbf{Step 3: Estimate Exchange}.
If $\actk_\dtimee = 1$: (i) Transmit $\boldsymbol{\psi}^\plyrind_\dtimee$ to neighbors $\neighborhoodk$ and collect $\lbrace \boldsymbol{\psi}^l_\dtimee\rbrace_{l\in\neighborhoodk}$; (ii) set
%
\begin{equation}
\textmd{a}^l_\dtime = \left\{\begin{array}{ll}
1, & \textmd{if node $\plyrind$ receieves node $l$'s estimate $\boldsymbol{\psi}_\dtime^l$,}\\
0, & \textmd{otherwise,}
\end{array}\right.
\end{equation}
and
\begin{equation}
\hat{\boldsymbol{\psi}}_\dtime^l = \left\{\begin{array}{ll}
\textmd{received estimate}, & \textmd{if } \textmd{a}^l_\dtimee = 1,\\
\boldsymbol{\psi}_{\dtime-1}^l, & \textmd{if } \textmd{a}^l_\dtimee = 0.
\end{array}\right.
\end{equation}
If $\actk_\dtimee = 0$, go to Step 5.

\textbf{Step 4: Fusion of Local Estimates}.
\begin{equation}
\boldsymbol{\psi}^\plyrind_\dtimee =  \sum_{l\in\cneighborhoodk} w_{\plyrind l} \hat{\boldsymbol{\psi}}^l_\dtimee.
\end{equation}

\textbf{Step 5: Regret Update}. Run `Step~3' in Algorithm~\ref{alg:social-clique}.

\textbf{Step 6: Recursion}.
Set $\dtimee\leftarrow \dtimee+1$, and go Step 1.
\end{algorithm}

\begin{figure}[!h]
\begin{center}
\subfloat[Local Adaptation and Learning]{\hspace{-0.2cm}
\includegraphics[width=0.95\linewidth]{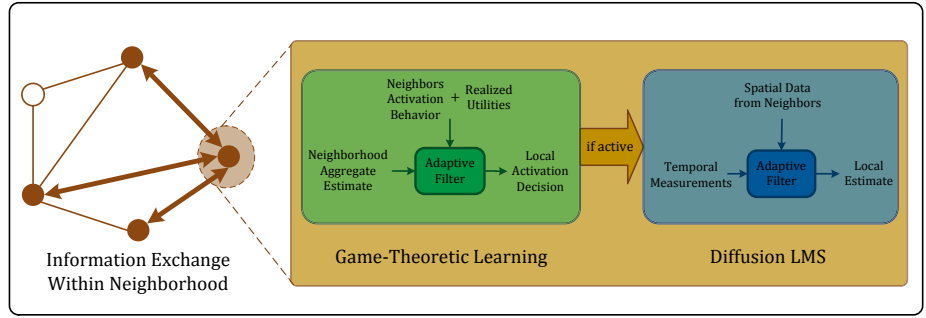}
\label{fig:LMS-local}
}\\
\vspace{-0.2cm}
\subfloat[Global Behavior]{
\includegraphics[width=0.6\linewidth]{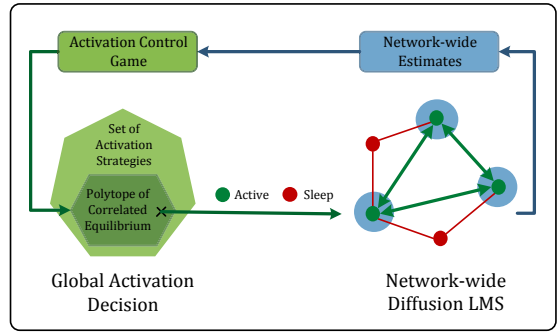}
\label{fig:LMS-global}
}
\end{center}
\caption{Energy-aware diffusion LMS is asymptotically consistent, yet the global activation behavior along the way
tracks the correlated equilibria set of the activation control game.}
\label{fig:diffusion-LMS}
\end{figure}

In such two timescale algorithms, the fast time-scale will see the slow component as quasi-static while the slow component sees the fast one near equilibrium. By virtue of the results of~\S\ref{sec:main-results-game}, consistency of the diffusion LMS~(\ref{eq:lms-convg}), and treatment of asynchronous stochastic approximation algorithms, it is shown in~\cite{NKY13b} that, if each sensor individually follows Algorithm~\ref{alg:social-clique}, the following results hold:
\begin{enumerate}
    \item[(i)] As $\mu\to 0$, $\boldsymbol{\psi}^{\plyrind,\mu}(t)$ converges weakly to $\boldsymbol{\psi}^o$ across the network. That is, the energy-aware diffusion LMS is consistent.
    \item[(ii)] As $\stepsize\to 0$, $\globbehav^\stepsize\cd$ converges weakly to $\CEe(\boldsymbol{\Psi}\cd)$. That is, the global activation behavior across the network belongs to the correlated $\epsilon$-equilibria set of the underlying activation control game.
\end{enumerate}
The local and global behavior of the energy-aware diffusion LMS in illustrated in Figure~\ref{fig:diffusion-LMS}. Note that, since the utility function~(\ref{eq:utility-function-lms}) is a function of the estimates across the network, so is the correlated equilibria set of the underlying game---it slowly evolves as the parameter estimates across the network $\boldsymbol{\Psi}_\dtimee$ change over time.
%


\subsection{Numerical Study}

\begin{figure}[!t]
     \centering
     \subfloat{
          \label{fig:net-fig}
          \hspace{-0.4cm}\includegraphics[width=0.45\textwidth]{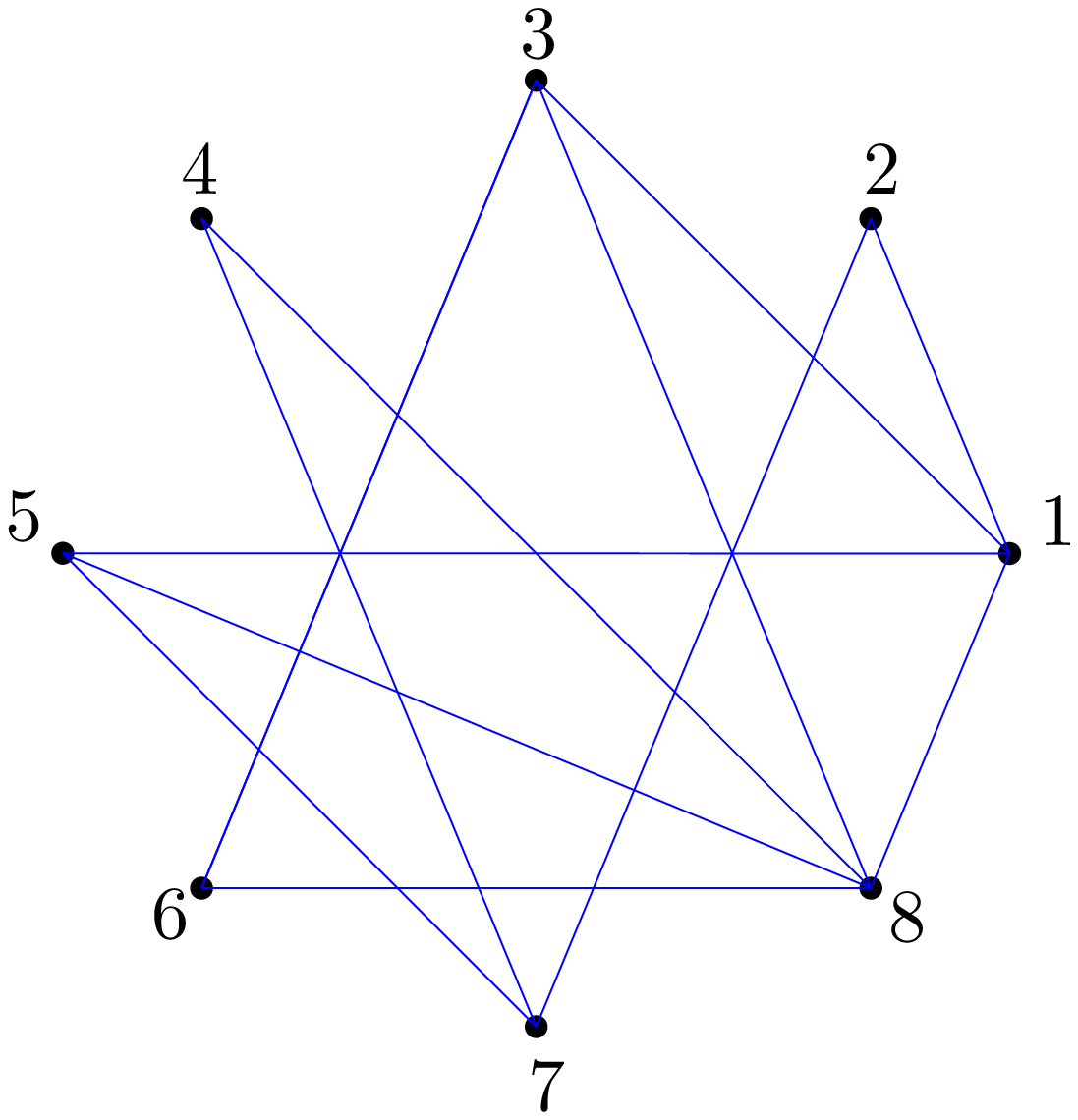}}
     \hspace{-.1in}
     \subfloat{
          \label{fig:h-statistics}
          \includegraphics[width=0.53\textwidth]{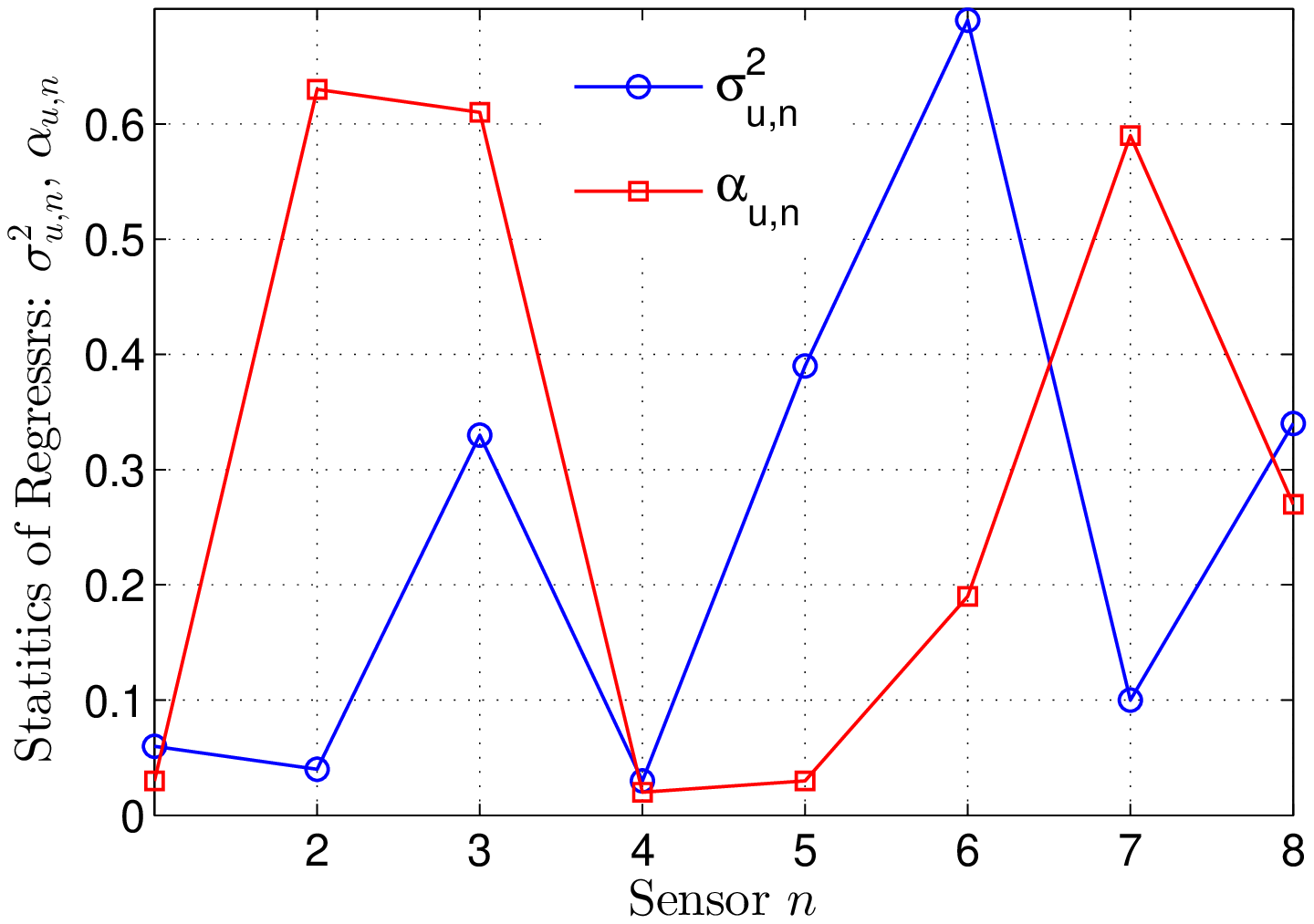}}
     \caption{Network topology and sensors' regressors statistics.}
     \label{fig:network}
\end{figure}

Fig.~\ref{fig:network} depicts the network topology that we study in this example. It further illustrates the statistics of each sensor's sequence of regressors that is generated by a Gaussian Markov source with local correlation function of the form $r(\tau) = \sigma_{\boldsymbol{u},\plyrind}^2 \alpha_{\boldsymbol{u},\plyrind}^{|\tau|}$, where $\alpha_{\boldsymbol{u},\plyrind}$ is the correlation index. We define $Q = [q_{kl}]$ in~(\ref{eq:C-1}) as
\begin{equation}
q_{\plyrind l} = \left\lbrace
\begin{array}{ll}
\disp \frac{1}{\big|\cneighborhoodk\big|}, & l\in\cneighborhoodk,\\
0, & \textmd{otherwise}.
\end{array}
\right.
\end{equation}
%
It is further assumed that sensors are equipped with Chipcon CC2420 transceiver chipset which implements CSMA/CA protocol for exchanging estimates with neighbors\footnote{The ZigBee/IEEE 802.15.4 standard is currently a leading choice for low-power communication in wireless sensor networks. It employs a CSMA/CA scheme for multiple access data transmission. In networks with tight energy constraints, the non-beacon-enabled (unslotted CSMA/CA) mode is more preferable as the node receivers do not need to switch on periodically to synchronize to the beacon.}. For brevity of presentation, we skip the detailed model description and expressions for $E_{\textmd{Tx}}(\cdot)$
and $E_{\textmd{Act}}$ in~(\ref{eq:local-utility}); the interested reader is referred to~\cite{NKY13b},~\cite[Appendix]{KMY08} for the details. We further assume the noise $v_n^k$ is i.i.d. with $\sigma_{v,k}^2 = 10^{-3}$,
and $\mu =  0.01$.

\begin{figure}[p]
     \begin{center}
     \includegraphics[width=0.8\textwidth]{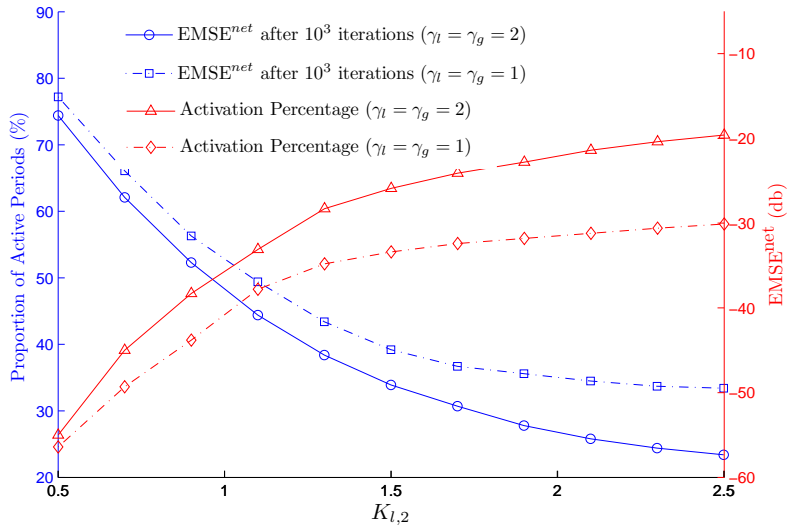}
     \end{center}
     \caption{Energy expenditure vs. estimation accuracy after $10^4$ iterations.}
     \label{fig:trade-off}
\end{figure}

\begin{figure}[p]
     \begin{center}
     \includegraphics[width=0.8\textwidth]{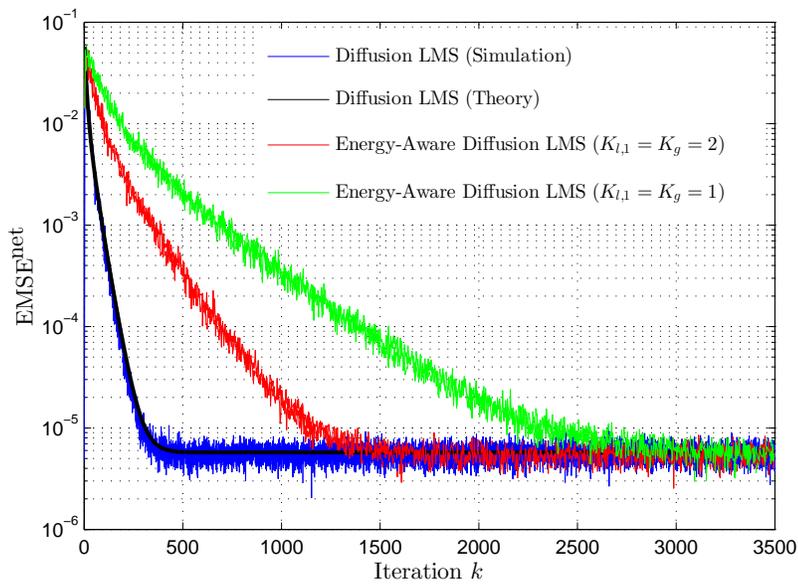}
     \end{center}
     \caption{Network excess mean square error $\textmd{EMSE}^{\textmd{net}}$.}
     \label{fig:emse-net}
\end{figure}

Define the network excess mean square error:
\begin{equation}
\textmd{EMSE}^\textmd{net} = \frac{1}{\fplyr} \sum_{\plyrind=1}^\fplyr \textmd{EMSE}^\plyrind
\end{equation}
where $\textmd{EMSE}^\plyrind$ denotes the EMSE for sensor $\plyrind$. $\textmd{EMSE}^\textmd{net}$ is simply obtained by averaging $\E |\boldsymbol{u}_\dtimee^\plyrind( \boldsymbol{\psi}^o - \boldsymbol{\phi}^{k}_{\dtimee-1})|^2$ across all sensors in the network. Figure~\ref{fig:trade-off} demonstrates the trade-off between energy expenditure in the network and the rate of convergence of the diffusion LMS algorithm in terms of $\textmd{EMSE}^{\textmd{net}}$ after $10^4$ iterations. Sensors become more conservative by increasing the energy consumption parameter $K_{l,2}$ and, accordingly, activate less frequently due to receiving lower utilities; see (\ref{eq:local-utility}). This reduces the average proportion of active sensors and increases $\textmd{EMSE}^{\textmd{net}}$ due to recording fewer measurements and less frequent fusion of neighboring estimates. Increasing the pricing parameters $\gamma_l$, in~(\ref{eq:local-utility}), and $\gamma_g$, in (\ref{eq:global-utility}), has the same effect as can be observed in Figure~\ref{fig:trade-off}.
The global performance of Algorithm~\ref{alg:diffusion-lms} is further compared with the standard diffusion LMS~\cite{LS08} in Figure~\ref{fig:emse-net}. As the pricing parameters $K_{l,1}$ in~(\ref{eq:local-utility}) (corresponding to the contribution of sensors in local parameter estimation) and $K_{g}$ in~(\ref{eq:global-utility}) (corresponding to the contribution of sensors in local diffusion) increase, sensors
activate more frequently. As shown in Figure~\ref{fig:emse-net}, this improves the rate of convergence of the energy-aware diffusion LMS algorithm.


\section{Closing Remarks}
\label{sec:conclusion-game}

In this chapter, we used non-cooperative game-theoretic learning as a model for interactive sensing with social sensors.
Besides the competitive situations,
equally abundant are interactions wherein multiple agents
cooperate to reach a common goal. For instance, family members and friends register to the same long-distance service to take advantage of more affordable rates. Similarly, multiple airlines may form alliances to provide convenient connections between as many airports as possible so as to increase profit~\cite{SV01}. Associated with each formed group of cooperating parties, namely, \emph{coalitions}, there corresponds a productive value which captures how fruitful their collaboration is in fulfilling the coalition's goal. Such a value may depend not only on parties that form a coalition and their connection (influence), but also on those who are excluded from the coalition. For instance, registering to a particular long-distance service clearly has advantages to the members of a family; however, due to the associated overhead costs, it may limit communication with friends who are registered to a different service. Therefore, any competition (if existent at all) will be among coalitions, rather than parties.

In such situations the main issue is to obtain a fair distribution of the joint revenues from cooperation so as to form stable coalitions. For instance, the productive value may be split based on the relative contributions or relations (influence) of different parties within each coalition. In the presence of multiple coalitions, each agent naturally favours the one that ensures the highest payoff. Allocation of shares from the joint revenues, thus, impacts the structure of formed coalitions and their stability.

\index{game!cooperative game}
Formation of such stable coalitions can be naturally modelled and analysed in a particular strain of the cooperative game framework, namely, \emph{coalition formation games}. Coalition formation is of particular significance in several economic, political and social contexts such as cartel formation, lobbies, customs unions, conflict, public goods provision, political party formation, etc.~\cite{R08,YZ13}. The single-valued \emph{Shapley value}~\cite{S53b,H89}, named in honour of Lloyd Shapley, and the set-valued \emph{core}\footnote{The formal concept of the core for cooperative games in characteristic form was first introduced by Gillies~\cite{G53} and Shapley~\cite{S53a}; see~\cite[Ch.~6]{M86} for additional background.} are the most prominent solution concepts in the context of cooperative games.

An extension of the game-theoretic learning approach to interactive sensing is to model formation of stable coalitions of sensors that collaborate to gather data about the same phenomenon of interest; cf.~\cite{NAK13,NK10} for two instances of cooperative game-theoretic learning approach to target localization and resource allocation problems.

Another direction for further research is to explore the adaptive learning algorithms of game-theoretic nature in large-scale discrete optimization problems, where the optimal solution is obtained by collecting observations from many separate sources via a social network~\cite{GPS07,NKY14}. One such problem is choosing influential sets of
individuals to maximize the spread of influence through a social network, which was first formulated as a discrete optimization in the seminal work~\cite{KKT03}: ``if we can try to convince a subset of individuals to adopt a new product or innovation, and the goal is to trigger a large cascade of further adoptions, which set of individuals should we target?'' For more recent works in this area, the reader is referred to~\cite{CWW10,CWY09}.

\chapter{Summary}\label{chapter:conclusion}

The motivation for this monograph stems from understanding how individuals interact in a social network and how simple local behavior can result in complex global behavior. The underlying tools employed in this monograph are widely used by the electrical engineering research community in the areas of signal processing, control, information theory and network communications. In what follows, we summarize the main results discussed in detail in this monograph:

Chapter~\ref{Chapter:SL} considered  social learning models for interaction among sensors where agents use their private observations along with actions of other agents to estimate an underlying state of nature.
We considered extensions of the basic social learning paradigm to online reputation systems in which agents communicate over a social network.  Despite the apparent simplicity in these information flows, the systems exhibit unusual behavior such as herding and data incest.  Also, an example of social-learning for change detection was considered.

Chapter~\ref{Chapter:tracking} analyzed the dynamics of a duplication-deletion graph where at each time instant, one node can either join or leave the graph (an extension to the duplication model of~\cite{CLD03,PSS03}). The power law component for such graph was computed.
 Also, a Markov-modulated random graph was proposed to model the social networks whose dynamics evolve over time according to a finite-state Markov chain.
 Using the stochastic approximation algorithms, the probability mass function of degree of each node (degree distribution) was estimated. Then, an upper bound was derived for
the distance between the estimated and the expected degree distribution. As a result of this bound, we showed that the scaled tracking error between the expected degree distribution and the estimated one weakly converges to a diffusion process. From
that, the covariance of this error can be computed.  Finally, we presented a discussion on simulating graphical realizations of a degree sequence which is a problem of much interest in numerical studies of social networks.

Chapter~\ref{Chapter:diffusion} discussed the Susceptible-Infected-Susceptible (SIS) model  for  diffusion of  information in social networks. By using the mean field dynamics, the degree distribution was shown to satisfy an ordinary differential equation. As a result, the probability of a link pointing to an infected node satisfies a scalar differential equation. Given these dynamics for the state of information diffusion, sampling of the nodes in the network can be viewed as a noisy measurements of the state. We outlined two types of sampling individuals in a network that are of recent interest---namely, social sampling and respondent-driven sampling. Given these noisy measurements, estimation of the underlying state was formulated as a Bayesian filtering problem.

Chapter~\ref{chapter:noncooperative} discussed, at length, a non-Bayesian formulation, where agents seek to achieve coordination in decision making by optimizing their own utility functions. We adopted a non-cooperative game-theoretic formulation and considered three protocols for information exchange among interacting agents. The well-known game-theoretic learning algorithm---namely, the Hart \& Mas-Colell's ``regret-matching'' procedure---was presented and extended to accommodate the information flow protocols that are typical in interactions among agents in social networks. Interestingly enough, we have shown that, while each individual has limitations in sensing the environment and communicating with other agents, the coordinated behavior among agents can lead to the manifestation of sophisticated behavior at the network level. More precisely, the global behavior of agents turn out to be distributed according to randomized strategies drawn from the convex polytope of correlated equilibria. Instances of such emergent global behavior are even observed in nature, e.g., fish schooling and birds flight in formation~\cite{CS11,TS11}.

In this monograph, to give a flavor of the various averaging analysis methods, we used a variety of  analysis tool  to analyse the asymptotic properties of the presented stochastic approximation algorithms: (a) Mean square convergence: Chapter~\ref{Chapter:tracking} used this method to derive bounds on the difference between the estimated and the expected degree distribution of a Markov-modulated random graph; (b) Maximum deviation bounds: Chapter~\ref{Chapter:diffusion} employed this method to derive an exponential bound (in terms of the number of agents) for the probability of deviation of the sample path of the infected distribution from the mean field dynamics in finite time interval; (c) Weak convergence:
Chapter~\ref{Chapter:tracking}  and
Chapter~\ref{chapter:noncooperative} use weak convergence methods to analyze the convergence of stochastic approximation algorithms. In Chapter~\ref{Chapter:tracking} these
methods were used to specify how the empirical distribution of a Markov chain can be tracked and how the tracking errors satisfy a switched Markovian diffusion process.
Finally Chapter~\ref{chapter:noncooperative} used weak convergence  methods to derive the limit system representing the asymptotics of the game-theoretic decision making procedure. This method was shown to be strong enough to tackle Markovian evolution of the parameters underlying the non-cooperative game model.

There exist other important problems in the context of social networks that have been extensively studied in the literature, hence, were not included in this monograph. Examples include consensus formation~\cite[Chapter~8]{J08},~\cite{KB08,TJ10}, metrics for measuring networks (other than degree distribution)~\cite[Chapter~2]{J08}, \cite{WKV13}, small world~\cite{Kle00,Wat99,WS98}, cooperative models of network formation~\cite[Chapter~1]{DW05},~\cite[Chapter~12]{J08},~\cite{SP09}, behavior dynamics in peer-to-peer media-sharing social networks~\cite{HK07,ZLL11}, and privacy and security modeling~\cite{Lan07,LYM03}, to name a few. The interested reader can find detailed discussions on these topics in the above references and the references therein. 


\appendix
\chapter{Appendix to Chapter~3}
\section{Proof of Theorem~3.3.1}
\label{ap:mu}
The proof is based on the proof of Lemma~4.1 in~\cite[Chapter~4, p79]{CL06}. To compute the expected degree distribution of the Markov-modulated random graph, we obtain a relation between the number of nodes with specific degree at time $\tim$ and the degree distribution of the graph at time $\tim-1$. For the sake of notational convenience, we drop the dependency of probabilities of connection and deletion on $\mc$ and we denote $\pdup(\mc)$ and $\pdel(\mc)$ by $\pdup$ and $\pdel$. Given the resulting graph at time $\tim$, the aim is to find the expected number of nodes with degree $i+1$ at time $\tim+1$. The following events can occur that result in a node with
degree $i+1$ at time $\tim+1$:
\begin{itemize}[leftmargin=12pt]
\item{\em Degree of a node with degree $i$ increments by one in the duplication step (Step~2 of Algorithm~\ref{alg:duplication}) and remains unchanged in the deletion step (Step~3):}\begin{itemize}[leftmargin=12pt]
     \item A node with degree $i$ is chosen at the duplication step as a parent node and remains unchanged in the deletion step. The probability of occurrence of such an event is $$\left(1-\frac{\pdel(i+1)+\pdel (1+pi)- \pdel (1+\pdup i)(i+1)/\sizegraph_\tim}{\sizegraph_\tim}\right)\frac{\degree_\tim(i)}{\sizegraph_\tim};$$ the probability of choosing a node with degree $i$ is $\frac{\degree_\tim(i)}{\sizegraph_\tim}$ and the probability of the event that this node remains unchanged in the deletion step is\footnote{The deletion step (Step~3 of Algorithm~\ref{alg:duplication}) comprises an edge-deletion step and a duplication step. The probability that the degree of node with degree $i$ changes in the edge-deletion step is $\frac{\pdel(i+1)}{\sizegraph_\tim}$; either this node or one of its neighbors should be selected in the edge-deletion step. Also given that the degree of this node dose not change in the edge-deletion step, if either this node or one of its neighbor is selected in the duplication step (within Step~3) then the degree of this node increments by one with probability $\frac{1+\pdup i}{\sizegraph_\tim}$. Therefore, the probability that the degree of a node of degree $i$ remains unchanged in Step~3 is $$1-\frac{\pdel(i+1)+\pdel (1+pi) - \pdel (1+\pdup i)(i+1)/\sizegraph_\tim}{\sizegraph_\tim}.$$ Note that for simplicity in our analysis, we assumed that the nodes whose degrees are changed in the edge-deletion part of Step~3, remain unchanged in the duplication part of Step~3 at that time instant. Also, the new node, which is generated in the vertex-duplication step of Step~2, remains unchanged in Step~3. Therefore, although the degree of parent node becomes $i+1$ in this case but this node is treated as a node with degree $i$ in Step~3 of Algorithm~\ref{alg:duplication}.} $$1-\frac{\pdel(i+1)+\pdel (1+pi)- \pdel (1+\pdup i)(i+1)/\sizegraph_\tim}{\sizegraph_\tim}.$$
    \item One neighbor of a node with degree $i$ is selected as a parent node; the parent node connects to its neighbors (including the node with degree $i$) with probability $\pdup$ in the edge-duplication part of Step~2. The probability of such an event is {\footnotesize$$\frac{\degree_\tim(i)pi}{\sizegraph_\tim}\left(1-\frac{\pdel(i+2)+\pdel (1+p(i+1))- \pdel (1+\pdup( i+1))(i+2)/\sizegraph_\tim}{\sizegraph_\tim}\right).$$} Note that the node whose degree is incremented by one in this event should remain unaffected in Step~3; the probability of being unchanged in Step~3 for such a node  is
        $$1-\frac{\pdel(i+2)+\pdel (1+p(i+1))- \pdel (1+\pdup( i+1))(i+2)/\sizegraph_\tim}{\sizegraph_\tim}.$$
\end{itemize}
\item{ \em A node with degree $i+1$ remains unchanged in both Step~2 and Step~3 of Algorithm~\ref{alg:duplication}:}
\begin{itemize}[leftmargin=12pt]
\item Using the same argument as above,  the probability of such an event is {\small $$f_\tim(i+1)\left(1-
\pdel\frac{i+3+ \pdup(i+1) -\frac{(1+\pdup(i+1))(i+2)}{\sizegraph_\tim}}{\sizegraph_\tim}\right)\left(1-
\frac{\pdup(i+1)+1}{\sizegraph_\tim}\right).$$}
\end{itemize}

\item{\em A new node with degree $i+1$ is generated in Step~2:}
\begin{itemize}
[leftmargin=12pt]
\item The degree of the most recently generated node (in the vertex-
duplication part of Step~2) increments to $i+1$; the new node connects to $``i"$  neighbors of the parent node and
remains unchanged in Step~3. The probability of this scenario
is

{\small $$\left(1-
\pdel\frac{i+3+ \pdup(i+1) -\frac{(1+\pdup(i+1))(i+2)}{\sizegraph_\tim}}{\sizegraph_\tim}\right)\sum_{j\geq
i}\frac{f_\tim(j)}{\sizegraph_\tim}{{j}\choose{i}}\pdup^i(1-\pdup)^{j-i}.$$}

\end{itemize}
\item{\em Degree of a node with degree $i+2$ decrements by one in Step~3:}
\begin{itemize}[leftmargin=12pt]
\item A node with degree $i+2$ remains unchanged in the duplication step
and one of its neighbors is eliminated in the deletion step. The
probability of this event is  $$\pdel\left(\frac{i+2}{\sizegraph_\tim}\right)\left(1-
\frac{\pdup(i+2)+1}{\sizegraph_\tim}\right).$$
\end{itemize}
\item{ \em A node with degree $i+1$ is generated in Step~3:}
\begin{itemize}[leftmargin=12pt]
\item The degree of the node generated in the vertex-duplication part of duplication step within Step~3 increments to
$i+1$. The probability of this event is $$\pdel\sum_{j\geq
i}\frac{1}{\sizegraph_\tim}f_\tim(j){{j}\choose{i}}\pdup^i(1-\pdup)^{j-i}.$$
\end{itemize}
\item{\em Degree of a node with degree $i$ increments by one in Step~3:}
\begin{itemize}[leftmargin=12pt]
\item  A node with degree $i$ remains unchanged in Step~2
and its degree increments by one in the duplication part of Step~3. The corresponding probability is
$$\frac{\pdel(1 + pi)}{\sizegraph_\tim}\left(1 - \frac{1+pi}{\sizegraph_\tim}\right).$$
\end{itemize}
\end{itemize}
 Let $\Omega$ denote the set of all arbitrary graphs and
$\mathcal{F}_\tim$ denote the sigma algebra generated by graphs
$\graph_\tau, \tau \leq \tim$. Considering the above events that result
in a node with degree $i+1$ at time $\tim+1$, the following recurrence
formula can be derived for the conditional expectation of
$f_{\tim+1}(i+1)$:

\vspace{-0.2cm}
{\small \begin{align}
\label{eq1}
&\mathbf{E}\{f_{\tim+1}(i+1)|\mathcal{F}_\tim\}=\nonumber\\
&\left(1-\pdel\frac{i+3+\pdup(i+1) -\frac{(1+\pdup(i+1))(i+2)}{\sizegraph_\tim}}{\sizegraph_\tim}\right)\left(1-
\frac{\pdup(i+1)+1}{\sizegraph_\tim}\right)f_\tim(i+1)\nonumber\\
&+ \left(1-\frac{\pdel(i+1)+\pdel (1+pi)- \pdel (1+\pdup i)(i+1)/\sizegraph_\tim}{\sizegraph_\tim}\right)\left(\frac{1+\pdup i}{\sizegraph_\tim}\right)f_\tim(i)\nonumber\\
& \ +\left(1-
\pdel\frac{i+3+ \pdup(i+1) -\frac{(1+\pdup(i+1))(i+2)}{\sizegraph_\tim}}{\sizegraph_\tim}\right)\sum_{j\geq
i}\frac{f_\tim(j)}{\sizegraph_\tim}{{j}\choose{i}}\pdup^i(1-\pdup)^{j-i}\nonumber\\
& \ +\pdel\sum_{j\geq i}\frac{f_\tim(j)}{\sizegraph_\tim}{{j}\choose{i}}\pdup^i(1-\pdup)^{j-
i} +\pdel\left(\frac{i+2}{\sizegraph_\tim}\right)\left(1-\frac{\pdup(i+2)+1}{\sizegraph_\tim}\right)
f_\tim(i+2)\nonumber\\
& \ +\frac{\pdel(1 + pi)}{\sizegraph_\tim}\left(1 -
\frac{1+pi}{\sizegraph_\tim}\right)f_\tim(i).
\end{align}}

\vspace{-0.2cm}
\noindent
Let $\baf^\mc_\tim(i)= \mathbf{E}\{f_\tim(i)|\mc_\tim = \mc\}$. By taking expectation of
both sides of (\ref{eq1}) with respect to trivial sigma algebra
$\{\Omega,\emptyset\}$, the smoothing property of conditional
expectations yields:

\vspace{-0.2cm}
{\small \begin{align}
\label{eq:1}
&\baf^\mc_{\tim+1}(i+1)=\nonumber\\&\quad\left(1-
\pdel\frac{i+3+\pdup(i+1) -\frac{(1+\pdup(i+1))(i+2)}{\sizegraph_\tim}}{\sizegraph_\tim}\right)\left(1-
\frac{\pdup(i+1)+1}{\sizegraph_\tim}\right)\baf^\mc_\tim(i+1)\nonumber\\\
& \ +\left(1-\frac{\pdel(i+1)+\pdel (1+pi)- \frac{\pdel (1+\pdup i)(i+1)}{\sizegraph_\tim}}{\sizegraph_\tim}\right)\left(\frac{1+\pdup i}{\sizegraph_\tim}\right)\baf^\mc_\tim(i)\nonumber\\
&\ +\left(1-
\pdel\frac{i+3+ \pdup(i+1) -\frac{(1+\pdup(i+1))(i+2)}{\sizegraph_\tim}}{\sizegraph_\tim}\right)\sum_{j\geq
i}\frac{\baf^\mc_\tim(j)}{\sizegraph_\tim}{{j}\choose{i}}\pdup^i(1-\pdup)^{j-i}\nonumber\\
&\  +\pdel\sum_{j\geq i}\frac{1}{\sizegraph_\tim}\baf^\mc_\tim(j){{j}\choose{i}}\pdup^i(1-
\pdup)^{j-i} +\pdel\left(\frac{i+2}{\sizegraph_\tim}\right)\left(1-\frac{\pdup(i+2)+1}{\sizegraph_\tim}\right)
\baf^\mc_\tim(i+2)\nonumber\\
&\ +\frac{\pdel(1 + pi)}{\sizegraph_\tim}\left(1 -
\frac{1+pi}{\sizegraph_\tim}\right)\baf^\mc_\tim(i).
\end{align}}

\vspace{-0.2cm}
\noindent
Assuming that size of the graph is sufficiently large,  each term like
$\frac{\baf_\tim(i')}{\sizegraph_\tim^2}$ can be neglected. Also, taking functional dependencies of $\pdup$ and $\pdel$ on $\mc$ into account, Eq. (\ref{eq:1}) can be written as
\begin{align}
\label{eq:barf}
\baf^{\mc}_{\tim+1}(i+1)& = \left(1- \frac{\pdel(\mc)(i+2)+
\pdel(\mc)\big(\pdup(\mc)(i+1)+1\big)}{\sizegraph_\tim}\right)\baf^{\mc}_\tim(i+1)
\nonumber\\
&\ +\left(\frac{(1+\pdup(\mc)
i)\pdel(\mc)}{\sizegraph_\tim}\right)\baf^{\mc}_\tim(i)+\pdel(\mc)\left(\frac{i+2}
{\sizegraph_\tim}\right)\baf^{\mc}_\tim(i+2)\nonumber\\
&\ + \pdel(\mc)\sum_{j\geq
i}\frac{1}{\sizegraph_\tim}\baf^{\mc}_\tim(\mc,j){{j}\choose{i}}\pdup(\mc_{\tim+1})^i
(1-\pdup(\mc_{\tim+1}))^{j-i}.
\end{align}

 Using (\ref{eq:1}), we can write the following recursion for the
$(i+1)$-th element of $\bg^{\mc}({\tim+1})$:
\begin{align}
\label{eq:barg}
\bg^{\mc}_{\tim+1}(i+1)& = \left(\frac{\sizegraph_\tim-\left(
\pdel(\mc)(i+2)+
\pdel(\mc)\big(\pdup(\mc)(i+1)+1\big)\right)}{\sizegraph_{\tim+1}}
\right)\bg^{\mc}_\tim(i+1)\nonumber\\
&\ +\left(\frac{(1+\pdup(\mc)
i)\pdel(\mc)}{\sizegraph_{\tim+1}}\right)\bg^{\mc}_\tim(i)+\pdel(\mc)\left(\frac{i
+2}{\sizegraph_{\tim+1}}\right)\bg^{\mc}_\tim(i+2)\nonumber\\
&\ +
\pdel(\mc)\sum_{j\geq
i}\frac{1}{\sizegraph_{\tim+1}}\bg^{\mc}_\tim(j){{j}\choose{i}}\pdup(\mc)^i(1-
\pdup(\mc))^{j-i}.
\end{align}
Since the probability of duplication step $\pdupstep = 0$, the number of vertices does not increase. Thus, $\sizegraph_\tim = \sizegraph_0$ and (\ref{eq:barg}) can be written as
\begin{align}
\label{eq:barg2}
\bg^{\mc}_{\tim+1}(i+1)=& \Big(1 - \frac{1}{\sizegraph_0}\left(\pdel(\mc)(i+2)+
\pdel(\mc)\big(\pdup(\mc)(i+1)+1\big)\right)\Big)\bg^{\mc}_\tim(i+1)
\nonumber\\
&+\frac{1}{\sizegraph_0}\Big((1+\pdup(\mc)
i)\pdel(\mc)\bg^{\mc}_\tim(i)+\frac{1}{\sizegraph_0}\pdel(\mc)(i+2)\bg^{\mc}_\tim(i+2)
\Big)\nonumber\\&+\frac{1}{\sizegraph_0} \pdel(\mc)\sum_{j\geq
i}\bg^{\mc}_\tim(j){{j}\choose{i}}\pdup(\mc)^i(1-\pdup(\mc))^{j-i}.
\end{align}
It is clear in~(\ref{eq:barg2}) that the vector
$\bg^{\mc}_{\tim+1}$ depends on elements of $\bg^{\mc}_\tim$. In a
matrix notation, (\ref{eq:barg2}) can be expressed as
\beq
\label{eq:true}
\bg^{\mc}_{\tim+1} = \left(\identity + \frac{1}{\sizegraph_0} \transition(\mc)\right)\bg^{\mc}_\tim
\eeq
where $\transition(\mc)$ is defined as (\ref{eq:L}).

To prove that $\transition(\mc)$ is a generator, we need to show
that $\transitionelem_{ii} < 0$ and $\sum_{i = 1}^{\sizegraph_0}\transitionelem_{ki} = 0$. Accordingly,
\begin{align}\label{temp}
\sum_{i = 1}^{\sizegraph_0}\transitionelem_{ki} &= -\left(\pdel(\mc)(k+1) +
\pdel(\mc)(1 + \pdup(\mc)k)\right) + (1 +
\pdup(\mc)k)\pdel(\mc) \nonumber\\&+ \pdel(\mc)k +
\pdel(\mc)\sum_{k\leq i-1} {{k}\choose{i-1}} \pdup(\mc)^{i-1} (1-
\pdup(\mc))^{k-i+1}\nonumber\\
& = -\pdel(\mc) + \pdel(\mc)\sum_{k\leq i-1} {{k}\choose{i-1}}
\pdup(\mc)^{i-1} (1-\pdup(\mc))^{k-i+1}.
\end{align}
Let $m = i - 1$. Then, (\ref{temp}) can be rewritten as
\begin{align}\label{temp2}
\sum_{i = 1}^{\sizegraph_0}\transitionelem_{ik} =& -\pdel(\mc) +
\pdel(\mc)\sum_{m=0}^k {{k}\choose{m}} \pdup(\mc)^{m} (1-
\pdup(\mc))^{k-m}\nonumber\\
=&  -\pdel(\mc) + \pdel(\mc)(1- \pdup(\mc))^k\sum_{m=0}^k
{{k}\choose{m}} \left(\frac{\pdup(\mc)}{1-\pdup(\mc)}\right)^{m}.
\end{align}
Knowing that $\sum_{m=0}^k{{k}\choose{m}} a^m = \left(1 + a\right)^k$,
(\ref{temp2}) can be written as
\begin{align}\label{temp3}
\sum_{i = 1}^{\sizegraph_0}\transitionelem_{ik}  =   -\pdel(\mc) +
\pdel(\mc)(1- \pdup(\mc))^k\left(\frac{1}{1 -
\pdup(\mc)}\right)^k = 0.
\end{align}
Also, it can be shown that $\transitionelem_{ii} < 0$. Since $\pdup(\mc)^{i-1} \leq 1$, $\pdup(\mc)^{i-1} < 1 + \frac{2}{i} + \pdup(\mc) + \pdup(\mc)^i$. Consequently, $i\pdel(\mc)\pdup(\mc)^{i-1}(1-\pdup(\mc))- \pdel(\mc)(i+2 + i\pdup(\mc)) < 0$. Therefore, $\transitionelem_{ii}< 0$ and the desired result follows. It is straightforward to show that all elements of  $\ttrue(\mc)^{N_0}$ are strictly greater than zero. Therefore, $\ttrue(\mc)$ is irreducible and aperiodic. (Recall that $\ttrue(\mc) = \identity + \frac{1}{\sizegraph_0} \transition(\mc)$.)

\section{Proof of Theorem~3.5.1}
\label{ap:bound}

 Define the Lyapunov function\index{Lyapunov function} $V(x) = (x'x)/2$ for  $x \in \mathbb{R}^\s$. 
Use $\Et$ to denote the conditional expectation
with respect to the $\sigma$-algebra $\salgebH_\tim$ generated by
$\{\obs_j,\mc_j, \quad j\leq \tim\}$. Then,
\begin{align}
\label{eq17}
\Et\{V&(\tg_{\tim+1})-V(\tg_\tim)\} = \nonumber\\&\Et\Big\{\tg'_\tim[-\esa\tg_\tim+\esa\left({\obs_{\tim+1}}-\bg(\mc_\tim)\right)
+\bg(\mc_\tim) - \bg(\mc_{\tim+1})]\Big\}\nonumber\\
&+\Et\Big\{|-\esa\tg_\tim+\esa\left(\obs_{\tim+1}-
\bg(\mc_\tim)\right)+\bg(\mc_\tim) -
\bg(\mc_{\tim+1})|^2\Big\}
\end{align}
where $\obs_{\tim+1}$ and $\bg(\mc_\tim)$ are  vectors in
$\mathbb{R}^\s$ with elements $\obs_\tim(i)$ and $\bg(\mc_\tim,i)$,\quad
$1 \leq i \leq \s$, respectively. Due to the Markovian assumption and the structure of the transition matrix of $\mc_\tim$, defined in~(\ref{eq:A}),
\begin{align}
\Et\{\bg&(\mc_\tim) - \bg(\mc_{\tim+1})\}=\E\{\bg(\mc_\tim) - \bg(\mc_{\tim+1})|\mc_\tim\}\nonumber\\
&=\sum_{i=1}^{\sizemc}\E\{\bg(i) - \bg(\mc_{\tim+1})|\mc_\tim = i\}\indicatorit\lbr\mc_\tim = i\rbr\nonumber\\
&=\sum_{i=1}^{\sizemc}\left[\bg(i) - \sum_{j=1}^{\sizemc}\bg(j)\A^\emc_{ij}\right]\indicatorit\lbr\mc_\tim = i\rbr\nonumber\\
&=-\emc\sum_{i=1}^{\sizemc}\sum_{j=1}^{\sizemc}\bg(j)\generatorelement_{ij}\indicatorit\lbr\mc_\tim = i\rbr\nonumber\\
&=O(\emc)
\end{align}
where $\indicatorit(\cdot)$ denotes the indicator function. Similarly, it is easily seen that
\beq
\Et\{|\bg(\mc_\tim) - \bg(\mc_{\tim+1})|^2\} = O(\emc).
\eeq

Using $K$ to denote a generic positive value
(with the notation $KK=K$  and $K+K =K$),
a familiar inequality $ab \leq \frac{a^2+b^2}{2}$ yields
\beq\label{eq:elem}  O(\esa\emc) = O(\esa^2 + \emc^2).\eeq  Moreover,
 we have $|\tg_\tim| = |\tg_\tim| \cdot 1 \leq (|\tg_\tim|^2
+1 )/2$. Thus, \beq\label{eq:pro} O(\emc)|\tg_\tim|\leq
O(\emc)\left(V(\tg_\tim) + 1\right). \eeq
Then, detailed estimates lead to
\begin{align}
\label{eq21}
\Et&\Big\{\Big|-\esa\tg_\tim+\esa\left(\obs_{\tim+1}-
\bg(\mc_\tim)\right)+\bg(\mc_\tim) -
\bg(\mc_{\tim+1})\Big|^2\Big\}
 \nonumber\\
&\le K \Et\Bigg\{\esa^2|\tg_\tim|^2+\esa^2|(\obs_{\tim+1}-
\bg(\mc_\tim)|^2 +\esa^2
\big|\tg'_\tim\left(\obs_{\tim+1}-
\bg(\mc_{\tim+1})\right)\big|\nonumber\\
&+ \esa |\tg'_\tim\left(\bg(\mc_\tim) - \bg(\mc_{\tim+1})\right)| +
\esa|\left(\obs_{\tim+1}-
\bg(\mc_\tim)\right)'\left(\bg(\mc_\tim)
- \bg(\mc_{\tim+1})\right)|\Bigg\}\nonumber\\ &+ \Et\{|\bg(\mc_\tim) -
\bg(\mc_{\tim+1})|\}^2.
\end{align}
It follows that
\begin{equation}
\label{eq21-2}
\begin{split}
&\Et\Big\{\Big|-\esa\tg_\tim+\esa\left(\obs_{\tim+1}-
\bg(\mc_\tim)\right)+\bg(\mc_\tim) -
\bg(\mc_{\tim+1})\Big|^2\Big\}\\
&\hspace{2cm}= O(\esa^2+\emc^2)(V(\tg_\tim)+1).
\end{split}
\end{equation}
Furthermore,
\begin{equation}
\label{eq23}
\begin{split}
\Et&\{V(\tg_{\tim+1})-V(\tg_\tim)\} =
-2\esa V(\tg_\tim) + \esa \Et\{\tg'_\tim[\obs_{\tim+1} -
\bg(\mc_\tim)]\} \\&+
\Et\{\tg'_\tim[\bg(\mc_{\tim+1}) - \bg(\mc_\tim)]\}
+ O(\esa^2 +\emc^2)(V(\tg_\tim)+1).
\end{split}
\end{equation}

To obtain the desired bound, define $V^{\emc}_1$ and $V^{\emc}_2$ as follows:
\begin{align}
V^{\emc}_1(\tg,\tim) &= \esa\sum_{j=\tim}^{\infty}\tg'\Et\{\obs_{j+1} -
\bg(\mc_j)\},\nonumber\\
V^{\emc}_2(\tg,\tim) &= \sum_{j=\tim}^{\infty}\tg'\mathbf{E}_\tim\{\bg(\mc_j)-
\bg(\mc_{j+1})\}.
\end{align}
It can be shown that
\bq{eq:order-vp} \barray
\ad |V^\rho_1(\tilde g, n)| = O(\e) (V(\tilde g)+ 1),\\
\ad |V^\rho_2(\tilde g, n)| = O(\rho) (V(\tilde g)+ 1).\earray\eq
Define $ W(\tg,\tim)$ as
\begin{equation}
 W(\tg,\tim) = V(\tg) +  V^{\emc}_1(\tg,\tim)+V^{\emc}_2(\tg,\tim).
 \end{equation}
This leads to
\begin{align}
\label{eq27}
\Et\{W(&\tg_{\tim+1},\tim+1) - W(\tg_\tim,\tim)\} = \Et\{V^{\emc}_1(\tg_{\tim+1},\tim+1)-
V^{\emc}_1(\tg_\tim,\tim)\}\nonumber\\&\hspace{5mm}+\Et\{V(\tg_{\tim+1})-
V(\tg_\tim)\} +\Et\{V^{\emc}_2(\tg_{\tim+1},\tim
+1)- V^{\emc}_2(\tg_\tim,\tim)\}.
\end{align}
Moreover,
\begin{equation}
\label{eq33}
\begin{split}
&\Et\{W(\tg_{\tim+1},\tim+1) - W(\tg_\tim,\tim)\} \\ &\qquad\quad=-2\esa V(\tg_\tim)+
O(\esa^2 +\emc^2)(V(\tg_\tim)+1).
\end{split}
\end{equation}
Equation~(\ref{eq33}) can be rewritten as
\begin{equation}
\label{eq30}
\begin{split}
\Et\{W(\tg_{\tim+1},\tim+1) - W(\tg_\tim,\tim)\} \leq  &O(\esa^2 +\emc^2)(W(\tg_\tim,\tim)+1)\\&-2\esa
W(\tg_\tim,\tim).
\end{split}
\end{equation}
If $\esa$ and $\emc$ are chosen small enough, then there exists a small
$\lambda$ such that $-2\esa + O(\emc^2)+ O(\esa^2) \leq -\lambda\esa$. Therefore, (\ref{eq30}) can be rearranged as
\begin{align}
\Et\{W(\tg_{\tim+1},\tim+1)\}\leq (1-\lambda\esa) W(\tg_\tim,\tim)+
O(\esa^2 +\emc^2).
\end{align}
Taking expectation of both sides yields
\begin{align}
\label{eq32}
\mathbf{E}\{W(\tg_{\tim+1},\tim+1)\}\leq
(1-\lambda\esa) \mathbf{E}\{W(\tg_\tim,\tim)\}+O(\esa^2 +\emc^2).
\end{align}
Iterating on~(\ref{eq32}) then results
\begin{equation}
\begin{split}
\mathbf{E}\{W(\tg_{\tim+1},\tim+1)\}\leq&
(1-\lambda\esa)^{\tim-N_\emc}\mathbf{E}\{W(\tg_{N_\emc},N_\emc)\}\\&+\sum_{j
= N_\emc}^{\tim}O(\esa^2 +\emc^2)(1-\lambda\esa)^{j-N_\emc}.
\end{split}
\end{equation}
As the result,
\begin{equation}
\begin{split}
\mathbf{E}\{W(\tg_{\tim+1},\tim+1)\}&\leq
(1-\lambda\esa)^{\tim-
N_\emc}\mathbf{E}\{W(\tg_{N_\emc},N_\emc)\}\\&\qquad+O\left(\esa
+\emc^2/\esa\right).
\end{split}
\end{equation}
If $\tim$ is large enough, one can approximate $(1-\lambda\esa)^{\tim-
N_\emc} = O(\esa)$. Therefore,
\begin{equation}
\mathbf{E}\{W(\tg_{\tim+1},\tim+1)\}\leq
O\left(\esa+\frac{\emc^2}{\esa}\right)
\end{equation}
Finally, using (\ref{eq:order-vp}) and replacing
$W(\tg_{\tim+1},\tim+1)$ with $V(\tg_{\tim+1})$, we obtain
\begin{equation}
\mathbf{E}\{V(\tg_{\tim+1})\}\leq
O\left(\emc+\esa+\frac{\emc^2}{\esa}\right).
\end{equation}

\section{Sketch of the Proof of Theorem~3.5.2}
\label{ap:conv-pf}
\noindent
The proof uses weak convergence\index{weak convergence} techniques \cite{EK86}.

Let $Z_k$ and $Z$ be ${\mathbb R}^r$-valued random vectors. We say $Z_k$ converges weakly to $Z$ ($Z_n \Rightarrow Z$) if for any bounded and continuous function $f(\cdot)$, $Ef(Z_k)\to Ef(Z)$ as $k\to \infty$. The sequence $\{Z_k\}$ is  tight if for each $\eta>0$, there is a compact set $K_\eta$ such that $P(Z_n\in K_\eta)\ge 1-\eta$ for all $k$. The definitions of weak convergence and tightness extend to random elements in more general metric spaces. On a complete separable metric space,  tightness\index{tightness} is equivalent to relative compactness, which is known as the Prohorov's Theorem~\cite{EK86}. By virtue of this theorem, we can extract convergent subsequences when tightness is verified. In what follows, we use a martingale problem formulation to establish the desired weak convergence. This usually requires to first prove tightness. The limit process is then characterized using certain operator related to the limit process. We refer the reader to \cite[Chapter 7]{KY03} for further details on weak convergence and related matters.

Since the proof is similar to \cite[Theorem 4.5]{YKI04}, we only
indicate the main steps in what follows and omit most of the verbatim details.

{\em Step 1:} First, we show that the two component process $(\hat
g^\e\cd,\theta^\e\cd)$ is tight in $D([0,T]:
\rr^{N_0}\times \sizemc)$, which is the space of functions defined on $[0,T]$ taking values in $\rr^{N_0}\times \sizemc$ that are right continuous, have left limits, and endowed with the Skorohod topology.
Using techniques similar to~\cite[Theorem 4.3]{YinZ05}, it can be shown that
$\theta^\e\cd$ converges weakly to a continuous-time Markov chain
generated by $\generatormatrix$. Thus, we mainly need to consider $\hat g^\e\cd$.
We show that
\begin{equation}
\lim_{\Delta\to 0}\limsup_{\e\to 0} \bE \lb\sup_{0\le s \le
\Delta} \bE^\e_t \left| \hat
g^\e(t+s)-  \hat g^\e(t)\right|^2\rb =0
\end{equation}
where $\bE^\e_t$ denotes the conditioning on the past information up to $t$.
Then, the tightness follows from the criterion \cite[p. 47]{Kush84}.

{\em Step 2:} Since $(\hat g^\e\cd,\theta^\e\cd)$ is tight, we can extract
weakly convergent subsequence according to the Prohorov theorem; see
\cite{KY03}.
To determine the limit, we show that $(\hat g^\e\cd, \theta^\e\cd)$ is a
solution of the martingale problem with operator
$L_0$. For each $i\in \M$ and
continuously differential function
with compact support $f(\cdot, i)$, the operator is given by
\bq{l1-def}
L_0 f(\hat g,i)= \nabla f'(\hat g,i) [- \hat g
 +\bar g(i)] + \sum_{j\in \M} \generatorelement_{ij}
f(\hat g,j), \ i\in \M.\eq
We can further demonstrate the martingale problem with operator $L_0$
has a unique weak solution. Thus, the desired
convergence property follows.\\

\section{Sketch of the Proof of Theorem~3.5.3}
\label{ap:sde-pf}
The proof comprises of four steps as described below:

{\em Step 1:} First, note
\bq{nu-defn}\nu_{n+1}=\nu_n - \e \nu_n +\sqrt \e (y_{n+1}-\bE \bar
g(\theta_n) )
+ { \bE [\bar g(\theta_n)- \bar g(\theta_{n+1}]\over \sqrt \e }.\eq
The approach is similar to that of~\cite[Theorem 5.6]{YKI04}. Therefore, we will be brief.

{\em Step 2:} Define the  operator
 \bq{op-def-sde-0}
 {\cal L} f(\nu,i)=-\nabla f'(\nu,i) \nu  + {1\over 2}
 \tr [\nabla ^2
 f(\nu,i)
 \Sigma(i) ] +\sum_{j\in \M} \generatorelement_{ij}
f(\nu,j), \ i\in \M,\eq
 for function $f(\cdot,i)$ with compact support that has continuous partial derivatives
 with respect to $\nu$ up to the second order. It can be shown that the associated martingale problem has
 a unique solution (in the sense of in distribution).

{\em Step 3:} It is natural now to work with a truncated process.
For a fixed, but otherwise arbitrary  $r_1>0$,
 define a truncation function
$$q^{r_1}(x)=\left\{ \barray 1, & \hbox{ if } x \in S^{r_1},\\
0, & \hbox{ if } x \in \rr^{N_0}- S^{r_1},\earray \right.$$
where $S^{r_1}= \{ x \in \rr^{N_0}: |x| \le r_1\}$.
Then, we obtain the truncated iterates
\begin{equation}
\label{trun-it}
\begin{split}
\nu^{r_1}_{n+1}&=\nu^{r_1}_n - \e \nu^{r_1}_n q^{r_1}(\nu^{r_1}_n)
 +\sqrt \e (y_{n+1}-\bE \bar g(\theta_n) )\\
 &\quad+ { \bE [\bar g(\theta_n)- \bar g(\theta_{n+1}]\over \sqrt \e }q^{r_1}(\nu^{r_1}_n).
 \end{split}
 \end{equation}
 Define $\nu^{\e,r_1}(t)= \nu^{r_1}_n$ for $t\in [\e n, \e n+\e)$. Then,
 $\nu^{\e,r_1}\cd$ is an $r$-truncation of
 $\nu^\e\cd$; see~\cite[p. 284]{KY03} for a definition.
 We then show the truncated process $(\nu^{\e,r_1}\cd, \theta^\e\cd)$
 is
 tight. Moreover, by Prohorov's theorem, we can extract a convergent
 subsequence with limit $(\nu^{r_1}\cd, \theta\cd)$
 such that
 the limit $(\nu^{r_1}\cd,\theta\cd)$ is the solution of the martingale
 problem with operator
 ${\cal L}^{r_1}$ defined by
 \bq{op-def-sde}
 {\cal L}^{r_1} f^{r_1}(\nu,i)=-\nabla\prime f^{r_1}(\nu,i) \nu  + {1\over 2}
 \tr [\nabla ^2
 f^{r_1}(\nu,i)
 \Sigma(i) ]+\sum_{j\in \M} \generatorelement_{ij}
f^{r_1}(\nu,j)\eq
for $i\in \M$, where $f^{r_1}(\nu,i)= f(\nu,i) q^{r_1}(\nu)$.

{\em Step 4:} Letting $r_1\to \infty$, we show that the un-truncated process also
converges and the limit, denoted by $(\nu\cd,\theta\cd)$, is precisely
the martingale problem with operator ${\cal L}$ defined in~(\ref{op-def-sde-0}). The limit covariance can further be
evaluated as in~\cite[Lemma 5.2]{YKI04}.

\chapter{Appendix to Chapter~5}
\label{sec:theorem-proofs-game}
This appendix starts with a formal definition of differential inclusions that capture the asymptotic local behavior of individual agents that  follow the algorithms of \S\ref{sec:local-behavior-game}.
\begin{definition}
\label{def:diff-inclusion}
Differential inclusions \index{differential inclusion} are generalizations of ordinary differential equations. In this monograph, we consider differential inclusions of the form
\begin{equation}
\label{eq:F_x}
\frac{d}{dt}X\in \mathcal{F}\left(X\right)
\end{equation}
where $X\in\mathbb{R}^r$ and $\mathcal{F}:\mathbb{R}^r\rightarrow\mathbb{R}^r$ is a Marchaud map~\cite{A97}. That is: (i) the graph and domain of $\mathcal{F}$ are nonempty and closed, (ii) the values $\mathcal{F}\left(X\right)$ are convex, and (iii) the growth of $\mathcal{F}$ is linear, i.e., there exists $\eta >0$ such that for every $X\in\mathbb{R}^r$:
\begin{equation}
\label{eq:linear_growth}
\sup_{Y\in \mathcal{F}\left(X\right)} \left\| Y\right\|\leq\eta\left(1 + \left\|X\right\|\right)
\end{equation}
where $\|\cdot\|$ denotes any norm on $\mathbb{R}^r$.
\end{definition}

The rest of this appendix presents details on the proofs of theorems presented in~\S\ref{sec:main-results-game}.  The key tools we use are weak convergence and Lyapunov stability methods. We refer to~\cite{EK86,KY03} for excellent textbook treatments of weak convergence theory.

\section{Proof of Theorem~5.4.2}
\label{sec:app_B_1}
Close scrutiny of the action selection strategy~(\ref{eq:strategy-MC}) reveals that $\big\lbrace \act^\plyrind_\dtimee,\regmatplyr_\dtimee\big\rbrace$ is a Markov chain with the transition matrix \index{matrix!transition matrix} $P\left(\regmatplyr\right) = \left[P_{ij}\left(\regmatplyr\right)\right]$ given by
\begin{align}
\label{eq:app_B_1_2}
P_{ij}\big(\regmatplyr_\dtimee\big) &= \mathbb{P}\big(\act^{\plyrind}_\dtimee = j|\act^{\plyrind}_{k-1} = i,\regmatplyr_\dtimee\big) \nonumber\\
& = \left\{
\begin{array}{ll}
\frac{1}{\inertia^\plyrind} \big|r^\plyrind_\dtimee(i,j)\big|^{+}, & i\neq j,\\
1-\sum_{j\neq i} P_{ij}\big(\regmatplyr_\dtimee\big), & i = j.
\end{array}
\right.
\end{align}
That is, the transition matrix is a function of the regret matrix $\regmatplyr$.
It is straightforward to show that the transition matrix $P\left(\regmatplyr\right)$ satisfies the following conditions:
\begin{enumerate}
    \item[(i)]  it is continuous in $\regmatplyr$;
    \item[(ii)] for each $\regmatplyr$, it is irreducible and aperiodic.
\end{enumerate}
Condition~(ii) essentially indicates that the Markov chain is ergodic in the sense that there is a unique stationary distribution $\statdistk\big(\regmatplyr\big)=\big(\statdistind^\plyrind_1\big(\regmatplyr\big),\ldots, \statdistind^\plyrind_{A^k}\big(\regmatplyr\big)\big)$ such that $\big[P\big(\regmatplyr\big)\big]^n \to \mathbf{1} \statdistk\big(\regmatplyr\big)$, a matrix with identical rows consisting of
the stationary distribution as $n\to \infty$. (The convergence in fact takes place exponentially fast.)

Further, the sequence of opponents' action profile $\big\lbrace \actprof^{-\plyrind}_\dtimee\big\rbrace$ is independent of $\big\lbrace\act^{\plyrind}_\dtimee\big\rbrace$, and satisfies the following condition:
\begin{equation}
\label{eq:app_B_1_3}
d \lb {1\over n} \sum^{n+m-1}_{\ell=m} \E_m \lbr B^{\plyrind}\big(i,\actprof^{-\plyrind}_\ell\big)\rbr, \mathcal{S}^{\plyrind}(i)\rb \to 0 \ \hbox{ in probability}
\end{equation}
where $ d[\cdot,\cdot]$ denotes the usual distance function, $\mathbf{E}_m$ denotes conditional expectation given the $\sigma$-algebra generated by $\big\{\regmatplyr_\ell, \act^{\plyrind}_{\ell-1}, \actprof^{-\plyrind}_{\ell-1}: \ell \le m\big\}$, and $\mathcal{S}^{\plyrind}(i) = \big[\mathcal{S}^{\plyrind}_{pq}(i)\big]$ is a set of $A^\plyrind\times A^\plyrind$ matrices, where
%
\begin{equation}
\label{eq:app_B_1_10}
\mathcal{S}^{\plyrind}_{pq}(i) = \left\lbrace
\begin{array}{ll}
\lbr\utilityk \big(q, \mixedstrat^{-\plyrind}\big) - \utilityk \big(i, \mixedstrat^{-\plyrind}\big); \mixedstrat^{-\plyrind} \in \Delta\actset^{-\plyrind}\rbr & p = i,\\
0 & p \neq i.
\end{array}
\right.
\end{equation}
%
It is then straight forward to show that, for any stationary distribution $\statdistk\big(\regmatplyr\big)$,
the set
\begin{equation}
\label{eq:app_B_1_5}
H^\plyrind\big(\regmatplyr\big) = \sum_{i\in\actset^{\plyrind}} \statdistindi\big(\regmatplyr\big)\mathcal{S}^{\plyrind}(i)
\end{equation}
is closed, convex, and upper semi-continuous; see~\cite[pp.108--109]{KY03}.

To proceed, recall the continuous-time interpolated process $\regmatplyrinterpol = \regmatplyr_\dtimee$ for $t\in[k\stepsize,(k+1)\stepsize)$. We first prove the tightness \index{tightness}. Consider~(\ref{eq:reg-update}) for the sequence of
$(A^k)^2$-valued vectors resulted after rearranging the elements of the regret matrix $\regmatplyr$ into a column vector.
Noting the boundedness of payoff functions,
and using H\"older's and Gronwall's inequalities, for any $0<T<\infty$, we obtain
\begin{equation}
\sup_{k \le T/\stepsize} \E  \big\| \regmatplyr_\dtimee \big\|^2<\infty
\end{equation}
where in the above and hereafter $t/\stepsize$ is understood to be the integer part of $t/\stepsize$ for each $t>0$. Next, considering $\regmatplyrinterpolcdot$, for any $t,s>0$, $\delta>0$, and $s<\delta$, it is fairly easy to verify that
\begin{equation}
R^{\plyrind,\stepsize}(t+s) -  R^{\plyrind,\stepsize}(t) =   \stepsize \sum^{(t+s)/\stepsize-1}_{\ell=t/\stepsize}\lb B^{\plyrind}\big(\act^\plyrind_\ell,\actset^{-\plyrind}_\ell\big)-R^\plyrind_\ell\rb.
\end{equation}
As a result,
\begin{equation}
\label{eq:app_B_1_6}
\lim_{\delta\to 0}
\limsup_{\mu\to 0}
\left\{
 \E
 \left[\sup_{0\le s \le \delta} \E^\stepsize_t\left \|
 R^{\plyrind,\stepsize}(t+s)
 -
 R^{\plyrind,\stepsize}(t)
 \right\|^2
 \right]
 \right\}  = 0.
\end{equation}
Thus, $\regmatplyrinterpolcdot$ is tight in the space of functions that are defined in $[0,\infty)$ taking values in $\mathbb{R}^{(A^k)^2}$; see~\cite[Chapter 7]{KY03} or~\cite[Theorem 3, p. 47]{Kush84}.

By Prohorov theorem~\cite{KY03}, one can extract a convergent subsequence. For notational simplicity, we still denote this convergent subsequence by $\regmatplyrinterpolcdot$ with limit $\regmatplyr(\cdot)$. Thus, $\regmatplyrinterpolcdot$ converges weakly to $\regmatplyr(\cdot)$. By Skorohod representation theorem~\cite{KY03}, with a slight abuse of notation, one can assume $\regmatplyrinterpolcdot \rightarrow \regmatplyr(\cdot)$ with probability one and the convergence is uniform on any finite interval. Next, using martingale averaging methods, we need to characterize the limit process.
Normally, one uses a smooth function $f\cd$ with compact support to carry out the analysis. Here, for simplicity, we suppress the dependence of $f\cd$ and proceed directly; see~\cite{KS84} for a similar argument.
Choose a sequence of integers $\lbrace k_\stepsize\rbrace$ such that $k_\stepsize\rightarrow \infty$ as $\stepsize\rightarrow 0$, but $\delta_\stepsize = \stepsize k_\stepsize \rightarrow 0$. Note
\begin{align}
R^{\plyrind,\stepsize}(t+s) - \regmatplyrinterpol &= \sum_{\ell:\ell \delta_\stepsize = t}^{t+s} \delta_\stepsize \sum_{i\in\actset^{\plyrind}} \frac{1}{k_\stepsize} \sum_{\tau = \ell k_\stepsize}^{\ell k_\stepsize + k_\stepsize - 1} B^\plyrind\big(i,\actprof^{-\plyrind}_\tau\big) I\left\{ \act^{\plyrind}_\tau = i\right\}\nonumber\\
&\quad - \sum_{\ell:\ell \delta_\stepsize=t}^{t+s} \delta_\stepsize \frac{1}{k_\stepsize} \sum_{\tau = \ell k_\stepsize}^{\ell k_\stepsize+k_\stepsize-1} \regmatplyr_\tau
\end{align}
where $\sum_{\ell:\ell \delta_\stepsize=t}^{t+s}$ denotes the sum over integers $\ell$ in the range $t\leq \ell \delta_\stepsize \leq t+s$.

We shall use the techniques in~\cite[Chapter 8]{KY03} to prove the desired result. Let $h\cd$ be any bounded and continuous function, $t,s>0$, $\kappa_0$ be an arbitrary positive integer, and $t_\iota \le t$ for all $\iota \le \kappa_0$. It is readily seen that, by virtue of the weak convergence and the Skorohod representation,
as $\stepsize\to 0$,
\begin{equation}
\begin{split}
\E h(R^{\plyrind,\stepsize}&(t_\iota): \iota\le \kappa_0)
\lb R^{\plyrind,\stepsize}(t+s)- R^{\plyrind,\stepsize}(t)\rb \\
&\to \E h(\regmatplyr(t_\iota): \iota\le \kappa_0) \lb \regmatplyr(t+s)- \regmatplyr(t) \rb.
\end{split}
\end{equation}
By using the technique of stochastic approximation (see, e.g.,~\cite[Chapter 8]{KY03}), we also have
\begin{equation}
\label{eq-X}
\begin{split}
\E
h\big(&R^{\plyrind,\stepsize}(t_\iota): \iota\le \kappa_0\big)\sum_{\ell \delta_\stepsize=t}^{t+s} \delta_\stepsize \frac{1}{k_\stepsize} \sum_{\tau = \ell k_\stepsize}^{\ell k_\stepsize+k_\stepsize-1} \regmatplyr_\tau\\
& \to \E h\big(R^k(t_\iota): \iota\le \kappa_0\big) \left[ \int^{t+s}_t \regmatplyr(u) du\right] \hbox{ as } \stepsize\to 0.
\end{split}
\end{equation}
Moreover, by the independence of $\act^{\plyrind}_\dtimee$ and $\actprof^{-\plyrind}_\dtimee$,
\begin{align}
\label{eq:A-21}
&\lim_{\stepsize \to 0} \ \E
h\big(R^{\plyrind,\stepsize}(t_\iota): \iota\le \kappa_0\big)\nonumber\\
&\qquad\times\left[\sum_{\ell \delta_\stepsize = t}^{t+s} \delta_\stepsize \sum_{i\in\actset^\plyrind} \frac{1}{k_\stepsize} \sum_{\tau = \ell k_\stepsize}^{\ell k_\stepsize + k_\stepsize - 1} B^\plyrind\big(i,\actprof^{-\plyrind}_\tau\big) I\left\{ \act^{\plyrind}_\tau = i\right\}\right]\nonumber\\
&\; = \lim_{\stepsize \to 0} \ \E
h\big(R^{\plyrind,\stepsize}(t_\iota): \iota\le \kappa_0\big)\nonumber\\
&\qquad\qquad\times\left[\sum_{\ell \delta_\stepsize = t}^{t+s} \delta_\stepsize \sum_{i\in\actset^\plyrind} \frac{1}{k_\stepsize} \sum_{\tau = \ell k_\stepsize}^{\ell k_\stepsize + k_\stepsize - 1} \mathbf{E}_{\ell k_\stepsize} \lbr B^{\plyrind}\big(i,\actprof^{-\plyrind}_\tau\big)\rbr
\E_{\ell k_\stepsize} I\left\{ \act^{\plyrind}_\tau = i\right\}\right]\nonumber\\
&\; = \lim_{\stepsize \to 0} \ \E
h\big(R^{\plyrind,\stepsize}(t_\iota): \iota\le \kappa_0\big)\Bigg[\sum_{\ell \delta_\stepsize = t}^{t+s} \delta_\stepsize \sum_{i\in\actset^{\plyrind}} \frac{1}{k_\stepsize} \sum_{\tau = \ell k_\stepsize}^{\ell k_\stepsize + k_\stepsize - 1}\sum_{j\in\actset^{\plyrind}}
I\left\{\actk_{\ell k_\stepsize}=j\right\}\nonumber\\
&\qquad\qquad \times \bigg[  \mathbf{E}_{\ell k_\stepsize}\lbr B^{\plyrind}\big(i,\actprof^{-\plyrind}_\tau\big) \rbr \varphi_i\big(\regmatplyr_{\ell n_\stepsize})
+  \mathbf{E}_{\ell k_\stepsize} \lbr B^{\plyrind}\big(i,\actprof^{-\plyrind}_\tau\big)\rbr\nonumber\\
&\qquad\qquad\quad\qquad
\times\Big[P_{j i}^{\{\tau-\ell k_\stepsize\}}\big(\regmatplyr_{\ell k_\stepsize}\big) -\varphi_i\big(\regmatplyr_{\ell k_\stepsize}\big) \Big] \bigg] \Bigg].
\end{align}
In~(\ref{eq:A-21}), $P_{ji}^{\{\tau- \ell k_\stepsize\}}\big(\regmatplyr\big)$ denotes the $(\tau-\ell k_\stepsize)$-step transition probability. Except the indicator function on the l.h.s. of~(\ref{eq:new}), the rest of the terms in the summand are independent of $j$, therefore,
\begin{equation}
\label{eq:new}
\begin{split}
&\sum_{j\in\actset^{\plyrind}}
I\left\{\act^{\plyrind}_{\ell k_\stepsize}=j\right\} \mathbf{E}_{\ell k_\stepsize} \lbr B^{\plyrind}\big(i,\actprof^{-\plyrind}_\tau\big)\rbr \varphi_i\big(\regmatplyr_{\ell k_\stepsize}\big)\\
&\qquad\quad = \mathbf{E}_{\ell k_\stepsize} \lbr B^{\plyrind}\big(i,\actprof^{-\plyrind}_\tau\big) \rbr \varphi_i\big(\regmatplyr_{\ell k_\stepsize}\big).
\end{split}
\end{equation}
Recalling the properties of the transition matrix $P\big(\regmatplyr\big)$, regardless of the choice of $j$, as $\stepsize\to 0$ and $\tau-\ell k_\stepsize\to \infty$, for each fixed $\regmatplyr$,
\begin{equation}
\left[P_{j i}^{\{\tau-\ell k_\stepsize\}}\big(\regmatplyr\big) -\varphi_i\big(\regmatplyr\big) \right]\to 0
\end{equation}
exponentially fast. As the result, as $\stepsize\to 0$,
\begin{align}
\label{eq:A-3}
&\E
h\big(R^{\plyrind,\stepsize}(t_\iota): \iota\le \kappa_0\big)\nonumber\\
&\;\times\Bigg[\sum_{\ell \delta_\stepsize = t}^{t+s} \delta_\stepsize \sum_{i\in\actset^{\plyrind}} \frac{1}{k_\stepsize} \sum_{k = \ell k_\stepsize}^{\ell k_\stepsize + k_\stepsize - 1} \sum_{j\in\actset^{\plyrind}} I\left\lbrace \act^{\plyrind}_{\ell k_\stepsize} = j\right\rbrace \ \mathbf{E}_{\ell k_\stepsize} \lbr B^{\plyrind}\big(i,\actprof^{-\plyrind}_\tau\big)\rbr\nonumber\\
&\quad\;\times
\Big[P_{j i}^{\{\tau-\ell k_\stepsize\}}\big(\regmatplyr_{\ell k_\stepsize}\big) -\varphi_i\big(\regmatplyr_{\ell k_\stepsize}\big) \Big]\Bigg]
 \to 0 \ \hbox{ in probability}.
\end{align}
Therefore,
\begin{align}
\label{eq:app_B_1_11}
&\lim_{\stepsize \to 0} \ \E
h\big(R^{\plyrind,\stepsize}(t_\iota): \iota\le \kappa_0\big)\nonumber\\
&\qquad\times\left[\sum_{\ell \delta_\stepsize = t}^{t+s} \delta_\stepsize \sum_{i\in\actset^\plyrind} \frac{1}{k_\stepsize} \sum_{\tau = \ell k_\stepsize}^{\ell k_\stepsize + k_\stepsize - 1} B^\plyrind\big(i,\actprof^{-\plyrind}_\tau\big) I\left\{ \act^{\plyrind}_\tau = i\right\}\right]\nonumber\\
& = \lim_{\stepsize \to 0} \ \E
h\big(R^{\plyrind,\stepsize}(t_\iota): \iota\le \kappa_0\big)\nonumber\\
&\qquad\times\Bigg[\sum_{\ell \delta_\stepsize = t}^{t+s} \delta_\stepsize \sum_{i\in\actset^{\plyrind}} \frac{1}{k_\stepsize} \sum_{\tau = \ell k_\stepsize}^{\ell k_\stepsize + k_\stepsize - 1} \mathbf{E}_{\ell k_\stepsize} \lbr B^{\plyrind}\big(i,\actprof^{-\plyrind}_\tau\big)\rbr \varphi_i\big(\regmatplyr_{\ell n_\stepsize}\big)\Bigg].
\end{align}
Using~(\ref{eq:app_B_1_3}) together with~(\ref{eq:app_B_1_11}), we obtain that, as $\stepsize\to 0$,
\begin{align}
\label{eq:app_B_1_12}
&\E
h\big(R^{\plyrind,\stepsize}(t_\iota): \iota\le \kappa_0\big)\left[\sum_{\ell \delta_\stepsize = t}^{t+s} \delta_\stepsize \sum_{i\in\actset^\plyrind} \frac{1}{k_\stepsize} \sum_{\tau = \ell k_\stepsize}^{\ell k_\stepsize + k_\stepsize - 1} B^\plyrind\big(i,\actprof^{-\plyrind}_\tau\big) I\left\{ \act^{\plyrind}_\tau = i\right\}\right]\nonumber\\
&\quad \to \E
h\big(R^{\plyrind,\stepsize}(t_\iota): \iota\le \kappa_0\big)\lb \int_{t}^{t+s} \sum_{i\in\actset^{\plyrind}} \varphi_i\big(\regmatplyr(u)\big)\mathcal{S}^{\plyrind}(i) du\rb.
\end{align}
Finally, combining~(\ref{eq-X}) and~(\ref{eq:app_B_1_12}), and using~(\ref{eq:app_B_1_5}) yields
\begin{equation}
\frac{d \regmatplyr(t)}{dt} \in  H^\plyrind\big(\regmatplyr\big) - \regmatplyr(t)
\end{equation}
as desired.

\section{Proof of Theorem~5.4.5}
\label{sec:app_B_2}

Define the Lyapunov function \index{Lyapunov function}
\begin{equation}
\label{eq:thrm-4-2}
\lyap\big(\regmatplyr\big) = \frac{1}{2}\lb \textmd{dist}\big[\regmatplyr,\mathbb{R}^-\big]\rb^2= \frac{1}{2}\sum_{i,j\in\actset^\plyrind} \lb\big| r^\plyrind(i,j)\big|^+\rb^2.
\end{equation}
For the sake of notational convenience, we drop the $\regmatplyr$ dependency of the invariant measure, and denote it simply by $\statdistk$ hereafter.

{\em Case I:} Taking the time-derivative of~(\ref{eq:thrm-4-2}), and substituting for $dr^\plyrind(i,j) / dt$ from~(\ref{eq:thrm-1-1})--(\ref{eq:thrm-1-2}), we obtain
\begin{align}
\label{eq:thrm-4-3}
{d\over dt}\lyap\big(\regmatplyr\big) &= \sum_{i,j\in\actset^\plyrind} \big| r^\plyrind(i,j)\big|^+ \cdot {d\over dt} r^\plyrind(i,j) \nonumber\\
& = \sum_{i,j\in\actset^\plyrind} \big| r^\plyrind(i,j)\big|^+ \Big[ \lb \utilityk\big(j,\mixedstrat^{-\plyrind}\big) - \utilityk\big(i,\mixedstrat^{-\plyrind}\big)\rb \statdistindi - r^\plyrind(i,j) \Big]\nonumber\\
& = \underbrace{\sum_{i,j\in\actset^\plyrind} \big| r^\plyrind(i,j)\big|^+ \lb \utilityk\big(j,\mixedstrat^{-\plyrind}\big) - \utilityk\big(i,\mixedstrat^{-\plyrind}\big)\rb \statdistindi}_{= 0\;\textmd{(See Lemma~\ref{lemma:first-term-zero} below)}} \nonumber\\
&\quad- \sum_{i,j\in\actset^\plyrind} \big| r^\plyrind(i,j)\big|^+ r^\plyrind(i,j)\nonumber\\
& = - \lyap\big(\regmatplyr\big).
\end{align}
In the last equality we used
\begin{equation}
\label{eq:thrm-4-4}
\sum_{i,j\in\actset^\plyrind} \big| r^\plyrind(i,j)\big|^+ r^\plyrind(i,j) = \sum_{i,j\in\actset^\plyrind} \lb \big| r^\plyrind(i,j)\big|^+ \rb^2 = \lyap\big(\regmatplyr\big).
\end{equation}
This completes the proof.

The following lemma shows that the first term in the r.h.s. of~(\ref{eq:thrm-4-3}) equals zero.

\vspace{0.2cm}
\begin{lemma}
\label{lemma:first-term-zero}
Let $\mixedstrat^{-\plyrind}$ denote a probability measure over the joint action space $\actset^{-\plyrind}$ of all agents excluding agent $k$. Then,
\begin{equation}
\label{eq:lemma-1-1}
\sum_{i,j\in\actset^\plyrind} \big| r^\plyrind(i,j)\big|^+ \lb \utilityk\big(j,\mixedstrat^{-\plyrind}\big) - \utilityk\big(i,\mixedstrat^{-\plyrind}\big)\rb \statdistindi = 0
\end{equation}
where $\statdistk = \big( \statdistind^\plyrind_1,\ldots,\statdistind^\plyrind_{\fact}\big)$ represents the invariant measure of transition probabilities~(\ref{eq:strategy-MC}).
\begin{proof}
Separating the summation over the two terms in ~(\ref{eq:lemma-1-1}) yields
\begin{align}
\label{eq:lemma-1-2}
&\sum_{i,j\in\actset^\plyrind} \big| r^\plyrind(i,j)\big|^+  \utilityk\big(j,\mixedstrat^{-\plyrind}\big) \statdistindi - \sum_{i,j\in\actset^\plyrind} \big| r^\plyrind(i,j)\big|^+  \utilityk\big(i,\mixedstrat^{-\plyrind}\big) \statdistindi\nonumber\\
& = \sum_{j,i\in\actset^\plyrind} \big| r^\plyrind(j,i)\big|^+  \utilityk\big(i,\mixedstrat^{-\plyrind}\big) \statdistindj - \sum_{i,j\in\actset^\plyrind} \big| r^\plyrind(i,j)\big|^+  \utilityk\big(i,\mixedstrat^{-\plyrind}\big) \statdistindi \nonumber\\
& = \sum_{i\in\actset^\plyrind} \utilityk\big(i,\mixedstrat^{-\plyrind}\big) \lb \sum_{j\in\actset^\plyrind} \statdistindj \big| r^\plyrind(j,i)\big|^+ - \sum_{j\in\actset^\plyrind} \statdistindi \big|r^\plyrind(i,j)\big|^+\rb.
\end{align}
In the second line we used $\sum_{i,j} a_{ij} = \sum_{j,i} a_{ji}$, and in the last line we changed the order of summation both applied to the first term. Finally, applying $\big|r^\plyrind(i,i)\big|^+ = 0$, for all $i\in\actset^\plyrind$, and substituting~(\ref{eq:stat-distr-prop}) into~(\ref{eq:lemma-1-2}) completes the proof.
\end{proof}
\end{lemma}

{\em Case II:} Define the Lyapunov function \index{Lyapunov function}
\begin{align}
\label{eq:thrm-4-5}
\lyap\big(\regmatplyrL,\regmatplyrG\big) &= \lb \textmd{dist}\Big[\regmatplyrL+\regmatplyrG,\mathbb{R}^-\Big]\rb^2\nonumber\\
& = \sum_{i,j\in\actset^\plyrind} \lb\left| r^{L,\plyrind}(i,j)+r^{G,\plyrind}(i,j)\right|^+\rb^2.
\end{align}
Taking the time-derivative of~(\ref{eq:thrm-4-5}) yields
\begin{align}
\label{eq:thrm-4-6}
{d\over dt}\lyap\big(\regmatplyrL,\regmatplyrG\big) = 2\sum_{i,j\in\actset^\plyrind} & \left| r^{L,\plyrind}(i,j)+r^{G,\plyrind}(i,j) \right|^+\nonumber\\
&\times
\lb {d\over dt} r^{L,\plyrind}(i,j) + {d\over dt} r^{G,\plyrind}(i,j)\rb.
\end{align}
Next, replacing for $d \regplyrLij/dt$ and $d \regplyrGij/dt$ from~(\ref{eq:thrm-2-1})--(\ref{eq:thrm-2-2}),
\begin{align}
\label{eq:thrm-4-7}
{d\over dt}\lyap\big(\regmatplyrL,\regmatplyrG\big) &= 2\sum_{i,j\in\actset^\plyrind} \left| r^{L,\plyrind}(i,j)+r^{G,\plyrind}(i,j) \right|^+ \nonumber\\
&\hspace{1.7cm}\times \Big[\big[\utilitykL\big(j,\neibmixedstrat\big) - \utilitykL\big(i,\neibmixedstrat\big)\big]\statdistindgroupnor \nonumber\\
&\hspace{2.4cm}+ \big[ \utilitykGt(j) - \utilitykGt(i)\big]\statdistindgroupnor \nonumber\\
&\hspace{2.4cm}- \big[ r^{L,\plyrind}(i,j)+r^{G,\plyrind}(i,j)\big]
\Big].
\end{align}
Substituting for $\statdistgroupnor$ from~(\ref{eq:statdistr-groups}) in~(\ref{eq:thrm-4-7}) yields
\begin{align}
\label{eq:lemma-2-2}
{d\over dt}&\lyap\big(\regmatplyrL,\regmatplyrG\big) = 2(1-\explor)\sum_{i,j} \statdistindi\left|r^{L,\plyrind}(i,j)+r^{G,\plyrind}(i,j)\right|^+\nonumber\\
& \times \lb \utilitykL\big(j,\neibmixedstrat\big) + \utilitykGt(j) - \big[ \utilitykL\big(i,\neibmixedstrat\big) + \utilitykGt(i)\big]\rb\nonumber\\
&\quad\;+{2\explor \over \fact} \sum_{i,j\in\actset^\plyrind} \left|r^{L,\plyrind}(i,j)+r^{G,\plyrind}(i,j)\right|^+ \nonumber\\
& \times\lb\utilitykL\big(j,\neibmixedstrat\big) + \utilitykGt(j) - \big[ \utilitykL\big(i,\neibmixedstrat\big) + \utilitykGt(i)\big]\rb\\
&\quad\;- 2\sum_{i,j\in\actset^\plyrind} \left|r^{L,\plyrind}(i,j)+r^{G,\plyrind}(i,j)\right|^+ \lb r^{L,\plyrind}(i,j)+r^{G,\plyrind}(i,j)\rb.\nonumber
\end{align}
Similar to Lemma~\ref{lemma:first-term-zero}, it is straightforward to show that the first term in the r.h.s. of~(\ref{eq:lemma-2-2}) is zero.
Therefore,
%
\begin{align}
\label{eq:lemma-2-4}
{d\over dt}\lyap\big(\regmatplyrL,\regmatplyrG\big) &= - 2 \lyap\big(\regmatplyrL,\regmatplyrG\big) \nonumber\\
&\quad+ {2\explor\over\fact} \sum_{i,j\in\actset^\plyrind} \left|r^{L,\plyrind}(i,j)+r^{G,\plyrind}(i,j)\right|^+ \nonumber\\
& \qquad\qquad\quad\times\Big[\utilitykL\big(j,\neibmixedstrat\big) + \utilitykGt(j) \nonumber\\
&\qquad\qquad\qquad\; - \big[ \utilitykL\big(i,\neibmixedstrat\big) + \utilitykGt(i)\big]\Big].
\end{align}
Finally, noting that the payoff function $\utilitykL\big(\cdot,\neibmixedstrat\big) + \utilitykGt(\cdot)$ is bounded, one can find a constant $\alpha$ such that
\begin{align}
\label{eq:thrm-4-8}
{d\over dt}\lyap\big(\regmatplyrL,\regmatplyrG\big) \leq &{2\alpha\explor\over\fact} \sum_{i,j\in\actset^\plyrind} \left| r^{L,\plyrind}(i,j)+r^{G,\plyrind}(i,j)\right|^+ \nonumber\\
&- 2\lyap\big(\regmatplyrL,\regmatplyrG\big).
\end{align}
In~(\ref{eq:thrm-4-8}), $\alpha > 0$ depends on the differences in payoffs for different actions $i$ and $j$ and can be specified based on the particular choice of the payoff function. In this lemma, however, we only require existence of such a bound, which is guaranteed since the payoff functions are bounded.

Now, assuming $\big|r^{L,\plyrind}(i,j)+r^{G,\plyrind}(i,j)\big|^+ >\epsilon>0$ for all $i,j\in\actset^\plyrind$, one can choose $\explor$ small enough such that
\begin{equation}
\label{eq:thrm-4-9}
{d\over dt}\lyap\big(\regmatplyrL,\regmatplyrG\big) \leq -\lyap\big(\regmatplyrL,\regmatplyrG\big).
\end{equation}
This proves global asymptotic stability of the limit system~(\ref{eq:thrm-2-1}) and
\begin{equation}
\label{eq:thrm-4-10}
\lim_{t\to\infty} \textmd{dist}\lb\regmatplyrL+\regmatplyrG,\mathbb{R}^-\rb = 0.
\end{equation}

{\em Case III:} Define the Lyapunov function \index{Lyapunov function}
\begin{align}
\label{eq:thrm-4-11}
\lyap\big(\globalregret\big) &= \lb \textmd{dist}\big[\globalregret,\mathbb{R}^-\big]\rb^2\nonumber\\
& = \sum_{k\in\communitys} \sum_{i,j\in\actset^\plyrind} \lb\big|r^{\plyrind}(i,j)\big|^+\rb^2.
\end{align}
Taking the time-derivative of~(\ref{eq:thrm-4-11}) and using a similar argument to Lemma~\ref{lemma:first-term-zero}, we obtain
\begin{equation}
\label{eq:thrm-4-12}
{d\over dt}\lyap\big(\globalregret\big) =  \sum_{k\in\communitys} \sum_{i,j\in\actset^\plyrind} \left|r^{\plyrind}(i,j)\right|^+ \lb \sum_{l\in\cneighborhoodk} \Cmatkl r^l(i,j) - r^{\plyrind}(i,j)\rb.
\end{equation}
Recall that $\globalregret$ represents the global regret matrix for social group $\communitys$ rearranged into a vector. We consider the following cases:

(i) $\big|\globalregret\big|^+ > 0$. This is the worst case scenario where all agents' regrets for switching from action $i$ to $j$ is positive. It is straightforward to rewrite~(\ref{eq:thrm-4-12}) as follows
\begin{equation}
\label{eq:thrm-4-13}
{d\over dt}\lyap\big(\globalregret\big) =  \lb\left|\globalregret\right|^+\rb^\top (\Cmatkron - I) \globalregret.
\end{equation}
Noting that, when $\big|\globalregret\big|^+ > 0$, $\big|\globalregret\big|^+  = \globalregret$, we obtaine
\begin{align}
\label{eq:thrm-4-14}
{d\over dt}\lyap\big(\globalregret\big) &=  \lb\globalregret\rb^\top (\Cmatkron - I) \globalregret\nonumber\\
&= \lb\globalregret\rb^\top \Cmatkron \globalregret - \lb\globalregret\rb^\top \globalregret.
\end{align}
Since $\Cmatkron\mathbf{1} = \mathbf{0}$, it is weakly diagonally dominant. Since $\Cmatkron$ is further symmetric, it is well-known from linear algebra that $\Cmatkron$ is negative semi-definite. That is, for all vectors $\mathbf{x}\in\mathbb{R}^{J_s}$, $\mathbf{x}^T\Cmatkron \mathbf{x} \leq 0$. This, together with~(\ref{eq:thrm-4-14}), concludes $d\lyap\big(\globalregret\big)/dt < 0$.

(ii) $\big|r^{\plyrind}(i,j)\big|^+ >0$ and $\big|r^l(i,j)\big|^+ = 0$ for some $l\in\communitys$. In~(\ref{eq:thrm-4-12}), looking at the term
\begin{equation}
\big|r^{\plyrind}(i,j)\big|^+ \lb \sum_{l\in\cneighborhoodk} \Cmatkl r^l(i,j) - r^{\plyrind}(i,j)\rb
\end{equation}
it can be clearly seen that $r^l(i,j) < 0 $ actually helps pulling the regrets to the negative orthant more forcefully since $c^s_{nl} r^{l}(i,j) - r^{\plyrind}(i,j) < 0$. This concludes the limit system~(\ref{eq:thrm-3-1}) is globally asymptotically stable and~(\ref{eq:thrm-4-1}) holds.

\section{Proof of Theorem~5.4.6}

The proof relies on how the ``regret'' measure is defined. For brevity, we present the proof for ``Case I''. Similar argument can be used to prove this result for the other cases. Recall, for $t\in[k\stepsize,(k+1)\stepsize)$, $r^{\plyrind,\varepsilon}(i,j)\ctimeet = r^\plyrind_\dtimee(i,j)$; hence, substituting~(\ref{eq:regret-def}) for $r^\plyrind_\dtimee(i,j)$ on the r.h.s., we obtain
\begin{align}
\label{eq:thrm-5-2}
r^{\plyrind,\varepsilon}(i,j)\ctimeet &= \stepsize \sum_{\tau \leq k} (1-\stepsize)^{k-\tau} \lb\utilityk\big(j,\actprof^{-\plyrind}(\tau)\big) - \utilityk \big(\actprof(\tau)\big)\rb \indicatoritau\nonumber\\
& = \sum_{\actprof^{-\plyrind}} \globinterpolind \lb\utilityk\big(j,\actprof^{-\plyrind}\big) - \utilityk \big(i,\actprof^{-\plyrind}\big)\rb
\end{align}
where $\globinterpolind$ denotes the interpolated empirical distribution of agent $\plyrind$ choosing action $i$ and the rest playing $\actprof^{-\plyrind}$. On any convergent subsequence $\lbrace\globbehav_{\underline{k}}\rbrace_{\underline{k}\geq 0}\rightarrow \boldsymbol{\pi}$, with slight abuse of notation, let $\globinterpol\ctimeet = \globbehav_{\underline{k}}$ and $\regmatplyrinterpol = \regmatplyr_{\underline{k}}$ for $t\in[\underline{k}\stepsize,\underline{k}\stepsize+\stepsize)$. Then,
\begin{equation}
\label{eq:thrm-5-3}
\lim_{t\to\infty} r^{\plyrind,\stepsize}(i,j)\ctimeet = \sum_{\actprof^{-\plyrind}} \pi^{\plyrind}\big(i,\actprof^{-\plyrind}\big) \lb\utilityk\big(j,\actprof^{-\plyrind}\big) - \utilityk \big(i,\actprof^{-\plyrind}\big)\rb
\end{equation}
where $\pi^{\plyrind}\big(i,\actprof^{-\plyrind}\big)$ denotes the probability of agent $\plyrind$ choosing action $i$ and the rest playing $\actprof^{-\plyrind}$. Next, combining~(\ref{eq:thrm-5-3}) with~(\ref{eq:corollary-1}) in Corollary~\ref{corollary-game}, and comparing with~(\ref{eq:CE_defn}) in Definition~\ref{def:CE}, the desired result follows.

To prove~(\ref{eq:thrm-5-1}) for ``Case II,'' one needs to look at the combined regrets $\regplyrLijinterpol\ctimeet + \regplyrGijinterpol\ctimeet$ instead. We first prove that the sequence $\big\{r^{G,\plyrind}_\dtimee(i,j)\big\}$ is an unbiased estimator of the sequence of regrets if agents were to observe the action profile of those outside their social group. The rest of the proof is similar to the above proof. The interested reader is referred to~\cite[Sec.~IV]{NKY13a} for the detailed proof. 


\addcontentsline{toc}{chapter}{References}
\bibliographystyle{plain}
\bibliography{references}
\addcontentsline{toc}{chapter}{Index}
\printindex

\end{document}